\documentclass[iop]{emulateapj}
\usepackage{graphicx}
\usepackage{hyperref}
\shorttitle{GC ILS Abundances}
\shortauthors{COLUCCI, BERNSTEIN, \& MCWILLIAM}

\begin{document}

\newcommand{\kms}{km~s$^{-1}$}
\newcommand{\rkms}{km~s$^{-1}$ \enskip}
\newcommand{\rAA}{{\AA \enskip}}
\defcitealias{m31p1}{C09}
\defcitealias{m31p2}{C14}
\defcitealias{paper3}{Paper III}
\defcitealias{paper4}{Paper IV}
\defcitealias{mb08}{Paper I}
\defcitealias{harris}{HGCC}
\defcitealias{m31inprep}{Colucci et. al, in preparation}

\title{Globular Cluster Abundances from High-Resolution,
  Integrated-Light Spectroscopy. II.  Expanding the Metallicity Range for Old Clusters and Updated Analysis Techniques \footnotemark[1]}
  \footnotetext[1]{This paper includes data gathered with the 6.5 meter Magellan 
Telescopes located at Las Campanas Observatory, Chile.}
\footnotetext[2]{NSF Astronomy and Astrophysics Postdoctoral Fellow}

\author{Janet E. Colucci\footnotemark[2]}
\affil{The Observatories of the Carnegie Institution for Science, 813 Santa Barbara St., Pasadena, CA 91101}

\author{Rebecca A. Bernstein}
\affil{The Observatories of the Carnegie Institution for Science, 813 Santa Barbara St., Pasadena, CA 91101} 

\author{Andrew McWilliam}
\affil{The Observatories of the Carnegie Institution for Science, 813 Santa Barbara St., Pasadena, CA 91101}

\begin{abstract}
We present abundances of globular clusters in the Milky Way and Fornax from integrated light spectra. Our goal is to evaluate the consistency of the integrated light analysis relative to standard abundance analysis for individual stars in those same clusters. This sample includes an updated analysis of 7 clusters from our previous publications and  results for 5 new clusters that expand the metallicity range over which our technique has been tested.  We find that the  [Fe/H] measured from integrated light spectra agrees to $\sim$0.1 dex for globular clusters with metallicities as high as [Fe/H]=$-0.3$, but the abundances measured for more metal rich clusters may be underestimated.   In addition we systematically evaluate the accuracy of abundance ratios, [X/Fe], for Na I, Mg I, Al I, Si I, Ca I, Ti I, Ti II, Sc II, V I, Cr I, Mn I, Co I, Ni I, Cu I, Y II, Zr I, Ba II, La II, Nd II, and Eu II.  The elements for which the integrated light analysis gives results that are most similar to analysis of individual stellar spectra are Fe I, Ca I, Si I, Ni I, and Ba II. The elements that show the greatest differences include Mg I and Zr I.  Some elements show good agreement only over a limited range in metallicity.  More stellar abundance data in these clusters would enable more complete evaluation of the integrated light results for other important elements.

 \end{abstract}


\section{Introduction}
\label{sec:intro}
\setcounter{footnote}{1}

Abundance analysis of high resolution integrated light (IL) spectra of globular clusters (GCs) holds significant
potential for galaxy evolution constraints. Due to the high luminosity
of GCs, high-resolution spectra of sufficient quality can
be obtained for GCs at extragalactic distances and, because they represent old stellar populations, GCs can be  used to probe
the chemical evolution of distant galaxies. The luminosities of
the brightest GCs are comparable to young supergiant stars,  for
which abundances have been measured in Local Group galaxies at distances of $\sim$1 Mpc \citep[e.g.][]{venn2001,kaufer2004, ic1613}
using high resolution spectra from large telescopes. However, unlike short lived supergiant stars
that reveal only recent gas compositions, GC ages cover the full
range of galactic history so that GCs can be used to probe the full 
formation history of the parent galaxy.

In a series of papers that demonstrated and developed our techniques, we have explored the use of high-resolution, integrated light spectra for the abundance analysis of globular clusters  \citep{bernstein02,mb08,scottphd,paper3,paper4}. Due to the low velocity
dispersions of GCs ($< 30$ \kms), the line widths of
the integrated light spectra are narrow enough that individual lines are well resolved and blending is not much more problematic than for
individual red giant branch (RGB) stars.  We have demonstrated that it is possible to use weak lines of numerous elements in the integrated light  spectra of
GCs to reveal a wealth of abundance information that is lost in low-resolution, low signal-to-noise ratio (S/N) spectra.

Our IL analysis method has been developed primarily using a ``training set" of Milky Way (MW) and Large Magellanic Cloud (LMC) GCs.
Our method was first explored in \cite[][``Paper I"]{mb08}  using the NGC 104 (47 Tuc) spectrum from that training set.  The method was further developed in \citet[][PhD thesis]{scottphd}, which focused on the old clusters available in the MW sample, and published in \citet[][``Paper III"]{paper3} and \citet[][``Paper IV"]{paper4}, which focused on the more varied abundance patterns and  mixed-age range clusters of the LMC sample.  Since then, we have published a number of papers in which we have both applied the basic analysis methodology and also refined many of the spectral measurement analysis techniques.  We are now in a position to publish a more complete sample from the MW than was presented in \citet[][PhD thesis]{scottphd}. This sample is particularly useful in establishing the accuracy of our analysis methods for clusters with a wide range of abundances. In addition, we can now perform our analysis using the refined methods that have been developed in our papers since \citetalias{paper4}. 

In \citetalias{mb08}, we demonstrated the core strategies of our method: the use of equivalent widths in spectra of unresolved GCs  to obtain chemical abundances in the same manner as is done using spectra of RGB stars.  In \cite{scottphd},  development
 of the technique  and a  detailed analysis was performed of the 7 MW training set GCs (NGC 104, NGC 362, NGC 2808, NGC 6093, NGC 6388 NGC 6752, NGC 6397). These ``old" ($>$10 Gyr) GCs   spanned a range in [Fe/H] that is most typical of GCs, ie.  $-2.2< $[Fe/H]$< -0.5$.  
The development of the technique was performed in two stages.  In the first stage, resolved high resolution photometry of the MW GCs was used to empirically determine the stellar population --- the color magnitude diagram (CMD) --- of the GCs.  The virtue of analyzing the spectra first with the empirical information was that it allowed us to develop  our basic integrated light line synthesis techniques and routines.  
As the resolved CMDs are obviously not available to use in the analysis of unresolved, extragalactic clusters, we then moved on to a second stage of analysis in which we developed a strategy for using the Fe lines to identify  best-fitting populaions
using theoretical isochrones from stellar evolutionary models to approximate the CMDs. In this analysis, we iteratively constrain both the Fe abundance and the best-fitting CMD. This technique was first published in detail in \cite{m31p1} in a first application to extragalactic GCs, which included 5 GCs in M31, discussed further below.

\begin{deluxetable*}{lrrrrrrrrrrrrr}
\tablecolumns{15}
\tablewidth{0pc}
\tablecaption{Observations and Cluster Properties\label{tab:obs}}
\tablehead{
\colhead{Name}  &\colhead{RA} & \colhead{Dec}& \colhead{$\mu_{{\rm
      V}}$}           & \colhead{V} &   \colhead{M$_{{\rm Vtot}}$}&\colhead{E(B-V)} &
\colhead{ r$_{{\rm core}}$}         &
\colhead{Telescope} & \colhead{Scan}  &\colhead{T$_{exp}$} &
\colhead{S/N} \\ \colhead{} & \colhead{(J2000)} &
\colhead{(J2000)}&  \colhead{($mag/\square^{''}$)}  && & &  \colhead{
  ($'$) }    &&\colhead{Area}&\colhead{(s)} & \colhead{(pixel$^{-1}$)}  }

\startdata
 NGC 104 &  00 24 05.2   &  -72 04 51   &  14.36   &  3.95   &  -9.42
 & 0.04 &
   3.2   & DuPont & 32x32 &11030&100 \\
 NGC 2808 &  09 12 02.6   &  -64 51 47   &  14.50   &  6.20   &  -9.39 &0.22
 & 0.8   & DuPont & 32x32& 10730&81\\
 NGC 362  &  01 03 14.3   &  -70 50 54   &  14.66   &  6.40   &  -8.41&0.05
 & 0.8 & DuPont & 32x32& 11021&89\\
NGC 6093 &  16 17 02.5   &  -22 58 30   &  14.84   &  7.33   &  -8.23&0.18
& 0.6 & DuPont & 32x32&7350&52\\
NGC 6397 &  17 40 41.3   &  -53 40 25   &  15.47$^{a}$   &  5.73   &
-6.63 &0.18 & 2.9  & DuPont & 32x32 &18374&57\\
NGC 6752 &  19 10 52.0   &  -59 59 05   &  14.88$^{a}$   &  5.40   &
-7.73& 0.04&
1.9 & DuPont & 32x32&11021&130
\\
\cutinhead{New Clusters from \cite{zaritsky14}}

NGC 6388  &   17 36 17.0   &  -44 44 06   &  13.33   &  6.72   &  -9.42&0.37
& 0.5 & Magellan & 30x30  & 3600 & 200\\

 NGC 6440       &         17 48 52.7&  -20 21 37 & 13.78  & 9.20 &
 -8.75 & 1.07& 0.5   & Magellan & 10x10 & 25200 & 162 \\
  NGC 6441      &          17 50 13.1 & -37 03 05   & 13.49  &
  7.15&  -9.63&0.47 & 0.6 & Magellan & 10x10&3600& 119\\
 NGC 6528      &          18 04 49.6 & -30 03 23 &15.34   &  9.60
 &-6.57 & 0.54&0.4  & Magellan & 10x10&10800 & 104\\
 NGC 6553       &         18 09 17.6 & -25 54 31   & 16.24  &  8.06
 & -7.77 & 0.63&1.0 & Magellan & 30x30&2700 & 103\\

\cutinhead{New Fornax Cluster}
Fornax 3 &  02 39 52.5 &-34 16 08.0 & 17.00  & 12.56&-8.12  &  0.04 & 0.04 &
Magellan &\nodata &6000 & 70  \\

\enddata
\tablerefs{Milky Way Cluster identifications, positions, V magnitudes, E(B-V)
  and core radius in arcmin are from
     \cite[][2010 revision]{harris}.  Milky Way  central surface brightnesses in
     V are from \cite{2005ApJS..161..304M} except for a., which is
  from \cite[][2010 revision]{harris}.    Coordinates and core radius
  for Fornax 3 are from \cite{fornax-mackey}.  The  E(B-V) and
  Mv$_{tot}$ for Fornax 3 are from
  \cite{buonanno}, using their distance modulus of 20.68. Data will be available at  \url{https://zenodo.org/record/163464} upon publication.
 }
\end{deluxetable*}

In \citetalias{paper3},  we further developed our Fe and age analysis method on the LMC GCs in our training set, which are a crucial addition to the training set because the LMC GCs have ages ranging from 10's of Myr to $>$10 Gyr.  This population of massive, high surface brightness GCs with ages $<$10 Gyr is not available in the MW.  \citetalias{paper3} presented an important modification to our analysis technique, in which we assessed  the impact of stochastic stellar population fluctuations when using theoretical isochrones to represent the GC populations.  Stochastic fluctuations are a worry when the GC CMD is not fully populated, so we evaluated the impact of possible mismatch of the isochrone population with stochastically populated CMDs.  In general, large stochastic fluctuations in the CMD are expected for stellar populations with  young ages $<$ 1 Gyr, where the number and properties of luminous red supergiants or asymptotic giant branch (AGB) stars can have a large impact on the flux weighted IL spectra.  In practice, our training set IL spectra can also suffer from stochastic  issues because we are not able to observe the whole GC population in the MW and LMC, only the core regions are high enough surface brightness to obtain high S/N IL spectra.  

 In \citetalias{paper4},   we analyzed $\sim$20 additional elements in the LMC training set GCs.  \citetalias{paper4} also included development of an additional component of our technique for IL spectral synthesis, which allows us to measure abundances in circumstances where equivalent width analysis is not possible or is insufficient.  This can happen when the velocity dispersions of the GCs are large ($>$15 \kms), which means there is  greater line broadening, which results in more line blending and line blanketing makes determination  of the true continuum difficult or impossible. Similarly, at high metallicities the line blending and blanketing can lead to these problems.  In addition, IL line synthesis can provide abundance estimates when the S/N of the data is too low for accurate analysis with equivalent widths.

 Recently, \cite{sakari,sakari_ers} have used similar IL analysis techniques on a sample of 5 MW GCs   to confirm the reliability of IL abundance measurements of elements key in chemical tagging studies. They performed a detailed analysis of 
the potential systematics that can occur in IL studies, including how systematic errors can be reduced when line-by-line differential abundances are used to calculate abundance ratios.  The 5 GCs in their sample covered a metallicity range of $-2.4<$[Fe/H]$<-0.7$. In addition \cite{larsen12} used similar strategies and different measurement techniques to recover the abundances of a  metal-poor ([Fe/H]$\sim$-2.3) cluster in Fornax.

Our first application of the high resolution IL abundance analysis to extragalactic GC systems was for a pilot study of 5 GCs in M31 in \cite[][``C09"]{m31p1}.  This sample was increased to 31 GCs in \cite[][``C14"]{m31p2}, where Fe, alpha, and light elements were analyzed.  In addition to M31, we presented Fe and Ca abundances for 10 GCs in NGC 5128 in \cite{cent}.   In both the M31 and NGC 5128 works we found evidence for 10 Gyr age GCs with metallicities as high as [Fe/H]$\sim-0.2$.  While our training set included metallicities this high for LMC clusters, these clusters were all quite young, with ages of $<$1 Gyr.  Since abundance analysis of both stars and GCs becomes more difficult at the highest metallicities due to increased line blending and line blanketing, it was not clear that the IL abundance analysis technique would have the same accuracy for  GCs with metallicities approaching solar or above (that have much cooler stellar populations than young clusters).

 The primary motivation for this work is to further test the IL analysis techniques for ``old" ($>$10 Gyr) GCs with high metallicities, as this is a regime that has not been included in IL test studies by any group. We first present an updated analysis of the MW training set GCs  presented in \citetalias{mb08} and \cite{scottphd}.    The updates include the methods described in \citetalias{paper3}, \citetalias{paper4} and \citetalias{m31p2}.  Specifically, we now  include  assessment of the effect of stochastic stellar population fluctuations in the MW training set, which leads to a more reliable determination of the stellar population without {\it a priori} CMD information.   We also include line synthesis abundance measurements of  elements, which increases the precision of our measurements over those in \citetalias{mb08} and \cite{scottphd}.  We increase the metallicity range of the training set by  supplementing our original training set sample with IL spectra of four  high metallicity  bulge GCs (NGC 6440, NGC 6441, NGC 6553, NGC 6528) and one low metallicity GC in the Fornax dwarf spheroidal galaxy (Fornax 3).  In particular, we use the bulge GCs   to more effectively assess the potential complications of analysis of IL GC spectra at high metallicity.   
The paper is outlined as follows: in \textsection \ref{sec:obs} we present the targets, observations and data reduction.  In \textsection \ref{sec:analysis} we present the IL abundance analysis techniques employed in this work.  In \textsection \ref{sec:results} we present an in depth discussion of the age and [Fe/H] solutions for each GC, which includes comparisons to known values from standard abundance analysis of individual stars in these GCs.  Finally, in \textsection \ref{sec:others} we compare the abundance ratios for an additional 19 elements to those in the reference stellar abundance studies presented in \textsection \ref{sec:results}, including a comparison of abundances calculated differentially, line-by-line.  In  \textsection \ref{sec:summary} we present our summary and conclusions.

\section{Targets, Observations and Reductions}
\label{sec:obs}

\subsection{Original Training Set}
The observations and reductions of the original training set sample are described in detail in \citetalias{mb08} and \cite{scottphd}, and are  summarized here. 
The  data were obtained using the the echelle
spectrograph on the 2.5 m du Pont telescope at Las Campanas.   The
spectra cover the range 3700--7800 \AA, with declining sensitivity and
spectral resolution toward the blue end.  Telluric absorption features
limit our ability to analyze the spectra at the reddest end of this
range, so our analysis focuses primarily on the high quality spectra
that were obtained from 4100--7500 \AA.
During the observations we uniformly scanned the core regions of the training set
clusters to simulate IL spectra that can be obtained for unresolved extragalactic GCs.  To both
maximize the S/N ratio of the spectra and minimize the
relative contribution from sky, we limited the scanned regions of each cluster to  the central $32\times32$
arcsec$^{2}$. The cluster regions were scanned once per exposure, so clear
conditions were necessary to ensure an unbiased weighting of the
cluster light.  As the entire slit was filled with cluster light
during these scans, significant sky flux outside the telluric emission
lines was only detected near twilight. To be conservative, exposures
were also obtained off the cluster (i.e. on ``pure sky'') to allow
measurement and subtraction of the sky signal from the science
exposures. These sky exposures were taken throughout the night,
temporally mixed with the cluster observations.  All clusters,
excluding NGC 2808, were observed during lunar dark time in 2000
July. NGC 2808 was observed during lunar dark time in 2001
January. The total integration time for each cluster is summarized in
Table \ref{tab:obs}, along with the general properties of the clusters.

\begin{figure}
\centering
\includegraphics[trim = 17mm 0mm 2mm 0mm, clip,scale=0.53]{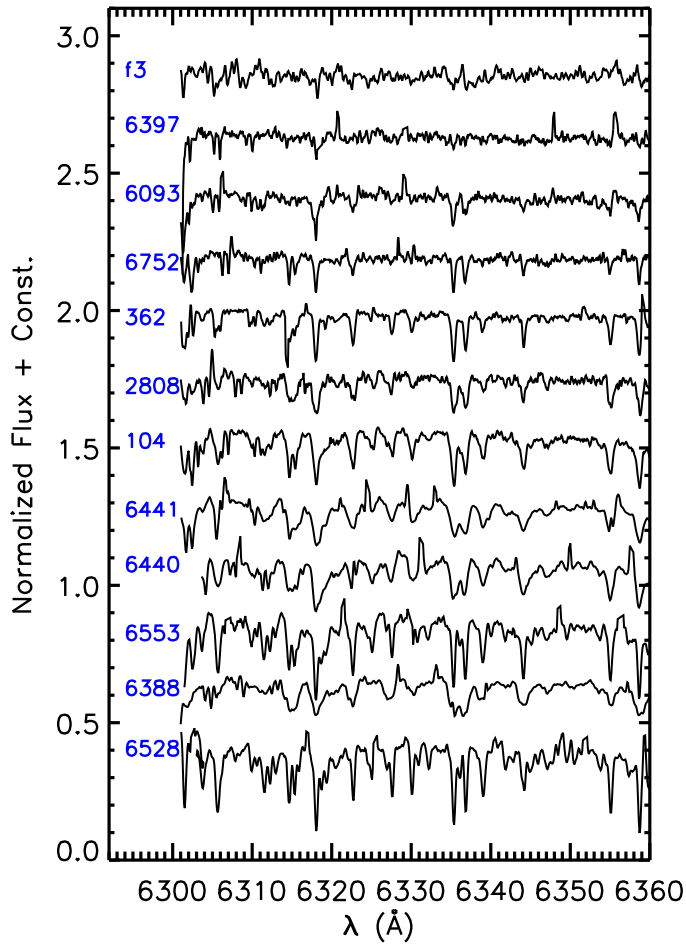}
\caption{ A section of the IL spectra for each cluster in the sample.  The spectra are normalized to 1.0, with a base at 0.0, and a constant offset has been applied for visual separation.  The spectra have been smoothed by 3 pixels for presentation. The clusters are presented top to bottom in order of increasing metallicity from our analysis. }
\label{fig:spec} 
\end{figure}

We performed the basic data reduction
steps using the echelle package in IRAF\footnote{IRAF is distributed
  by the National Optical Astronomy Observatories, which are operated
  by the Association of Universities for Research in Astronomy, Inc.,
  under cooperative agreement with the National Science
  Foundation.}, including the routines for overscan, bias subtraction,
and flat-field division.  Echelle orders were identified, traced, and
extracted using IRAF's APALL routine. This routine also allows for the
subtraction of inter-order, scattered-light, which is easy to measure
in this spectrograph for orders bluer than 6000 \AA, but difficult to
measure at redder wavelengths because the wings of adjacent orders
begin to overlap at those wavelengths. In order to measure the
scattered light despite this overlap, an empirical scattered-light
model was produced. For a detailed discussion of our scattered-light
subtraction see \citetalias{mb08}.  The sky spectra were scaled
and subtracted from the individual integrated-light exposures using
simple arithmetic routines. The extracted spectra were wavelength
calibrated using standard IRAF routines and the Th-Ar spectra taken
before and after each science spectrum.

In the analysis of the GCs' fully reduced spectra, we measure the
equivalent widths (EWs) of spectral absorption lines while
simultaneously fitting the continuum level in each order.  To do so,
it is only necessary to perform a relative flux calibration of the spectra by
normalizing out the strong echelle blaze function from each order.  An
adequate approximation of the blaze function of each order
was obtained using observations of a bright G-star,
which is roughly the color of our IL spectra.  The approximate 
blaze function was obtained by fitting a low order polynomial
to the G-star's continuum in each order.
No attempt was made to remove telluric absorption lines from the
spectra, however telluric template stars were compared to each order
of our final IL spectra to assure that no absorption lines were
measured near telluric lines.  Representative S/N values
for each cluster's final IL spectrum are listed in Table
\ref{tab:obs}, and Figure \ref{fig:spec} displays a
representative spectral region for each cluster.

\subsection{Supplemental Data}

High resolution IL spectra of five additional GCs (NGC 6388, NGC 6440, NGC 6441, NGC 6528, NGC 6553) were taken in collaboration with D. Zaritsky for a program focussed  on dynamical analysis \citep{zaritsky14}.  These GCs were observed with the MIKE spectrograph \citep{mike} on the Magellan Clay Telescope.  The telescope was set  to raster scan the slit (0.75"x5") across the central, high surface brightness  region of the cluster during each exposure. The area of each scan is listed in Table \ref{tab:obs} and approximately corresponds to the cluster half light radius.      The spectra have a wavelength coverage of approximately 3700 to 9800 \AA; we primarily use the    region between 4000 and 7600 \AA, which has  the highest S/N and least background contamination.  We reduced the spectra using the MIKE IDL pipeline \citep{mikeredux}.  More details about the observations can be found in \cite{zaritsky1, zaritsky14}.   NGC
6388 was observed with the duPont telescope for the original training set as well as with Magellan.  For the analysis in this work we use the higher S/N MIKE spectra.

We obtained IL spectra of Fornax 3 in March 2014 with the MIKE spectrograph on Magellan, using a slit size of 1.0"x5".  In this case, 
 due to the much greater distance to the Fornax galaxy,  traditional stationary (rather than scanned) exposures were more effective.   Fornax 3 is still quite extended with a core radius of 2.4" \citep{fornax-mackey}, so we split the observations into 3 exposures to cover as much of the cluster light as possible.  The first exposure was centered on the cluster, and the other two were offset by one slit width to the left and right of the central exposure.  Like the MW bulge GCs, the Fornax 3 spectrum was  reduced using the MIKE IDL pipeline. 

Upon publication the spectra of the training set clusters will be downloadable from \url{https://zenodo.org/record/163464}.

\section{Analysis Details}        
\label{sec:analysis}

To calculate integrated light spectra, we use the routine ILABUNDS \citep{mb08},  which employs the most up to date (2014) version of Moog  \citep{1973ApJ...184..839S}.  All calculations are performed under the assumption of local thermodynamic equilibrium (LTE).  ILABUNDS can be used for EW matching (\textsection \ref{sec:ews}), or for spectral synthesis (\textsection \ref{sec:syn}).  

In all of our analyses, theoretical CMDs are constructed to represent the cluster population; the stellar parameters from the CMDs are used to synthesize spectra for $\sim$25 representative stellar types, which are then combined into a flux-weighted integrated light spectrum.  The properties of the stars are taken from the isochrones used to construct the theoretical CMDs.  We use the  extended AGB  canonical isochrones  from the Teramo group with  mass$-$loss parameter of $\eta$=0.2
 \citep{2004ApJ...612..168P,2006ApJ...642..797P,2007AJ....133..468C}, and determine the number of stars of each type using the initial mass function of   \cite{2002Sci...295...82K}.  The total number of stars in the CMD is normalized to the absolute magnitude of the GC and the percentage of the total GC flux contained in the scanning observation. More details on sampling effects are discussed in   \textsection \ref{sec:results}.

The 1-D, plane parallel, ODFNEW and AODFNEW model grids of Kurucz\footnote{The models are available  from R. L. Kurucz's Website at  http://kurucz.harvard.edu/grids.html}\citep[e.g.][]{2004astro.ph..5087C} are used to interpolate stellar atmospheres for the 25 stellar types in our theoretical CMDs.

\begin{deluxetable}{lrr}
\tablecolumns{3}
\tablewidth{0pc}
\tablecaption{Velocity Dispersions\label{tab:sigv}}
\tablehead{
\colhead{Name}  &\colhead{v$_{\sigma,Z}$} &\colhead{v$_{\sigma}$}\\ & \colhead{\kms}&\colhead{\kms}}

\startdata
 NGC 104 &11.5$\pm$0.2 &12.6$\pm$1.2 \\
 NGC 2808 & 13.0$\pm$0.5 & 13.7$\pm$1.1\\
 NGC 362 &9.2$\pm$0.4& 9.1$\pm$1.2\\
NGC 6093 &  9.5$\pm$0.5& 11.9$\pm$3.3\\

NGC 6752 & 6.6$\pm0.4$ & 7.0$\pm$1.5\\

NGC 6388  &   18.4$\pm$0.6 &23.1$\pm$2.5\\

 NGC 6440     & 13.3$\pm$0.7& 15.9$\pm$1.7\\
  NGC 6441    & 16.5$\pm$0.7  & 18.1$\pm$2.8\\
 NGC 6528      &     5.8$\pm$0.5& 6.4$\pm$1.4\\
 NGC 6553      & 7.0$\pm$0.5 & 7.8$\pm$1.9\\

Fornax 3   &\nodata&9.9$\pm$2.6\\

\enddata
\tablerefs{ Column 2 contains measurements from \cite{zaritsky2, zaritsky14}.  Column 3 contains the measurements we made using cross correlation between template stars in this work. }
\end{deluxetable}

\begin{figure}
\centering
\includegraphics[scale=0.53]{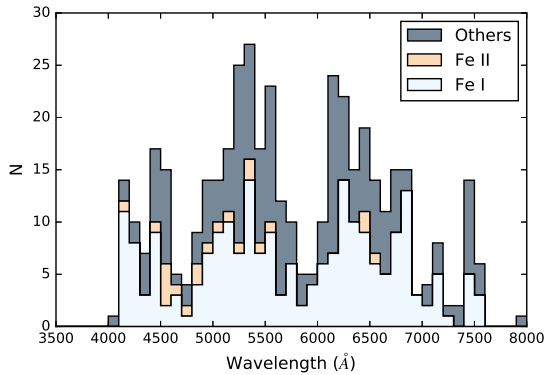}
\caption{ The wavelength distribution of lines used in this analysis. Each bin corresponds to 100 \AA. }
\label{fig:line_histo} 
\end{figure}

Line lists were taken from \cite{1994ApJS...91..749M}, \cite{1995AJ....109.2757M},
\cite{1998AJ....115.1640M},  \citetalias{mb08},  \cite{m31p2}, and references therein, with updates to log {\it gf} values from \cite{2009A&A...497..611M} (Fe II), \cite{sobeck} (Cr), \cite{lawlerla} (La), and \cite{hartognd} (Nd).
Our line lists are specifically chosen to include lines with the most accurate  {\it gf} values, and that
  give consistent and reliable results in our training set clusters.
Because each cluster has a unique combination of velocity dispersion,
metallicity, and systemic velocity, each cluster has a unique list of
detectable and trustworthy  lines.  Our Fe-line list is 
intended to  include Fe I lines that cover as wide a range as
possible in $\lambda$, excitation potential (EP), and EW, all of which are powerful
diagnostics for evaluating the quality of our abundance solutions.  We use a total of 217 different Fe I lines, 21 Fe II lines and 198 lines of other elements. A histogram showing the wavelength distribution of the lines in our analysis is shown in Figure \ref{fig:line_histo}, which  shows  the usefulness of the   extensive wavelength coverage  of the data.

We primarily use lines that have EWs or inferred EWs that are less than $\sim$150 m\AA, in order to avoid lines that may be saturated in a significant number of stars in the population.   For abundance analysis utilizing spectral synthesis, we supplement our primary line list with that of the Kurucz database.\footnote{http://kurucz.harvard.edu/linelists.html} In our default  lists we include  transitions for important molecules like C$_{2}$, CN, NH, CH, and  MgH (see \textsection \ref{sec:tio} for TiO).     Hyperfine splitting is included in the line synthesis for Sc II, V I, Mn I, Co I, Cu I,  Zr I, Ba II, La II, and  Eu II.  

 \begin{deluxetable*}{rrr|rrrrrrr}
\tablecolumns{10}
\tablewidth{0pc}
\tablecaption{GC Fe I EWs\label{tab:ew_stub}}
\tablehead{
 \colhead{$\lambda$}& \colhead{E.P.} &  \colhead{log{\it gf}} & \colhead{F3} & \colhead{N104} &  \colhead{N2808}&  \colhead{N362}  & \colhead{N6093} & \colhead{N6397} & \colhead{N6752} \\ \colhead{(\AA)}& \colhead{(eV)} && \colhead{(m\AA)} & \colhead{(m\AA)}& \colhead{(m\AA)}& \colhead{(m\AA)}& \colhead{(m\AA)}& \colhead{(m\AA)}& \colhead{(m\AA)} }
\startdata

5216.283 &  1.608 & -2.082 &  64.8 &  \nodata&  \nodata& 126.2 & 116.2 &  36.7 &  74.7 \\
5225.534 &  0.110 & -4.755 &  \nodata&  \nodata&  \nodata&  84.9 &  \nodata&  \nodata&  53.2 \\
5232.952 &  2.940 & -0.057 &  76.2 &  \nodata&  \nodata&  \nodata&  89.4 &  65.8 & 113.2 \\
5266.563 &  2.998 & -0.385 &  \nodata&  \nodata&  \nodata&  \nodata&  87.8 &  46.1 &  85.2 \\
5281.798 &  3.038 & -0.833 &  \nodata&  \nodata&  \nodata& 121.8 &  64.2 &  \nodata&  63.2 \\
5281.798 &  3.038 & -0.833 &  \nodata&  \nodata&  \nodata& 115.7 & \nodata&  \nodata&\nodata\\
5283.629 &  3.241 & -0.524 &  33.6 &  \nodata&  \nodata& 148.8 &  \nodata&  35.4 &  88.4 \\
5302.307 &  3.283 & -0.720 &  \nodata&  \nodata& 101.4 & 119.2 &  73.6 &  22.5 &  64.5 \\
5307.369 &  1.608 & -2.912 &  \nodata& 117.9 &  78.0 &  82.7 &  70.5 &  \nodata&  65.6 \\
5324.191 &  3.211 & -0.103 &  \nodata&  \nodata&  \nodata&  \nodata& 105.2 &  46.0 &  90.8 \\

\enddata
\tablecomments{EW measurements for Fornax 3, NGC 104, NGC 2808, NGC
  362, NGC 6093, NGC 6397 and NGC 6752. Lines listed twice correspond
  to measurements made in adjacent orders with overlapping
  spectral coverage. This table is presented in its entirety 
   in the electronic addition of the journal.   }
\end{deluxetable*}

\subsection{ Fe EW Analysis}
\label{sec:ews}

Our core analysis method allows us to measure the
abundance of the cluster using the observed EWs of absorption
features in a cluster's integrated light spectrum, analogous to standard EW
 techniques used for the analysis of single stars.

It is advantageous to use a standardized tool to ensure consistent measurements are made.
We measure  EWs of absorption lines in the IL spectra using GETJOB
(McWilliam et al. 1995). This software includes semi-automated
routines  to interactively fit  low-order
polynomials to the continuum level of each echelle order using
specified continuum regions.  Line profile fits are made
with  single, double, or triple Gaussian profiles to
isolate desired lines and obtain their EWs.  Particular care
is needed to set the continuum level in the IL spectra, especially when the velocity dispersion is large.

In \cite{m31p2}, with a sample of 31 GCs in M31, we performed tests to determine when EW analysis alone was insufficient for measuring an accurate [Fe/H].  We found that EW analysis and line synthesis analysis yielded consistent results for GCs with [Fe/H]$<$-0.3 and velocity dispersions ($\sigma_{v}$) less than 15 \kms.    Following that work, we use EWs to determine the [Fe/H] of the MW GCs here when those criteria are met.  For the original training set GCs included in  \cite{scottphd}, all are analyzed using EWs, with the exception of NGC 6388, which has a  velocity dispersion of $\sigma_{v}\sim$20 \kms.  For the supplemental high metallicity GCs, all [Fe/H] are measured with  spectral synthesis, as described in the next section.

\subsection{ Fe Spectral Synthesis Analysis}
\label{sec:syn}

The Fe spectral synthesis analysis is described in detail in \cite{paper3} and \cite{m31p2}. Briefly, we use a $\chi^{2}$ minimization scheme to fit the synthesized IL spectra to  the observed spectra. We calculate synthetic spectra for a 20 \rAA region centered on the Fe line of interest; this full region is used to set the continuum or ``pseudo-continuum" level.   The fitting of each line must be checked by eye to ensure that the continuum fit is accurate, that the line is not affected by bad blending or sky absorption lines, and that the region isn't affected by local noise.    The actual  $\chi^{2}$-minimization is performed in a smaller $\sim$0.5 \rAA region  around the line itself in order to obtain the most accurate measurement. 
 For the abundance matching,   spectra are calculated with abundances that are  $\pm$0.5 dex around the   average abundance, in steps of 0.1 dex.  A starting guess abundance and age are necessary to create an initial theoretical CMD for calculating the IL spectra.  This initial CMD and its spectra are used to identify the most trustworthy lines for the cluster and to establish proper continuum placement for each line.  First guess abundances in this work were taken from the catalog of  \cite[][2010 edition]{harris}; when extragalactic GCs are studied, low resolution spectroscopic metallicity estimates can be used as first guess abundances.   We use a first guess age of 10 Gyr for the initial CMD.    When the lines and continuum are satisfactorily identified, the $\chi^{2}$-minimization can be performed systematically  for  CMDs of any other age/metallicity combination, and the most self consistent age/metallicity solution can be identified.
 
 In the EW analysis we use the behavior of the EWs of the Fe lines as one of our diagnostics for isolating the best possible stellar population.  In the line synthesis analysis we calculate a ``pseudo-EW" by re-synthesizing the Fe line alone with the abundance we derived when nearby line blends were included.

 \begin{deluxetable}{r|rrrrr}
\tablecolumns{6}
\tablecaption{ Fe I Synthesis Abundances\label{tab:syn_stub}}
\tablehead{ &\multicolumn{5}{c}{12+log(X/H)}\\
 \colhead{$\lambda$} & \colhead{n6388} & \colhead{n6440} &  \colhead{n6441}&  \colhead{n6528}  & \colhead{n6553}  }
\startdata

5216.283 & \nodata&  6.88 &  6.51 &  6.82 & \nodata \\
5232.952 & \nodata&  6.88 &  6.71 &  7.02 & \nodata \\
5281.798 &  7.08 &  7.28 &  6.61 &  7.02 &  6.89  \\
5283.629 &  7.28 &  7.38 &  7.21 &  6.92 &  7.09  \\
5302.307 &  7.08 &  7.28 &  7.01 &  7.12 &  7.09  \\
5307.369 &  7.08 &  6.78 &  6.91 &  7.22 &  6.69  \\
5367.476 &  6.68 &  6.88 &  6.91 &  6.82 &  7.09  \\
5369.974 &  6.98 &  7.28 &  7.01 &  6.92 & \nodata \\
5383.380 & \nodata&  6.68 & \nodata&  6.72 &  6.89  \\
5389.486 & \nodata& \nodata&  6.71 & 7.22& 6.89 \\

\enddata
\tablecomments{Abundance measurements from $\chi^{2}$ fitting of
  spectral syntheses for NGC 6388, NGC 6440, NGC 6441, NGC 6528 and
  NGC 6553. Abundances for individual lines are measured in steps of 0.1 dex from the mean abundance obtained from all lines.  Abundances are quoted for the CMD solution with the
  oldest age for each GC.  The full table is available in the
  electronic edition of the journal.}
\end{deluxetable}

 \begin{deluxetable*}{lrrr|rrrrrrrrrrr}
\tablecolumns{15}
\tablewidth{0pc}
\tablecaption{Synthesized Fe II Abundances\label{tab:feii}}
\tablehead{\colhead{Species}& \colhead{$\lambda$}&\colhead{EP} & \colhead{log{\it gf}} & \multicolumn{10}{c}{12+log(X/H)} \\ &\colhead{(\AA)}& \colhead{(eV)} & &
 \colhead{F3} & \colhead{n104} &  \colhead{n2808}&  \colhead{n362}  & \colhead{n6093} & \colhead{n6388} & \colhead{n6440}& \colhead{n6441}& \colhead{n6528}& \colhead{n6553}& \colhead{n6752} }
\startdata
 Fe   II &  4122.67 &  2.58 &  -3.26 &   \nodata &   6.52 & \nodata &   6.17 & \nodata & \nodata &     \nodata & \nodata & \nodata & \nodata &   5.77 \\
     Fe   II &  4489.18 &  2.83 &  -2.96 &    5.31 & \nodata & \nodata & \nodata & \nodata & \nodata & \nodata & \nodata &   7.12 & \nodata &   5.97 \\
     Fe   II &  4508.29 &  2.86 &  -2.44 &      5.41 &   7.02 &   6.63 & \nodata & \nodata & \nodata &   7.28 & \nodata &   7.02 & \nodata &   5.87 \\
     Fe   II &  4520.22 &  2.81 &  -2.65 &    \nodata &   6.82 & \nodata & \nodata & \nodata & \nodata&   \nodata &   7.14 & \nodata & \nodata &   5.97 \\
     Fe   II &  4582.83 &  2.84 &  -3.18 &    \nodata &   6.62 & \nodata &   6.47 & \nodata & \nodata & \nodata & \nodata &   6.92 & \nodata & \nodata \\
     Fe   II &  4583.84 &  2.81 &  -1.93 &      5.31 &   6.92 & \nodata &   6.57 &   6.01 & \nodata   &   7.18 & \nodata &   7.22 &   7.49 & \nodata \\
     Fe   II &  4620.52 &  2.83 &  -3.21 &    \nodata &   6.52 & \nodata &   6.47 & \nodata & \nodata & \nodata & \nodata &   7.22 &   6.79 &   5.87 \\
     Fe   II &  4731.45 &  2.89 &  -3.10 &    \nodata &   6.82 &   6.63 &   6.47 & \nodata & \nodata   & \nodata & \nodata & \nodata & \nodata &   6.17 \\
     Fe   II &  4839.99 &  2.68 &  -4.75 &    \nodata & \nodata & \nodata & \nodata & \nodata & \nodata& \nodata & \nodata & \nodata & \nodata &   5.97 \\
     Fe   II &  4855.55 &  2.70 &  -4.46 &    \nodata & \nodata & \nodata &   6.37 & \nodata & \nodata& \nodata & \nodata & \nodata & \nodata & \nodata \\
     Fe   II &  4923.93 &  2.89 &  -1.26 &      5.51 &   7.02 &   6.63 &   6.37 &   5.41 & \nodata    & \nodata & \nodata &   6.62 & \nodata &   5.77 \\
     Fe   II &  5018.44 &  2.89 &  -1.10 &    \nodata & \nodata &   6.63 &   6.47 &   5.41 & \nodata   &   \nodata &   7.53 &   7.12 &   6.89 & \nodata \\
     Fe   II &  5197.58 &  3.23 &  -2.22 &      5.61 &   6.82 &   6.83 &   6.47 & \nodata & \nodata    &   7.18 & \nodata &   6.92 & \nodata &   5.77 \\
     Fe   II &  5234.62 &  3.22 &  -2.18 &    \nodata &   6.72 & \nodata & \nodata & \nodata & \nodata & \nodata & \nodata &   6.82 &   6.99 &   5.67 \\
     Fe   II &  5325.55 &  3.22 &  -3.16 &    \nodata &   6.92 & \nodata &   6.57 & \nodata & \nodata  & \nodata & \nodata & \nodata &   7.39 &   5.77 \\
     Fe   II &  5362.87 &  3.20 &  -2.57 &    \nodata & \nodata &   6.73 &   6.57 & \nodata & \nodata  &   \nodata & \nodata &   7.71 & \nodata &   5.97 \\
     Fe   II &  5425.26 &  3.20 &  -3.22 &    \nodata &   6.82 & \nodata & \nodata & \nodata & \nodata & \nodata & \nodata &   7.12 &   7.19 &   5.97 \\
     Fe   II &  5534.85 &  3.24 &  -2.75 &    \nodata & \nodata & \nodata &   6.37 & \nodata &   6.98  &   \nodata &   6.74 &   7.02 &   6.79 &   5.87 \\
     Fe   II &  6432.68 &  2.89 &  -3.57 &    \nodata & \nodata &   6.73 &   6.37 &   5.71 & \nodata   &   7.18 &   7.53 &   7.02 &   7.19 & \nodata \\
     Fe   II &  6456.38 &  3.90 &  -2.05 &    \nodata &   7.02 & \nodata &   6.57 & \nodata & \nodata  & \nodata & \nodata &   7.02 &   7.39 &   5.97 \\
     Fe   II &  6516.08 &  2.89 &  -3.31 &    \nodata &   7.02 &   6.43 &   6.47 & \nodata & \nodata   & \nodata & \nodata &   7.32 &   7.39 & \nodata \\

\enddata
\tablecomments{Abundance measurements for individual Fe II lines.  Abundances are presented in the same format as Table \ref{tab:syn_stub}. }
\end{deluxetable*}

To compare the synthesized spectra to the data,  we need to convolve the synthesis with the observed one dimensional velocity dispersion (v$_{\sigma}$) of the cluster, which dominates the broadening of the lines.   As in our previous work \citepalias[e.g.][]{m31p2}, we obtain the velocity dispersion of our IL spectra by cross correlation with Galactic template stars using the routine {\it fxcor} in IRAF.  For each cluster we used template stars that were observed with identical setups of the cluster.  For the clusters observed with the duPont telescope (NGC 104, NGC 2808, NGC 362, NGC 6093, NGC 6397, NGC 6752), we use a single template star,  HR805.   The velocity dispersion of NGC 6397 is comparable to or smaller than the resolution of the data of $\sim$3 \kms, so we do not report a measurement here and only use EW analysis for the abundances of all species of this cluster.    For the clusters observed in \cite{zaritsky14} with the Magellan telescope, we use the mean v$_{\sigma}$ obtained from three  different template stars: HD033771, HR914, and HD171391. Our measurements are given in Table \ref{tab:sigv}. For Fornax 3 we use the mean  v$_{\sigma}$  obtained from HD033771 and HD171391.  For comparison, we  show the velocity dispersion measurements made using the same data for the MW GCs but different techniques in \cite{zaritsky2} and \cite{zaritsky14}.  The two measurements agree for most clusters, although we note that small differences are expected because we have not corrected for aperture or 3 dimensional effects in this work, since our goal here is only to account for the line broadening of the spectra.

\begin{figure}
\centering
\includegraphics[scale=0.45]{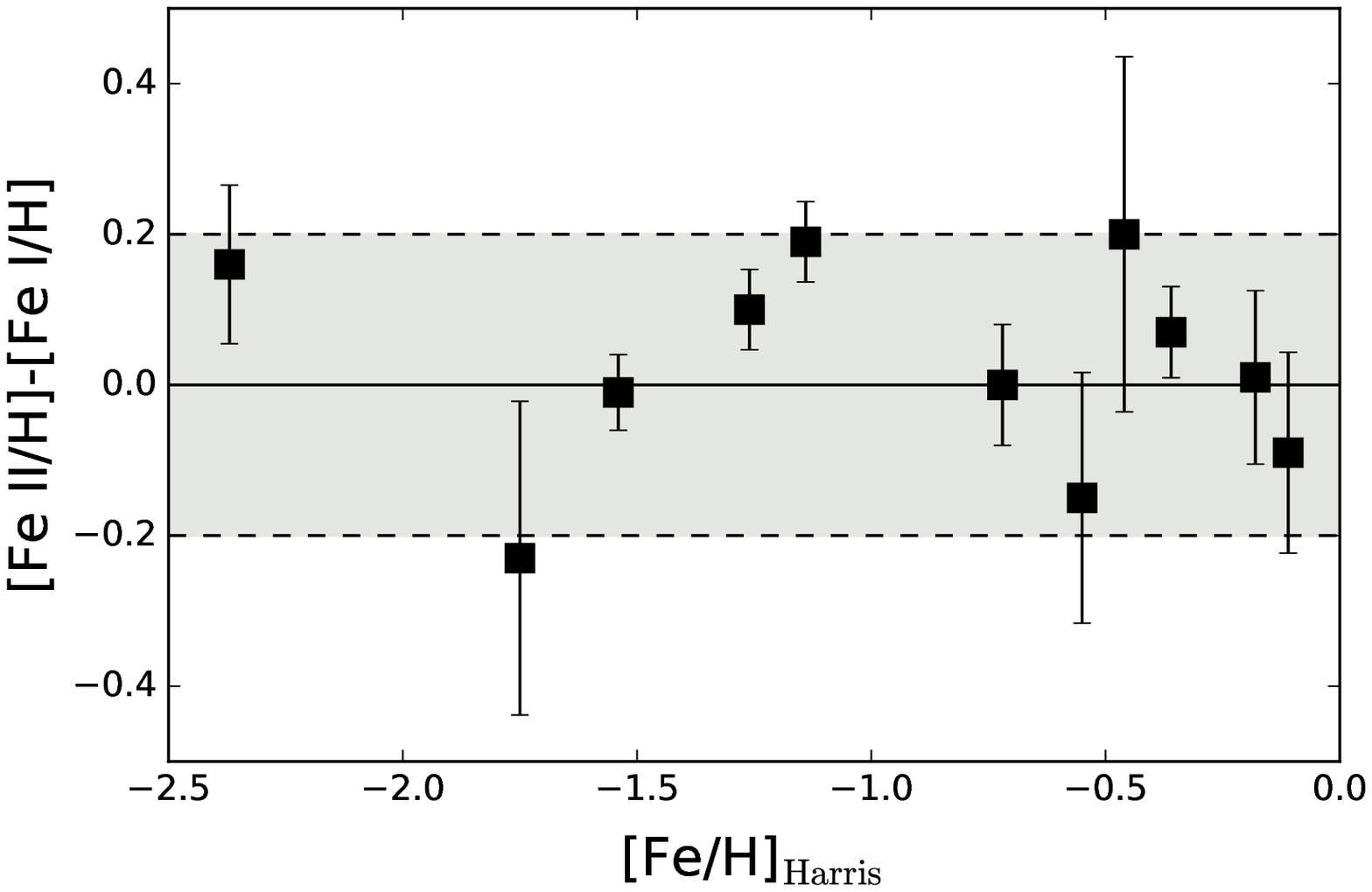}
\caption{ The difference between the abundances measured for Fe I and Fe II lines, as a function of the [Fe/H] from \cite[][2010 edition]{harris}, with the exception of Fornax 3 for which we use the reference abundance of \cite{letarte}. A solid line is shown at 0, which would equal perfect agreement.  The shaded gray region shows a range of $\pm$0.2 dex from perfect agreement to guide the eye, which encompasses the majority of points.  The error bars show the total uncertainties in Fe I and Fe II mean abundances added in quadrature. }
\label{fig:fe} 
\end{figure}

\subsection{Isolating the best population}
\label{sec:method}

Our goal for study of unresolved extragalactic GCs is to be able to identify a GC's age and metallicity using diagnostics in the IL spectra themselves.  Therefore, in analyzing the training set clusters we have allowed for CMD solutions with ages between 1-15 Gyr and   -$2.5<$ [Fe/H] $<+0.2$.   We isolate the most self consistent solution by first requiring that the input metallicity of the CMD is consistent with  the derived average abundance from the Fe I lines.    We then look for solutions with the smallest line-to-line statistical scatter ($\sigma_{N,FeI}$), and minimize any trends of the Fe abundance with excitation potential, wavelength and EW, similar to techniques used in analysis of individual stars.  For typical ``old" GCs, we can usually constrain the age to a range of 10-15 Gyr.  This age spread translates to a spread in associated metallicity; for typical old GCs the uncertainty in the [Fe/H] due to the age is $\sim$0.05 dex.   For final abundances we average the results obtained using the youngest and oldest CMDs that give self consistent results.
 For the total uncertainty in [Fe/H], we add the uncertainty due to age ($\sigma_{A,FeI}$) and the standard deviation of the mean abundance ($\sigma_{FeI}$=$\sigma_{ {N,FeI}} / \sqrt{{\rm N_{FeI}} -1} $) in quadrature.  We note that for abundances of elements besides Fe I, we can sometimes only measure one line of a given species.  In these cases we assign a typical line-to-line scatter uncertainty of 0.1 dex \citepalias[see][]{m31p2}, and add that in quadrature with the age uncertainty. 

\begin{deluxetable*}{l|lllll|lllll|c}
\tablecolumns{12}
\tablewidth{0pc}
\tablecaption{Fe Abundance and Age Results\label{tab:fe}}
\tablehead{
\colhead{Cluster}   & \multicolumn{5}{c}{[Fe I/H]} &
\multicolumn{5}{c}{[Fe II/H] } & Age  \\ & \colhead{[Fe I/H]}  &
\colhead{$\sigma$} &  \colhead{$\sigma_{A,FeI}$} &
\colhead{$\sigma_{FeI}$} & \colhead{N$_{FeI}$}  & \colhead{ [Fe II/H]} & \colhead{$\sigma$} &  \colhead{$\sigma_{A,FeII}$} &
\colhead{$\sigma_{FeII}$} & \colhead{N$_{FeII}$} & (Gyr) }

\startdata
 NGC 104 & -0.65 & 0.19 & 0.05 & 0.05 & 111 & -0.65  & 0.18 &0.04&
 0.06 &14   & 10.0 $\pm$ 3.0 \\
 NGC 2808 &  -1.04 & 0.22 & 0.02 & 0.04 & 58 & -0.85 & 0.12 & 0.01 & 0.04
 & 8 &
 11.5 $\pm$ 1.5  \\
 NGC 362  & -1.14  & 0.16 & 0.04 & 0.04 & 75 &-1.04 & 0.11 & 0.02 & 0.03 & 15  & 14.0
 $\pm$ 1.0\\
NGC 6093 &  -1.65 & 0.25 &0.08 & 0.09 & 65 &  -1.88 & 0.33 & 0.02 & 0.19 &
4 & 12.5 $\pm$ 2.5\\
NGC 6388 & -0.33 & 0.18 & 0.13 & 0.13 & 42 & -0.48 & 0.10 & 0.06 &0.10 & 1 & 9.0 $\pm$ 4.0 \\
NGC 6397 & -2.05 & 0.21 & 0.02 & 0.03 & 58 & \nodata & \nodata &
\nodata & \nodata  &  \nodata & 11.0 $\pm$ 4.0 \\
 NGC 6440       &  -0.34  & 0.22 & 0.08 & 0.08 & 58 & -0.27 & 0.10 &
 0.04 & 0.05 & 4 & 9.0 $\pm$ 4.0
 \\
  NGC 6441      &  -0.46 & 0.21 & 0.11 & 0.11 & 60 & -0.26 &0.34 & 0.02 &
  0.22 & 4 & 9.0 $\pm$ 6.0 \\
 NGC 6528      &  -0.31 & 0.22 & 0.05 & 0.06 & 65 & -0.40 & 0.18 &
 0.03 & 0.07 & 14 & 8.5 $\pm$ 1.5\\
 NGC 6553       &  -0.35 & 0.19 & 0.03 & 0.03 & 68 & -0.34 & 0.27 &
 0.01 & 0.10 & 10 & 10.0 $\pm$ 3.0 \\
NGC 6752 & -1.58 & 0.20 &  0.02 & 0.03 & 81 & -1.59 & 0.13 & 0.02 &
0.04 & 14 & 11.0 $\pm$ 4.0 \\
Fornax 3   & -2.27 & 0.27 & 0.03 & 0.05 & 36 & -2.11 & 0.15 & 0.06 &
0.09 & 5 & 14.0 $\pm$ 1.0  \\

\enddata
\tablerefs{Final results for Fe abundances and derived ages. Column 2 shows the mean abundance obtained from Fe I lines, and Column 7 shows the mean abundance obtained from Fe II lines.  Column 3 and Column 8 show $\sigma$, which corresponds to the standard devation
  in the abundance measured from the lines of each species.
Column 4 and Column 9 show the uncertainty in abundance due to the age range of the best solutions, $\sigma_{A,X}$.
   Column 5 and Column 10 show the total uncertainty for each species, which is defined as $\sigma_{X}=\sqrt(\sigma_{A,X}^{2} +
  \sigma_{N,X}^{2})$ where $\sigma_{N,X} = \sigma/
  \sqrt(N_{X}-1)$ and $\sigma$ is the standard deviation of the
  abundance of N$_{X}$ lines.  The number of lines measured for each species, N$_{X}$,  are listed in Columns 6 and 11.  }
\end{deluxetable*}

\begin{deluxetable*}{l|r|l|c}
\tablecolumns{4}
\tablewidth{0pc}
\tablecaption{Abundance Data from the Literature\label{tab:source}}
\tablehead{
\colhead{Cluster}  & & \multicolumn{2}{c}{Individual Sources} \\ & \colhead{[Fe/H]$^{1}$} & \colhead{ [Fe/H]$^{2}$}   }

\startdata
 NGC 104 &  -0.72& -0.75$\pm$0.05& \cite{koch2008},\cite{thyg2014},\cite{cordero2014},\cite{carretta47tuc}\\
 NGC 2808 & -1.14 & -1.18$\pm$0.04 &\cite{carretta2808}, \cite{carretta2808-06}, \cite{marino14}\\
 NGC 362  & -1.26 & -1.24$\pm$0.08&\cite{shetrone362}, \cite{worley2010}, \cite{carretta362}\\
NGC 6093 &  -1.75& -1.76$\pm$0.04 &\cite{carretta6093}, \cite{cavallo2004}\\
NGC 6388 & -0.55 & -0.62$\pm$0.25 &\cite{wallerstein07}, \cite{carretta6388}\\
NGC 6397 & -2.02&  -2.07$\pm$0.03 &\cite{koch2011}, \cite{gratton01}, \cite{lind11}\\
 NGC 6440       &     -0.36& -0.56$\pm$0.02$^{3}$ & \cite{origlia08}\\
  NGC 6441      &   -0.46& -0.45$\pm$0.08&  \cite{origlia08}, \cite{gratton6441} \\
 NGC 6528      &  -0.11& -0.07$\pm$0.13 &\cite{origlia05}, \cite{zoccali04}, \cite{carretta6528}\\
 NGC 6553       &     -0.18&  -0.28$\pm$0.18 & \cite{cohen99}, \cite{alves06},
 \cite{melendez03}, \cite{barbuy99}\\
NGC 6752 & -1.54&-1.56$\pm$0.10 &\cite{cavallo2004}, \cite{gruyters14},
\cite{yong2005}, \cite{gratton01}, \cite{james04}\\
Fornax 3   &\nodata  &  -2.38$\pm$0.03$^{3}$  &  \cite{letarte} \\

\enddata
\tablerefs{1.  [Fe/H] from \citet[][2010 revision]{harris}.  2. Mean and
   standard deviation of [Fe/H] from the
  references listed in column 4.  3. Because there is only one high
  resolution spectroscopic study
  of NGC 6440 and Fornax 3 we list the dispersion of abundance between stars given
  by \cite{origlia08} and \cite{letarte}, respectively  as the dispersion.}
\end{deluxetable*}

\section{Fe Analysis Tests and Results}
\label{sec:results}

First we summarize the analysis tests that we have performed, and then we discuss the individual solutions on a cluster by cluster basis.  Initial age and [Fe/H] abundance solutions are found using measurements of Fe I lines that are obtained with EW analysis (N104, N2808, N362, N6093, N6397, N6752, Fornax 3) or line synthesis analysis (N6388, N6440, N6441, N6528, N6553), following our standard method described in \citetalias{mb08}, \cite{scottphd},  \citetalias{paper3}, and \citetalias{m31p2}. The EWs measured for each Fe I line are given in Table \ref{tab:ew_stub}, and the abundances measured from synthesis for each Fe I line are given in Table \ref{tab:syn_stub}. The measurements of Fe II lines, which are synthesized in all GCs, are given in Table \ref{tab:feii}.  The final solutions for the 12 GCs are given in Table \ref{tab:fe}. We first discuss GCs measured using EWs, and then GCs measured using line synthesis. 

{\it  1. Stochastic Effects.} After the CMDs with the most consistent age and [Fe/H] combinations were determined in the standard way, we next performed tests with stochastically sampled CMDs as we did in \citetalias{paper3} to determine if stochasticity, which manifests as the presence or absence of short lived stars, was affecting our results, and whether more self-consistent solutions could be obtained with stochastically sampled CMDs.  We find that about half of the GCs have better  solutions with stochastically sampled CMDs, although in most cases the difference in the final mean [Fe/H] is small.  Note that stochasticity has an effect on many of the solutions for the training set GCs primarily because we are only able to sample the highest surface brightness part of the GC cores, and that this is not an issue when observing massive and distant extragalactic GCs.

{\it 2.  Constraints from Fe II lines.}   We have additionally performed tests to determine whether Fe II lines can be helpful in isolating the best stellar populations.  We ignored Fe II lines in our previous analyses because we were only able to measure a few Fe II lines using EWs with comparable precision to the Fe I lines.  At present, we are recovering more Fe II measurements by using line synthesis in all GCs, instead of EWs \citepalias[see ][]{m31p2}, and it is useful to determine if we can use the additional Fe II lines to our advantage. We find that in the majority of cases, adding Fe II lines can marginally improve the final abundance solutions as a function of wavelength, EP and EW.  However, the behavior pattern of the solutions overall remains very similar so that it is not necessarily more helpful in constraining the best CMD solutions.   In a couple of cases, as described in more detail below, we find a difference between the Fe I and Fe II abundances, which are shown in   Figure \ref{fig:fe}, although the measured abundances from Fe II lines are always within the line to line scatter of the larger set of Fe I lines.   The discrepancy could be  due to non-LTE (NLTE) effects, as investigated in detail by  \cite{lind2012}, although in this case we might expect that the Fe I abundances are underestimated compared to the Fe II abundances and perhaps for the problem to be exacerbated at low metallicity.  However, we don't find either of these things to be consistently true in the sample. It is likely that a large part of the discrepancy is due to the difficulty in making clean measurements of  Fe II lines as compared to the Fe I lines.  We find that the worst agreement between Fe I and Fe II abundances is found for GCs where we measure less than 10 Fe II lines, which also tend to be the GCs that have larger velocity dispersions (NGC 6388) and lower S/N spectra (NGC 6093, Fornax 3).  We note that all but 3 of the Fe II lines  have wavelengths $<$5500 \AA, which are the regions of the spectra with the lowest S/N.  We conclude that our standard analysis using Fe I lines alone is the most consistent method across the range of cluster properties and data quality.

{\it 3. Horizontal Branch Morphology Effects.}
 As shown in our previous work, when we derive a young or intermediate age ($<$10 Gyr) for a GC we must also determine whether the solution is also consistent with that for a CMD with  an old age and a very blue horizontal branch (HB), since both cases would require more flux in hot stars for the best overall solution.       
 Significant effort has been put into determining the potential abundance systematics introduced by mismatches between actual horizontal branch morphology and the horizontal branch morphology assumed in the theoretical CMDs.  \cite{sakari,sakari_ers}  found that for abundances of most elements, the upper limits on systematic offsets were between 0.05 - 0.2 dex, and that HB morphology had a much bigger effect on the Fe II abundances than the Fe I abundances.   \citetalias{m31p1}  found that completely replacing the original HB stars in the CMD with very hot blue HB stars  for a metal-poor M31 GC resulted in a difference of 0.05 dex.  \citetalias{m31p2} found that although abundance offsets were generally small, ages could be systematically underestimated when blue HBs were missing in theoretical CMDs for moderately high metallicity clusters ([Fe/H]$>-0.7$), and that for appropriate age uncertainties  it is necessary to test whether any cluster with a derived age of $<$10 Gyr could also be consistent with an age $>$10 Gyr and a blue HB.    In this work we find an  example of this effect  for our solution   for NGC 6441.

Because blue HB stars can have a noticeable effect on the abundances derived from bluer Fe I lines, we perform additional tests here to determine if the Fe lines can be used to infer the  HB morphology themselves.   This would potentially allow the HB morphology - metallicity relationship to be studied in more distant galaxies where it is difficult or impossible to determine the HB morphology photometrically.   
In our previous papers, we presented tests  for  evaluating if GCs with younger ages were also consistent  with older ages and blue  HB morphology by deriving abundances for an ``extreme'' CMD, where we replaced all of the red HB stars with blue HB stars while conserving the total V flux in the horizontal branch.  We then compared the behavior and self consistency of the abundance solutions  with the extreme CMD to see if the abundance diagnostics ($\sigma_{Fe}$, Fe vs. $\lambda$, Fe vs. EP, Fe vs. EW )  were improved over the initial solutions.      
Here we  test additional CMDs between the red and blue extremes, where we replace 25 \%, 50\%, 75\% or 100\% of the red HB stars with blue HB stars.  In this way we can test  for a missing amount of flux in hot stars by  trying to eliminate trends in the abundance diagnostics and minimizing the line-to-line scatter of Fe I abundances.

   We tested each GC with the four versions of blue HB CMDs, to see if the results were consistent with the actual HB morphology of the GCs from their photometric CMDs. Because convenient quantitative measurements of HB morphology aren't available for all the GCs in our sample (e.g. HB indexes based on star counts, median RGB/HB color difference, temperature limits, etc.),  we have only looked for qualitative agreement with the resolved CMDs.  In this sample, we do not find that only GCs with known blue HBs have better solutions when blue HBs were added. In fact, we found that a CMD with a 50\% blue HB improved the diagnostics for both NGC 104, which has red HB, and NGC 6752, which has a blue HB. In the original solutions (i.e. no blue HB added), both NGC 104 and NGC 6752 have small trends of increasing Fe abundance measured from redder Fe lines. By  adding blue HB stars to the CMD, the derived abundance from bluer lines is increased, thus removing the trend in the original solutions. Because NGC 104 doesn't have a blue HB there must be another explanation (be it some systematic error or population mismatch) for the observed trend in the original solution that could be falsely interpreted as a blue HB. Since HB morphology is not a unique solution to eliminating Fe abundance trends with wavelength, we conclude that the Fe I lines themselves can't be used to infer HB morphology in unresolved GCs without additional information.  Therefore in our final analysis strategy we only test for compatibility with an extreme blue HB when an intermediate or young age is derived.  This more accurately reflects the degeneracy present when constraining the age of an unresolved GC, and consequently the abundance uncertainties may be larger for some GCs.

{\it 4. Comparison to reference abundances from individual stars.}   For each GC we  compare the [Fe/H] we derive to [Fe/H] measured by other authors using standard high resolution analysis of individual cluster stars.   The sources for the reference studies are listed in Table \ref{tab:source}.  In column 1 we list the [Fe/H] from the Milky Way Globular Cluster Catalog \citep[][2010 edition, hereafter ``HGCC"]{harris}, which is based on the abundance scale of \cite{carrettascale}.    In column 2 we list the [Fe/H]  that we obtain from an unweighted mean of the individual works listed in column 3, which are the references we use to compare the additional elements we measure in \textsection \ref{sec:others}. We also give the standard deviation of the measurements to give a sense of the dispersion between recent studies.  We have limited these references to works published in approximately the last 15 years, which in most cases use similar techniques to the optical RGB star analysis the IL method is based on.  The exceptions are IR spectra measurements for the  bulge GCs NGC 6440, NGC 6441, and NGC 6528 performed by \cite{origlia05} and \cite{origlia08}, which are included because there is very little abundance information available for these highly reddened GCs. The IR comparison should be sufficient for most of the elements we are comparing for these two GCs (Fe, Mg, Si, Ti, Al, Ca, V), as recent studies have shown reasonable agreement between optical and IR abundances \citep[e.g.][]{2015AJ....150..148H,2015MNRAS.448...42L}. 
 We use the additional comparison to the [Fe/H] in column 2 because it includes abundance studies performed after the publication of the 2010 edition of \citetalias{harris}, but note that the average [Fe/H] in column 2 is usually very similar to the [Fe/H] of \citetalias{harris}.  In the following sections we give the IL abundance measurement with the uncertainty in [Fe/H] from Column 5 in Table \ref{tab:fe}, which is the combined uncertainty from the age of the CMD and  the line-to-line scatter  divided by $\sqrt{N-1}$ lines.

\begin{figure*}
\centering

\includegraphics[angle=90,scale=0.2]{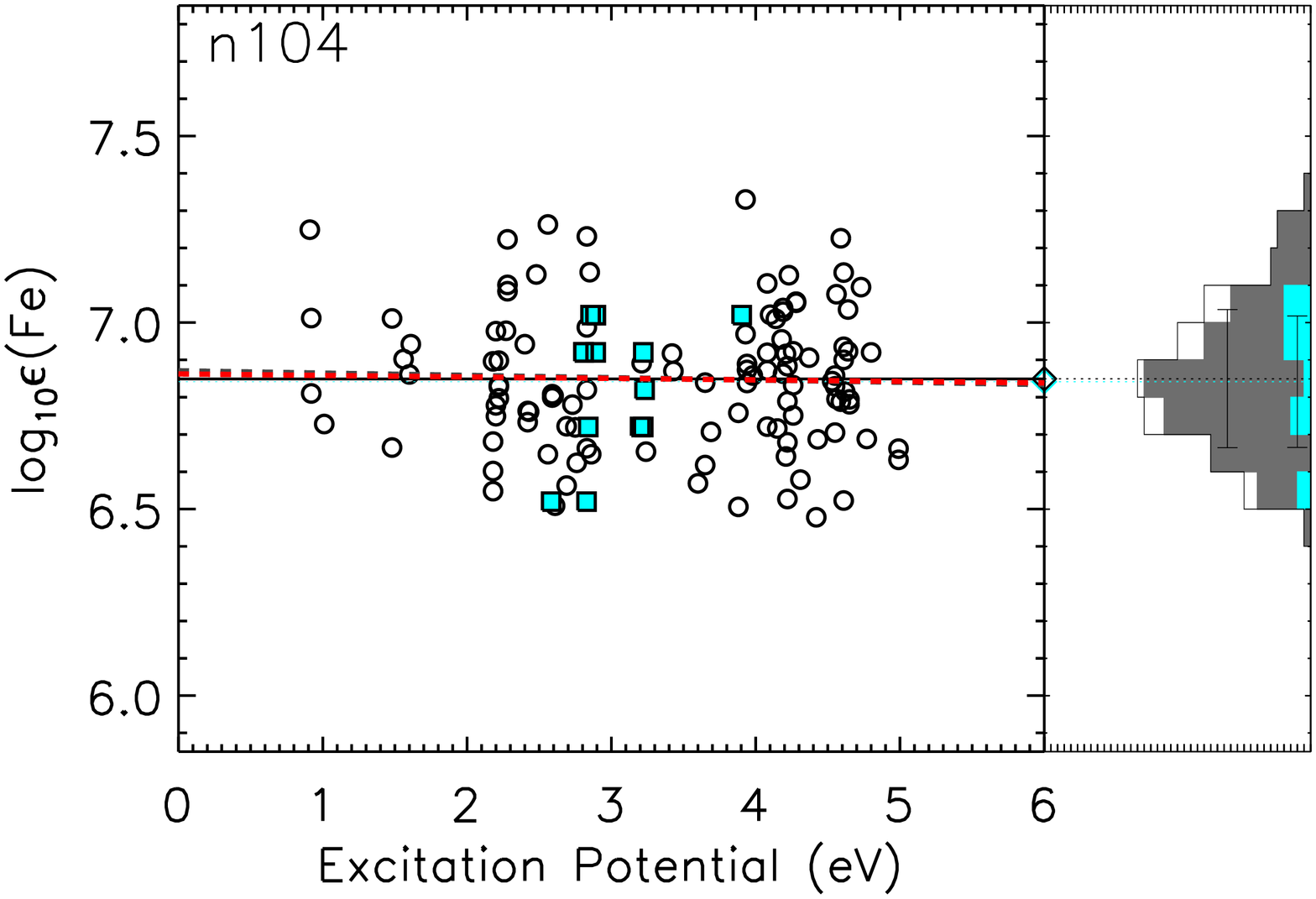}
\includegraphics[angle=90,scale=0.2]{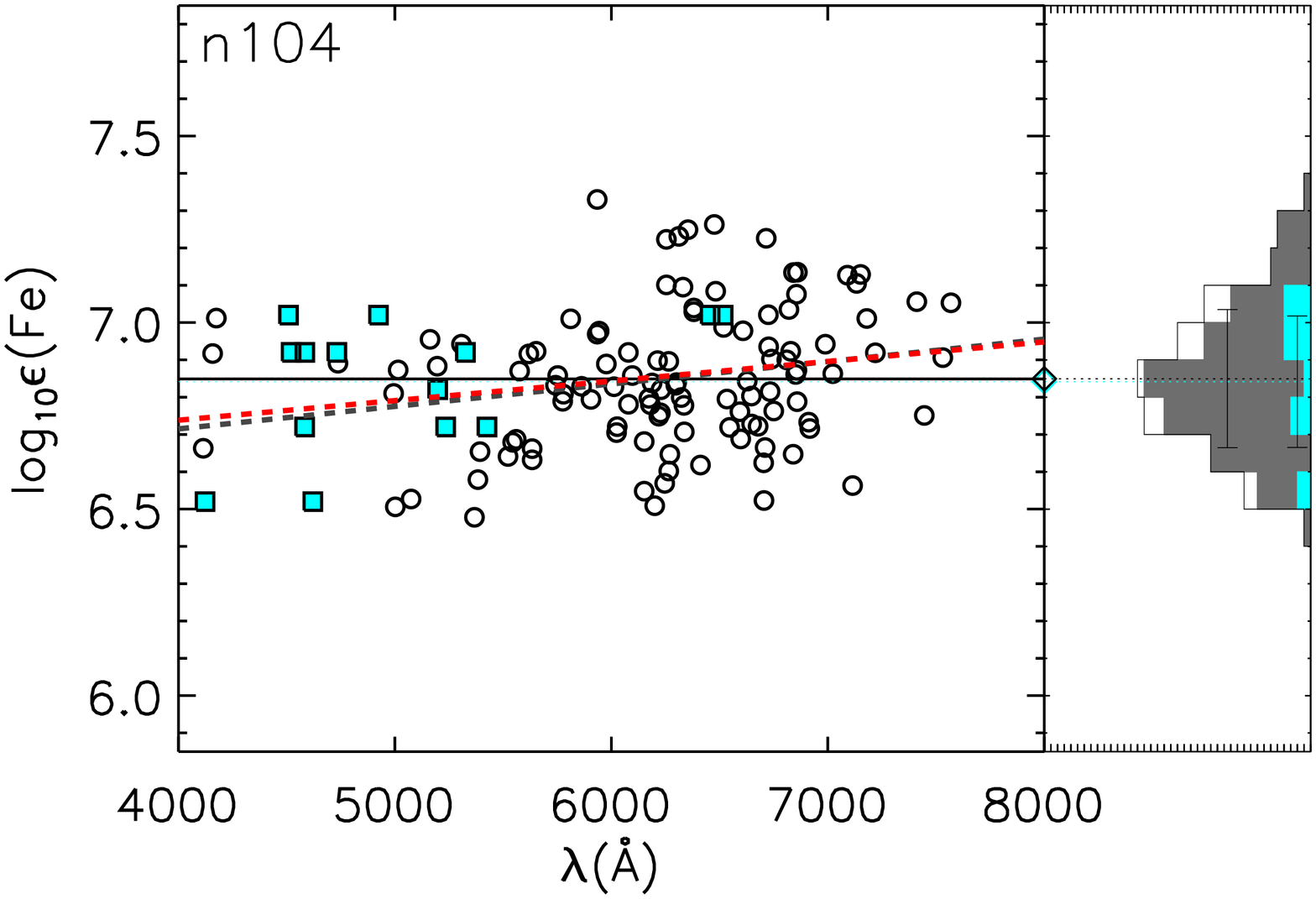}
\includegraphics[angle=90,scale=0.2]{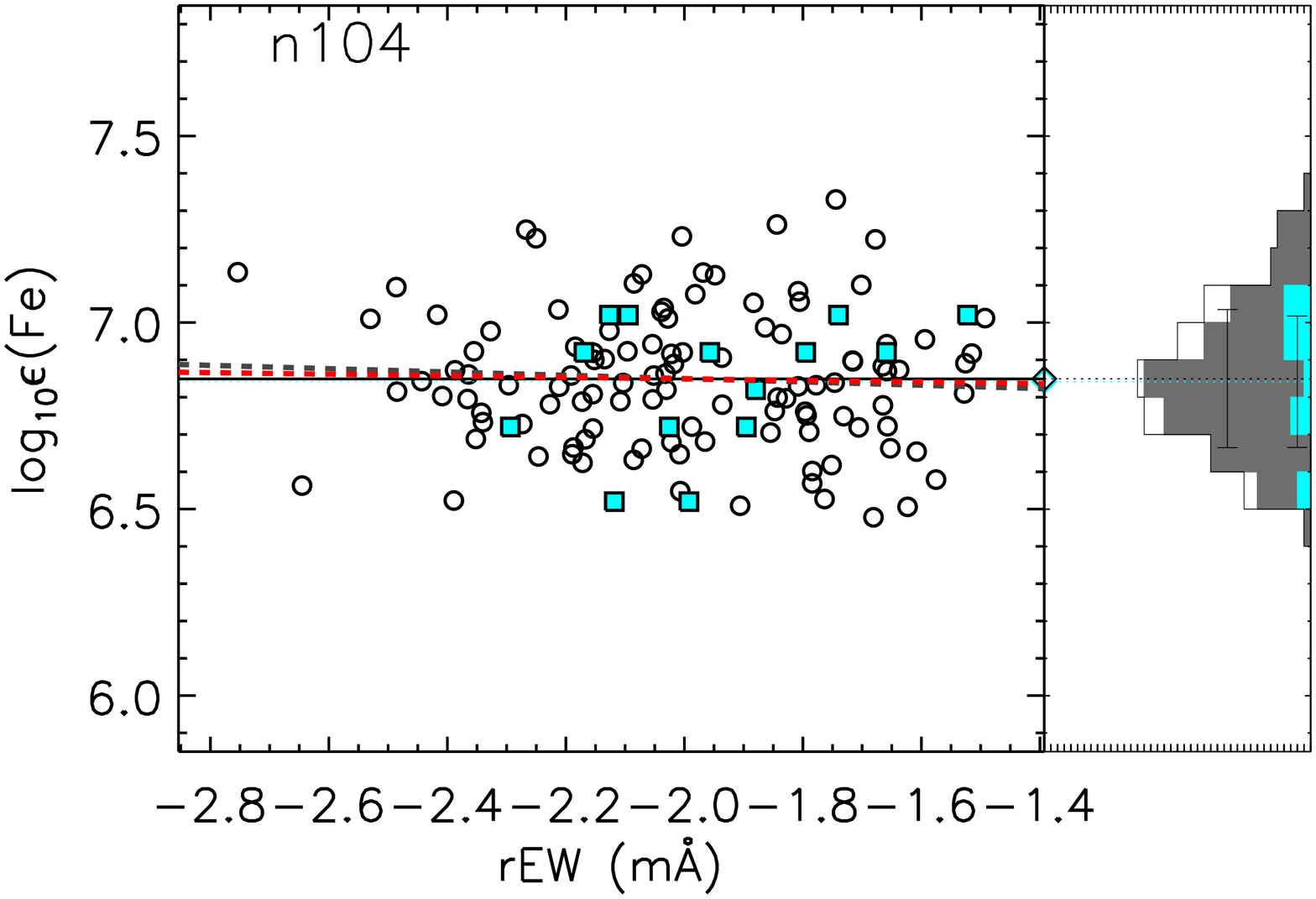}

\includegraphics[angle=90,scale=0.2]{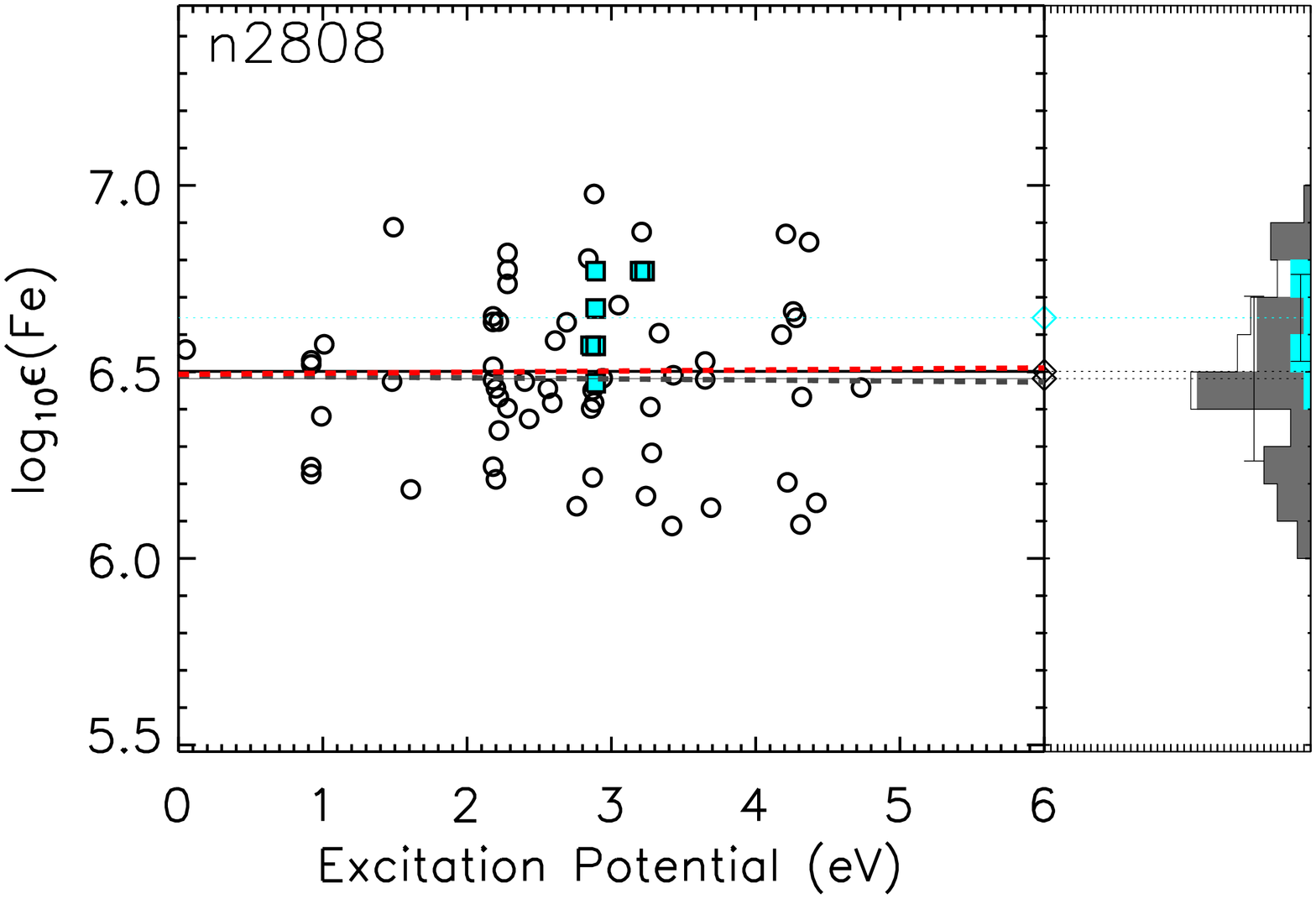}
\includegraphics[angle=90,scale=0.2]{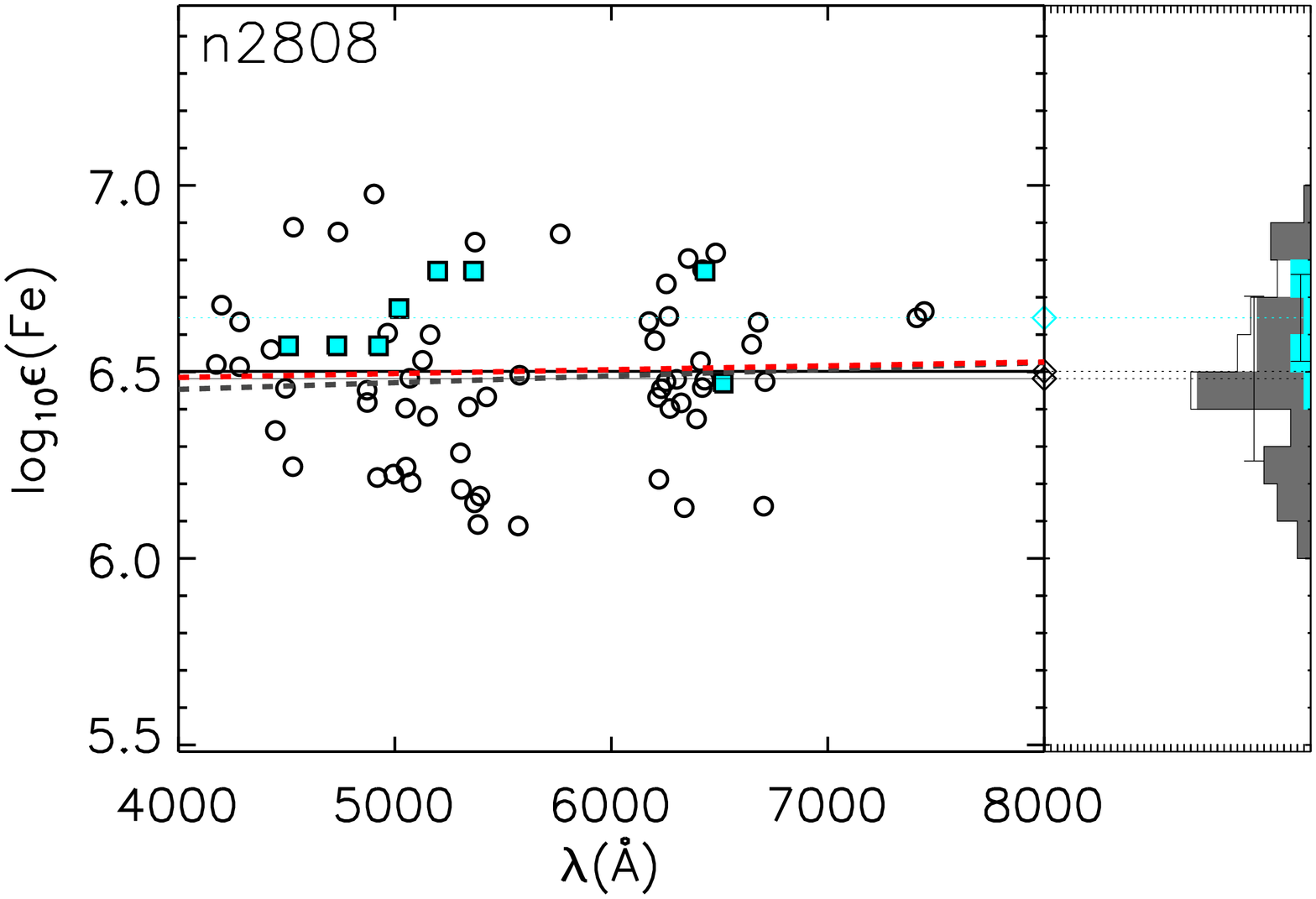}
\includegraphics[angle=90,scale=0.2]{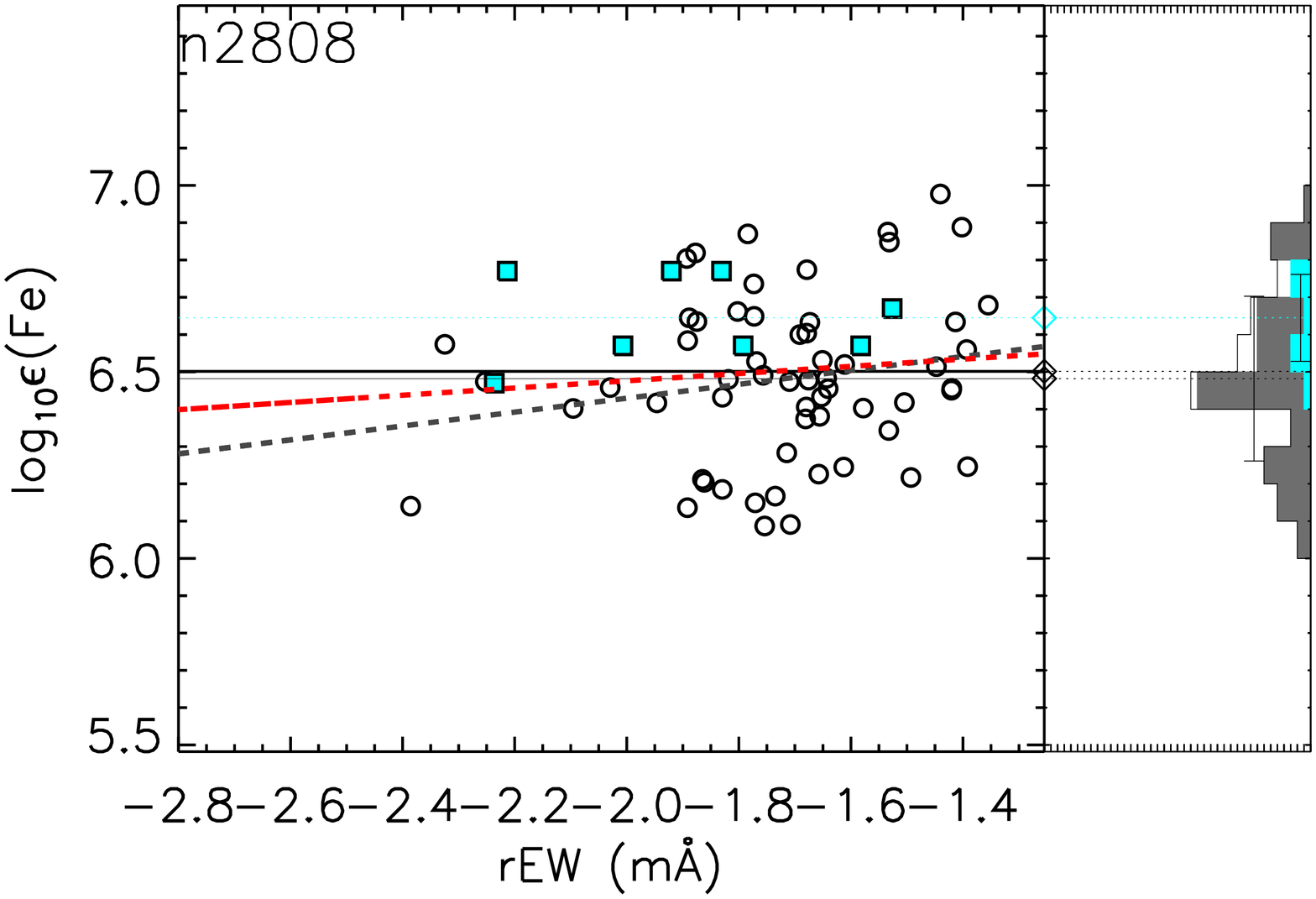}

\includegraphics[angle=90,scale=0.2]{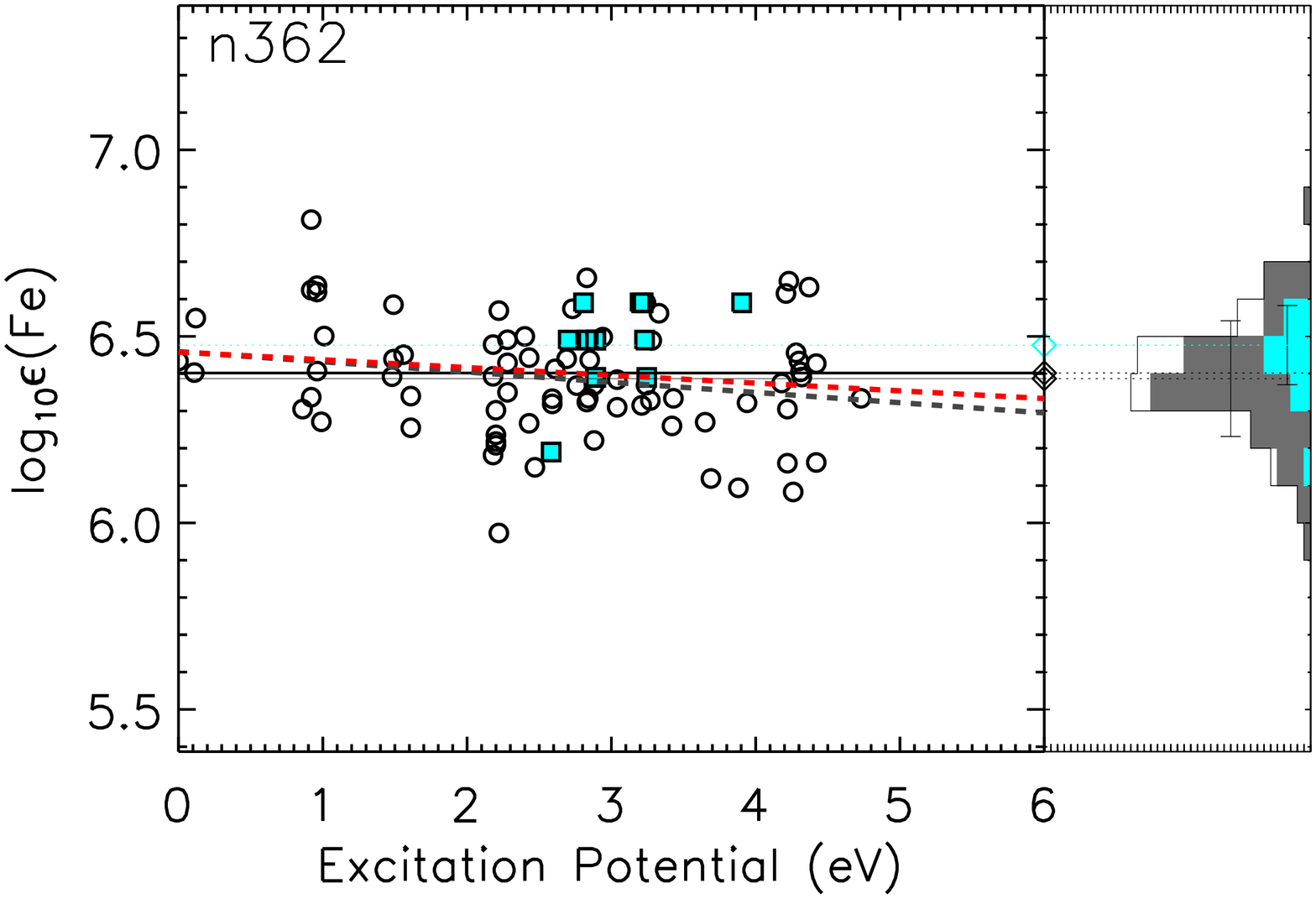}
\includegraphics[angle=90,scale=0.2]{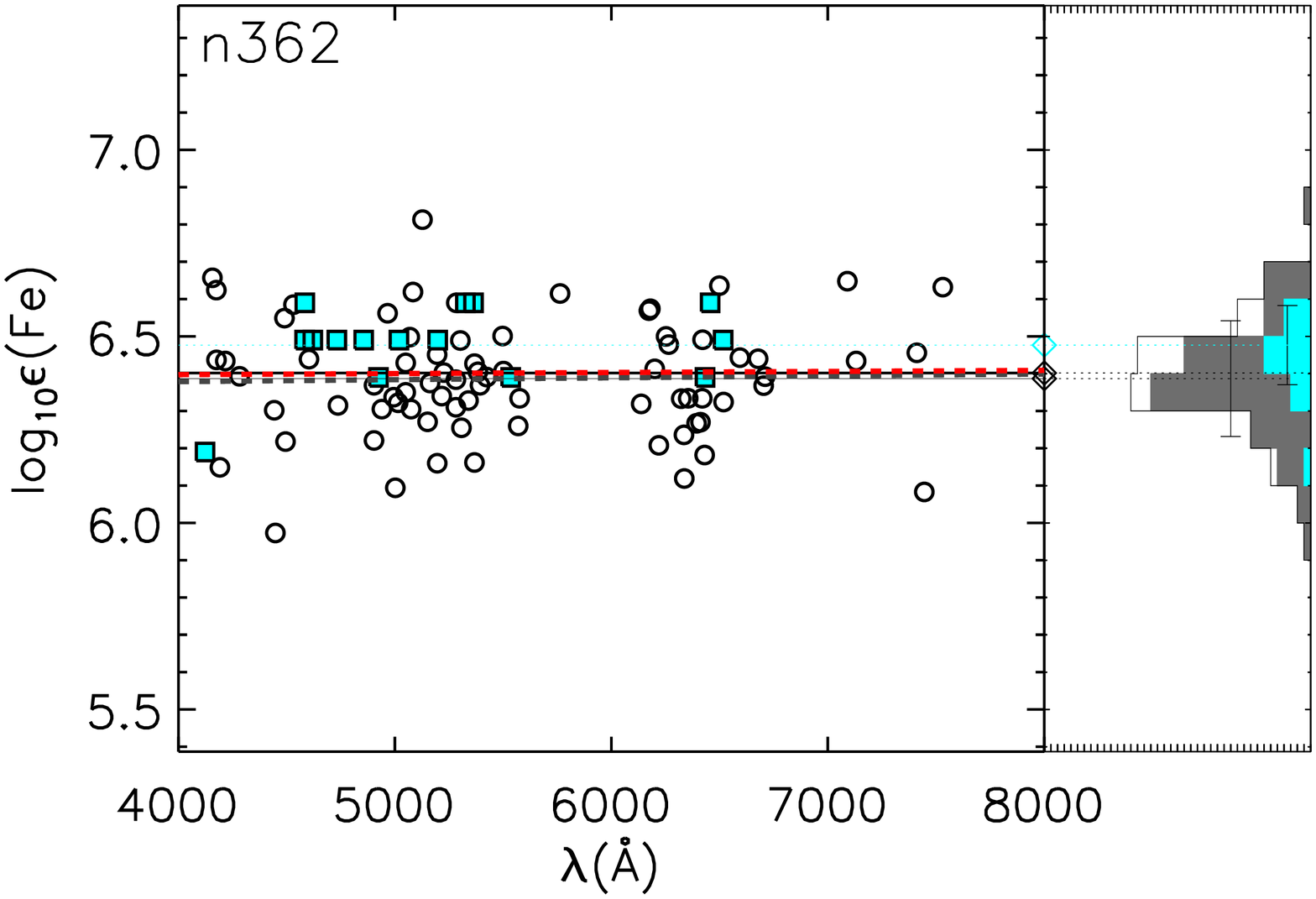}
\includegraphics[angle=90,scale=0.2]{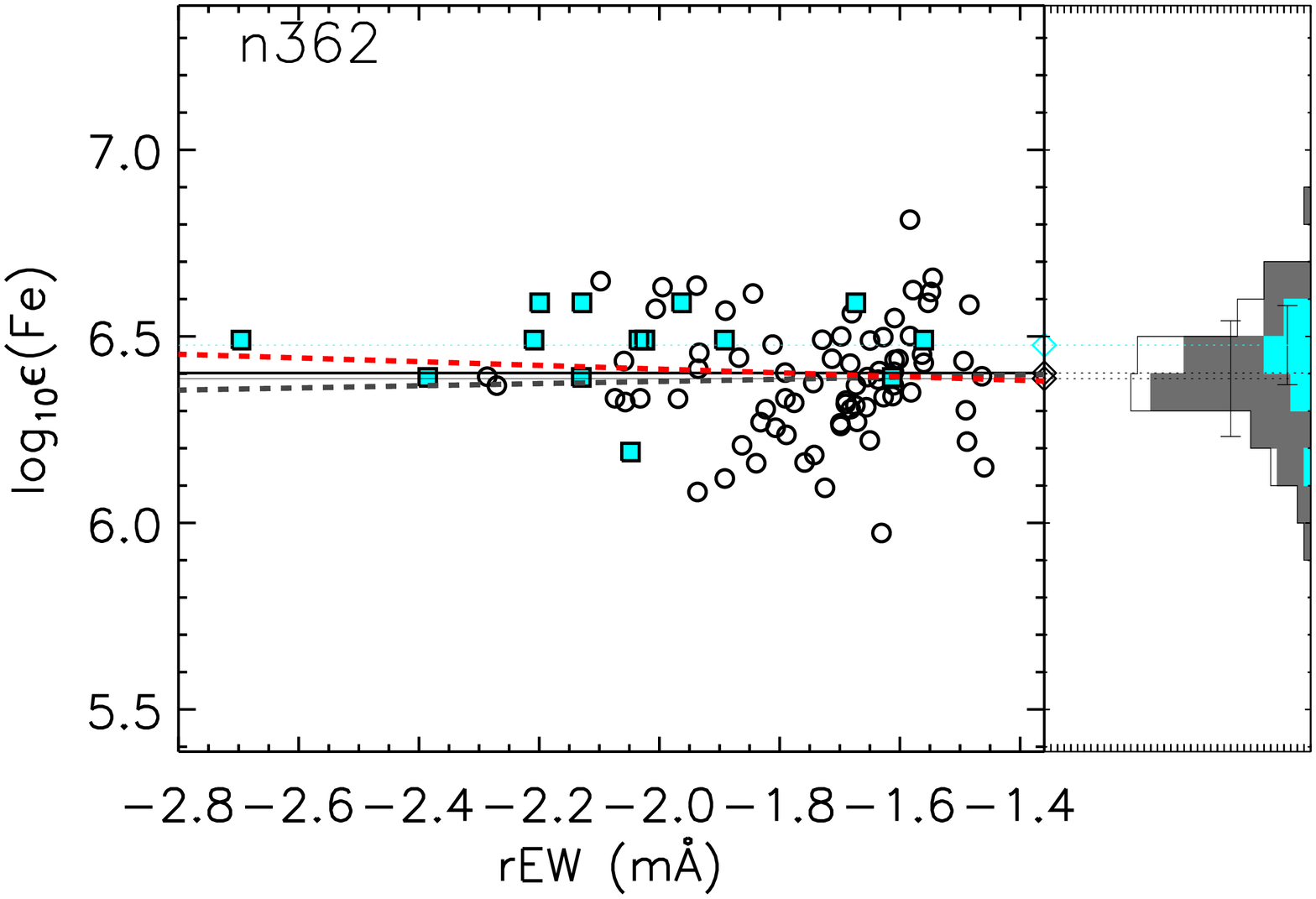}

\includegraphics[angle=90,scale=0.2]{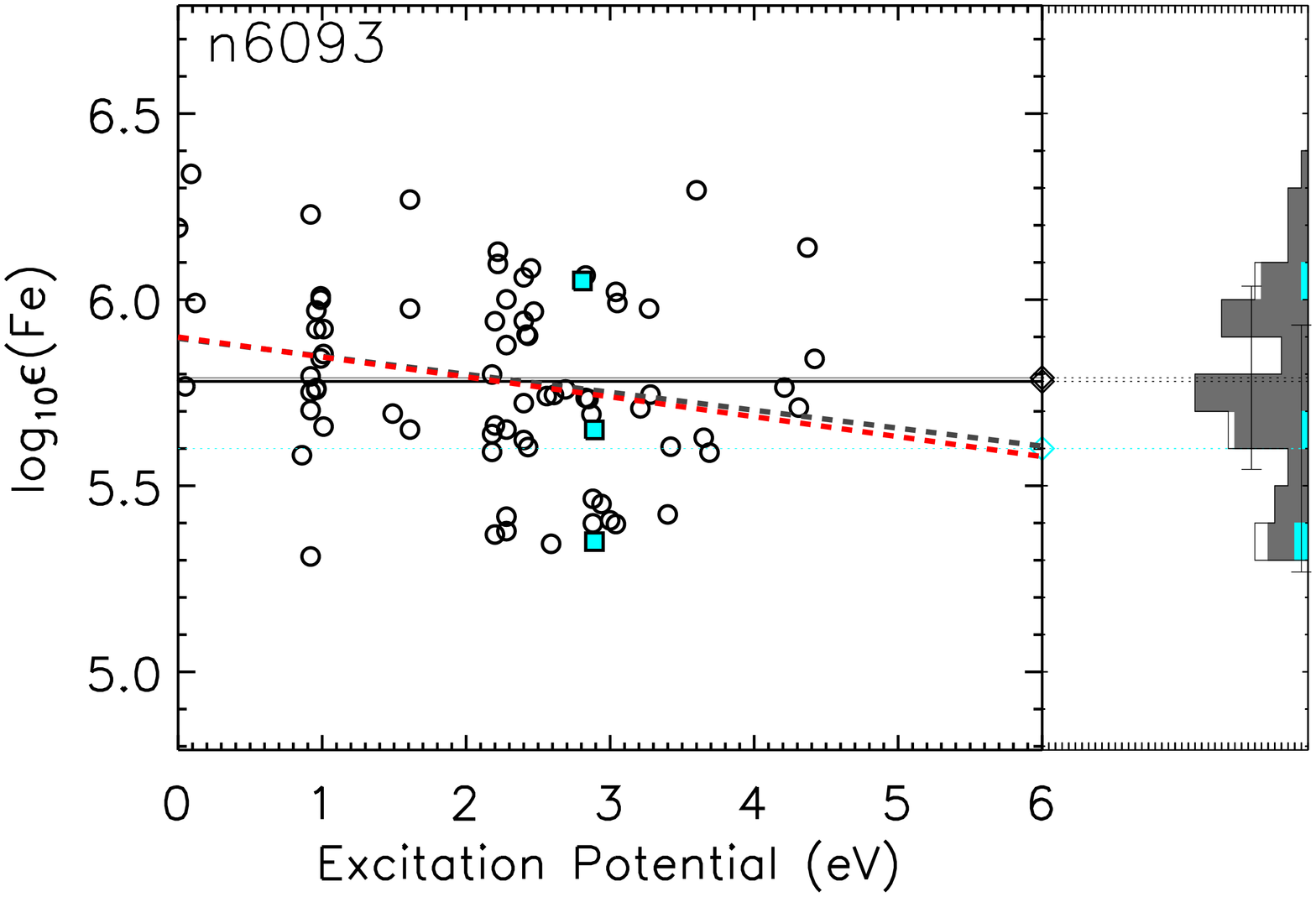}
\includegraphics[angle=90,scale=0.2]{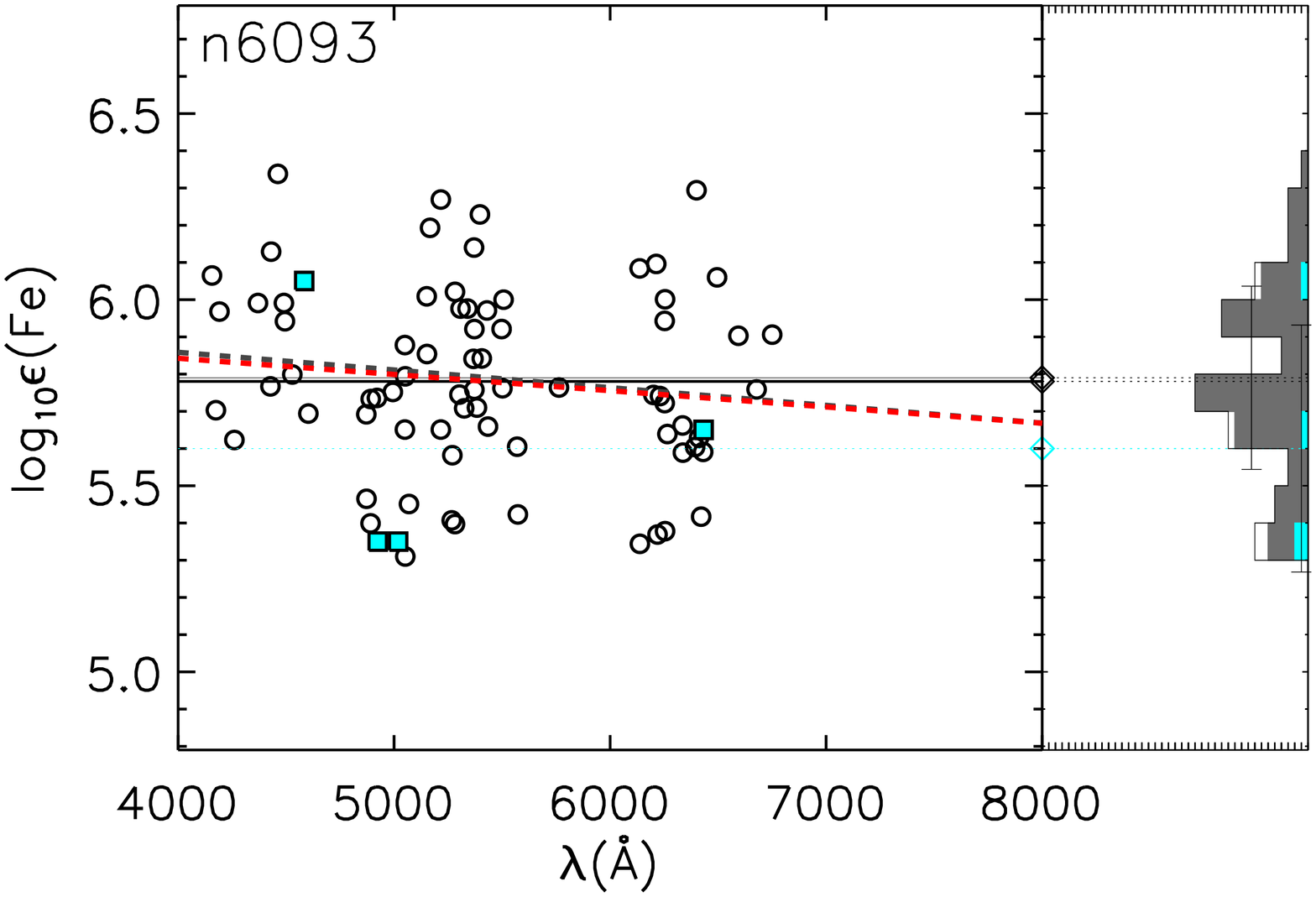}
\includegraphics[angle=90,scale=0.2]{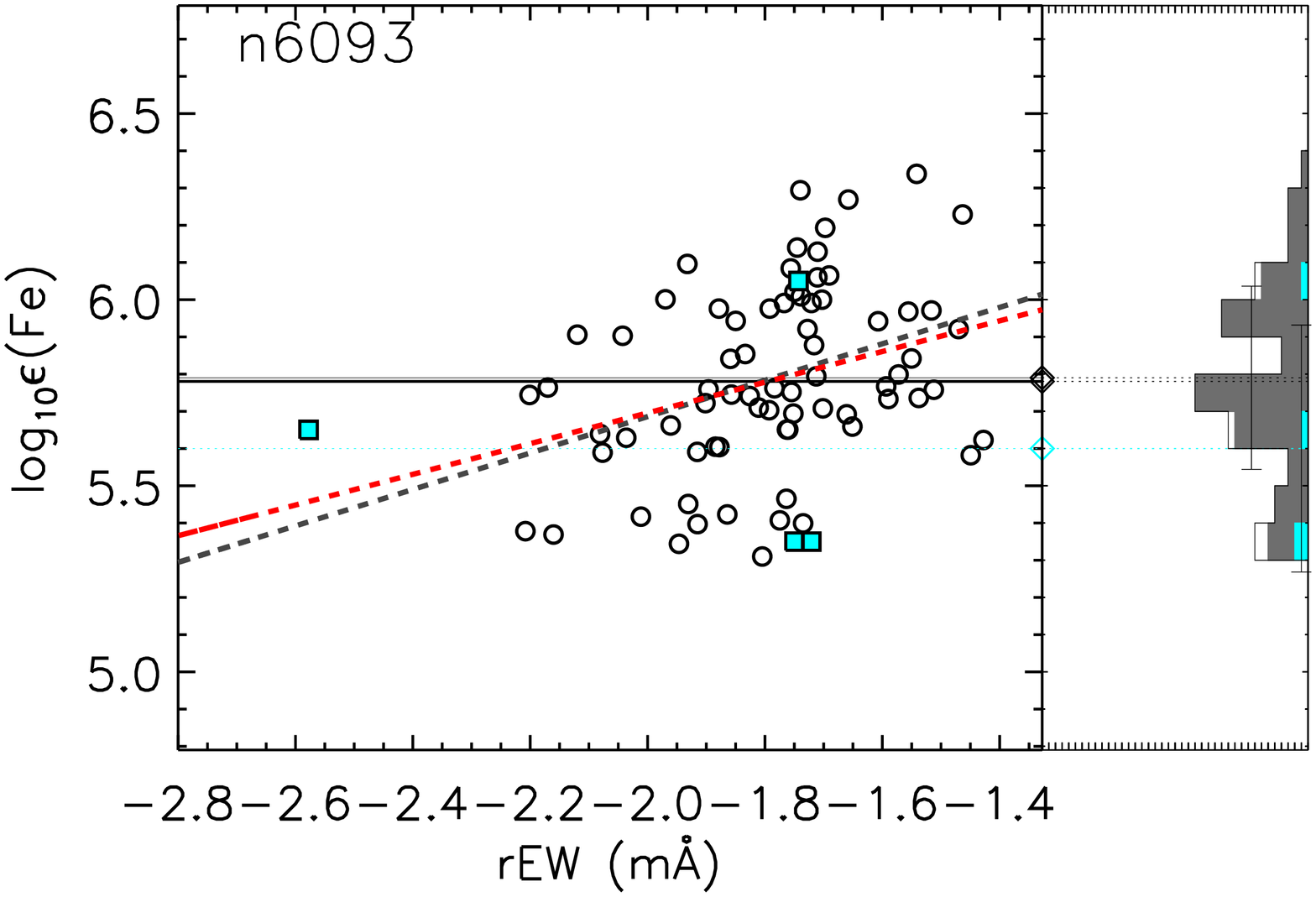}

\caption{   Fe line abundances as a function of  EP (left panels), wavelength (center panels), and reduced EW (rEW, right panels), where reduced EW is defined as  rEW=log$_{10}$(EW/$\lambda$).  From top to bottom the clusters shown are NGC 104, NGC 2808, NGC 362, and NGC 6093.  For each panel open circles show abundances measured for Fe I lines, and filled cyan squares show abundances measured for Fe II lines.  The solid horizontal black line shows the mean abundance from Fe I lines, and the dotted horizontal cyan line shows the mean abundance from Fe II lines.  The black dashed line shows a linear least squares fit to the Fe I abundances as a function of EP, wavelength or rEW.  The dashed red line shows a linear least squares fit of all Fe I and Fe II lines with EP, wavelength or rEW.   On the right side of each panel are histograms of the Fe I abundances in gray, Fe II abundances in cyan, and all lines together in white.    }
\label{fig:sol1} 
\end{figure*}

\subsection{NGC 104}

For the analysis of NGC 104, we use the Fe I EW measurements in \citetalias{mb08}, and supplement them with additional measurements of bluer lines.  The bluest line in \citetalias{mb08} is found at 5862 \AA, and here we  add an additional 57 Fe I lines with   4000 \AA $< \lambda <$ 5860 \rAA so that our analysis of NGC 104 covers a similar wavelength range to the other GCs in this work.  In addition, we measured 14 Fe II lines with spectral synthesis. The solutions with our standard CMDs are very self-consistent, and we obtain a best fit age of 7 - 15 Gyr.  The solution for an age of 10 Gyr is shown in Figure \ref{fig:sol1}; it shows no dependence of abundance with EP or EW, and a small dependence of abundance with wavelength. We obtain identical abundances using Fe I and Fe II lines, and including the Fe II lines in the solution makes a marginal improvement.  We did not find that stochastically sampled CMDs improve the solution in this case.   Our final measurement is [Fe/H]$=-0.65 \pm 0.05$, which is in good agreement with the IL measurement obtained using theoretical CMDs from \citetalias{mb08} of [Fe/H]$=-0.70 \pm 0.021 \pm 0.05$.   Our measurement is consistent, within the uncertainties, to the recent abundance studies in Table \ref{tab:source}, which range from [Fe/H]$=-0.67$ in \cite{carretta47tuc} to [Fe/H]$=-0.79$ in \cite{cordero2014}.  A more metal-poor IL abundance of [Fe/H]$=-0.81 \pm 0.02$ was  measured by \cite{sakari} using the same data and similar techniques with a resolved CMD instead of a theoretical CMD.  The two measurements are still reasonably close given the possible systematic errors in IL abundance analysis discussed in depth in \citetalias{mb08} and  \cite{sakari_ers}.  For example, \citetalias{mb08} find that the IL abundances derived with a photometric CMD instead of theoretical CMD  resulted in a change of $-0.05$ dex in [Fe/H], and \cite{sakari} attribute the 0.05 dex difference in [Fe/H] from photometric CMDs between their analysis and \citetalias{mb08} to differences in how the abundance ratio was calculated.   

\subsection{NGC 2808}
\begin{figure}
\centering
\includegraphics[angle=90,scale=0.35]{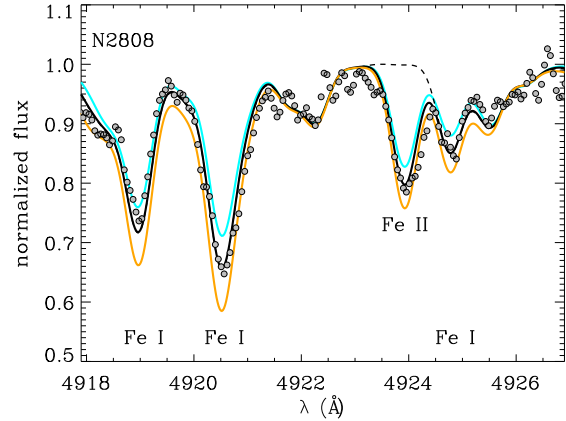}

\caption{  An example of Fe I and Fe II lines in the bluer region of the spectra for NGC 2808, where a spectrum synthesized with the cluster's  mean abundance consistently fits the features.  The data is shown in grey circles and has been smoothed by 3 pixels. The solid black line corresponds to synthesized spectra with the mean [Fe/H] abundance derived from all lines.  The cyan and orange lines correspond to synthesized spectra with abundances $-0.4$ dex and $+0.4$ dex from the mean abundance, respectively. As a consistency check, with the dashed black line we show the region synthesized without the Fe II line  to check for the influence of underlying blends.     }
\label{fig:2808feii} 
\end{figure}

We measure abundances for 59 Fe I lines and  8 Fe II lines for NGC 2808, with final abundances of [Fe I/H]$=-1.04\pm0.04$ and [Fe II/H]$=-0.85\pm0.04$, as well as an age constraint  of 10-13 Gyr.  The difference between the mean abundances from Fe I and Fe II lines is larger than in most cases, although  the Fe II abundance remains consistent with the line-to-line scatter of the Fe I abundances. There is no dependence of abundance with EP or wavelength, but some dependence on abundance vs. EW, which is alleviated somewhat by including abundances of Fe II lines.  We did not find better solutions with stochastically sampled CMDs. We show an example of the NGC 2808 data and synthesized spectra in Figure \ref{fig:2808feii}, where a region with both Fe I and Fe II lines can be seen. As a test we show the region synthesized without the Fe II line, and find this to be consistent with the pseudo-continuum level, which we require for a clean measurement.  Our Fe I measurement is $\sim$0.1 dex more metal rich than the reference abundances in Table \ref{tab:source}, which is a little larger than the formal 1-sigma errors. 

\subsection{NGC 362}
Our final abundance measurements for NGC 362 are [Fe I/H]$=-1.14\pm0.04$ and [Fe II/H]$=-1.04\pm0.03$, obtained from 75 Fe I and 15 Fe II lines, respectively.  We find that CMDs with ages between 13-15 Gyr give the most self consistent solutions, with only small dependences of abundance with EP and EW that marginally improve when including Fe II lines.  We did not find a significant improvement in the solution when using stochastically sampled CMDs.  Our IL measurement is consistent within the uncertainties with the reference abundances in Table \ref{tab:source}, which range from [Fe/H]$=-1.33$ in \cite{shetrone362} to [Fe/H]$=-1.17$ in \cite{carretta362}.

\subsection{NGC 6093}
\begin{figure}
\centering

\includegraphics[angle=90,scale=0.35]{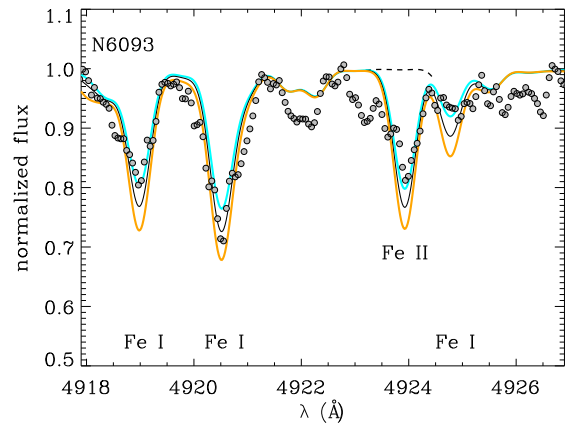}

\caption{  An example of the line-to-line scatter in Fe I and Fe II abundances in the lower quality data for NGC 6093.  The colors and symbols are the same as in Figure \ref{fig:2808feii}.   }
\label{fig:6093feii} 
\end{figure}

For NGC 6093 we measure final abundances of  [Fe I/H]$=-1.65\pm0.09$ from 73 Fe I lines, [Fe II/H]=$-1.88\pm0.19$ from  4 Fe II lines, and find an age of 10-15 Gyr.  The line-to-line scatter for this cluster is larger than in the previous cases, and there are moderate trends in the abundance diagnostics in Figure \ref{fig:sol1}. The worse quality of these solutions are not completely unexpected given that this GC has the lowest S/N in the  sample.  NGC 6093 also has one of the larger differences between the abundance derived from Fe I and that from Fe II, however we are only able to measure 4 Fe II lines with a significant line-to-line scatter. In Figure \ref{fig:6093feii} we show an example of the lower quality data and line-to-line scatter in abundance derived from several nearby lines.  The abundance derived from the Fe II line in Figure \ref{fig:6093feii} is one of the lowest we measure for this cluster.  As we did for NGC 2808, we show the region synthesized without the Fe II line and again find this to be consistent with the pseudo-continuum level.   For this cluster we found that stochastically sampled CMDs provided some improvement in the solution, although significant trends in the abundance diagnostics remain.  Our IL abundance measurements are consistent within the uncertainties with the reference abundances in Table \ref{tab:source}.

\subsection{NGC 6397}

\begin{figure*}
\centering

\includegraphics[angle=90,scale=0.2]{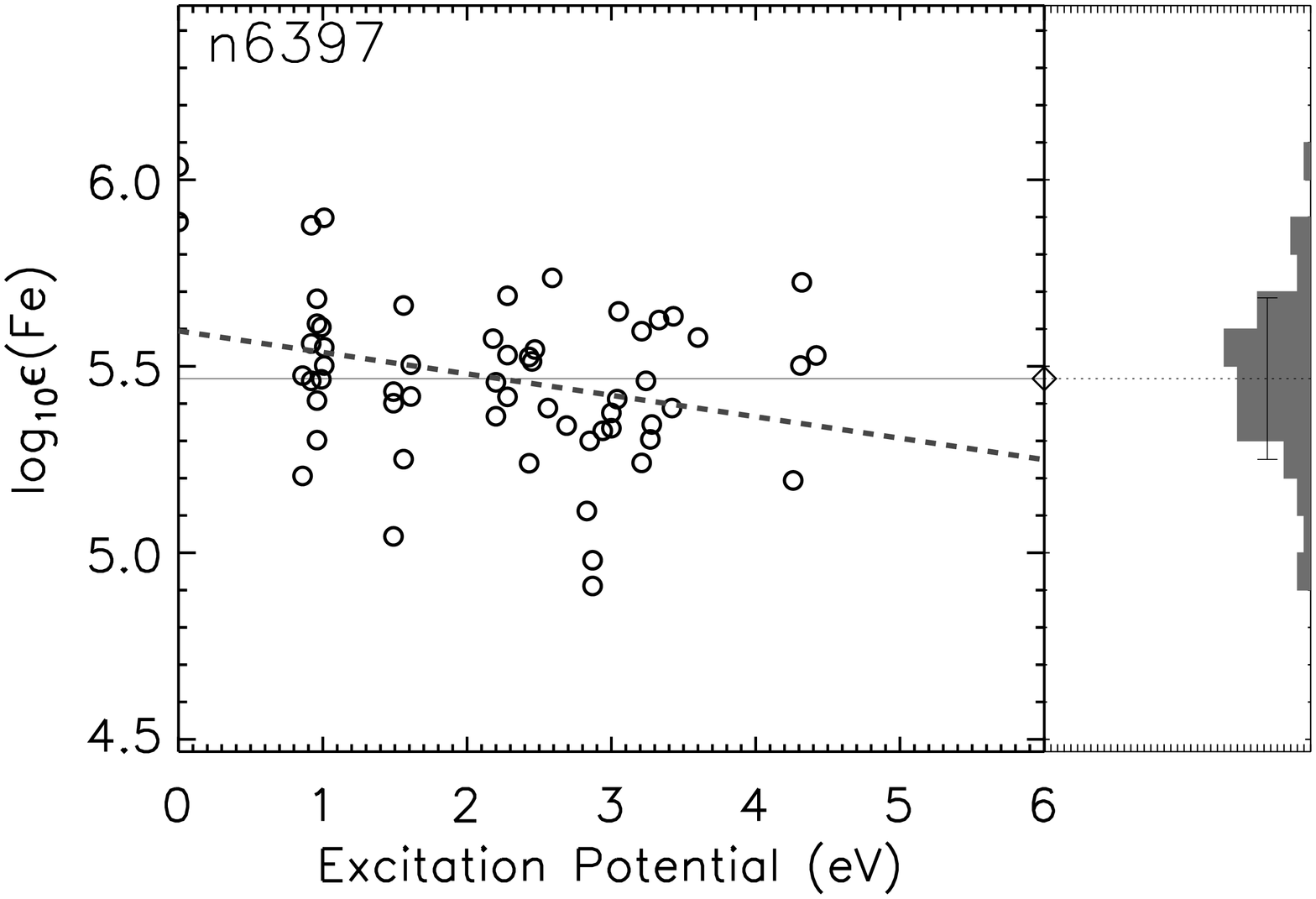}
\includegraphics[angle=90,scale=0.2]{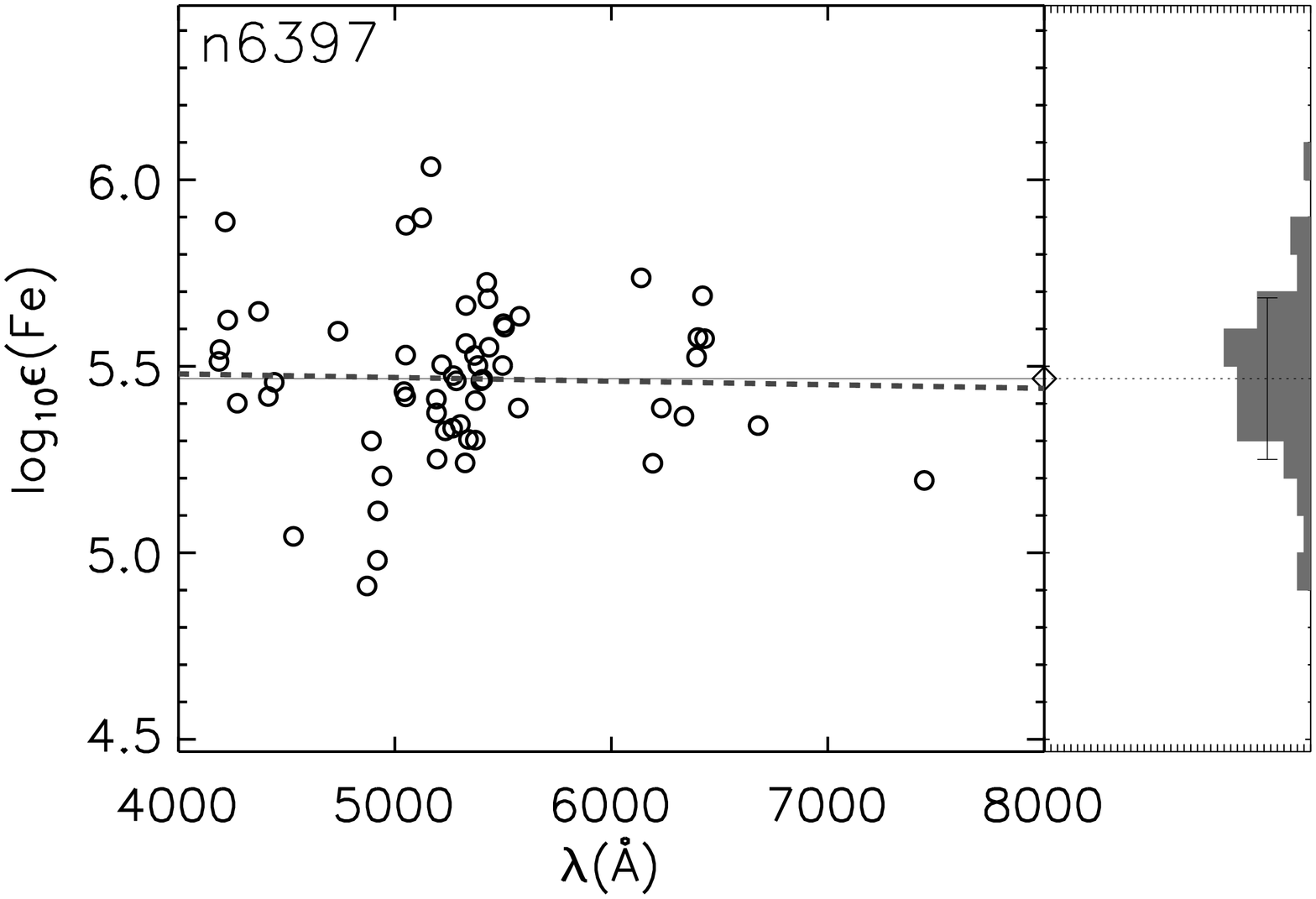}
\includegraphics[angle=90,scale=0.2]{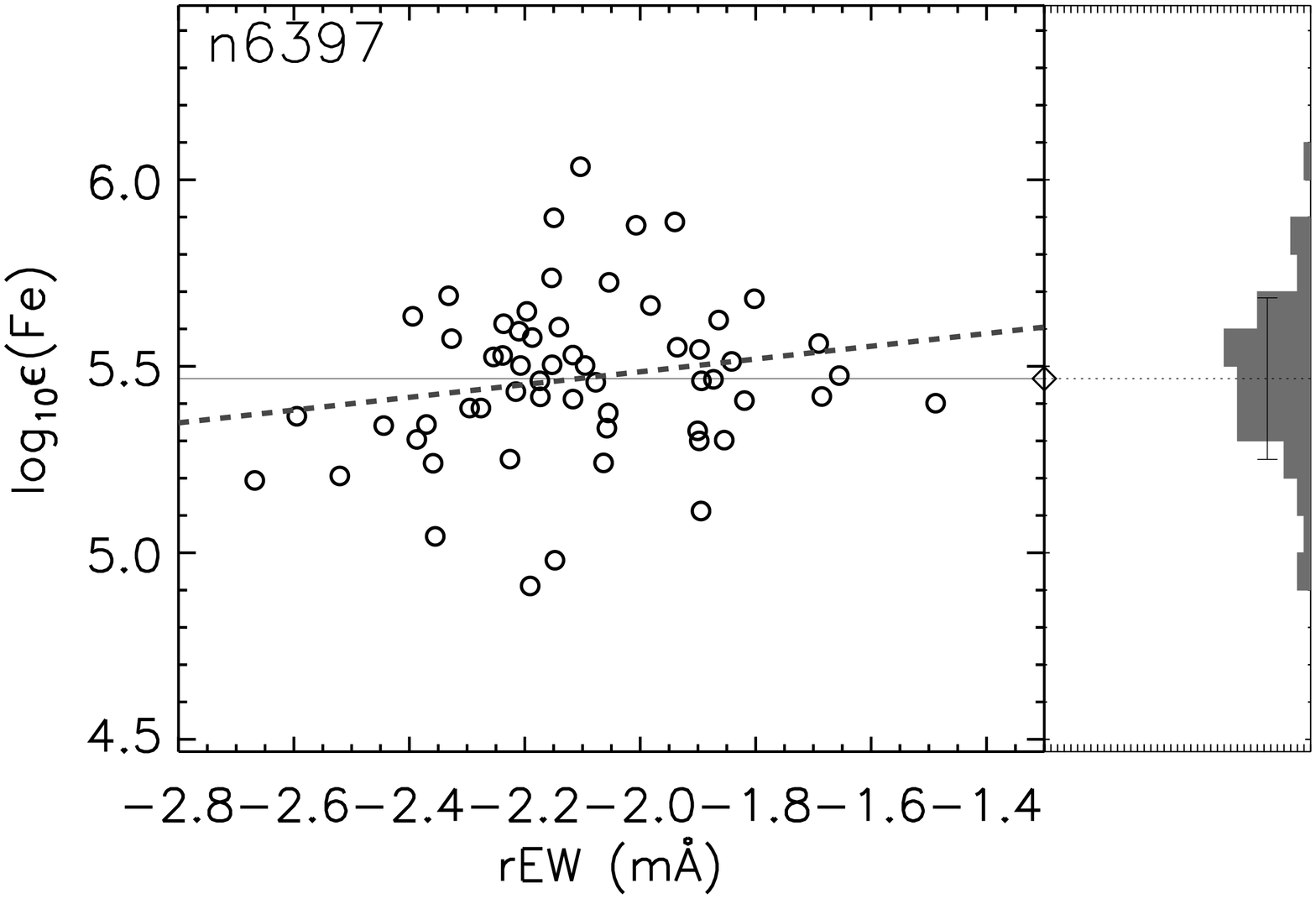}

\includegraphics[angle=90,scale=0.2]{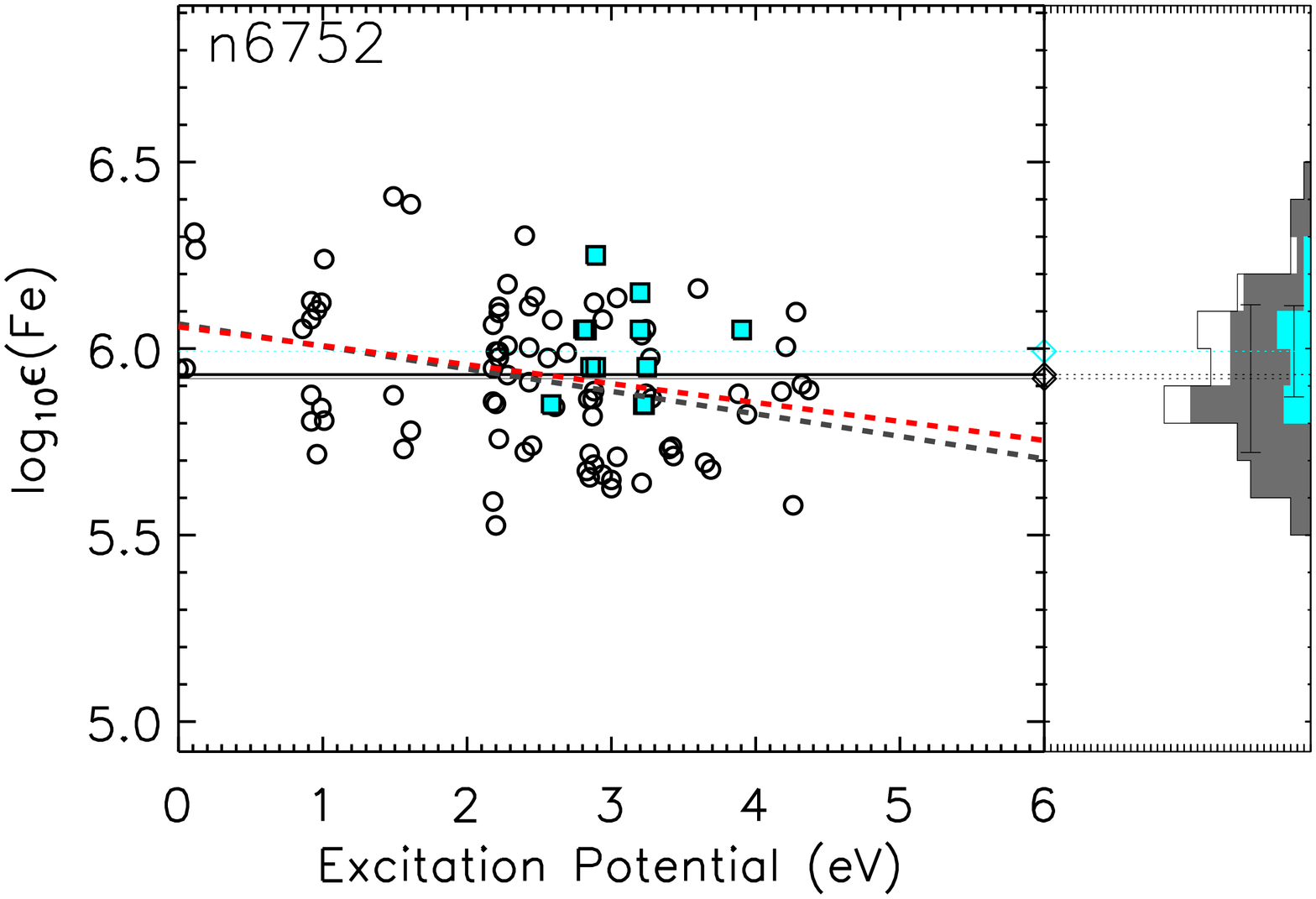}
\includegraphics[angle=90,scale=0.2]{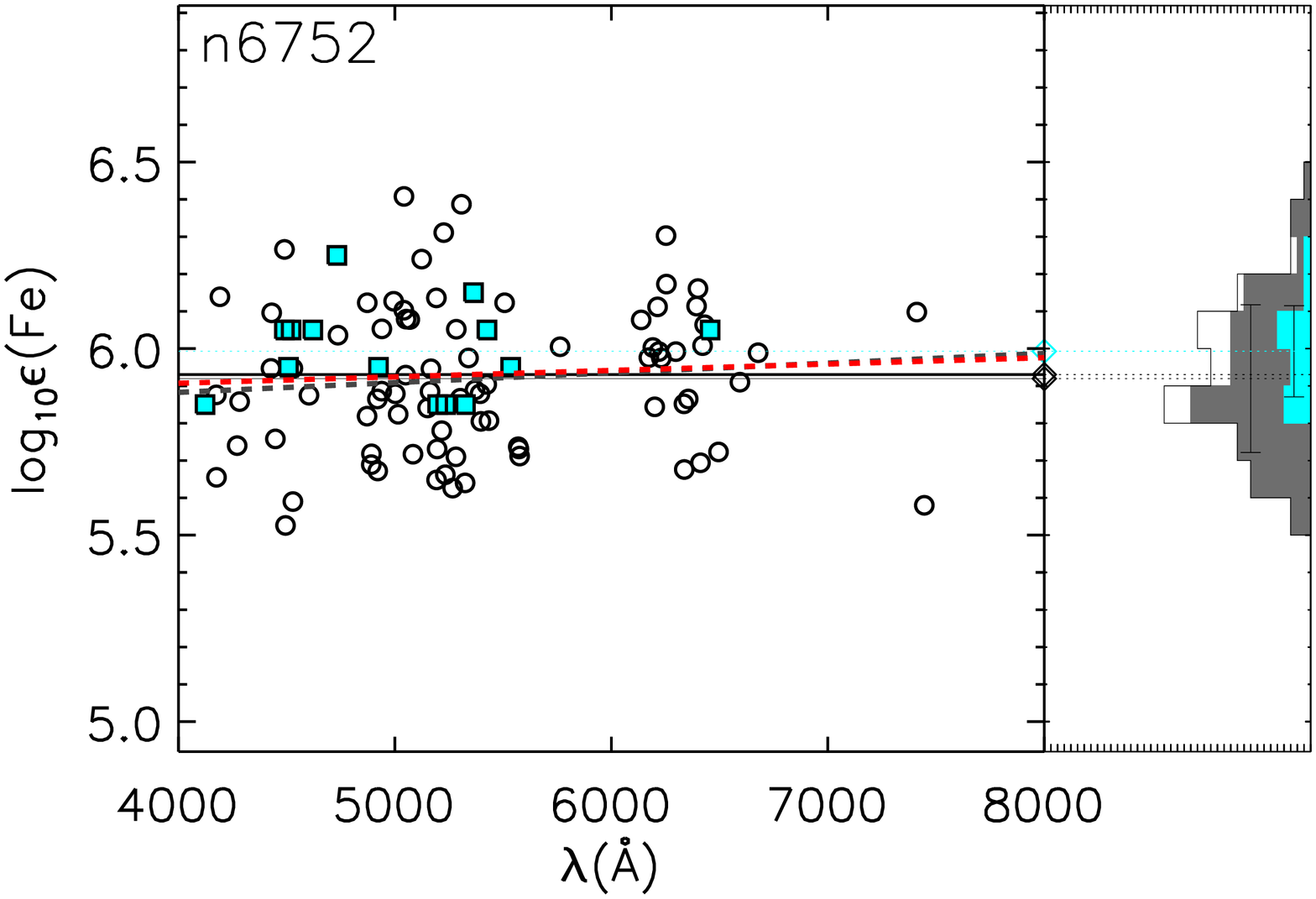}
\includegraphics[angle=90,scale=0.2]{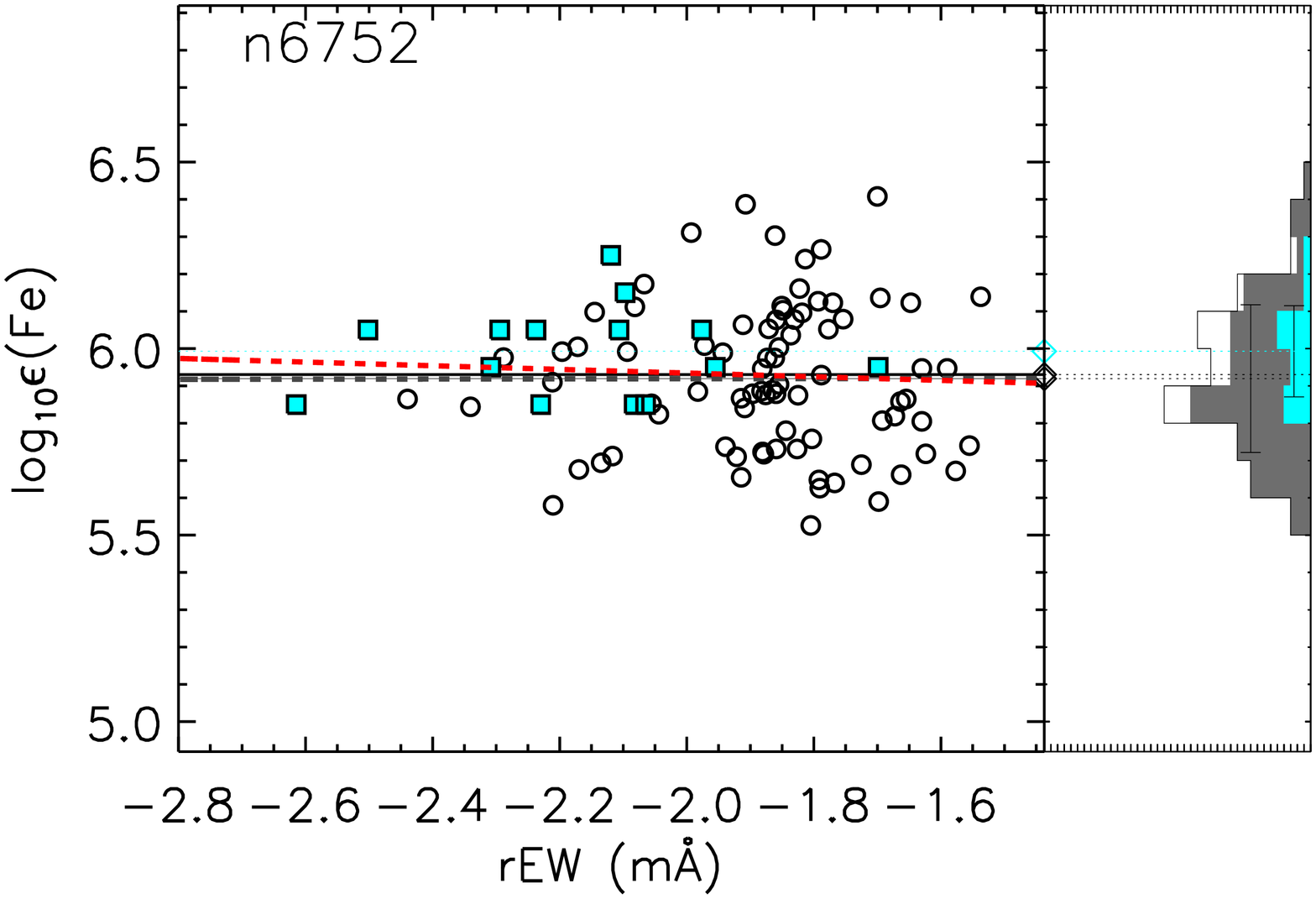}

\includegraphics[angle=90,scale=0.2]{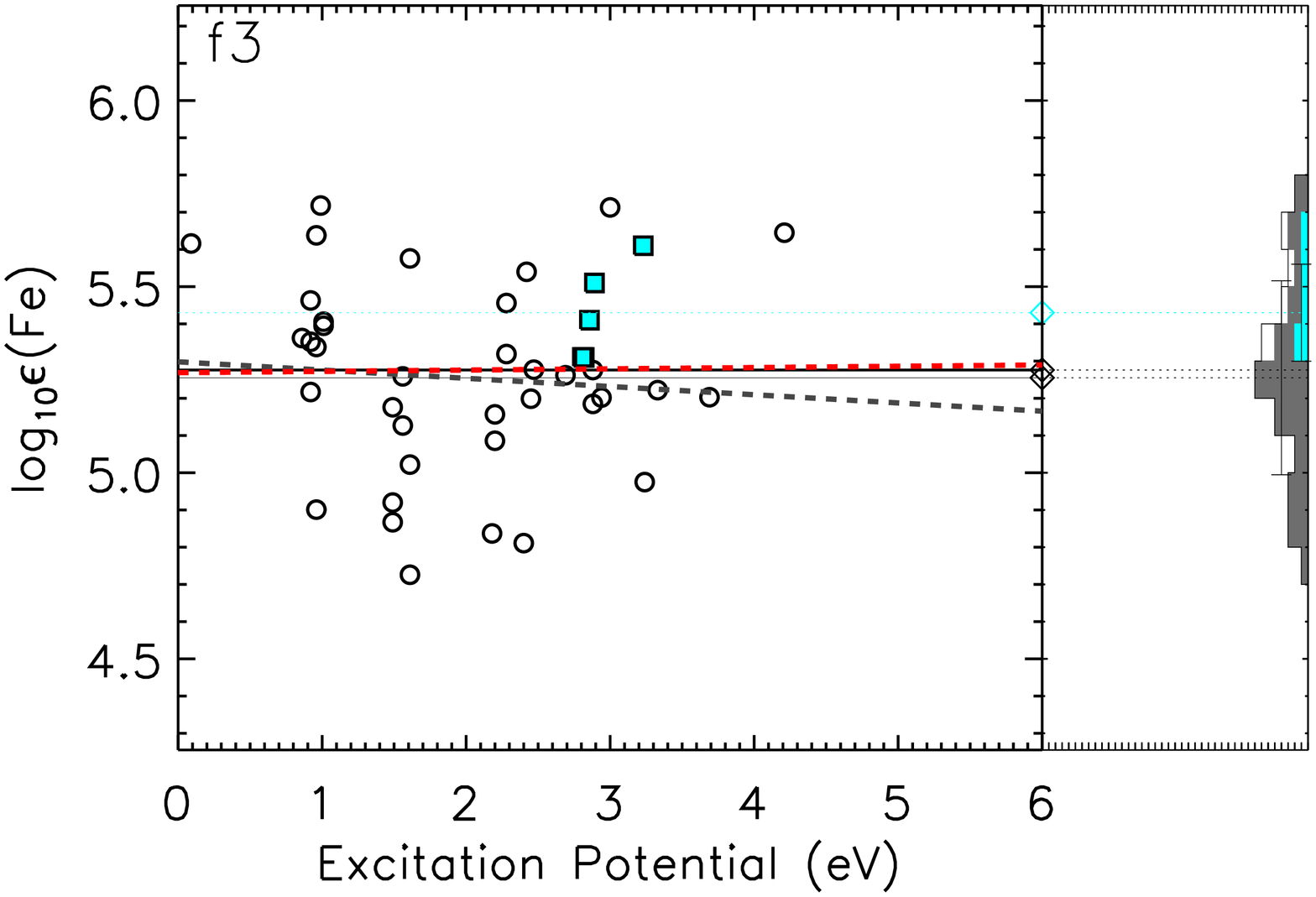}
\includegraphics[angle=90,scale=0.2]{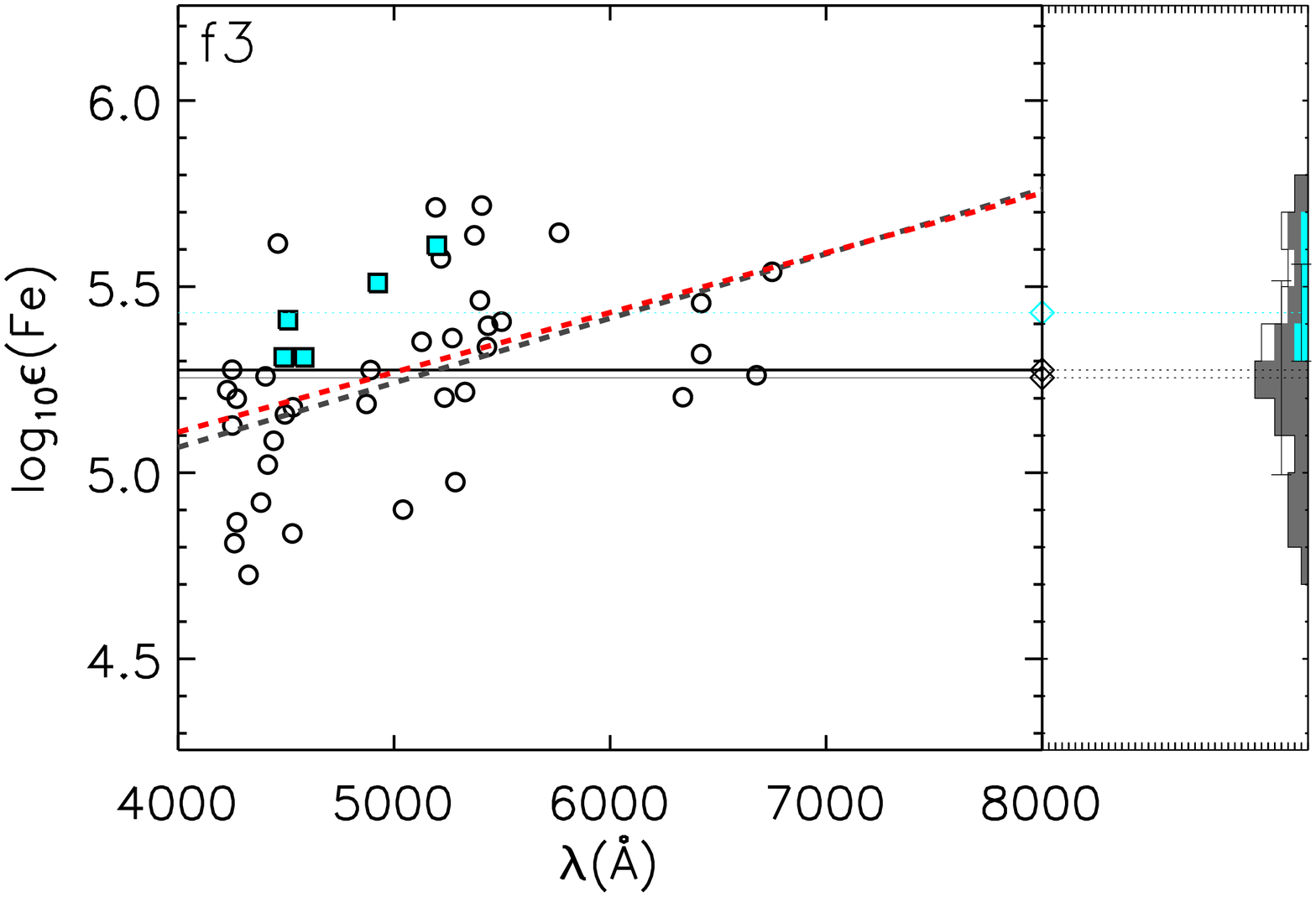}
\includegraphics[angle=90,scale=0.2]{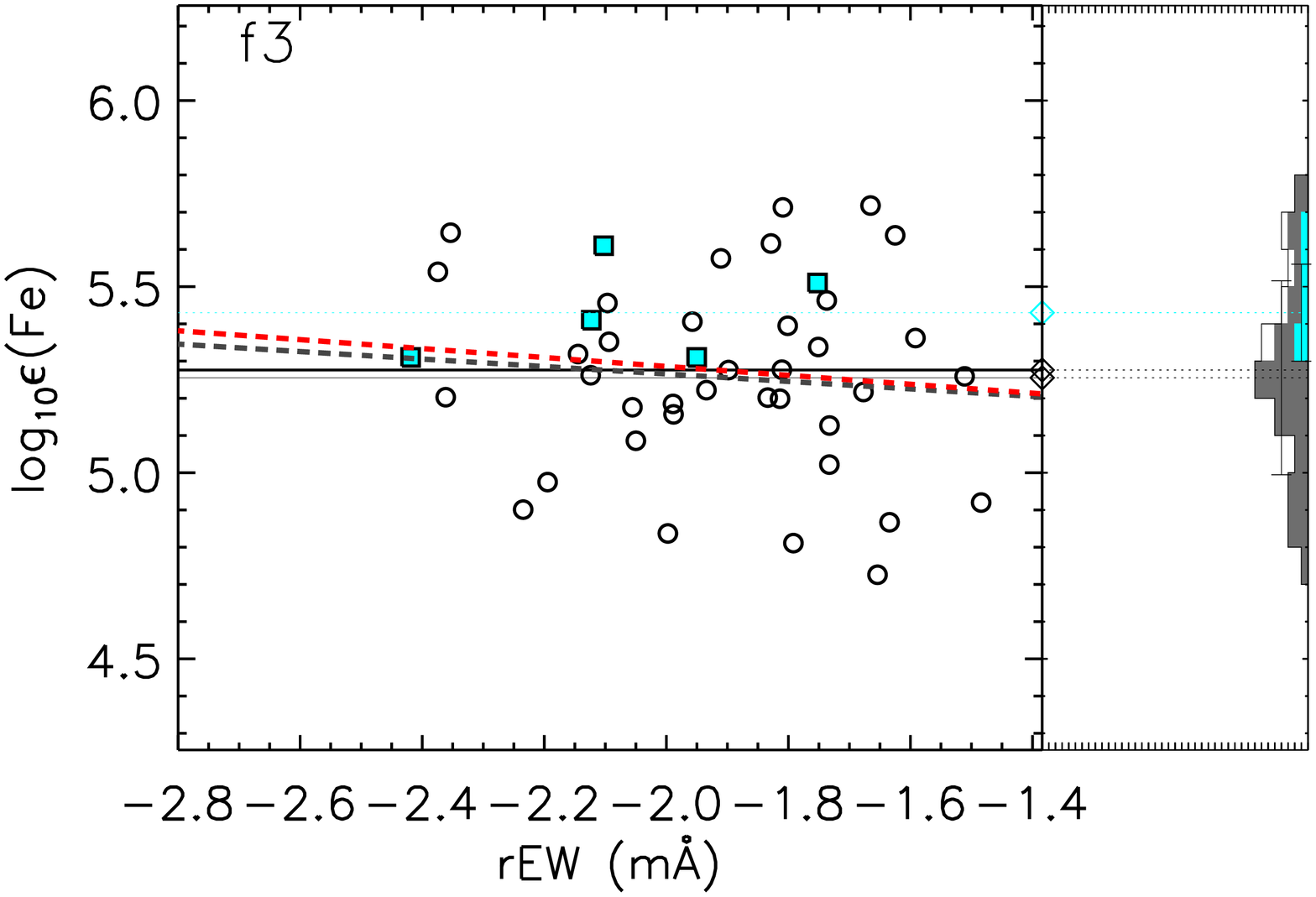}

\caption{  The same as Figure \ref{fig:sol1} for the clusters NGC 6397, NGC 6752 and Fornax 3, from top to bottom respectively.      }
\label{fig:sol2} 
\end{figure*}

NGC 6397 is a special case because the sampling of our data is poor, due to its nearby distance and low mass.  In \cite{scottphd} a comparison of resolved photometry to the corresponding scanned region of the IL spectra showed that the brightest stars (M$_{V} > 0$) were missing in the scanned area, as also shown in \cite{bernstein02}.   Therefore, in \cite{scottphd} all stars brighter than M$_{V} = 0$ were removed from the theoretical CMDs in order to provide a more accurate test of the IL technique.  We accordingly follow the same procedure in this work.   Our final solution is [Fe/H]$=-2.05\pm0.03$ from 58 Fe I lines, and an age between 7 - 15 Gyr.  We were unable to make a clean measurement of any Fe II line, which is likely partly due to the low S/N of the data and  also to the lack of luminous RGB stars, which are large contributors to the EWs. The final abundance diagnostics are shown in Figure \ref{fig:sol2}, where moderate dependences of abundance with EP and EW can be seen. The reference abundances in Table \ref{tab:source} for NGC 6397 are $-2.02$ and $-2.07\pm0.03$, which both agree within the uncertainties with our IL measurement.

\begin{figure*}
\centering
\includegraphics[angle=90,scale=0.2]{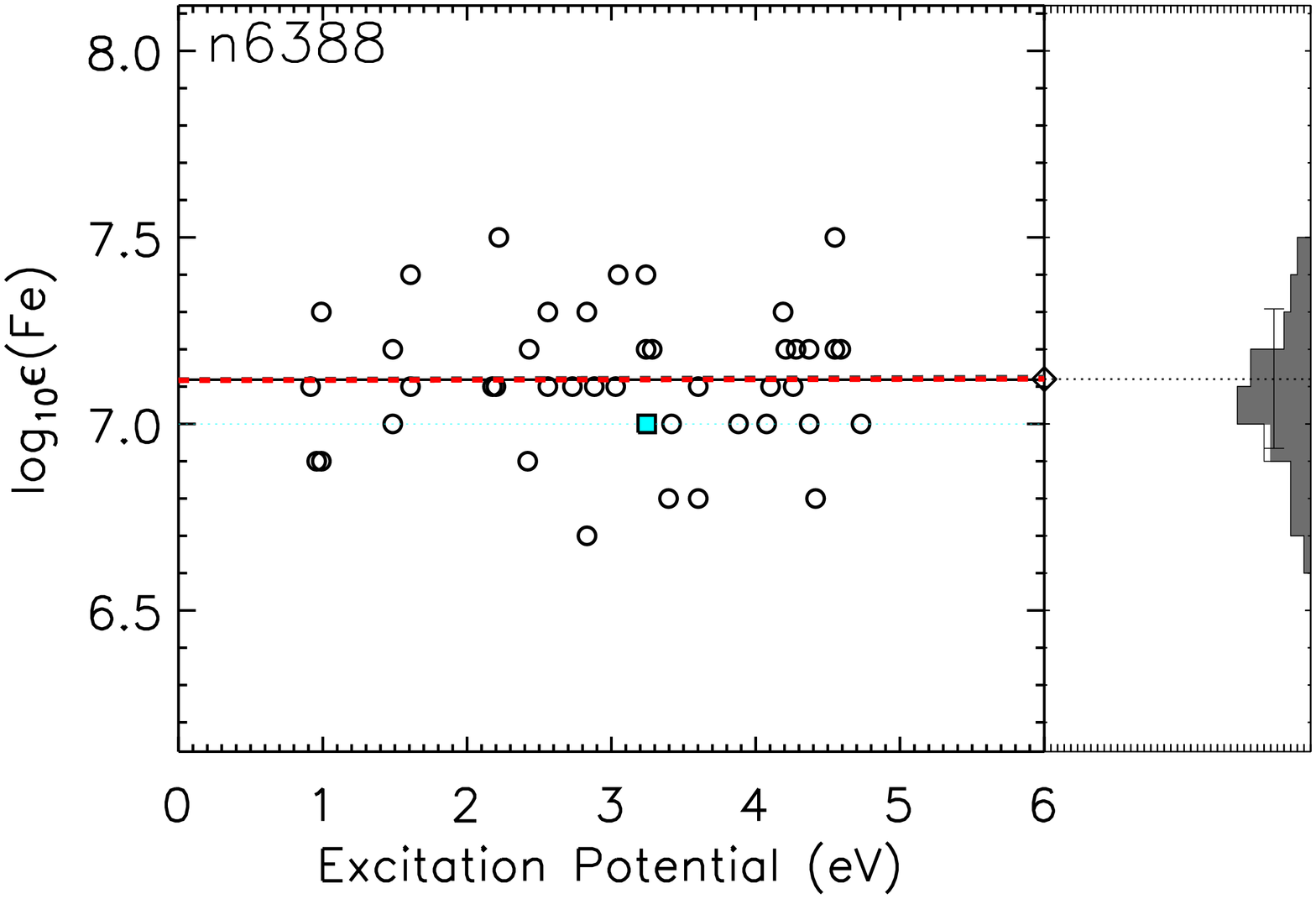}
\includegraphics[angle=90,scale=0.2]{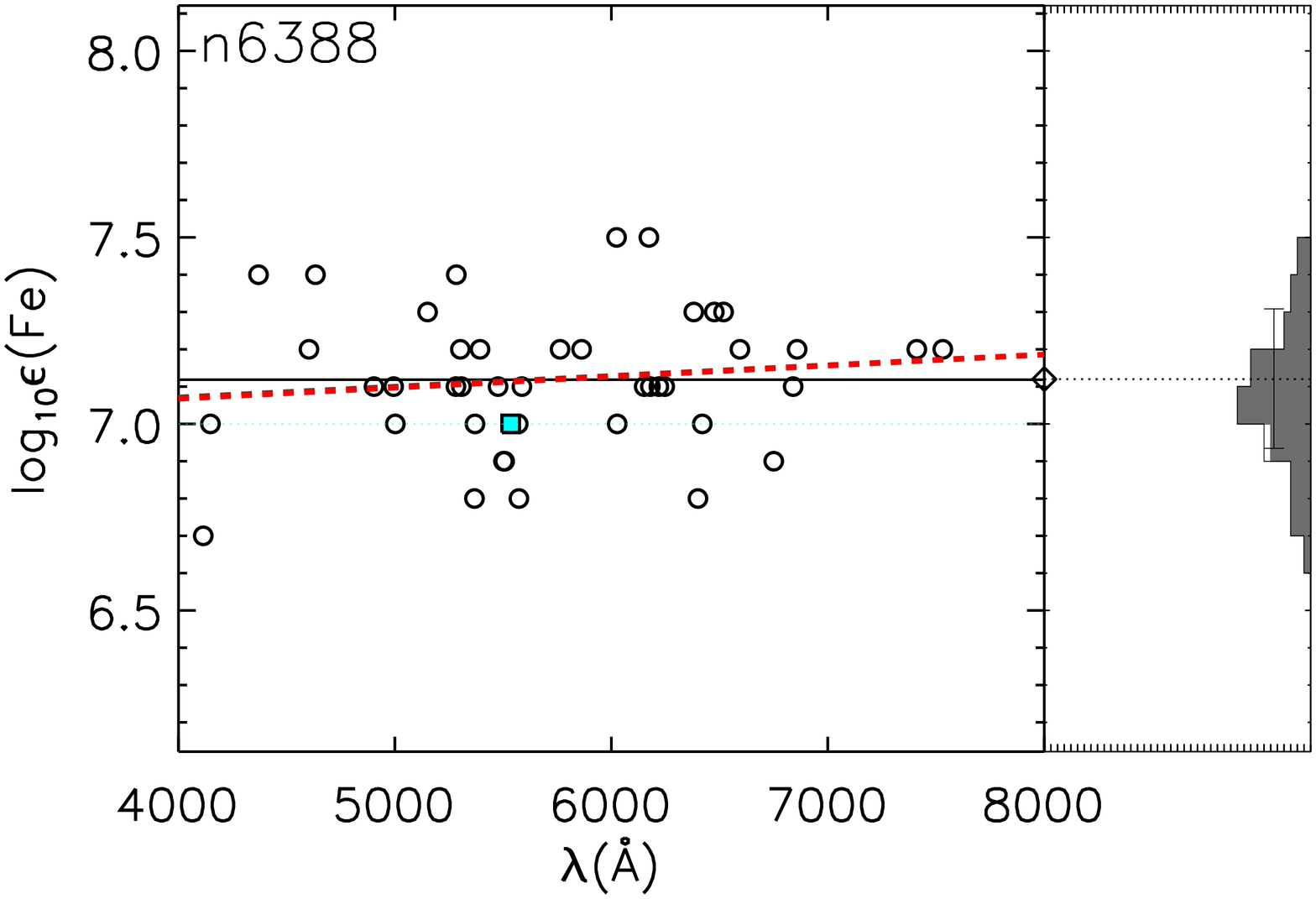}
\includegraphics[angle=90,scale=0.2]{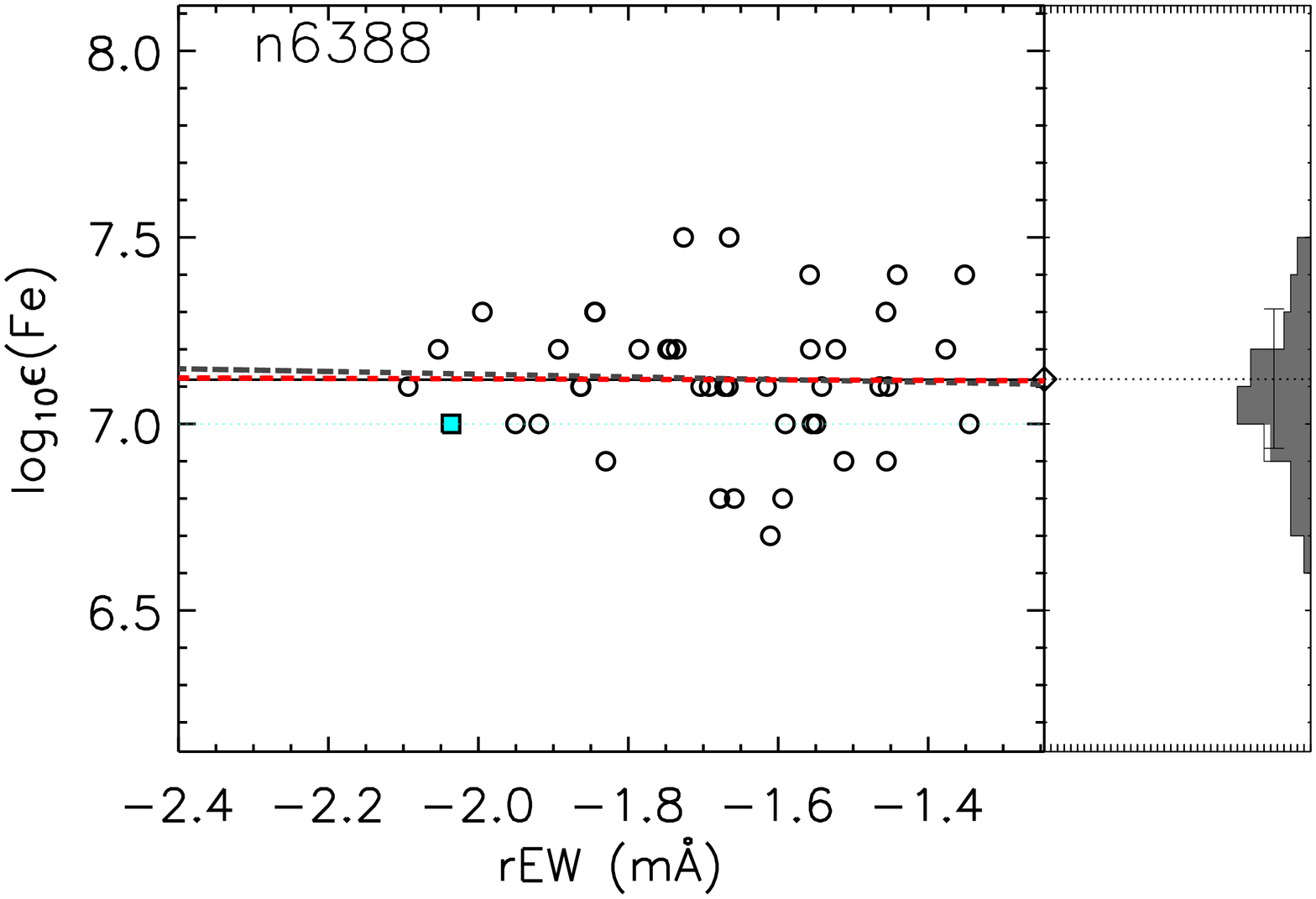}

\includegraphics[angle=90,scale=0.2]{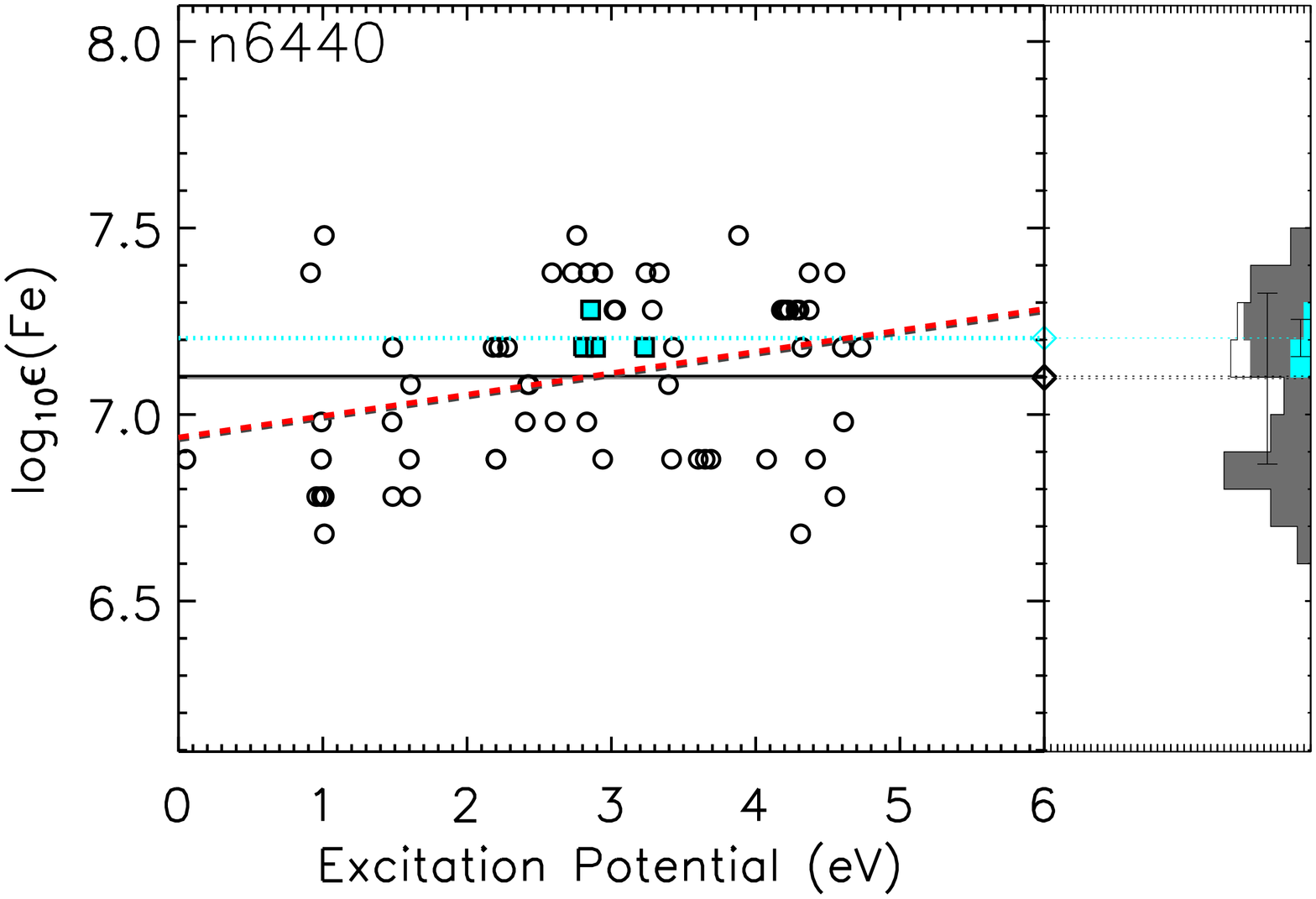}
\includegraphics[angle=90,scale=0.2]{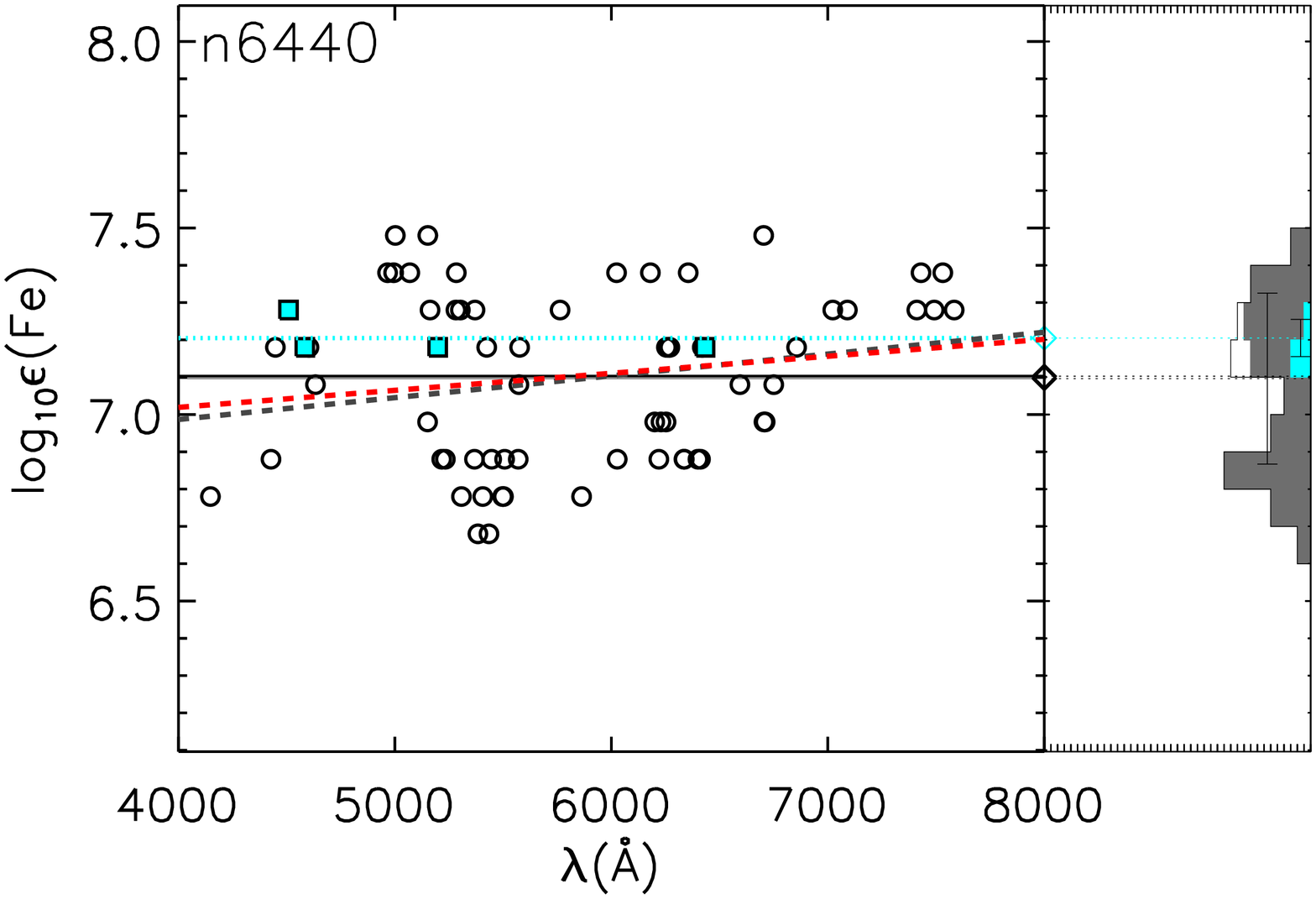}
\includegraphics[angle=90,scale=0.2]{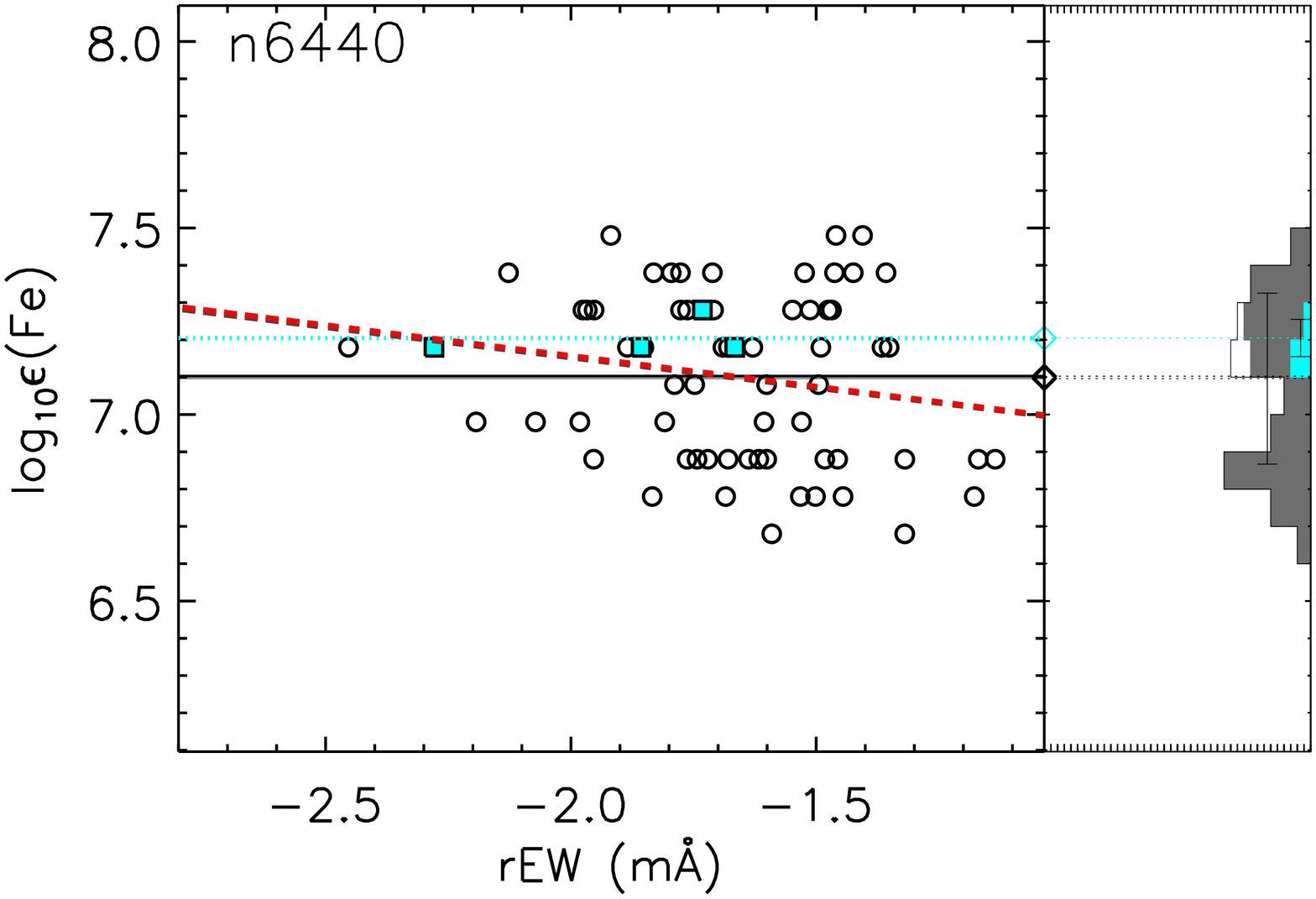}

\includegraphics[angle=90,scale=0.2]{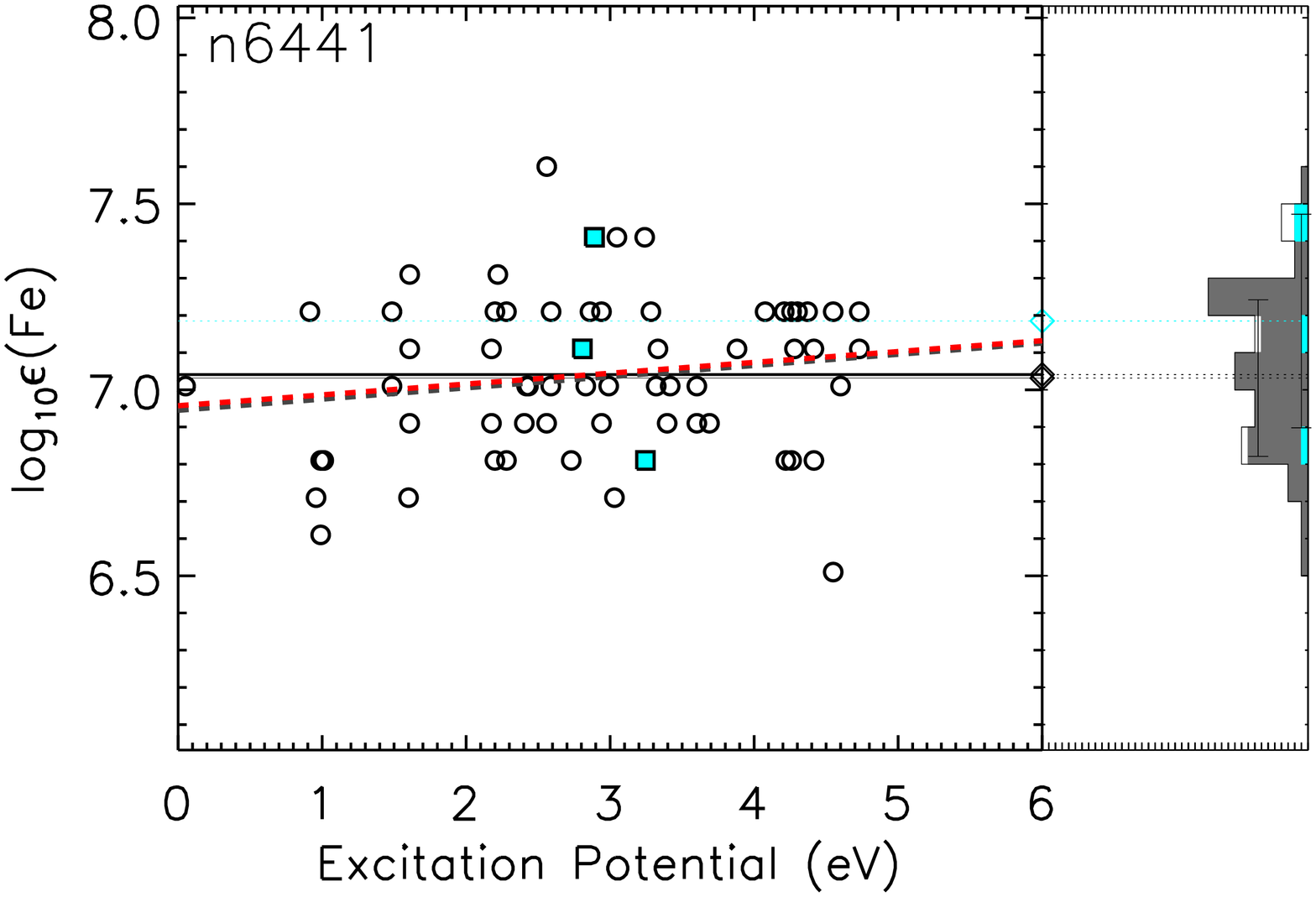}
\includegraphics[angle=90,scale=0.2]{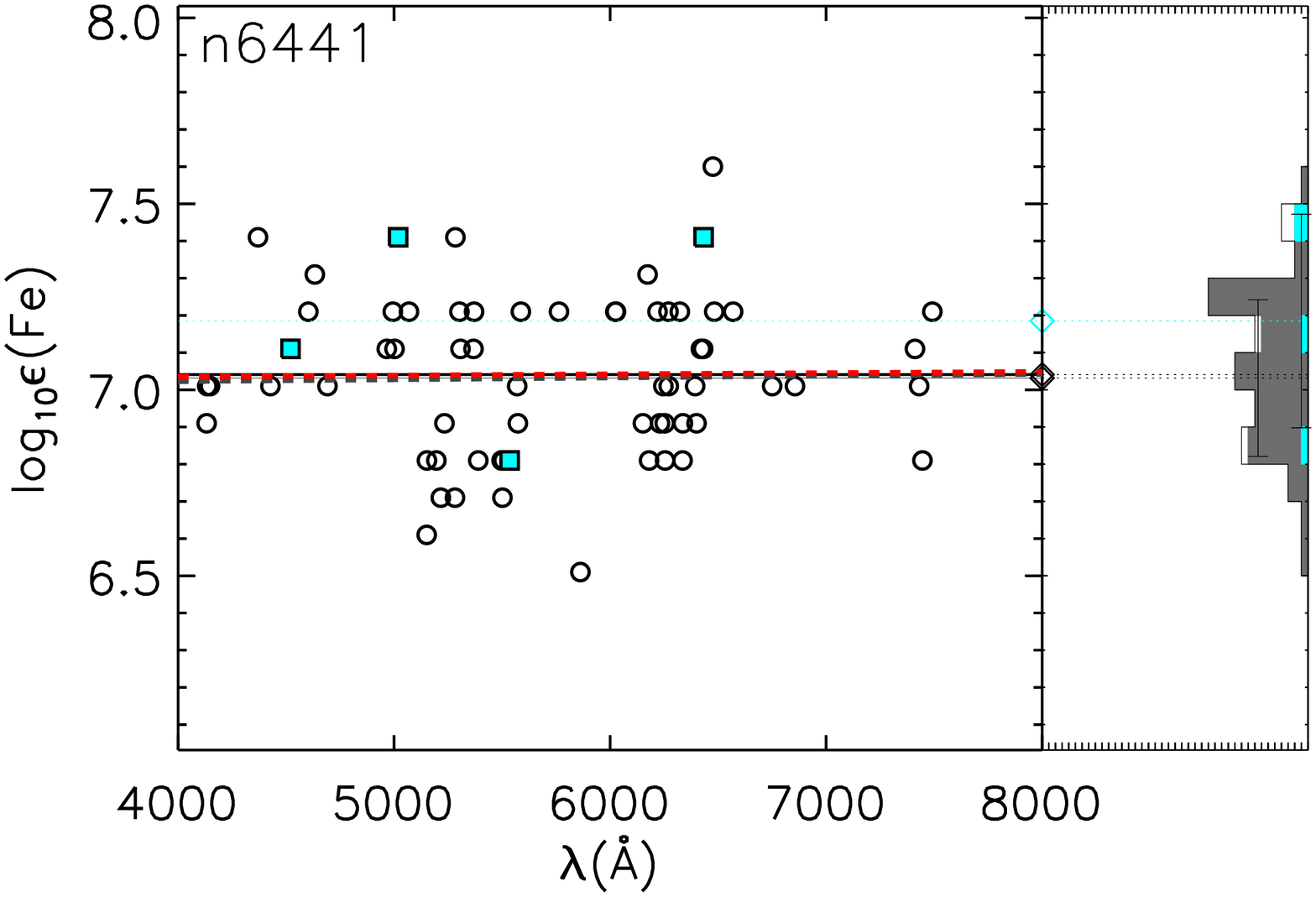}
\includegraphics[angle=90,scale=0.2]{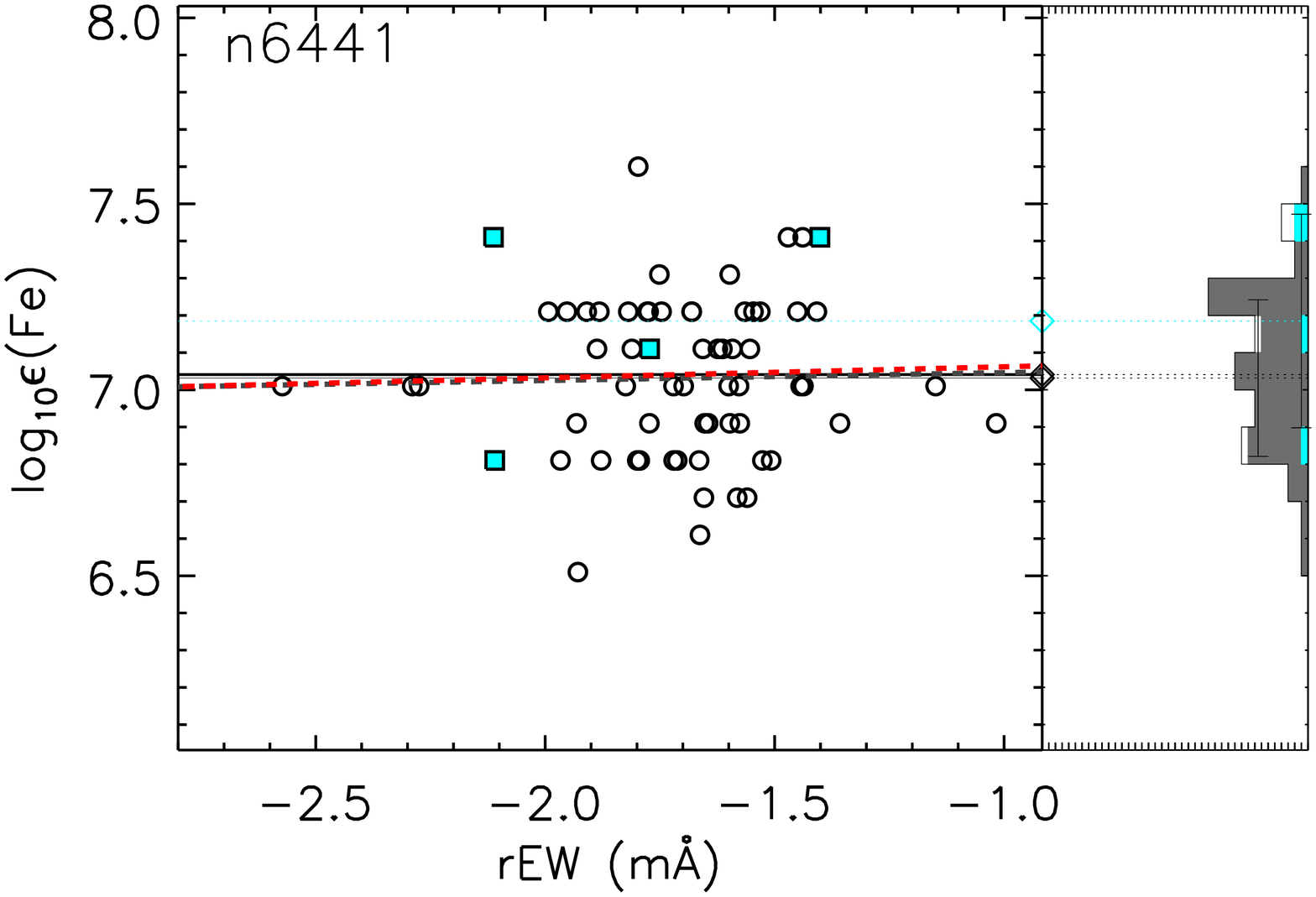}

\includegraphics[angle=90,scale=0.2]{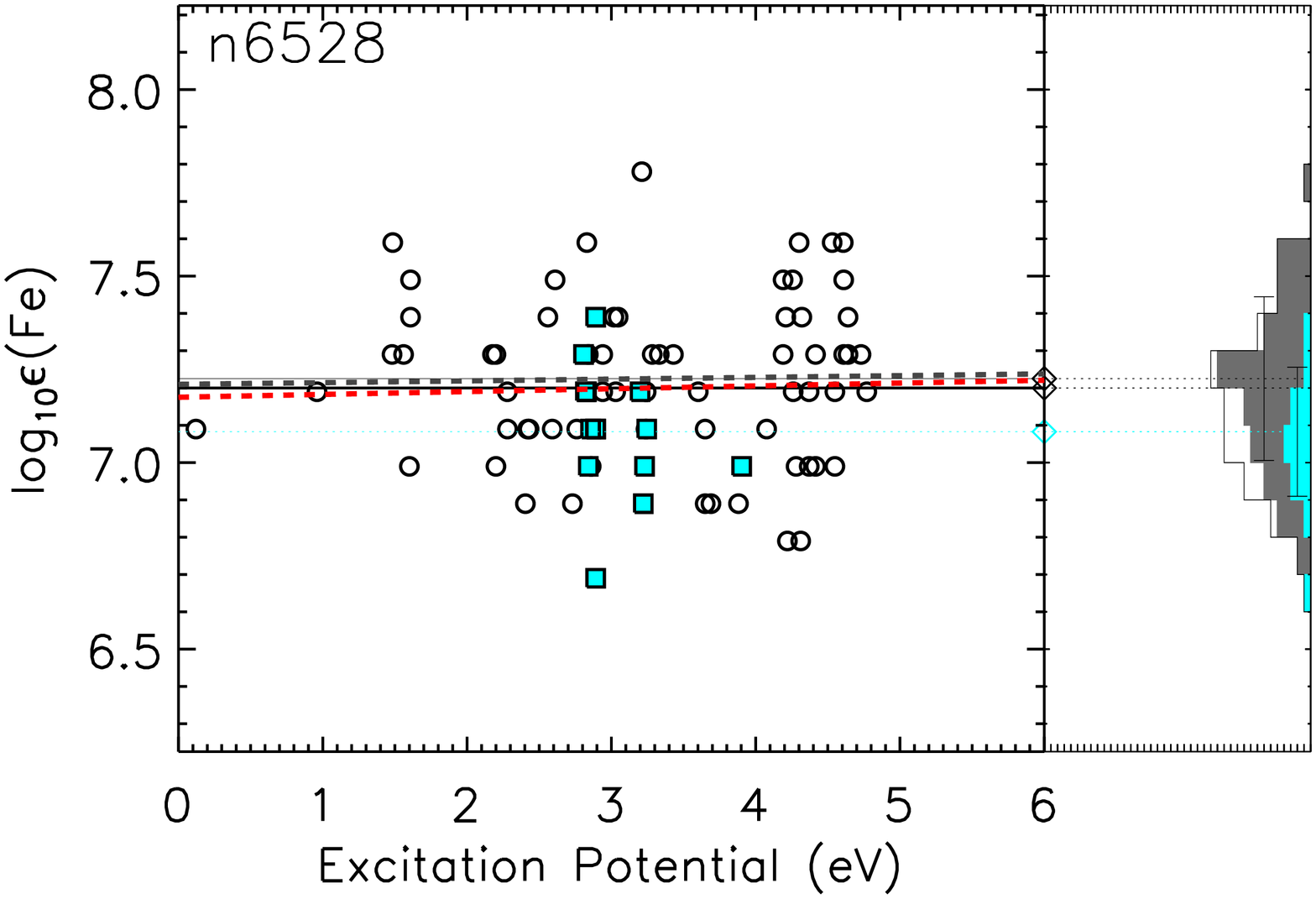}
\includegraphics[angle=90,scale=0.2]{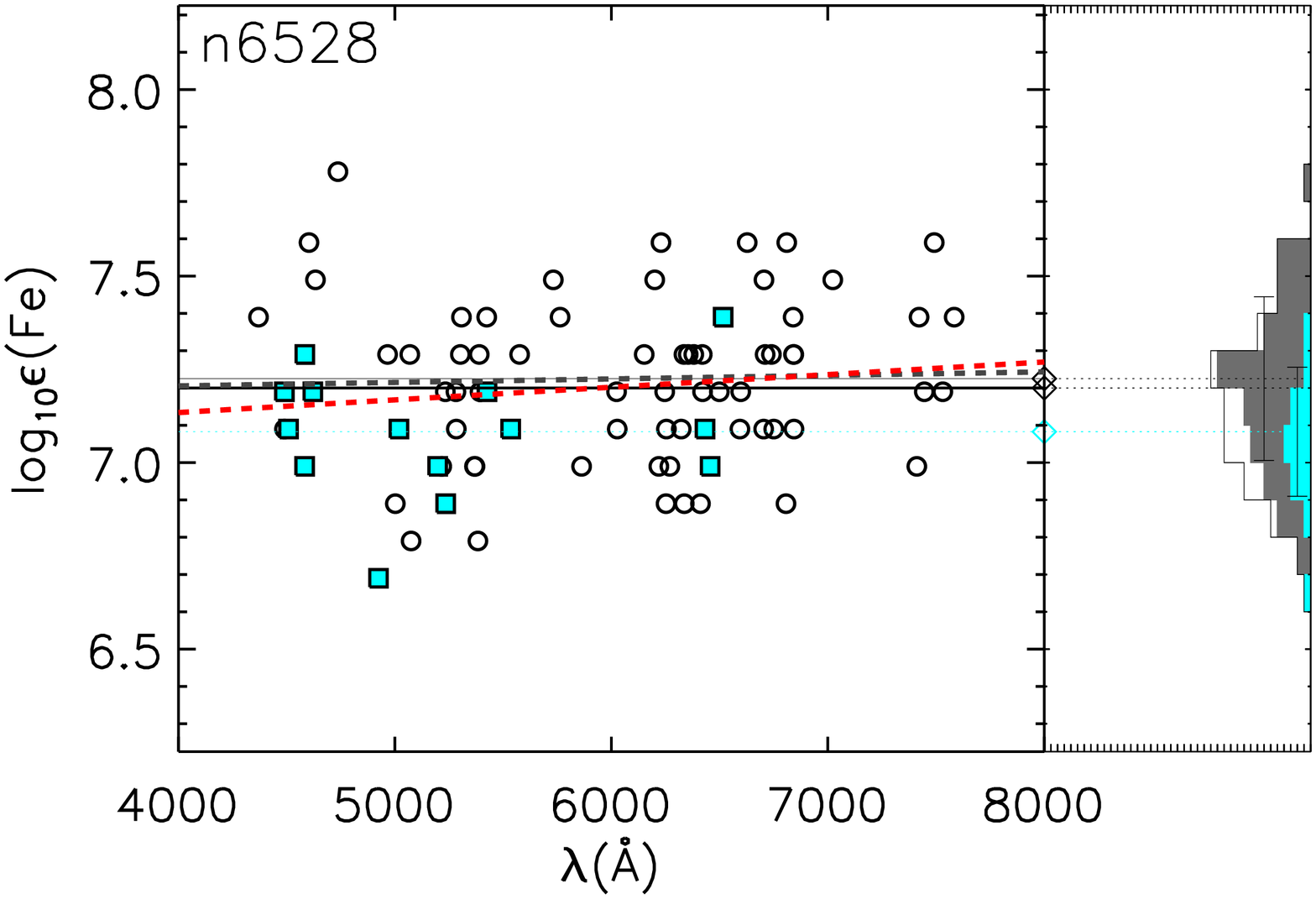}
\includegraphics[angle=90,scale=0.2]{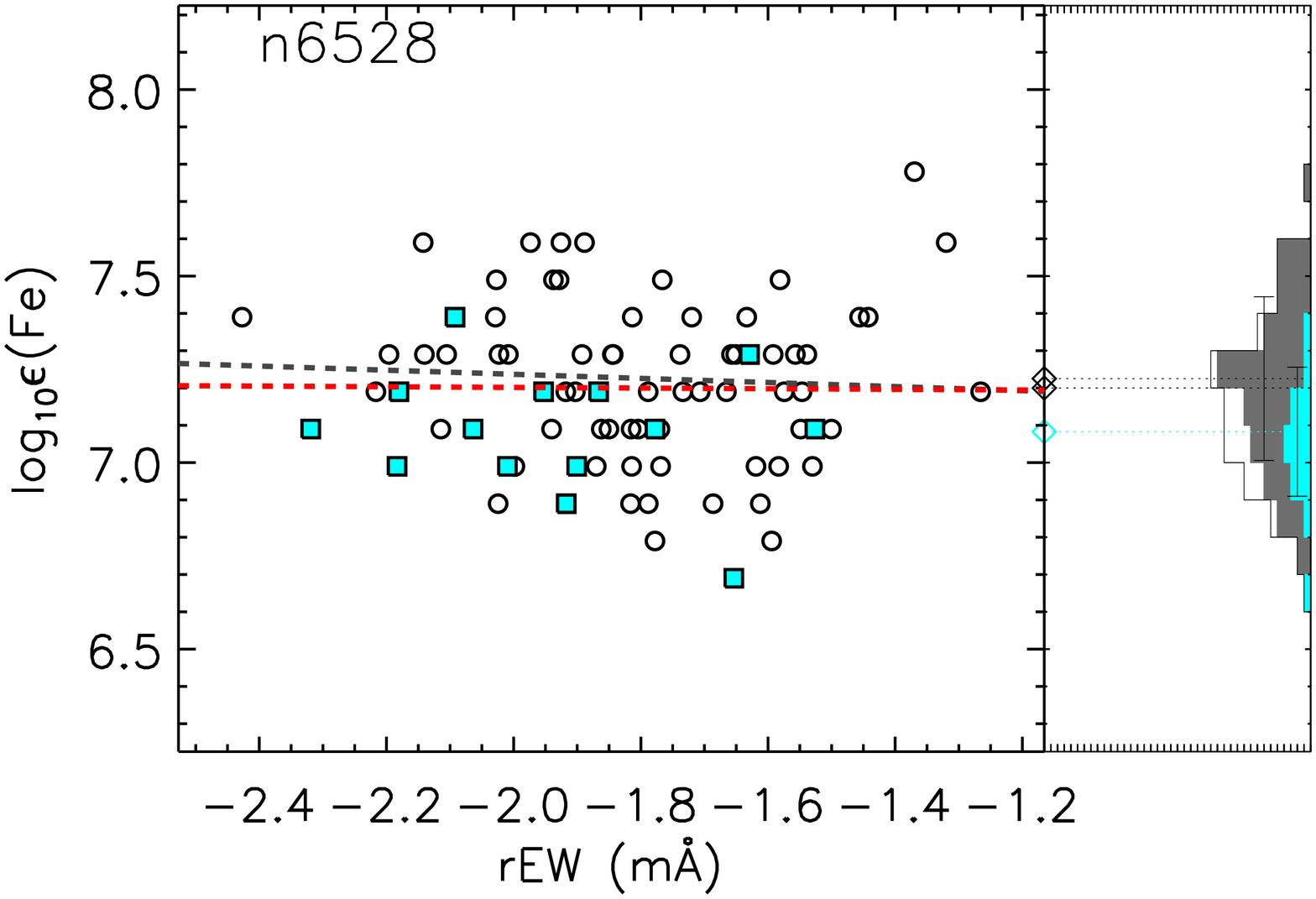}

\includegraphics[angle=90,scale=0.2]{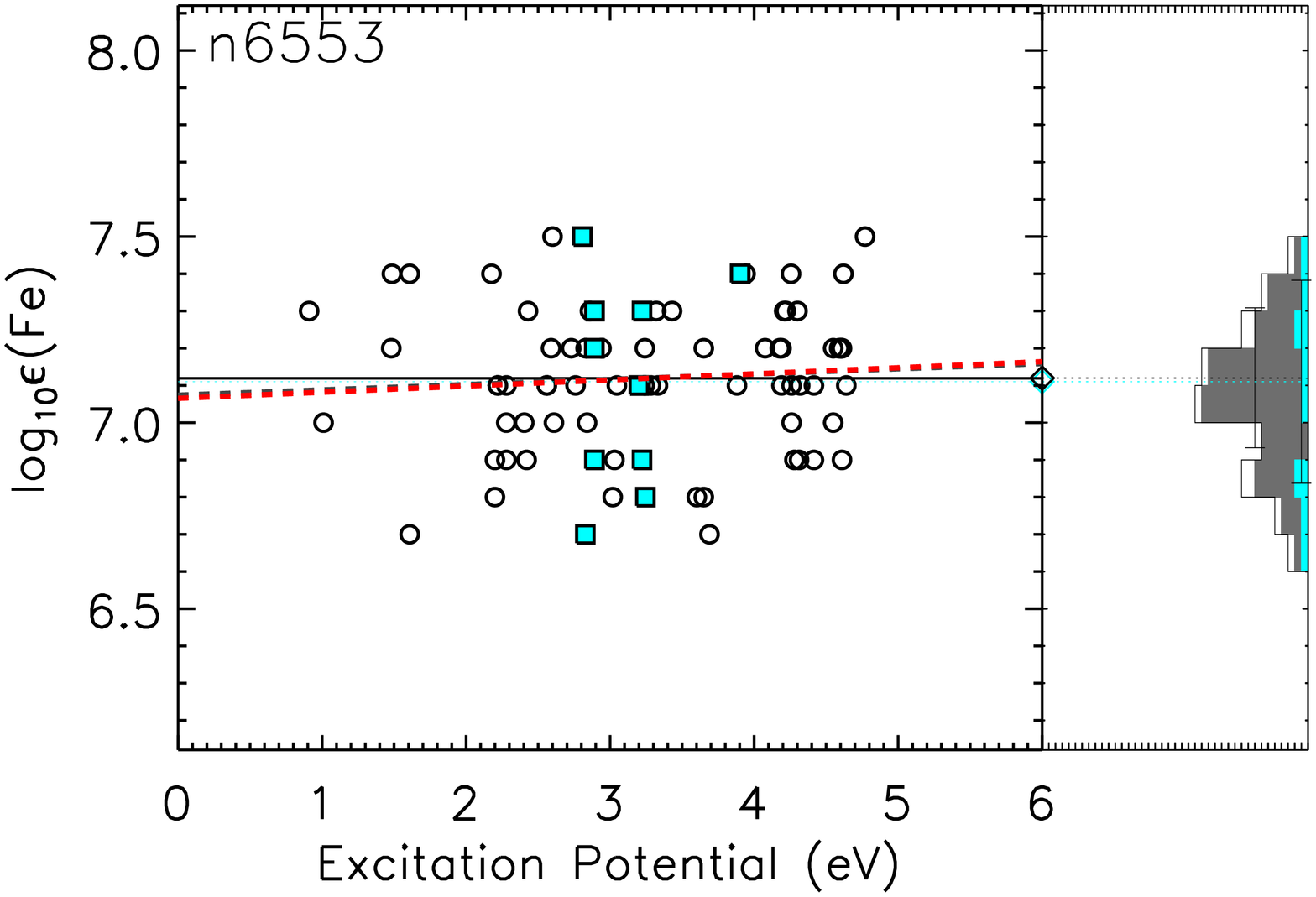}
\includegraphics[angle=90,scale=0.2]{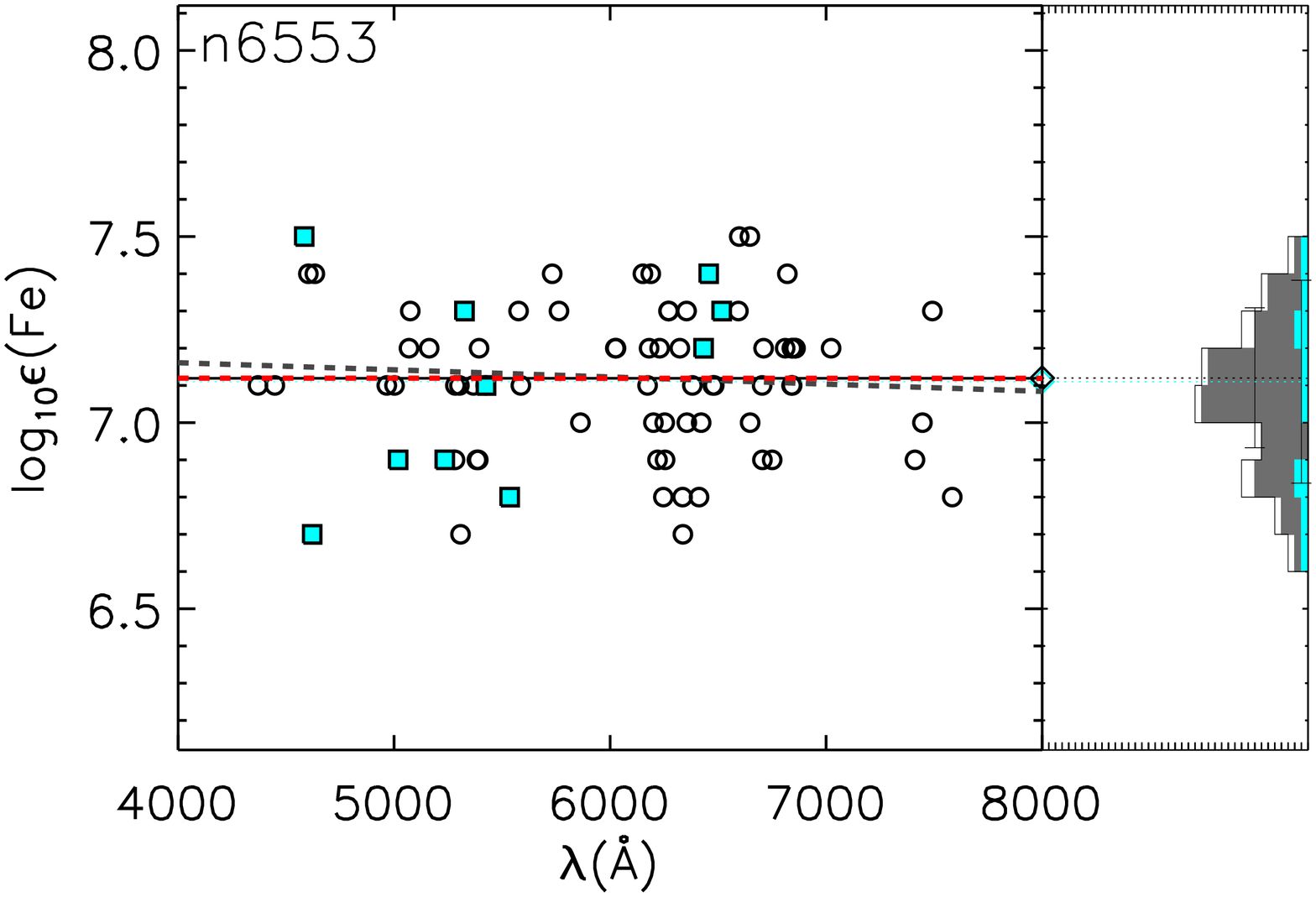}
\includegraphics[angle=90,scale=0.2]{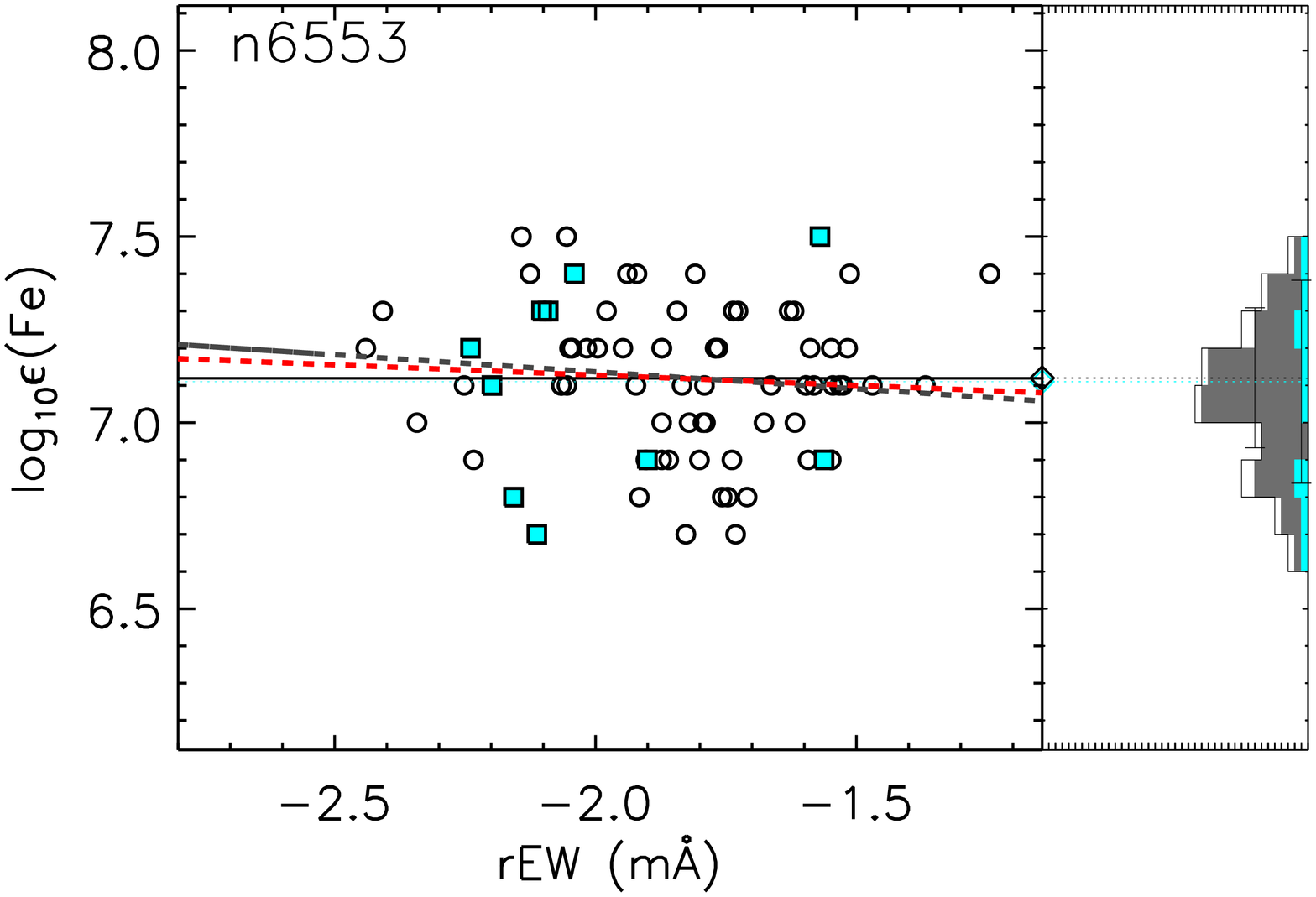}

\caption{ The same as Figure \ref{fig:sol1} for clusters measured with line synthesis.  From top to bottom are the clusters NGC 6388, NGC 6440, NGC 6441, NGC 6528 and NGC 6553.}
\label{fig:sol3} 
\end{figure*}

 We note that as can be expected,  performing the abundance analysis with the luminous RGB stars included in the theoretical CMD results in a lower derived abundance ([Fe/H]$\sim-2.2$), which is a good illustration of the possible systematic error that can occur from a poorly matched population.  Fortunately this is an extreme case, and IL spectra of old extragalactic clusters are unlikely to be this poorly sampled.

\subsection{NGC 6752}

The  results for NGC 6752 are  [Fe I/H]$=-1.58\pm0.03$ from 81 Fe I lines, [Fe II/H]$=-1.59\pm0.04$ from 15 Fe II lines, and an age of 7 to 15 Gyr.  We find that stochastically sampled CMDs improve the solution, although there is still some dependence of abundance with EP.  The inclusion of Fe II lines improves the diagnostics marginally.  Our result agrees well with the average reference abundances, which range from [Fe/H]$=-1.43$ in \cite{gratton01} to [Fe/H]$=-1.67$ in \cite{gruyters14}.

\subsection{Fornax 3}

Fornax 3 is the only cluster in this sample that is outside the Milky Way itself, but stellar abundance comparisons are available from \cite{letarte} and it allows us to extend the  metallicity range we sample to [Fe/H]$\sim -2.3$.    Our IL abundance measurement is [Fe I/H]$=-2.27\pm0.05$ from 36 Fe I lines and [Fe II/H]$=-2.11\pm0.09$ from 5 Fe II lines.  The average Fe II abundance is 0.2 dex higher than the Fe I abundance, but the dispersion in abundance between individual Fe II lines is fairly large.  We do not find that stochastically sampled CMDs improve the abundance diagnostics, which have a particularly large dependence on the wavelength of the line, as seen in Figure \ref{fig:sol2}.  The abundance of Fornax 3 from individual stars by \cite{letarte} is [Fe/H]$=-2.38\pm0.03$, which is within 2-sigma from our result for Fe I lines.  An abundance measurement from IL spectra was also measured by \cite{larsen12} of [Fe/H]=$-2.33$ using different measurement techniques and is consistent with our result.

\subsection{NGC 6388}

NGC 6388 has the largest velocity dispersion in the sample ($\sigma_{v}\sim$ 20 \kms), as well as high metallicity, and therefore it was more challenging to measure a precise abundance. Line synthesis analysis was used for both Fe I and Fe II lines, and we only include lines where we believe we are able to accurately identify the pseudo-continuum.  Our final result is [Fe I/H]$=-0.33\pm0.13$ from 42 Fe I lines, and [Fe II/H]$=-0.48\pm0.1$ from 1 Fe II line, where we have assigned a typical measurement uncertainty of 0.1 dex because the measurement is from a single line.  The most self consistent solutions for NGC 6388 were found when using stochastically sampled CMDs, and the final solutions have very little dependence on EP, wavelength or EW, as shown in Figure \ref{fig:sol3}.  CMDs with ages of 5 Gyr or greater gave very similar solutions, leading to a larger age uncertainty for this GC of 5-15 Gyr.     There is a large dispersion in reference abundances for NGC 6388, which is in part due to the high foreground extinction to this cluster (E(B-V)=0.37).  \cite{wallerstein07} measure [Fe/H]=$-0.79$ and \cite{carretta6388} measure [Fe/H]=$-0.44$; our IL measurement is consistent with the average of these two studies of [Fe/H]=$-0.62\pm0.25$, within the noted large uncertainties. 

\subsection{NGC 6440}

 For NGC 6440 we find [Fe I/H]=$-0.34\pm0.08$ from 62 Fe I lines, [Fe II/H]$=-0.26\pm0.04$ from 4 Fe II lines,   and an age of 5$-$13 Gyr.  This cluster has larger uncertainties in the Fe abundance than most of the other clusters in the sample, and consistent abundances from  Fe I and  Fe II lines.   We note that the uncertainty in the Fe I abundance is dominated by the age uncertainty.   We find that stochastically sampled CMDs improve the final solutions, although trends persist in the abundance diagnostics.   The comparison to reference abundances for NGC 6440 is difficult because there are few measurements available.  The only high resolution study that provides abundance measurements for elements other than Fe is the IR study of \cite{origlia08}, who find [Fe/H]$=-0.56\pm0.02$.   However a higher metallicity of [Fe/H]=$-0.36$ is given by \citetalias{harris}, which is based on the \cite{origlia08} measurement,  lower resolution index measurements of \cite{1995A&A...303..468M}, Ca II triplet measurements of \cite{armandroffzinn}, and the transformation of the  \cite{armandroffzinn} value by \cite{carrettascale}.  Our measurement agrees well with the value given in \citetalias{harris}, but is higher than that of \cite{origlia08}.

\subsection{NGC 6441}

We measure [Fe I/H]$=-0.46\pm0.11$ and [Fe II/H]$=-0.26\pm0.22$ for NGC 6441, using 60 Fe I and 4 Fe II lines, respectively.  While the difference between the mean abundances from Fe I and Fe II lines is large, they are consistent with each other given the large uncertainties. 
We initially found that the best solutions for NGC 6441 had a CMD age of 3 - 7 Gyr, even when stochastically sampled CMDs were used.  Because NGC 6441 is a resolved Milky Way GC, we know that it is in fact old \citep[e.g.][]{2009ApJ...694.1498M}, and one of the few high metallicity ``second parameter'' clusters with a predominantly blue HB \citep{1997ApJ...484L..25R}. In our previous work     \citepalias[e.g.][]{m31p2}, we have found that some clusters are consistent with both a younger age or an old age with a blue HB.  Because the set of isochrones we use do not include blue HBs for either intermediate or high metallicities by default, we perform ad hoc tests where we replace red HB stars with blue HB stars in the theoretical CMDs, as discussed earlier.  When we perform these tests for NGC 6441, we indeed find that when blue HBs are assumed, CMDs with old ages provide equally good solutions as the 3 -7 Gyr aged CMDs.   Therefore, our final age constraint for NGC 6441 is an age between 3 - 15 Gyr.  While this is a large range in age, the $\sigma_{age}$ for [Fe/H] we derive over this range is only slightly larger than for GCs with the best age constraints, since with the addition of blue horizontal branch stars the mean colors and temperatures of the CMDs at the two age extremes are similar.   Our final solution for [Fe/H]  is also  consistent with the reference values in Table \ref{tab:source}.

\begin{figure}
\centering
\includegraphics[trim = 0mm 0mm 50mm 0mm, clip,scale=0.45]{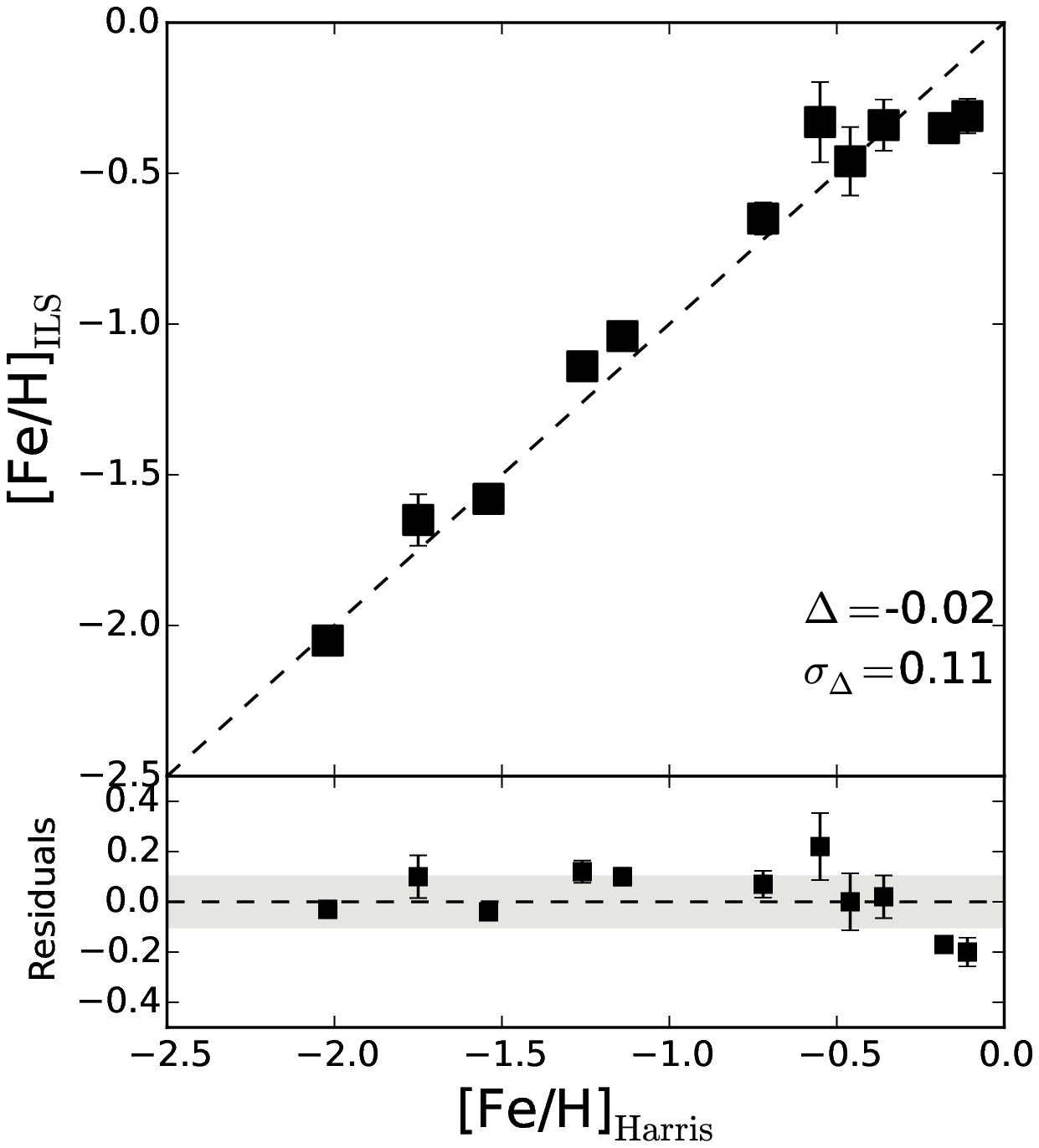}

\caption{  A  comparison of the abundance from IL spectra ([Fe/H]$_{ILS}$) and the abundance of \citet[][2010 edition]{harris} ([Fe/H]$_ {Harris}$), note that Fornax 3 is not included as it does not have an abundance from \citet[][2010 edition]{harris}.    In the top panel a dashed line shows a 1:1 line.  The bottom panel shows the residuals of   [Fe/H]$_{ILS}$- Fe/H]$_ {Harris}$.   A dashed line at 0 shows perfect agreement and the shaded gray area shows $\pm$0.1 dex around perfect agreement, which encompasses the majority of points.    The weighted mean of the residuals for the whole sample ($\Delta$) is shown in the lower right, as well as  the weighted standard deviation around the mean ($\sigma_{\Delta}$). }
\label{fig:fe_lit} 
\end{figure}

\begin{figure}
\centering
\includegraphics[trim = 0mm 0mm 50mm 0mm, clip,scale=0.45]{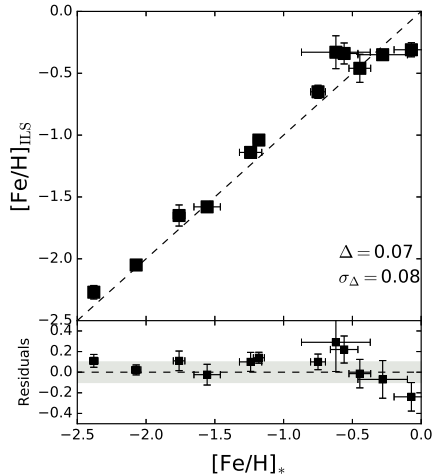}

\caption{ The same as Figure \ref{fig:fe_lit} for a comparison of the abundances from  IL spectra and the average  abundance of individual stars from the reference abundances in Table \ref{tab:source}.  }
\label{fig:fe_lit2} 
\end{figure}

\subsection{NGC 6528}
The final measurements for NGC 6528 are [Fe I/H]=$-0.31\pm0.06$ from 65 Fe I lines, and [Fe II/H]=$-0.40\pm0.07$ from 15 Fe II lines.   We find that the best matching ages are between 7-10 Gyr, and that the solution is improved when using stochastically sampled CMDs.  The final abundance diagnostics in Figure \ref{fig:sol3} show little to no dependence of abundance on EP, wavelength or EW.  Our result is more metal poor than the reference values, which range from [Fe/H]=$-0.17$ in \cite{origlia05} to [Fe/H]=$+0.07$ in \cite{carretta6528}. While there is considerable dispersion in the reference values, they are all systematically higher than the value we find with the IL technique. As we show in \textsection \ref{sec:others}, we also find that several of  the abundance ratios for NGC 6528 are  in poor agreement with the reference abundances.

 We  confirm that even rare, stochastically sampled CMDs with an [Fe/H] as high as solar do not result in self consistent solutions, so the discrepancy is not due to sampling issues.  We also recalculate the abundances using Barklem damping constants, but overall the effect is a lowered derived abundance. 
We next used NGC 6528 as a test case for the effect of TiO lines on the IL abundances, as discussed in more detail in \textsection \ref{sec:tio}, but this did not alleviate the discrepancy either.  In another test we find that it is possible to derive a higher abundance if we remove the bright, red stars in the same way we did for NGC 6397.  We can recover a more consistent [Fe/H] in this case, but find that this CMD results in  abundance ratios that  significantly deviate from the reference abundances. 

In conclusion, we are unable to determine if the  discrepancy in abundances for NGC 6528 is  due to a failure of our analysis method at solar metallicities or a particular issue affecting the IL spectra of NGC 6528 alone.

\subsection{NGC 6553}
For NGC 6553 we measure [Fe I/H]=$-0.35\pm0.03$ from 68 Fe I lines, [Fe II/H]=$-0.34\pm0.10$ from 10 Fe II lines, and an age of 7-13 Gyr.  
We also find that the solutions are improved when using stochastically sampled CMDs. The final abundance diagnostics show little to no dependence of abundance with EP and wavelength, and a small dependence of abundance with EW; including the Fe II lines in the diagnostics provides marginal improvement.  The reference abundances for NGC 6553 give a metallicity of  approximately [Fe/H]=$-0.2$, with the exception of \cite{barbuy99}, who find [Fe/H]=$-0.55$.   The   mean reference abundance in Table \ref{tab:source} is [Fe/H]=$-0.28\pm0.18$, where the large dispersion is driven by the much lower measurement of \cite{barbuy99}.     Our IL measurement is consistent with this mean given the large dispersion, but lower than most of the measurements and the value of [Fe/H]=$-0.18$ given by \citetalias{harris}.

\subsection{ILS method Performance over all [Fe/H]}

The updated Milky Way test sample extends the metallicity range to evaluate of the accuracy of the IL spectra analysis method.   First we briefly comment on the ages derived for the GCs using the ILS technique.  As discussed in our previous papers, the ILS method with Fe lines alone does not usually provide precise relative age measurements ($\pm$ 1 or 2 Gyr) for typical ``old" GCs, and the absolute ages given as solutions are dependent on the model isochrones used to construct the synthetic CMDs, which is one reason why we perform all of the ILS work with the same library of isochrones.  The subtle differences in the stellar populations for CMDs with ages between  $\sim$10-15 Gyr do not usually result in  changes in the Fe line strengths that are significant enough for us to confidently measure at this time.  However, we do sometimes see the effect of  hot blue HB stars on the Fe l line strengths, but this results in a larger age uncertainty, as we see for NGC 6441 in this sample. While it is disappointing that more precise age information cannot be obtained using this technique, the advantage is that the abundance measurements are usually robust to potential  errors in the age measurement of a few Gyr; more details can also be found in \citet{sakari_ers}.   With that in mind we look for broad agreement between our age constraints and the known ages of the training set GCs from resolved photometry.  In this sample, all of the GCs have previously measured ages of $>$10 Gyr \citep{2009ApJ...694.1498M, 6440age, 6528age,6553age,fornaxage}, and using the ILS technique we find that all of the GCs are consistent with having an age of at least 10 Gyr, as shown in Table \ref{tab:fe}.   NGC 6388, NGC 6440, NGC 6441, and NGC 6528 are also consistent with somewhat younger ages of $\sim$5-7 Gyr, and the uncertainties in the abundance due to the age are also given in Table \ref{tab:fe}.

In the previous section we discussed the [Fe/H] comparison for each GC in detail, and in this section we look at the behavior over [Fe/H] as a whole.  
In Figure \ref{fig:fe_lit} we compare the IL results to the \citetalias{harris} values.  We find that the IL results are accurate to within $\sim$0.1 dex until  [Fe/H]$\sim-0.3$. 
The IL results are offset by 0.2 dex to lower abundances for the highest metallicity clusters NGC 6553 and NGC 6528 which have values from \citetalias{harris} of $-0.18$ and $-0.11$, respectively.  To evaluate systematic behavior across the whole sample we calculate a mean residual ([Fe/H]$_{ILS}$- [Fe/H]$_{Harris}$) that is weighted by the total uncertainty, as well as a weighted standard deviation of the residuals.  For the comparison to \citetalias{harris} we find a systematic offset of $\Delta_{FeI}$=-0.02, with a dispersion of $\sigma_{\Delta}$=0.11.  We note that this is similar to what was found in \cite{scottphd} for the original 7 MW training set GCs, where a systematic of  $\Delta_{FeI}$=0.01 with $\sigma_{\Delta}$=0.09 was measured.  
 
\begin{figure}
\centering
\includegraphics[scale=0.3]{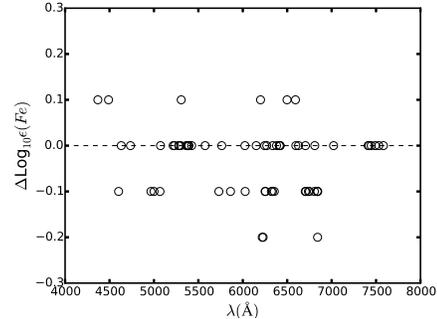}

\caption{ The difference in abundance ($\Delta Log_{10}\epsilon(Fe)= Log_{10}\epsilon(Fe)_{TiO}-Log_{10}\epsilon(Fe)_{NoTiO}$ ) derived for individual lines when low excitation potential TiO transitions are included in the analysis. }
\label{fig:tio} 
\end{figure}

 In  Figure \ref{fig:fe_lit2} we compare the IL results to the mean of the reference stellar abundances in Table \ref{tab:source}.   In this case we also find that the abundances for [Fe/H]$\leq-0.3$ are consistent, within the uncertainties, to $\sim$0.1 dex, and that NGC 6553 is in better agreement because of the lower reference value of [Fe/H]$_{*}=-0.28$.  The highest metallicity GC NGC 6528 is again offset to lower abundances.    We note that in both comparisons NGC 6388 is offset to higher abundances by $\sim$0.2 dex, but the uncertainties from both the IL and stellar techniques are large for this GC; this is likely a combination of the large velocity dispersion, high metallicity, and large extinction.  For the comparison to the stellar reference abundances to the whole sample of clusters we calculate a systematic offset of $\Delta_{FeI}$=0.07, with a dispersion of $\sigma_{\Delta}$=0.08.

Given these results we can conclude that the IL analysis method that we have developed, when applied to  typical ``old" GCs, is accurate to $\sim$0.1 dex for $-2.4<$[Fe/H]$<-0.3$, but may systematically underestimate the abundance of higher metallicity GCs. This conclusion is primarily based on the abundance discrepancy for NGC 6528 ([Fe/H]$\sim$0). It is less clear how well the method performs for NGC 6553 ([Fe/H]$\sim$-0.3), given the large dispersion in reference abundances, but to be conservative we recommend that IL abundance measurements for GCs with [Fe/H]$>-0.3$ be interpreted cautiously.   It is possible that because the highest metallicity GCs are in the bulge, that there could be a significant contribution from background stars to the IL spectra.   As these stars would have a range of velocities, the overall effect would be to dampen the continuum of the IL spectra further, which would result in a lower derived [Fe/H].

   There is the additional question of how we would know if we have a GC that truly has [Fe/H]$\sim-0.3$, or if it is a  GC  with solar [Fe/H] for which we are  underestimating the abundance. Unfortunately we only have one GC, NGC 6528, to base our conclusion on, and we cannot conclusively determine if there is an issue specific to this GC, or a failure in the IL analysis at high metallicity.  
    We note that we do find that the other abundance ratios (see \textsection \ref{sec:others}) for NGC 6528 are sometimes offset to higher values than the rest of the GCs in the sample, which may be a useful red flag.
Of course expanding the sample of $\sim$12 Gyr old test clusters with [Fe/H]$>-0.3$ to more than 1-2 clusters would be ideal, but unfortunately resolved clusters with metallicities this high (and better isolated from background stars) are rare, as are high resolution  abundance studies of their member stars.

\subsection{TiO Effects}
\label{sec:tio}

As discussed in \citetalias{mb08},  line blanketing from TiO molecular lines in M giant stars is a concern when deriving abundances for metal-rich clusters.  We note that TiO is not included in our standard synthesis line lists because it is computationally expensive. 
\citetalias{mb08} presented tests to evaluate the impact of  TiO line blanketing in detail for the metallicity of NGC 104 ([Fe/H]=$-0.7$).  The results were that the impact of TiO line blanketing was small in the V band ($\sim$5500 \AA) at this metallicity, and resulted in a change of the overall mean [Fe/H] of $\leq$0.02 dex, although it was speculated that the impact at higher metallicities could be more pronounced.   \citetalias{mb08} suggested that the wavelength interval of 7300-7600 \AA, which has lower TiO opacity and therefore less blanketing, could potentially be used to constrain the contribution of M giant stars when compared to bluer wavelengths that are more affected.

Since we now have clusters more metal rich than [Fe/H]=$-0.70$, we have performed some additional simple tests to investigate whether TiO line blanketing is having a significant effect on our results.  We note that since the IL abundance results show similar agreement to reference abundances for $-0.70<$[Fe/H]$<-0.3$ as they do for clusters with [Fe/H]$<-0.7$,  at first glance any systematics introduced by TiO line blanketing do not seem to be dominating the abundance measurement in this metallicity window.  However, since we have larger discrepancies  between the IL results and reference stellar results for [Fe/H]$>-0.3$, we investigate whether TiO features are affecting our results in this regime.    

 As an initial consistency check, for the metal rich clusters in our sample we can compare the abundances of Fe lines in the 7300-7600 \rAA window to those at bluer wavelengths to look for systematic differences. Unfortunately there are only a handful of line measurements in the 7300-7600 \rAA region, as can be seen in Figure \ref{fig:sol3}.  Nevertheless, for NGC 6441, NGC 6528, and NGC 6553 the Fe lines in the 7300-7600 \rAA region have a similar dispersion in [Fe/H] as the bluer lines.  For NGC 6440 the mean abundance of Fe I lines with wavelengths $>$7000 is higher than the mean abundance of bluer lines, which manifests as the trend of increasing Fe abundance with wavelength.  However we note that there are only  5 lines at wavelengths $>$7000,  the individual abundances of these lines still fall within the dispersion of the bluer lines, and the abundances of these lines are not the most extreme over the entire wavelength range.   For these reasons, and because we measure an IL [Fe/H] that agrees very well with the reference stellar abundances, we conclude that there is no obvious sign of the impact of M giant TiO blanketing from the behavior of the abundances of the Fe lines themselves.

Since NGC 6528 is the most metal-rich cluster in our sample, it is an interesting test case to perform an Fe line synthesis for each line which includes low EP TiO lines.    To have the maximum impact from TiO lines, M giant stars with temperatures of $\sim$3900 K must be present in the theoretical CMD.  Since these stars are rare in the CMDs with averaged stellar properties, we needed to use a stochastically sampled CMD that has these stars present.  We note that none of the  CMDs with these cool stars were included as best fit solutions when the stochastic analysis was initially done because none of them produced self consistent abundances--- we do not recover the input abundance of the CMD itself, which is a sign of a population mismatch.  As expected, when including these cool stars that produce strong features in the CMDs, we derive an abundance significantly lower than in the CMDs where they are not present.  

Whether the CMD is a mismatch or not, we can test the effect of including cool stars and TiO transitions on the abundances we derive.   The difference in the abundances derived for individual Fe I lines when TiO lines are and are not included in the synthesis is shown in Figure \ref{fig:tio}.  As already demonstrated in \citetalias{mb08}, including TiO reduces the derived abundance in some cases, but also increases the derived abundance in others.  Moreover, the abundances derived for  most of the lines that we use in our analysis are the same. Only a handful of lines result in abundance differences of $>\pm$0.1 dex (6219 \AA, 6229 \AA, and 6842 \AA).  For the most part it appears that the broadening from the velocity dispersion  makes the effect on the flux weighted spectra small.   The difference in the mean abundance for the two analyses is 0.03 dex, so we conclude that TiO lines are not having a significant impact on our analysis.  

We note that it is not necessarily the case that TiO features do not impact the IL spectra at these metallicities at all, but because we  exclude Fe lines that appear to be significantly blended or where the pseudo-continuum is not well identified, in our analysis we are already avoiding regions where TiO (or any other molecular) bands are significantly affecting the spectra.  We believe that this is evidence that our conservative selection of only the most trustworthy Fe lines for each individual cluster produces as robust a measurement as possible of the abundance from IL spectra using our technique. 

\begin{deluxetable*}{crrrrrrrrrrrr}
\tablecaption{ILS Abundance Results\label{tab:bigab1}}

\tablehead{
&\colhead{F3 }&\colhead{n104 }&\colhead{n2808 }&\colhead{n362 }&\colhead{n6093}&\colhead{n6388}&\colhead{n6397 }&\colhead{n6440}&\colhead{ n6441 }&\colhead{n6528 }&\colhead{n6553 }&\colhead{n6752 }}

\startdata
$\rm{[Na/Fe]}$ & \nodata & 0.24 & -0.17 & -0.02 & 0.24 & -0.17 & \nodata & 0.10 & 0.41 & 0.31 & 0.33 & 0.04 \\
$\sigma_{A,Na}$ & \nodata & 0.00 & 0.06 & 0.01 & 0.04 & 0.04 & \nodata & 0.04 & 0.01 & 0.02 & 0.01 & 0.01 \\
$\sigma_{N,Na}$ & \nodata & 0.08 & 0.00 & 0.10 & 0.10 & 0.35 & \nodata & 0.23 & 0.08 & 0.22 & 0.11 & 0.07 \\
$\sigma_{Na}$ & \nodata & 0.08 & 0.06 & 0.10 & 0.11 & 0.36 & \nodata & 0.23 & 0.08 & 0.22 & 0.11 & 0.07 \\
N$_{Na}$ & \nodata & 3 & 2 & 4 & 1 & 2 & \nodata & 3 & 3 & 3 & 3 & 2 \\
\\

$\rm{[Mg/Fe]}$ & 0.21 & 0.29 & -0.03 & 0.07 & 0.17 & -0.02 & -0.09 & 0.18 & 0.23 & 0.02 & 0.05 & 0.06 \\
$\sigma_{A,Mg}$ & 0.03 & 0.00 & 0.01 & 0.04 & 0.00 & 0.00 & 0.02 & 0.04 & 0.03 & 0.00 & 0.00 & 0.01 \\
$\sigma_{N,Mg}$ & 0.22 & 0.08 & 0.10 & 0.02 & 0.13 & 0.10 & 0.10 & 0.17 & 0.10 & 0.06 & 0.03 & 0.07 \\
$\sigma_{Mg}$ & 0.22 & 0.08 & 0.10 & 0.05 & 0.13 & 0.10 & 0.10 & 0.18 & 0.10 & 0.06 & 0.03 & 0.07 \\
N$_{Mg}$ & 2 & 3 & 1 & 3 & 3 & 1 & 1 & 2 & 1 & 3 & 2 & 4 \\
\\

$\rm{[Al/Fe]}$ & \nodata & 0.33 & \nodata & 0.15 & \nodata & \nodata & \nodata & 0.31 & 0.64 & 0.52 & 0.43 & \nodata \\
$\sigma_{A,Al}$ & \nodata & 0.03 & \nodata & 0.03 & \nodata & \nodata & \nodata & 0.08 & 0.00 & 0.00 & 0.04 & \nodata \\
$\sigma_{N,Al}$ & \nodata & 0.11 & \nodata & 0.10 & \nodata & \nodata & \nodata & 0.03 & 0.18 & 0.18 & 0.03 & \nodata \\
$\sigma_{Al}$ & \nodata & 0.12 & \nodata & 0.10 & \nodata & \nodata & \nodata & 0.08 & 0.18 & 0.18 & 0.05 & \nodata \\
N$_{Al}$ & \nodata & 2 & \nodata & 1 & \nodata & \nodata & \nodata & 2 & 2 & 2 & 2 & \nodata \\
\\

$\rm{[Si/Fe]}$ & \nodata & 0.43 & 0.32 & 0.29 & 0.35 & 0.27 & \nodata & 0.27 & 0.35 & 0.17 & 0.30 & 0.40 \\
$\sigma_{A,Si}$ & \nodata & 0.02 & 0.03 & 0.03 & 0.04 & 0.05 & \nodata & 0.08 & 0.13 & 0.04 & 0.01 & 0.04 \\
$\sigma_{N,Si}$ & \nodata & 0.07 & 0.02 & 0.07 & 0.10 & 0.16 & \nodata & 0.11 & 0.09 & 0.06 & 0.03 & 0.06 \\
$\sigma_{Si}$ & \nodata & 0.07 & 0.03 & 0.08 & 0.11 & 0.17 & \nodata & 0.14 & 0.16 & 0.07 & 0.03 & 0.07 \\
N$_{Si}$ & \nodata & 8 & 2 & 6 & 1 & 4 & \nodata & 7 & 2 & 9 & 7 & 5 \\
\\

$\rm{[Ca/Fe]}$ & 0.24 & 0.18 & 0.19 & 0.23 & 0.27 & 0.02 & 0.24 & 0.21 & 0.24 & 0.20 & 0.06 & 0.41 \\
$\sigma_{A,Ca}$ & 0.02 & 0.01 & 0.04 & 0.04 & 0.00 & 0.06 & 0.01 & 0.01 & 0.02 & 0.03 & 0.01 & 0.00 \\
$\sigma_{N,Ca}$ & 0.10 & 0.04 & 0.07 & 0.05 & 0.05 & 0.07 & 0.08 & 0.12 & 0.08 & 0.06 & 0.07 & 0.04 \\
$\sigma_{Ca}$ & 0.10 & 0.04 & 0.09 & 0.06 & 0.05 & 0.09 & 0.08 & 0.12 & 0.08 & 0.07 & 0.07 & 0.04 \\
N$_{Ca}$ & 6 & 12 & 9 & 11 & 14 & 8 & 7 & 5 & 8 & 10 & 10 & 17 \\
\\

$\rm{[Sc/Fe]}$ & 0.11 & 0.18 & 0.04 & 0.06 & -0.05 & -0.11 & \nodata & 0.17 & -0.20 & 0.14 & 0.20 & 0.06 \\
$\sigma_{A,Sc}$ & 0.04 & 0.01 & 0.01 & 0.03 & 0.06 & 0.00 & \nodata & 0.04 & 0.10 & 0.02 & 0.01 & 0.08 \\
$\sigma_{N,Sc}$ & 0.11 & 0.28 & 0.23 & 0.08 & 0.11 & 0.10 & \nodata & 0.07 & 0.10 & 0.09 & 0.10 & 0.12 \\
$\sigma_{Sc}$ & 0.12 & 0.28 & 0.23 & 0.09 & 0.13 & 0.10 & \nodata & 0.08 & 0.14 & 0.10 & 0.10 & 0.14 \\
N$_{Sc}$ & 3 & 2 & 2 & 6 & 3 & 1 & \nodata & 2 & 1 & 5 & 5 & 4 \\
\\

$\rm{[TiI/Fe]}$ & 0.29 & 0.05 & 0.17 & 0.13 & 0.30 & 0.19 & \nodata & 0.05 & 0.17 & 0.57 & 0.29 & 0.39 \\
$\sigma_{A,TiI}$ & 0.03 & 0.02 & 0.02 & 0.01 & 0.05 & 0.05 & \nodata & 0.01 & 0.04 & 0.08 & 0.02 & 0.00 \\
$\sigma_{N,TiI}$ & 0.05 & 0.05 & 0.09 & 0.08 & 0.08 & 0.18 & \nodata & 0.14 & 0.23 & 0.09 & 0.08 & 0.06 \\
$\sigma_{TiI}$ & 0.06 & 0.06 & 0.09 & 0.08 & 0.09 & 0.19 & \nodata & 0.14 & 0.24 & 0.12 & 0.08 & 0.06 \\
N$_{TiI}$ & 4 & 6 & 5 & 8 & 4 & 3 & \nodata & 4 & 3 & 11 & 6 & 8 \\
\\

$\rm{[TiII/Fe]}$ & 0.35 & 0.27 & 0.23 & 0.12 & 0.16 & 0.24 & 0.42 & -0.14 & 0.41 & 0.08 & 0.04 & 0.04 \\
$\sigma_{A,TiII}$ & 0.07 & 0.00 & 0.01 & 0.05 & 0.03 & 0.00 & 0.13 & 0.01 & 0.01 & 0.08 & 0.00 & 0.01 \\
$\sigma_{N,TiII}$ & 0.08 & 0.08 & 0.13 & 0.12 & 0.09 & 0.10 & 0.15 & 0.19 & 0.11 & 0.11 & 0.10 & 0.06 \\
$\sigma_{TiII}$ & 0.10 & 0.08 & 0.13 & 0.13 & 0.09 & 0.10 & 0.20 & 0.19 & 0.11 & 0.14 & 0.10 & 0.06 \\
N$_{TiII}$ & 8 & 3 & 4 & 7 & 9 & 1 & 8 & 2 & 2 & 5 & 4 & 10 \\
\\

$\rm{[V/Fe]}$ & \nodata & -0.01 & -0.00 & -0.14 & \nodata & 0.05 & \nodata & 0.12 & -0.10 & 0.40 & 0.18 & 0.26 \\
$\sigma_{A,V}$ & \nodata & 0.02 & 0.04 & 0.00 & \nodata & 0.07 & \nodata & 0.02 & 0.06 & 0.02 & 0.05 & 0.02 \\
$\sigma_{N,V}$ & \nodata & 0.04 & 0.10 & 0.04 & \nodata & 0.10 & \nodata & 0.06 & 0.07 & 0.05 & 0.06 & 0.04 \\
$\sigma_{V}$ & \nodata & 0.05 & 0.11 & 0.04 & \nodata & 0.12 & \nodata & 0.06 & 0.09 & 0.05 & 0.08 & 0.04 \\
N$_{V}$ & \nodata & 11 & 5 & 5 & \nodata & 1 & \nodata & 7 & 5 & 9 & 11 & 3 \\
\\

$\rm{[Cr/Fe]}$ & -0.14 & -0.07 & -0.04 & -0.13 & -0.18 & -0.17 & -0.34 & 0.07 & -0.07 & 0.10 & 0.12 & -0.09 \\
$\sigma_{A,Cr}$ & 0.00 & 0.00 & 0.02 & 0.00 & 0.00 & 0.04 & 0.03 & 0.03 & 0.03 & 0.01 & 0.02 & 0.04 \\
$\sigma_{N,Cr}$ & 0.18 & 0.08 & 0.11 & 0.07 & 0.09 & 0.14 & 0.21 & 0.11 & 0.08 & 0.05 & 0.06 & 0.07 \\
$\sigma_{Cr}$ & 0.18 & 0.08 & 0.11 & 0.07 & 0.09 & 0.15 & 0.21 & 0.11 & 0.09 & 0.05 & 0.06 & 0.09 \\
N$_{Cr}$ & 3 & 8 & 3 & 9 & 5 & 2 & 3 & 5 & 5 & 13 & 6 & 9 \\
\\

$\rm{[Mn/Fe]}$ & -0.24 & -0.26 & -0.42 & -0.36 & \nodata & \nodata & \nodata & -0.31 & -0.30 & 0.14 & -0.06 & -0.28 \\
$\sigma_{A,Mn}$ & 0.03 & 0.00 & 0.01 & 0.01 & \nodata & \nodata & \nodata & 0.02 & 0.04 & 0.04 & 0.05 & 0.01 \\
$\sigma_{N,Mn}$ & 0.10 & 0.08 & 0.15 & 0.07 & \nodata & \nodata & \nodata & 0.15 & 0.13 & 0.06 & 0.06 & 0.05 \\
$\sigma_{Mn}$ & 0.10 & 0.08 & 0.15 & 0.07 & \nodata & \nodata & \nodata & 0.15 & 0.13 & 0.07 & 0.07 & 0.05 \\
N$_{Mn}$ & 1 & 5 & 3 & 4 & \nodata & \nodata & \nodata & 4 & 2 & 4 & 4 & 4 \\

\enddata

\tablecomments{IL abundance ratio results. All ratios are taken with respect to the Fe I solutions and calculated with the solar differential abundances line by line. The solutions for the youngest and oldest CMDs for each cluster have been averaged, and  $\sigma_{A,X}$ corresponds to the uncertainy due to the assumed age of the CMD.  $\sigma_{N,X}$ corresponds to the error in the mean of the abundance of different lines,  and the number of lines measured for each species is listed as N$_{X}$.  The total error is listed as  $\sigma_{X}$, which corresponds to  $\sigma_{A,X}$ and $\sigma_{N,X}$ added in quadrature. For elements where only one line of a given species was measured we have assigned a typical error of  $\sigma_{N,X}$=0.1. }
\end{deluxetable*}

\begin{deluxetable*}{crrrrrrrrrrrr}
\tablecaption{ILS Abundance Results Continued\label{tab:bigab}}

\tablehead{
&\colhead{F3 }&\colhead{n104 }&\colhead{n2808 }&\colhead{n362 }&\colhead{n6093}&\colhead{n6388}&\colhead{n6397 }&\colhead{n6440}&\colhead{ n6441 }&\colhead{n6528 }&\colhead{n6553 }&\colhead{n6752 }}

\startdata
$\rm{[Co/Fe]}$ & \nodata & 0.17 & 0.09 & 0.11 & \nodata & -0.10 & \nodata & 0.20 & 0.24 & 0.28 & 0.09 & 0.01 \\
$\sigma_{A,Co}$ & \nodata & 0.02 & 0.01 & 0.00 & \nodata & 0.00 & \nodata & 0.00 & 0.04 & 0.01 & 0.05 & 0.04 \\
$\sigma_{N,Co}$ & \nodata & 0.02 & 0.05 & 0.05 & \nodata & 0.00 & \nodata & 0.09 & 0.11 & 0.10 & 0.14 & 0.10 \\
$\sigma_{Co}$ & \nodata & 0.03 & 0.05 & 0.05 & \nodata & 0.00 & \nodata & 0.09 & 0.12 & 0.10 & 0.15 & 0.11 \\
N$_{Co}$ & \nodata & 6 & 4 & 8 & \nodata & 2 & \nodata & 5 & 3 & 4 & 4 & 1 \\
\\

$\rm{[Ni/Fe]}$ & 0.13 & -0.08 & -0.05 & -0.13 & -0.03 & -0.05 & -0.06 & -0.15 & 0.01 & -0.04 & -0.10 & -0.01 \\
$\sigma_{A,Ni}$ & 0.03 & 0.01 & 0.04 & 0.00 & 0.05 & 0.02 & 0.10 & 0.03 & 0.04 & 0.00 & 0.01 & 0.04 \\
$\sigma_{N,Ni}$ & 0.02 & 0.04 & 0.05 & 0.03 & 0.11 & 0.11 & 0.10 & 0.06 & 0.06 & 0.05 & 0.05 & 0.05 \\
$\sigma_{Ni}$ & 0.04 & 0.05 & 0.06 & 0.03 & 0.12 & 0.11 & 0.14 & 0.07 & 0.07 & 0.05 & 0.05 & 0.06 \\
N$_{Ni}$ & 2 & 20 & 10 & 20 & 8 & 4 & 1 & 12 & 13 & 18 & 18 & 15 \\
\\

$\rm{[Cu/Fe]}$ & \nodata & -0.01 & -0.56 & -0.56 & \nodata & \nodata & \nodata & 0.58 & 0.31 & 0.30 & 0.23 & \nodata \\
$\sigma_{A,Cu}$ & \nodata & 0.01 & 0.06 & 0.01 & \nodata & \nodata & \nodata & 0.03 & 0.04 & 0.01 & 0.04 & \nodata \\
$\sigma_{N,Cu}$ & \nodata & 0.14 & 0.10 & 0.00 & \nodata & \nodata & \nodata & 0.10 & 0.28 & 0.35 & 0.21 & \nodata \\
$\sigma_{Cu}$ & \nodata & 0.14 & 0.12 & 0.01 & \nodata & \nodata & \nodata & 0.10 & 0.29 & 0.35 & 0.21 & \nodata \\
N$_{Cu}$ & \nodata & 2 & 1 & 2 & \nodata & \nodata & \nodata & 1 & 2 & 2 & 2 & \nodata \\
\\

$\rm{[Y/Fe]}$ & \nodata & -0.10 & -0.23 & -0.01 & -0.01 & \nodata & \nodata & \nodata & \nodata & -0.04 & \nodata & 0.09 \\
$\sigma_{A,Y}$ & \nodata & 0.01 & 0.02 & 0.01 & 0.04 & \nodata & \nodata & \nodata & \nodata & 0.01 & \nodata & 0.01 \\
$\sigma_{N,Y}$ & \nodata & 0.28 & 0.07 & 0.07 & 0.07 & \nodata & \nodata & \nodata & \nodata & 0.10 & \nodata & 0.06 \\
$\sigma_{Y}$ & \nodata & 0.28 & 0.07 & 0.07 & 0.08 & \nodata & \nodata & \nodata & \nodata & 0.10 & \nodata & 0.06 \\
N$_{Y}$ & \nodata & 2 & 3 & 2 & 2 & \nodata & \nodata & \nodata & \nodata & 1 & \nodata & 5 \\
\\

$\rm{[Zr/Fe]}$ & \nodata & -0.03 & \nodata & 0.17 & \nodata & \nodata & \nodata & -0.42 & \nodata & \nodata & 0.33 & \nodata \\
$\sigma_{A,Zr}$ & \nodata & 0.04 & \nodata & 0.04 & \nodata & \nodata & \nodata & 0.11 & \nodata & \nodata & 0.04 & \nodata \\
$\sigma_{N,Zr}$ & \nodata & 0.10 & \nodata & 0.10 & \nodata & \nodata & \nodata & 0.10 & \nodata & \nodata & 0.10 & \nodata \\
$\sigma_{Zr}$ & \nodata & 0.11 & \nodata & 0.11 & \nodata & \nodata & \nodata & 0.15 & \nodata & \nodata & 0.11 & \nodata \\
N$_{Zr}$ & \nodata & 1 & \nodata & 1 & \nodata & \nodata & \nodata & 1 & \nodata & \nodata & 1 & \nodata \\
\\

$\rm{[Ba/Fe]}$ & 0.12 & 0.32 & 0.11 & 0.25 & 0.03 & 0.00 & -0.04 & 0.18 & 0.20 & 0.35 & -0.04 & 0.08 \\
$\sigma_{A,Ba}$ & 0.00 & 0.11 & 0.04 & 0.02 & 0.03 & 0.00 & 0.15 & 0.04 & 0.04 & 0.01 & 0.03 & 0.04 \\
$\sigma_{N,Ba}$ & 0.08 & 0.00 & 0.00 & 0.07 & 0.11 & 0.10 & 0.18 & 0.07 & 0.07 & 0.07 & 0.08 & 0.04 \\
$\sigma_{Ba}$ & 0.08 & 0.11 & 0.04 & 0.07 & 0.11 & 0.10 & 0.24 & 0.08 & 0.08 & 0.07 & 0.09 & 0.06 \\
N$_{Ba}$ & 5 & 2 & 2 & 3 & 5 & 1 & 4 & 2 & 2 & 2 & 3 & 3 \\
\\

$\rm{[La/Fe]}$ & \nodata & 0.02 & \nodata & 0.47 & \nodata & \nodata & \nodata & 0.03 & 0.16 & 0.26 & 0.32 & \nodata \\
$\sigma_{A,La}$ & \nodata & 0.04 & \nodata & 0.04 & \nodata & \nodata & \nodata & 0.04 & 0.03 & 0.01 & 0.10 & \nodata \\
$\sigma_{N,La}$ & \nodata & 0.07 & \nodata & 0.00 & \nodata & \nodata & \nodata & 0.10 & 0.10 & 0.00 & 0.21 & \nodata \\
$\sigma_{La}$ & \nodata & 0.08 & \nodata & 0.04 & \nodata & \nodata & \nodata & 0.11 & 0.10 & 0.01 & 0.24 & \nodata \\
N$_{La}$ & \nodata & 2 & \nodata & 2 & \nodata & \nodata & \nodata & 1 & 1 & 2 & 2 & \nodata \\
\\

$\rm{[Nd/Fe]}$ & \nodata & 0.10 & 0.28 & 0.39 & \nodata & \nodata & \nodata & \nodata & \nodata & 0.59 & 0.29 & 0.61 \\
$\sigma_{A,Nd}$ & \nodata & 0.00 & 0.01 & 0.03 & \nodata & \nodata & \nodata & \nodata & \nodata & 0.01 & 0.05 & 0.04 \\
$\sigma_{N,Nd}$ & \nodata & 0.14 & 0.10 & 0.02 & \nodata & \nodata & \nodata & \nodata & \nodata & 0.15 & 0.07 & 0.10 \\
$\sigma_{Nd}$ & \nodata & 0.14 & 0.10 & 0.04 & \nodata & \nodata & \nodata & \nodata & \nodata & 0.15 & 0.09 & 0.11 \\
N$_{Nd}$ & \nodata & 3 & 1 & 6 & \nodata & \nodata & \nodata & \nodata & \nodata & 3 & 4 & 1 \\
\\

$\rm{[Eu/Fe]}$ & 0.79 & 0.37 & \nodata & 0.40 & \nodata & \nodata & \nodata & 0.63 & \nodata & \nodata & 0.42 & \nodata \\
$\sigma_{A,Eu}$ & 0.01 & 0.04 & \nodata & 0.00 & \nodata & \nodata & \nodata & 0.04 & \nodata & \nodata & 0.04 & \nodata \\
$\sigma_{N,Eu}$ & 0.10 & 0.10 & \nodata & 0.10 & \nodata & \nodata & \nodata & 0.10 & \nodata & \nodata & 0.10 & \nodata \\
$\sigma_{Eu}$ & 0.10 & 0.11 & \nodata & 0.10 & \nodata & \nodata & \nodata & 0.11 & \nodata & \nodata & 0.11 & \nodata \\
N$_{Eu}$ & 1 & 1 & \nodata & 1 & \nodata & \nodata & \nodata & 1 & \nodata & \nodata & 1 & \nodata \\

\enddata
\tablecomments{Measurements given in the same format as Table \ref{tab:bigab1}.}
\end{deluxetable*}

\begin{deluxetable*}{lrrrr|rrrrrrrrrrr}
\tablecolumns{16}
\tablewidth{0pc}
\tablecaption{Individual Line Abundances\label{tab:nf_stub}}
\tablehead{
\colhead{Species}& \colhead{$\lambda$}&\colhead{EP} & \colhead{log{\it gf}}& \colhead{Sun}& \colhead{F3} & \colhead{n104} &  \colhead{n2808}&  \colhead{n362}  & \colhead{n6093} & \colhead{n6388} & \colhead{n6440}& \colhead{n6441}& \colhead{n6528}& \colhead{n6553}& \colhead{n6752}\\ &\colhead{(\AA)}& \colhead{(eV)} }
\startdata

 Ca    I &  5581.98 &  2.52 &  -0.56 &  6.23 & \nodata &   5.53 & \nodata & \nodata &   5.02 & \nodata & \nodata & \nodata &   6.02 & \nodata &   4.96 \\
     Ca    I &  5588.76 &  2.53 &   0.36 &  5.88 & \nodata & \nodata & \nodata & \nodata &   4.82 & \nodata & \nodata & \nodata & \nodata & \nodata &   4.96 \\
     Ca    I &  5590.13 &  2.52 &  -0.57 &  6.21 & \nodata &   5.73 &   5.54 &   5.08 &   5.02 & \nodata & \nodata & \nodata &   5.92 &   5.90 &   5.06 \\
     Ca    I &  5601.29 &  2.53 &  -0.69 &  6.27 & \nodata &   5.83 &   5.74 &   5.48 &   4.92 &   6.09 &   6.39 &   6.25 &   5.82 &   6.30 &   5.16 \\
     Ca    I &  5857.46 &  2.93 &   0.24 &  6.20 & \nodata &   5.83 &   5.34 &   5.38 & \nodata &   5.99 & \nodata &   6.05 &   6.12 & \nodata &   5.36 \\
     Ca    I &  6102.73 &  1.88 &  -0.79 &  6.45 &   4.22 &   5.83 &   5.44 &   5.38 &   5.22 & \nodata & \nodata & \nodata & \nodata & \nodata &   5.16 \\
     Ca    I &  6122.23 &  1.89 &  -0.32 &  6.42 &   4.42 & \nodata &   5.44 &   5.38 &   5.02 &   5.89 & \nodata & \nodata & \nodata & \nodata &   5.06 \\
     Ca    I &  6162.18 &  1.90 &  -0.09 &  6.40 &   4.32 & \nodata & \nodata &   5.48 &   4.92 & \nodata & \nodata & \nodata & \nodata & \nodata &   5.06 \\
     Ca    I &  6166.44 &  2.52 &  -1.14 &  6.36 & \nodata &   5.83 &   5.24 &   5.28 & \nodata &   6.19 &   6.29 &   5.75 & \nodata &   5.70 &   5.16 \\
     Ca    I &  6169.04 &  2.52 &  -0.80 &  6.34 & \nodata & \nodata & \nodata & \nodata & \nodata & \nodata & \nodata &   5.95 &   6.32 &   5.90 & \nodata \\
     Ca    I &  6439.08 &  2.52 &   0.39 &  6.02 &   4.42 &   5.63 & \nodata &   5.38 &   4.92 &   5.79 &   5.69 &   5.65 & \nodata &   5.80 &   4.96 \\
     Ca    I &  6449.81 &  2.52 &  -0.50 &  6.29 & \nodata &   5.63 & \nodata & \nodata &   4.92 & \nodata & \nodata & \nodata & \nodata & \nodata &   5.06 \\
     Ca    I &  6455.60 &  2.52 &  -1.29 &  6.29 & \nodata &   5.73 & \nodata &   5.58 & \nodata & \nodata &   5.79 &   6.15 &   5.92 &   6.10 &   5.06 \\
     Ca    I &  6462.68 &  2.52 &   0.26 &  6.27 &   4.22 & \nodata &   5.44 & \nodata &   4.72 & \nodata & \nodata & \nodata & \nodata & \nodata &   4.86 \\

\enddata
\tablecomments{Abundance measurements for individual lines. The reference solar abundance for each line is given in Column 5, see text for details. The full table is available in the
  electronic edition of the journal. }
\end{deluxetable*}

\begin{deluxetable}{lrr}
\tablecolumns{3}
\tablewidth{0pc}
\tablecaption{NGC 6397 EWs \label{tab:n6397_abund}}
\tablehead{
\colhead{Species}& \colhead{$\lambda$}&\colhead{EW}   \\ &\colhead{(\AA)}& \colhead{m\AA}  }
\startdata

Mg I & 5528.418 & 51.7 \\
Ca I & 5588.764 & 34.0 \\
Ca I & 5857.459 & 20.3 \\
Ca I & 6102.727 & 35.8 \\
Ca I & 6122.226 & 39.1 \\
Ca I & 6162.180 & 58.0 \\
Ca I & 6439.083 & 45.1 \\
Ca I & 7148.150 & 43.7 \\

Ti II & 4395.040 & 82.9 \\
Ti II & 4399.778 & 64.0 \\
Ti II & 4443.812 & 51.9 \\
Ti II & 4450.491 & 53.5 \\
Ti II & 4468.500 & 75.3 \\
Ti II & 4501.278 & 57.3 \\
Ti II & 4563.766 & 39.9 \\
Ti II & 5188.698 & 43.9 \\

Cr I & 5204.470 & 45.5 \\
Cr I & 5206.044 & 34.9 \\
Cr I & 5409.799 & 15.8 \\
Ni I & 5476.921 & 38.2 \\

Ba II & 4554.036 & 38.8 \\
Ba II & 4934.095 & 52.6 \\
Ba II & 6141.727 & 40.0 \\
Ba II & 6141.727 & 38.1 \\

\enddata
\end{deluxetable}

\begin{deluxetable*}{lrrrrrrrrrrrr}
\tablecaption{Abundance Ratio Data from the Literature\label{tab:biglit}}
\tablehead{
\colhead{[X/Fe] }&\colhead{F3 }&\colhead{n104 }&\colhead{n2808 }&\colhead{n362 }&\colhead{n6093}&\colhead{n6388}&\colhead{n6397 }&\colhead{n6440}&\colhead{ n6441 }&\colhead{n6528 }&\colhead{n6553 }&\colhead{n6752 }}

\startdata
$[{\rm Na/Fe}]_{*}$ & 0.09 & 0.25 & 0.28 & 0.07 & 0.44 & 0.45 & 0.19 & \nodata & \nodata & 0.41 & 0.65 & 0.27 \\
$\sigma_{Na,*}$ & 0.37 & 0.06 & 0.23 & 0.05 & 0.22 & 0.20 & 0.04 & \nodata & \nodata & 0.02 & 0.10 & 0.05 \\
Ref & 1&4& 3& 2& 1& 2& 3& \nodata & \nodata& 2& 2& 3\\

$[{\rm Mg/Fe}]_{*}$ & 0.07 & 0.43 & 0.27 & 0.34 & 0.45 & 0.24 & 0.26 & 0.33 & 0.33 & 0.19 & 0.34 & 0.25 \\
$\sigma_{Mg,*}$ & 0.37 & 0.03 & 0.07 & 0.02 & 0.10 & 0.04 & 0.14 & 0.03 & 0.06 & 0.15 & 0.07 & 0.22 \\
Ref & 1&3& 2& 2& 1& 2& 3& 1& 2& 3&  3& 3\\

$[{\rm Al/Fe}]_{*}$ & \nodata & 0.33 & 0.45 & 0.28 & 0.42 & 0.44 & 0.54 & 0.46 & 0.53 & \nodata & 0.34 & 0.46 \\
$\sigma_{Al,*}$ & \nodata & 0.13 & 0.10 & 0.05 & 0.08 & 0.21 & 0.36 & 0.05 & 0.10 & \nodata & 0.23 & 0.31 \\
Ref & \nodata&4& 1& 2& 2& 2& 3& 1&1 & \nodata & 2 &4\\

$[{\rm Si/Fe}]_{*}$ & \nodata & 0.33 & 0.39 & 0.30 & 0.34 & 0.30 & 0.34 & 0.32 & 0.33 & 0.25 & 0.23 & 0.29 \\
$\sigma_{Si},*$ & \nodata & 0.05 & 0.04 & 0.08 & 0.10 & 0.02 & 0.14 & 0.10 & 0.08 & 0.15 & 0.11 & 0.05 \\
Ref & \nodata&4& 3& 2& 1& 2& 2& 1& 2& 3& 3& 2\\

$[{\rm Ca/Fe}]_{*}$ & 0.24 & 0.27 & 0.42 & 0.25 & 0.30 & 0.01 & 0.33 & 0.37 & 0.18 & 0.07 & 0.21 & 0.27 \\
$\sigma_{Ca,*}$ & 0.03 & 0.06 & 0.10 & 0.11 & 0.08 & 0.08 & 0.12 & 0.04 & 0.13 & 0.41 & 0.14 & 0.04 \\
Ref & 1&4& 3& 2& 2& 2& 2& 1& 2& 3& 3& 3\\

$[{\rm Sc/Fe}]_{*}$ & \nodata & 0.17 & -0.04 & -0.07 & -0.03 & -0.01 & 0.11 & \nodata & 0.11 & -0.05 & -0.12 & 0.00 \\
$\sigma_{Sc,*}$ & \nodata & 0.06 & 0.10 & 0.02 & 0.10 & 0.09 & 0.10 & \nodata & 0.06 & 0.10 & 0.10 & 0.06 \\
Ref & \nodata&2& 1& 2& 1& 2& 1& \nodata &  1& 1&1 & 2\\

$[{\rm TiI/Fe}]_{*}$ & -0.04 & 0.32 & 0.03 & 0.24 & 0.15 & 0.21 & 0.28 & 0.33 & 0.31 & 0.08 & 0.23 & 0.15 \\
$\sigma_{TiI,*}$ & 0.12 & 0.04 & 0.07 & 0.09 & 0.05 & 0.22 & 0.02 & 0.08 & 0.02 & 0.21 & 0.26 & 0.01 \\
Ref & 1&4&  2& 2& 2& 2& 2&  1 & 2& 3& 3&2\\

$[{\rm TiII/Fe}]_{*}$ & 0.18 & 0.39 & 0.20 & 0.21 & 0.32 & 0.30 & 0.31 & \nodata & \nodata & -0.12 & 0.12 & 0.40 \\
$\sigma_{TiII,*}$ & 0.13 & 0.02 & 0.10 & 0.10 & 0.20 & 0.10 & 0.10 & \nodata & \nodata & 0.1 & 0.19 & 0.20 \\
Ref & 1&3&  1& 1& 2& 1& 1& \nodata & \nodata& 1 & 2&1\\

$[{\rm V/Fe}]_{*}$ & \nodata & 0.11 & 0.04 & -0.03 & -0.04 & 0.30 & \nodata & \nodata & 0.15 & -0.20 & \nodata & -0.28 \\
$\sigma_{V,*}$ & \nodata & 0.08 & 0.10 & 0.02 & 0.10 & 0.13 & \nodata & \nodata & 0.20 & 0.10 & \nodata & 0.10 \\
Ref & \nodata&2& 1& 2& 1& 2& \nodata& \nodata & 1& 1& \nodata&1\\

$[{\rm Cr/Fe}]_{*}$ & -0.36 & 0.04 & 0.03 & -0.03 & -0.04 & -0.06 & 0.02 & \nodata & 0.13 & 0.00 & 0.04 & -0.07 \\
$\sigma_{Cr,*}$ & 0.12 & 0.10 & 0.10 & 0.10 & 0.02 & 0.03 & 0.10 & \nodata & 0.03 & 0.10 & 0.10 & 0.03 \\
Ref & 1&2& 1& 1& 2& 2& 1&\nodata & 1& 1& 1&2\\

$[{\rm Mn/Fe}]_{*}$ & 0.01 & -0.24 & -0.34 & -0.33 & -0.48 & -0.10 & -0.54 & \nodata & \nodata & -0.37 & \nodata & -0.45 \\
$\sigma_{Mn,*}$ & 0.03 & 0.06 & 0.11 & 0.10 & 0.10 & 0.21 & 0.10 & \nodata & \nodata & 0.10 & \nodata & 0.10 \\
Ref & 1&2& 2& 1& 1& 2& 1& \nodata & \nodata &1 & \nodata &1\\

$[{\rm Co/Fe}]_{*}$ & \nodata & -0.00 & \nodata & -0.11 & -0.22 & 0.07 & 0.10 & \nodata & \nodata & \nodata & \nodata & -0.02 \\
$\sigma_{Co,*}$ & \nodata & 0.10 & \nodata & 0.13 & 0.10 & 0.04 & 0.10 & \nodata & \nodata & \nodata & \nodata & 0.10 \\
Ref & \nodata&1& \nodata& 1 & 1& 2& 1&\nodata &\nodata & \nodata & \nodata &1\\

$[{\rm Ni/Fe}]_{*}$ & 0.05 & -0.02 & -0.07 & -0.08 & -0.09 & 0.04 & -0.14 & \nodata & \nodata & 0.10 & -0.06 & -0.08 \\
$\sigma_{Ni,*}$ & 0.14 & 0.09 & 0.10 & 0.02 & 0.06 & 0.02 & 0.10 & \nodata & \nodata & 0.10 & 0.09 & 0.04 \\
Ref & 1&3&  1& 2& 2& 2& 1&\nodata &\nodata &1 & 1 &3\\

$[{\rm Cu/Fe}]_{*}$ & \nodata & -0.14 & \nodata & -0.49 & -0.52 & \nodata & -0.89 & \nodata & \nodata & \nodata & \nodata & -0.61 \\
$\sigma_{Cu,*}$ & \nodata & 0.10 & \nodata & 0.12 & 0.10 & \nodata & 0.10 & \nodata & \nodata & \nodata & \nodata & 0.10 \\
Ref & \nodata&1& \nodata& 1& 1& \nodata& 1&\nodata &\nodata & \nodata & \nodata  &1\\

$[{\rm Y/Fe}]_{*}$ & -0.22 & 0.07 & \nodata & 0.18 & -0.07 & \nodata & -0.26 & \nodata & \nodata & \nodata & 0.07 & -0.01 \\
$\sigma_{Y,*}$ & 0.03 & 0.10 & \nodata & 0.16 & 0.10 & \nodata & 0.10 & \nodata & \nodata & \nodata & 0.10 & 0.01 \\
Ref & 1&1& \nodata& 2& 1& \nodata& 1&\nodata &\nodata & \nodata & 1&2\\

$[{\rm Zr/Fe}]_{*}$ & \nodata & 0.41 & \nodata & 0.42 & -0.03 & -0.12 & \nodata & \nodata & \nodata & \nodata & -0.54 & 0.18 \\
$\sigma_{Zr,*}$ & \nodata & 0.1 & \nodata & 0.11 & 0.10 & 0.05 & \nodata & \nodata & \nodata & \nodata & 0.19 & 0.10 \\
Ref & \nodata&\nodata & 1& 2& 1& 2&\nodata &\nodata & \nodata & \nodata & 2&1\\

$[{\rm Ba/Fe}]_{*}$ & 0.21 & 0.25 & 0.25 & 0.35 & 0.16 & 0.18 & -0.08 & \nodata & 0.14 & 0.14 & -0.19 & 0.06 \\
$\sigma_{Ba,*}$ & 0.10 & 0.10 & 0.01 & 0.19 & 0.16 & 0.04 & 0.10 & \nodata & 0.05 & 0.10 & 0.13 & 0.17 \\
Ref & 1&1& 2 & 3 & 1& 2& 1&\nodata & 1&  1& 2&2\\

$[{\rm La/Fe}]_{*}$ & 0.79 & 0.14 & \nodata & 0.35 & 0.33 & 0.22 & \nodata & \nodata & \nodata & \nodata & 0.01 & 0.12 \\
$\sigma_{La,*}$ & 0.23 & 0.10 & \nodata & 0.16 & 0.07 & 0.20 & \nodata & \nodata & \nodata & \nodata & 0.17 & 0.02 \\
Ref & 1&2& \nodata& 2 & 2& 2& \nodata &\nodata &\nodata & \nodata& 2&2\\

$[{\rm Nd/Fe}]_{*}$ & 0.61 & 0.04 & \nodata & 0.46 & 0.22 & 0.33 & 0.18 & \nodata & \nodata & \nodata & \nodata & 0.22 \\
$\sigma_{Nd,*}$ & 0.15 & 0.10 & \nodata & 0.15 & 0.10 & 0.10 & 0.10 & \nodata & \nodata & \nodata & \nodata & 0.10 \\
Ref & 1&1& \nodata&  2& 1&1 & 1&\nodata &\nodata & \nodata & \nodata &1\\

$[{\rm Eu/Fe}]_{*}$ & 0.88 & 0.38 & \nodata & 0.68 & 0.66 & 0.24 & 0.40 & \nodata & \nodata & \nodata & 0.05 & 0.43 \\
$\sigma_{Eu,*}$ & 0.10 & 0.08 & \nodata & 0.11 & 0.20 & 0.07 & 0.10 & \nodata & \nodata & \nodata & 0.07 & 0.12 \\
Ref & 1&2&  \nodata& 3 & 2& 2& 1&\nodata &\nodata & \nodata & 2&3\\

\enddata
\tablecomments{ Reference abundances used in the comparisons in \textsection \ref{sec:others}. Each ratio corresponds to the mean of the ratios measured in studies of individual stars from the references listed in Table \ref{tab:source}.  $\sigma_{X,*}$ corresponds to the standard deviation of the abundance ratio from different studies. Note that different sets of elements are given in each reference, for this reason we explicitly give the number of studies used in the mean for each ratio. For ratios where only one study was used we assign a typical dispersion of 0.1 dex.
For NGC 6440 and Fornax 3 only one reference was available and the $\sigma_{X,*}$ given corresponds to the standard deviation in abundances between individual stars in \cite{origlia08} and \cite{letarte}, respectively.   }
\end{deluxetable*}

\section{Results for Abundance Ratios}
\label{sec:others}

In this section we compare the cluster IL abundance ratios, which are given in Table \ref{tab:bigab1} and Table \ref{tab:bigab}, to the abundances from individual GC stars from the references listed in Table \ref{tab:source}.  Our primary goal is to determine if there are any large systematic offsets between the IL abundances and the abundances measured from individual stars.    With the exception of NGC 6397, all of the abundances in this section are calculated from line synthesis matching in the same way as the line synthesis matching for Fe discussed above, and the synthesized spectra include nearby blends of other atoms and molecular lines (without TiO). 
The abundance measurements obtained via line synthesis matching for the individual lines are given in Table \ref{tab:nf_stub}, and the EWs measured for NGC 6397 are given in Table \ref{tab:n6397_abund}.

 As in our previous work, all of the abundance measurements are calculated in LTE with no corrections for NLTE effects  applied, although NLTE effects can be significant for some elements \citep[e.g.][]{nlte1,mgnlte,tinlte}.  We have tried to limit our comparisons to abundances from the literature to studies that also use the LTE assumption.   We note that NLTE abundance corrections could potentially improve comparisons for some elements.  Since the synthetic stellar populations include a large range in stellar types, more detailed modeling is necessary to attempt NLTE corrections for the IL abundances, which is beyond the scope of the current paper, but may prove helpful in the future.

Another goal in this section is to evaluate the impact of line-by-line differential abundances, similar to that performed  by \cite{sakari}, who found that line-by-line differential abundances could reduce systematic offsets, particularly for Mg I.  To that end, we calculate abundances in two ways: first in an "absolute" sense, by which we mean abundance ratios are calculated with a standard reference solar abundance distribution  from \cite{asplundreview}, and second in a "differential" sense,  where we use our own solar abundance analysis to calculate ratios relative to the solar abundance we derive for each line.   We use a solar spectrum from the \cite{2005MSAIS...8..189K} solar flux atlas and like \cite{sakari} we use solar atmospheric parameters of T$_{\rm{eff}}$=5777 K, log {\it g}=4.44 dex, $\chi$=0.85 km s$^{-1}$ and [M/H]=0.0 \citep{2005AJ....130..597Y}. We derive Fe$_ {\odot}$=7.50 from an EW analysis.  For other elements in the Sun we primarily use EW derived abundances, but we use abundances derived from synthesis for those requiring hyperfine splitting (HFS), which include Sc II, V I, Mn I, Co I, Cu I,  Zr I, Ba II, La II, and  Eu II, as well as Y II and Nd II, for which abundances are determined from very few lines, and Mg I, which has strong wings in the solar spectrum.  Like \cite{sakari}, we eliminate Na I from the differential abundance analysis due to complications from non-LTE effects.  The solar abundance for each line is listed in  Table \ref{tab:nf_stub} along with the cluster IL measurements.  We note that it is not always possible to derive a solar abundance for a line we use in the IL analysis if the line is too strong or too blended in the solar spectrum.  For these lines we use the mean solar abundance derived from the lines we do measure, with the exception of  Nd II, La II, Y II and Eu II,  where we use the \cite{asplundreview} value because we only have solar measurements for one or two lines in common with the ILS analysis.

\begin{figure*}
\centering
\includegraphics[scale=0.4]{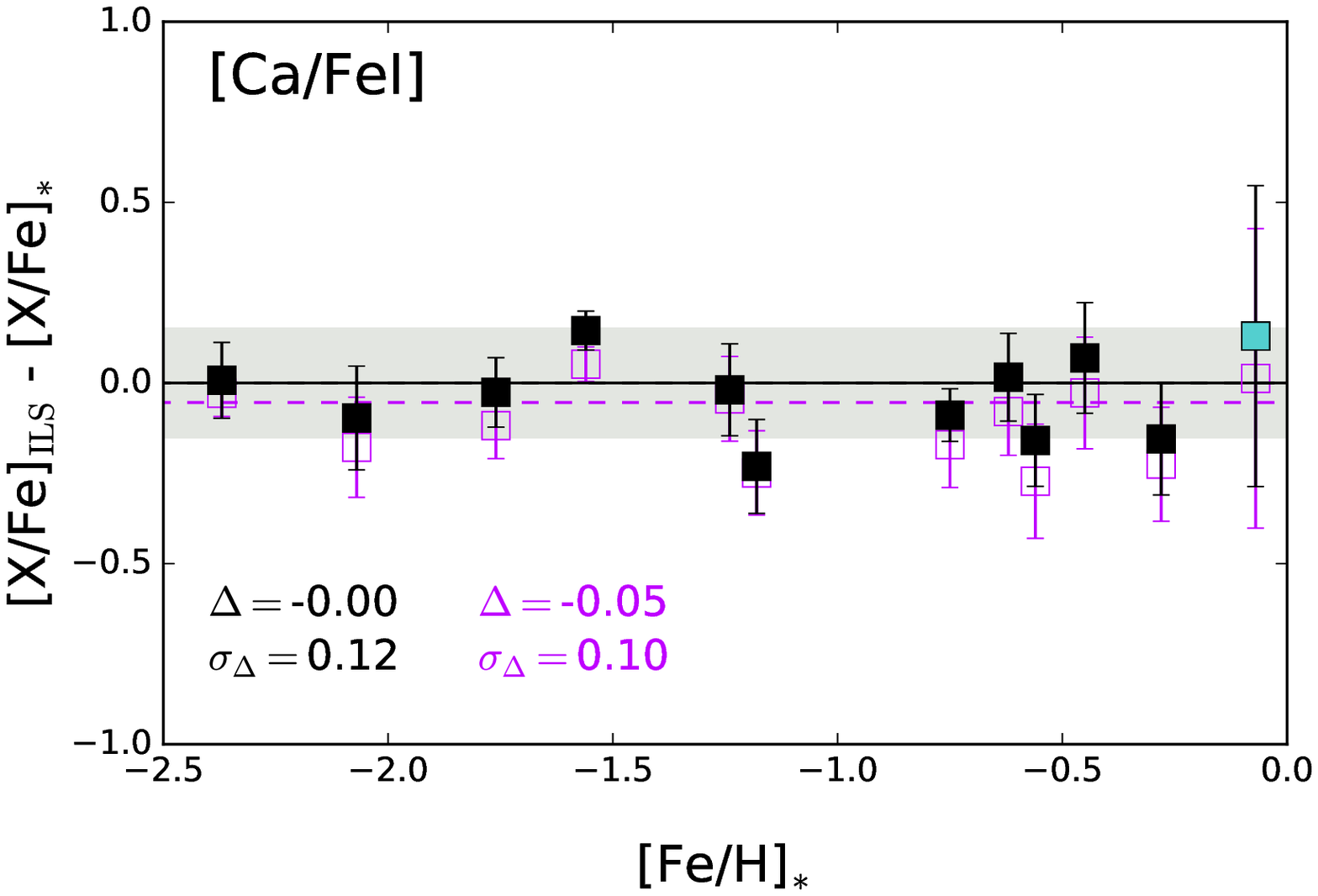}
\includegraphics[scale=0.4]{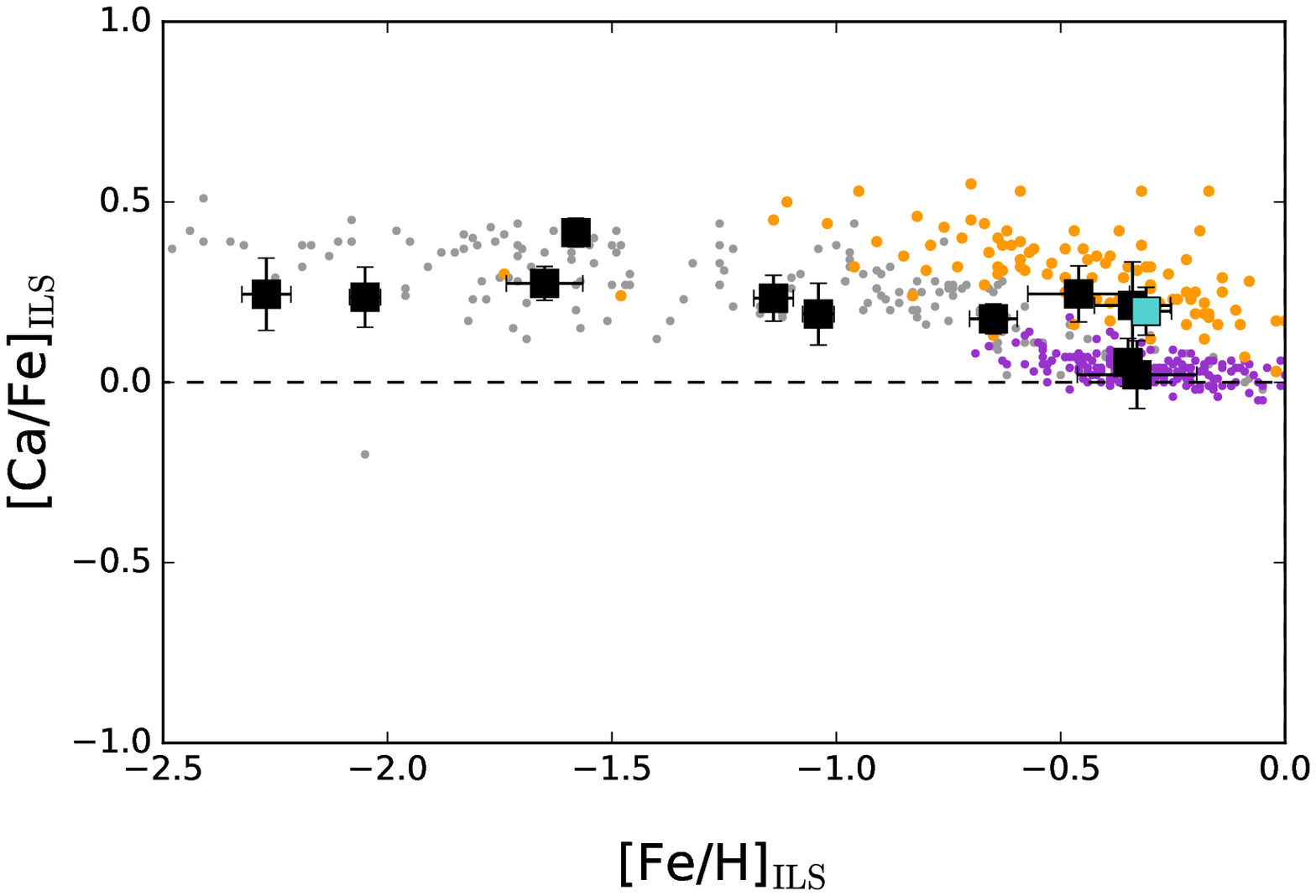}
\includegraphics[scale=0.4]{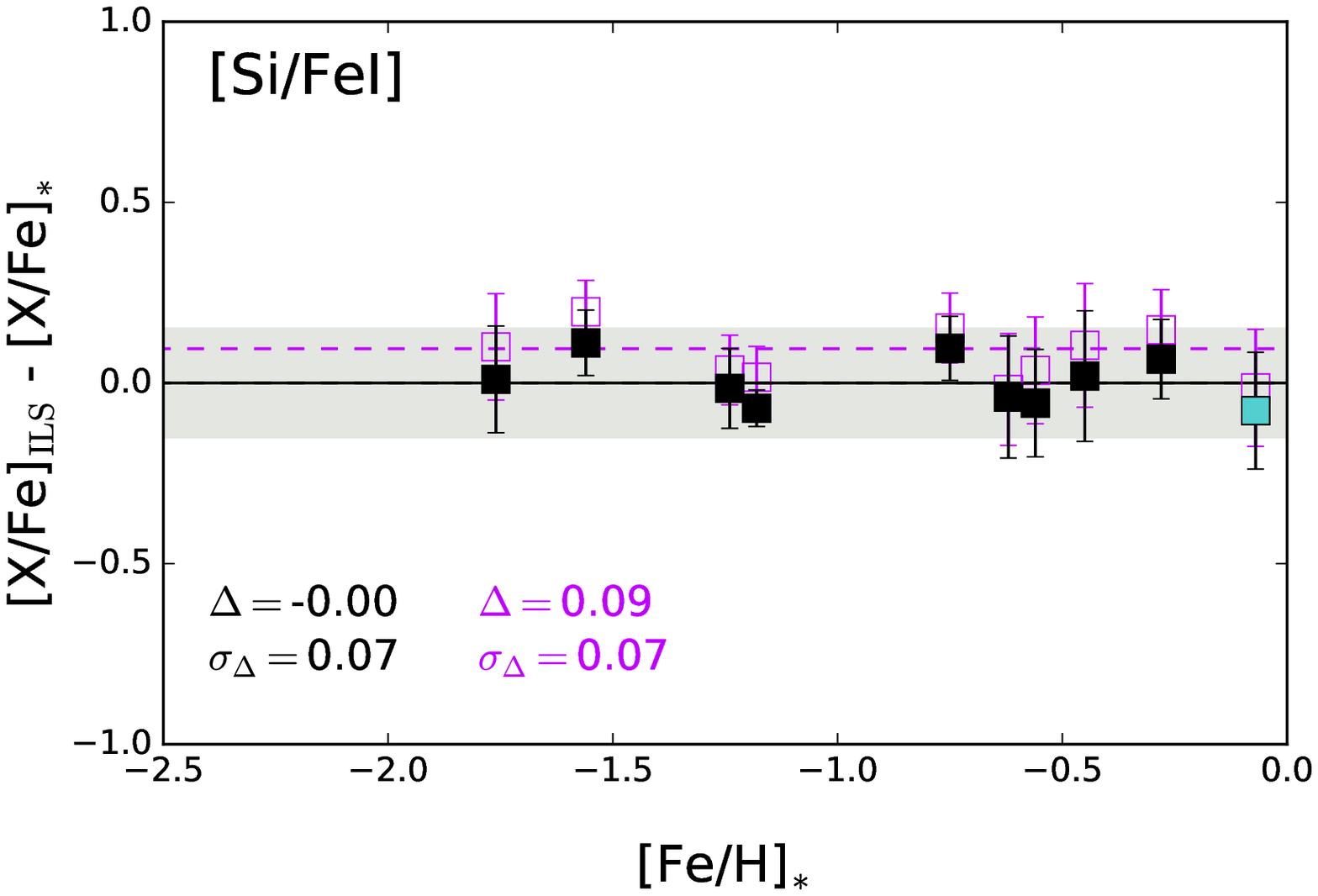}
\includegraphics[scale=0.4]{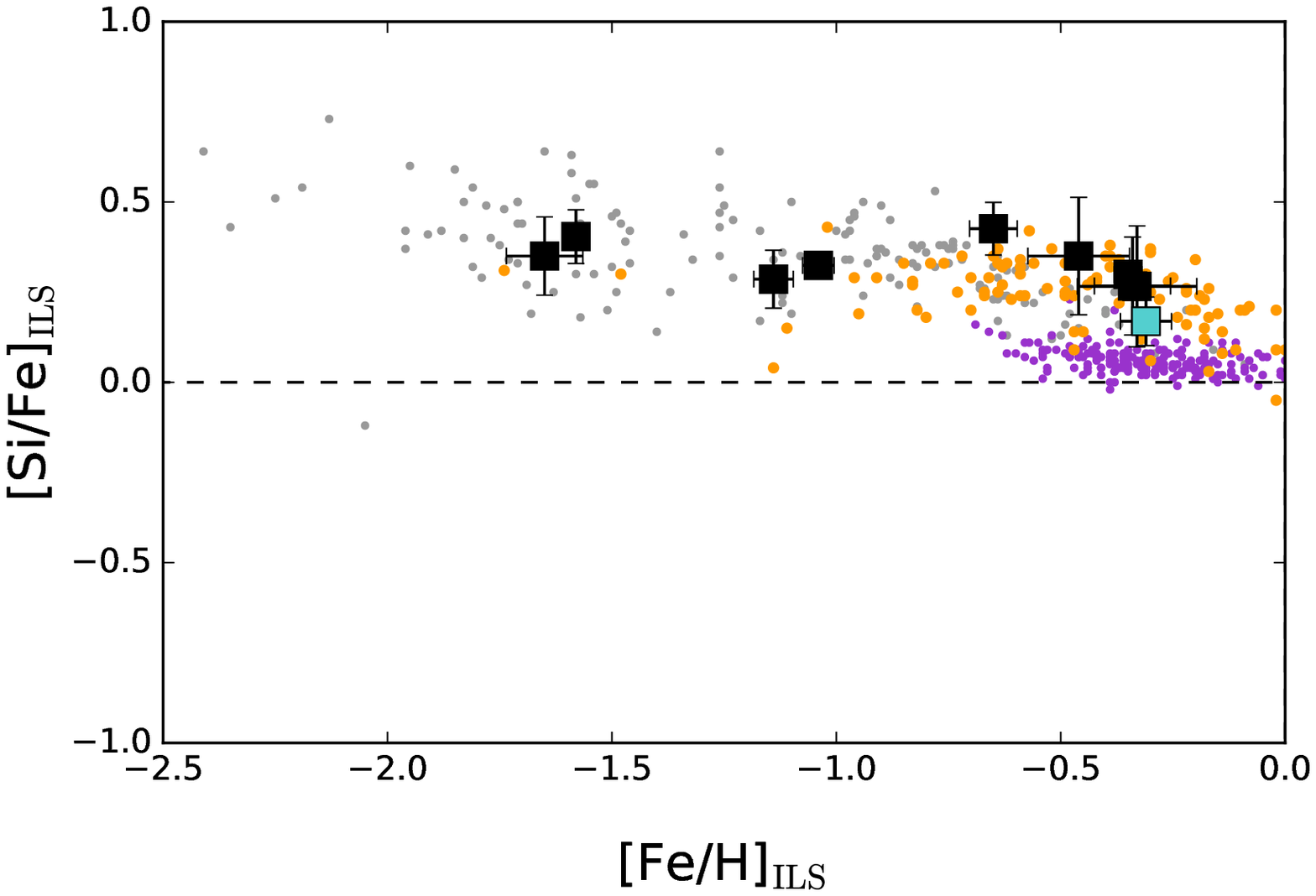}
\includegraphics[scale=0.4]{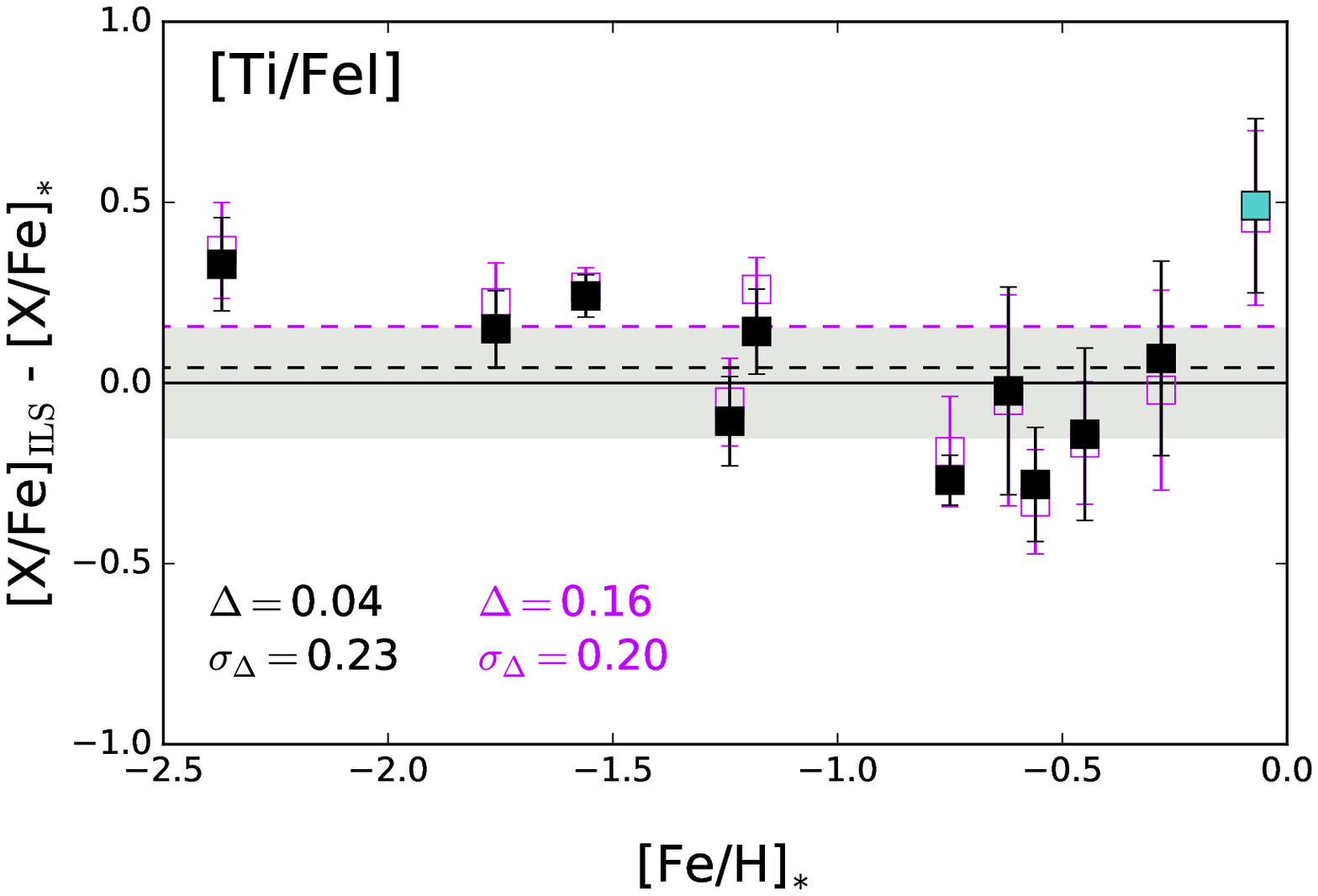}
\includegraphics[scale=0.4]{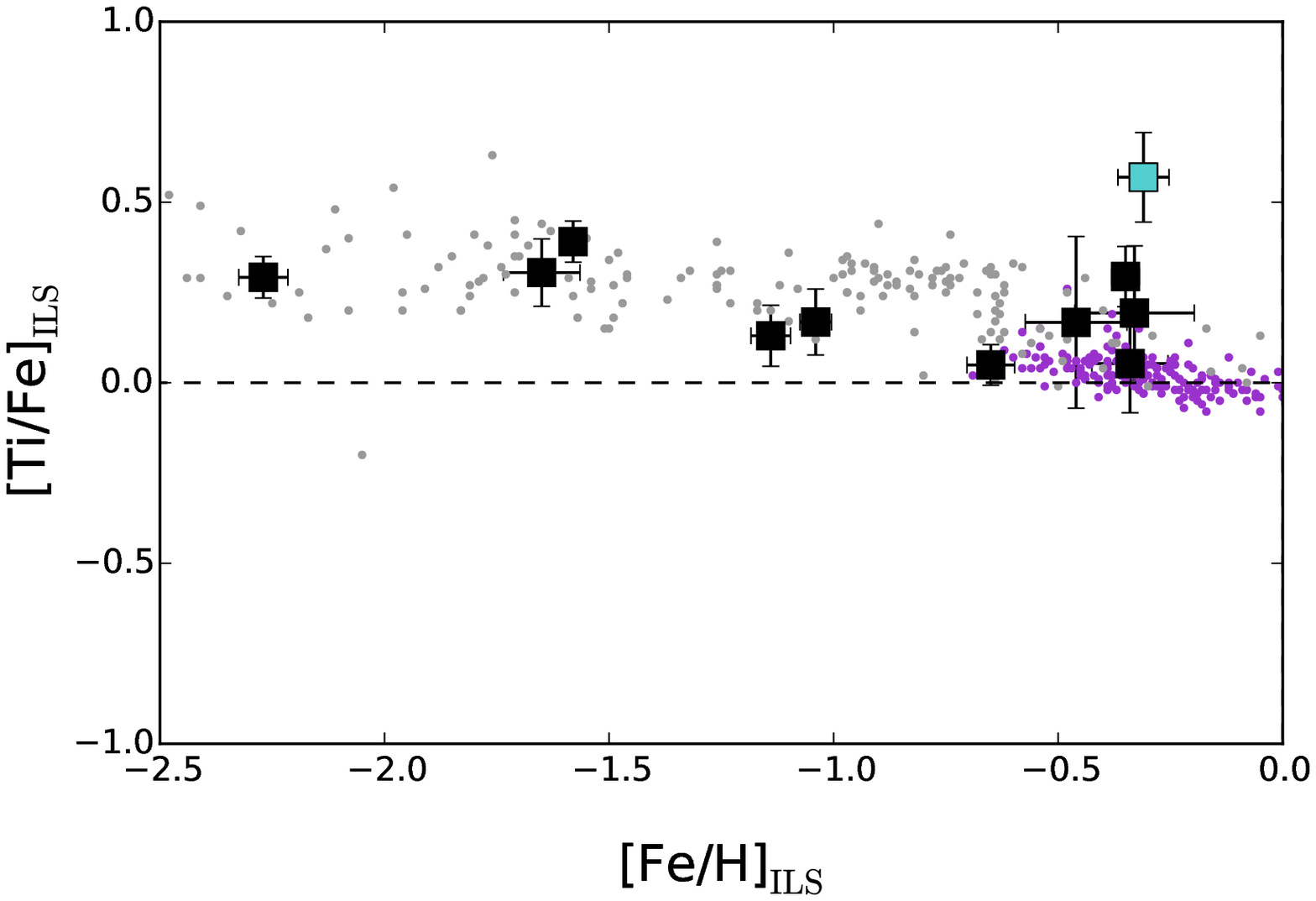}

\caption{ Results for the abundances of [Ca/Fe] (top panels), [Si/Fe] (middle panels), and [TiI/Fe] (bottom panels).   Left panels compare the abundances derived from the IL spectra in this work, [X/Fe]$_{ILS}$, to the mean abundance for each cluster from the stellar abundance studies  listed in Table \ref{tab:source}, [X/Fe]$_{*}$, shown as a function of the stellar abundance.  Error bars in the left panels correspond to the uncertainty in the IL abundance, $\sigma_{Tot,X}$, and the standard deviation of the stellar abundance studies added in quadrature.  The residuals for absolute abundances are shown in purple and the residuals for differential abundances are shown in black.
The corresponding dashed lines are drawn at the offset calculated from the weighted mean of the residuals, where the weights correspond to the error bars shown for each point.  The value of the offset is shown as $\Delta$ in the lower part of the panels, and the weighted standard deviation of the residuals is shown as $\sigma_{\Delta}$.  Note that abundances for NGC 6528 are shown by the cyan square, but are not used in the calculation of  $\Delta$ or $\sigma_{\Delta}$, as explained in the text.
  To guide the eye, a solid line is shown at [X/Fe]$_{ILS}$-[X/Fe]$_{*}$=0 for perfect agreement, and a shaded region is shown in grey that corresponds to  $\pm$0.15 dex from perfect agreement, which is the range including most of the offsets for all elements.        The right panels show the behavior of the differential abundance ratios as a function of the ILS abundance with Milky Way field star abundances shown for comparison.  Note that the right panels include all the abundances measured in this work, while the left panels include only the abundances for which there are reference stellar values.   The error bars in the right panels correspond to $\sigma_{Tot,X}$ for each species. Field star abundances are taken from   \cite{reddy03} (purple), \cite{fulbright2000} (gray), and \cite{ishigaki13} (blue).  MW Bulge field star abundances of \cite{johnson14} are shown as orange circles. } 
\label{fig:alpha} 
\end{figure*}

\begin{figure}
\centering
\includegraphics[scale=0.4]{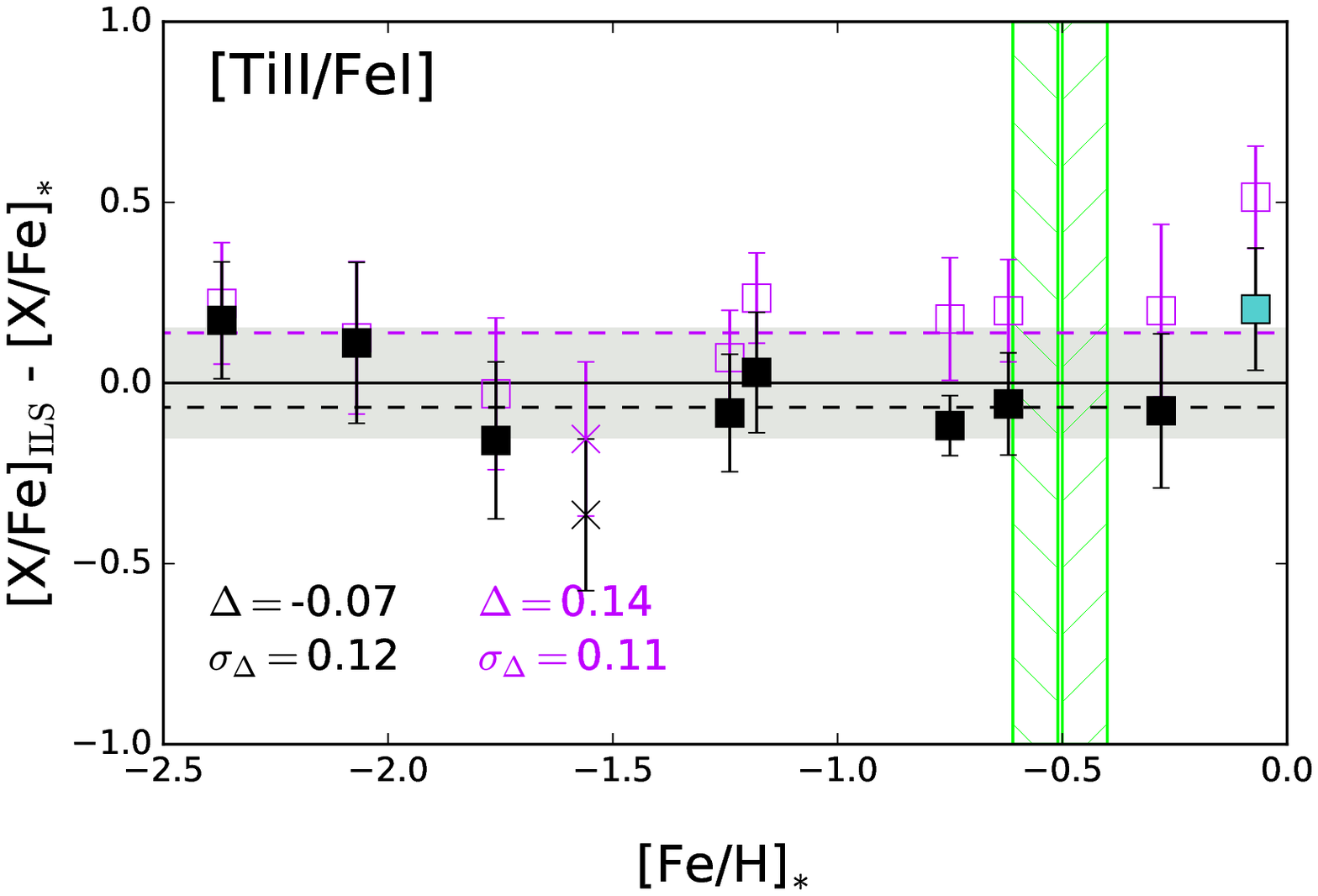}

\includegraphics[scale=0.4]{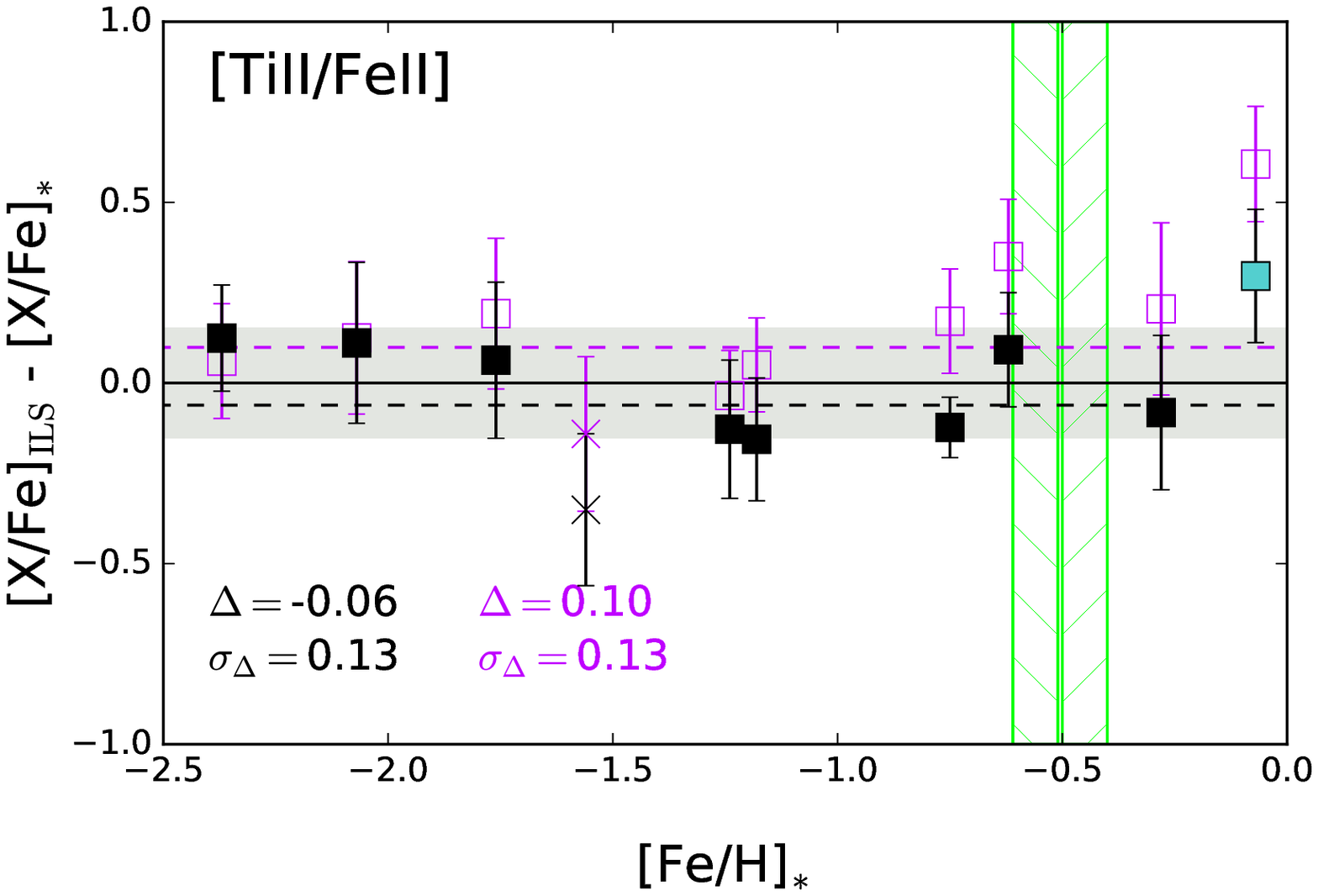}
\includegraphics[scale=0.4]{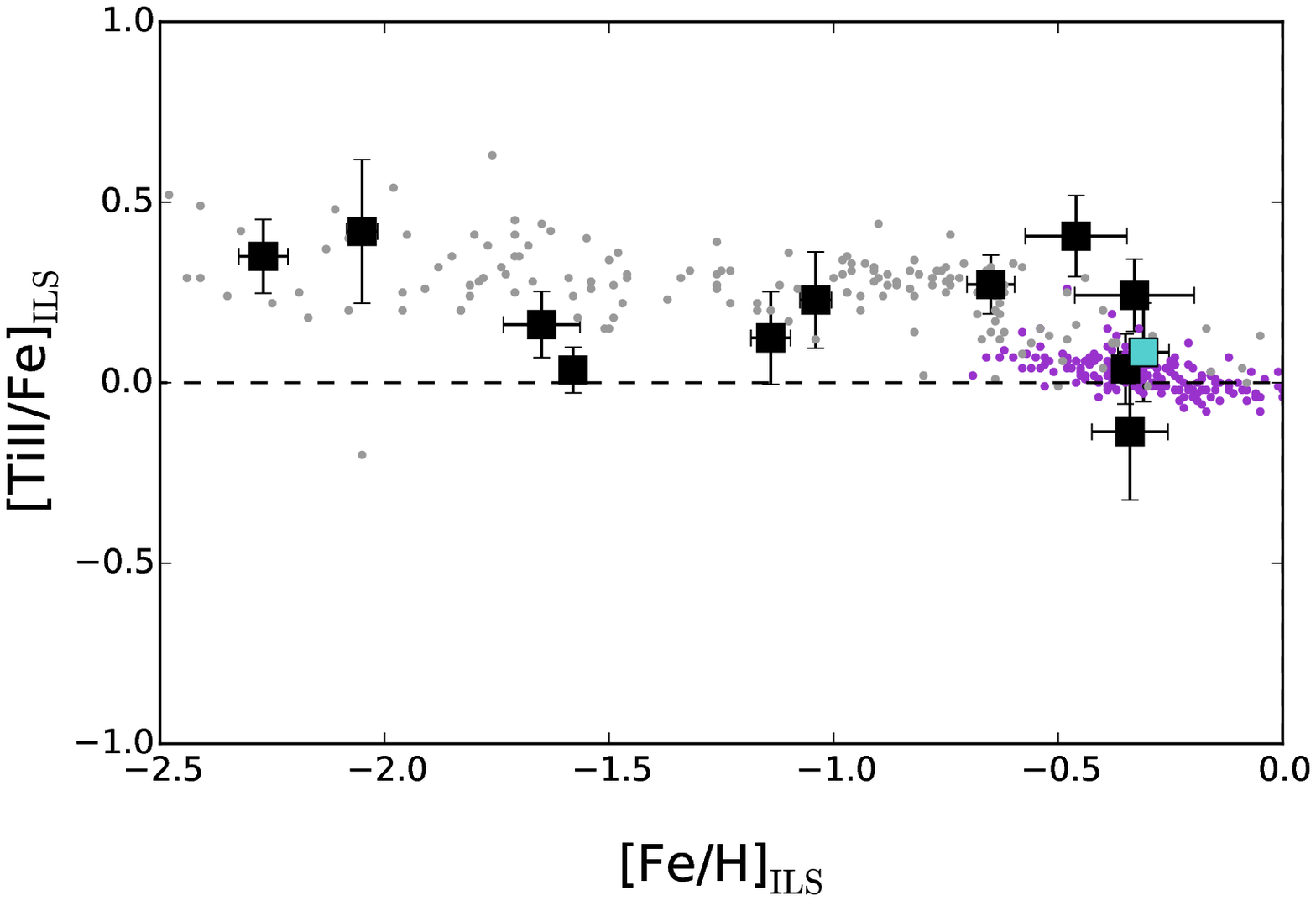}

\caption{IL  and stellar comparison results for [Ti II/Fe].  The top panel shows the Ti II ratio taken with respect to Fe I, and the middle panel shows the Ti II ratio taken with respect to Fe II.   The bottom panel shows the differential [TiII/Fe] results for all clusters compared to MW field stars; in this panel the the ratio has been taken with respect to Fe I.   Symbols and lines in all panels are the same as in Figure \ref{fig:alpha}.  The measurement for NGC 6752 is shown as a cross as a reminder that there may be an issue with the reference abundance. Green hatched regions show areas where the GC IL measurement exists, but no stellar data is available for comparison.  }
\label{fig:ti2} 
\end{figure}

\begin{figure*}
\centering
\includegraphics[scale=0.4]{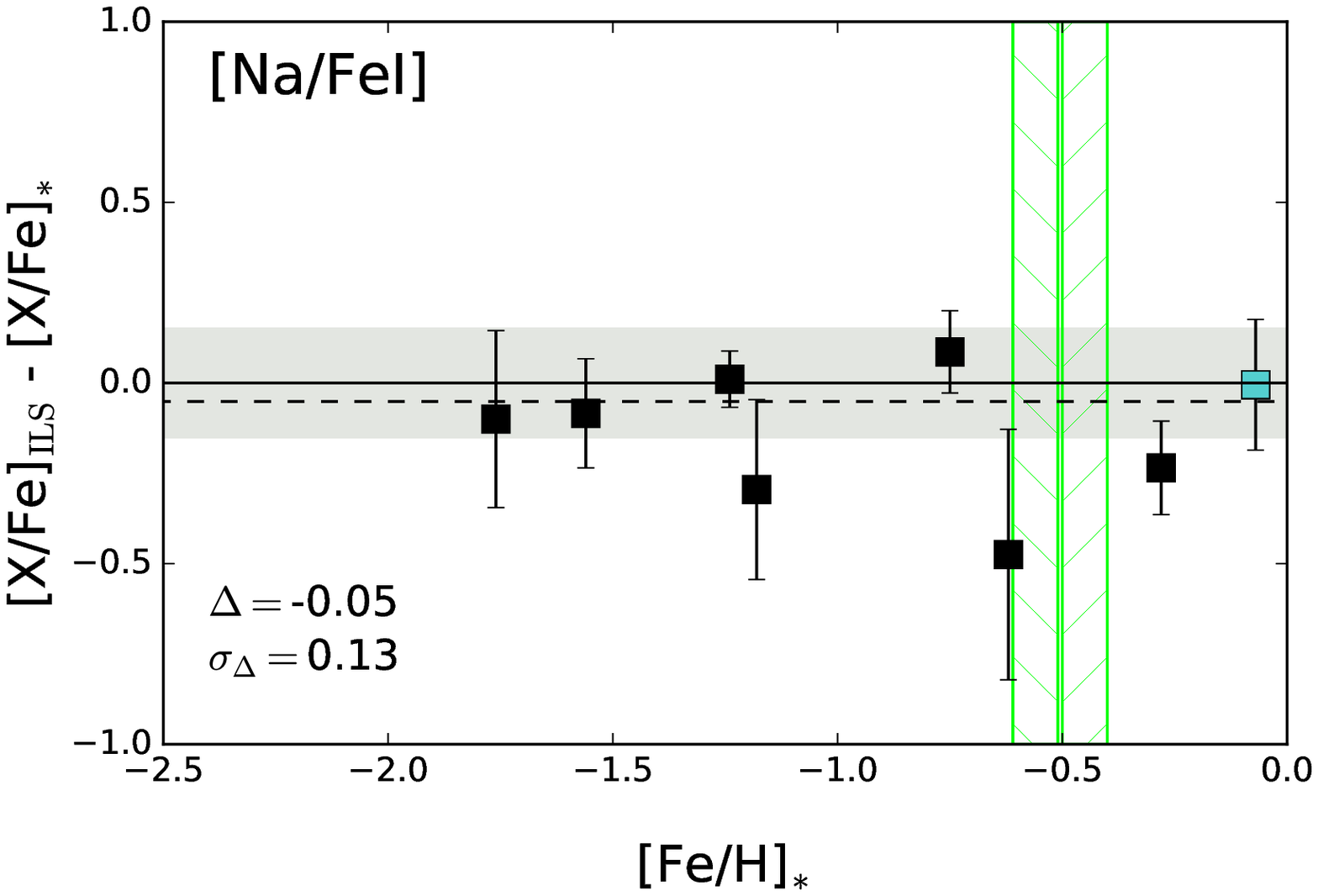}
\includegraphics[scale=0.4]{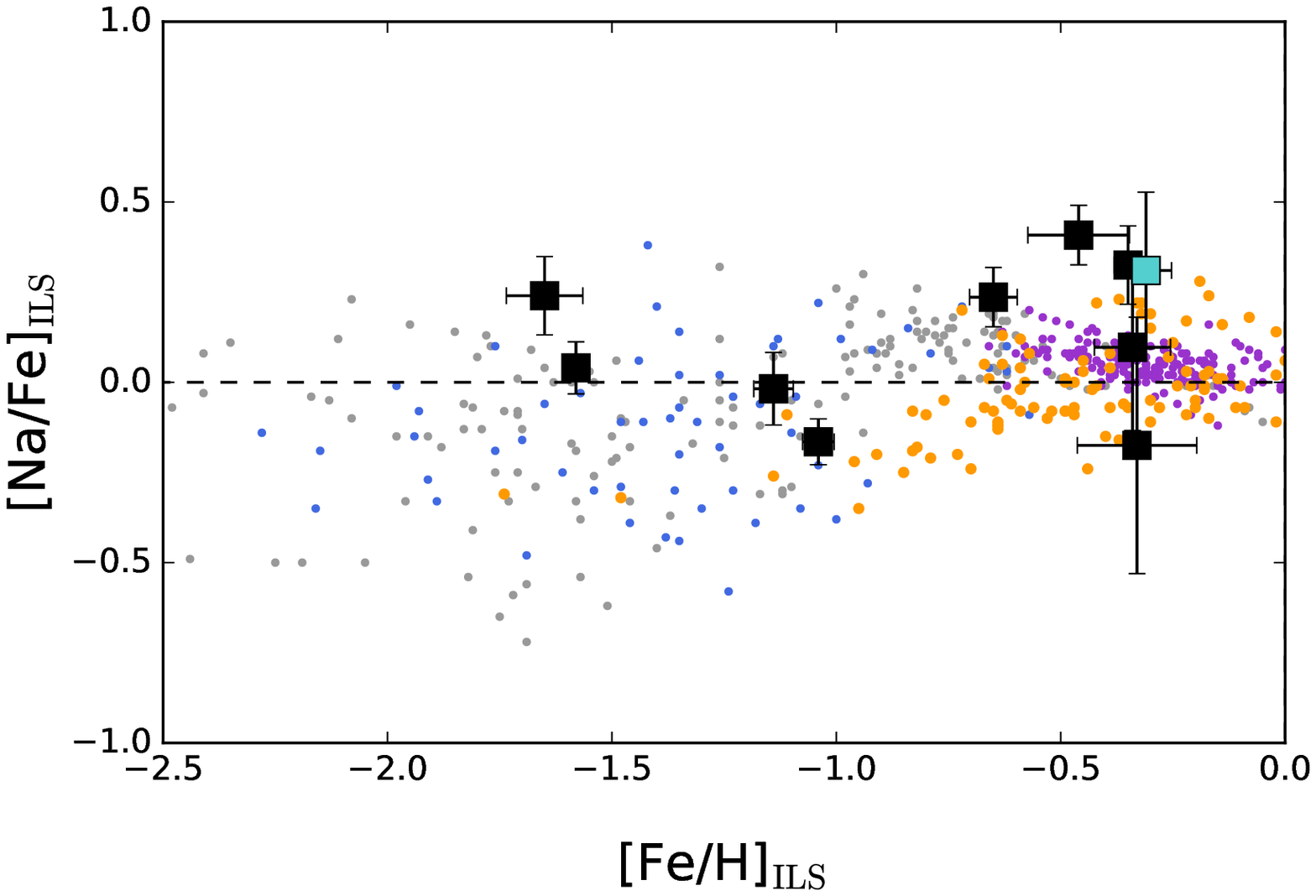}

\includegraphics[scale=0.4]{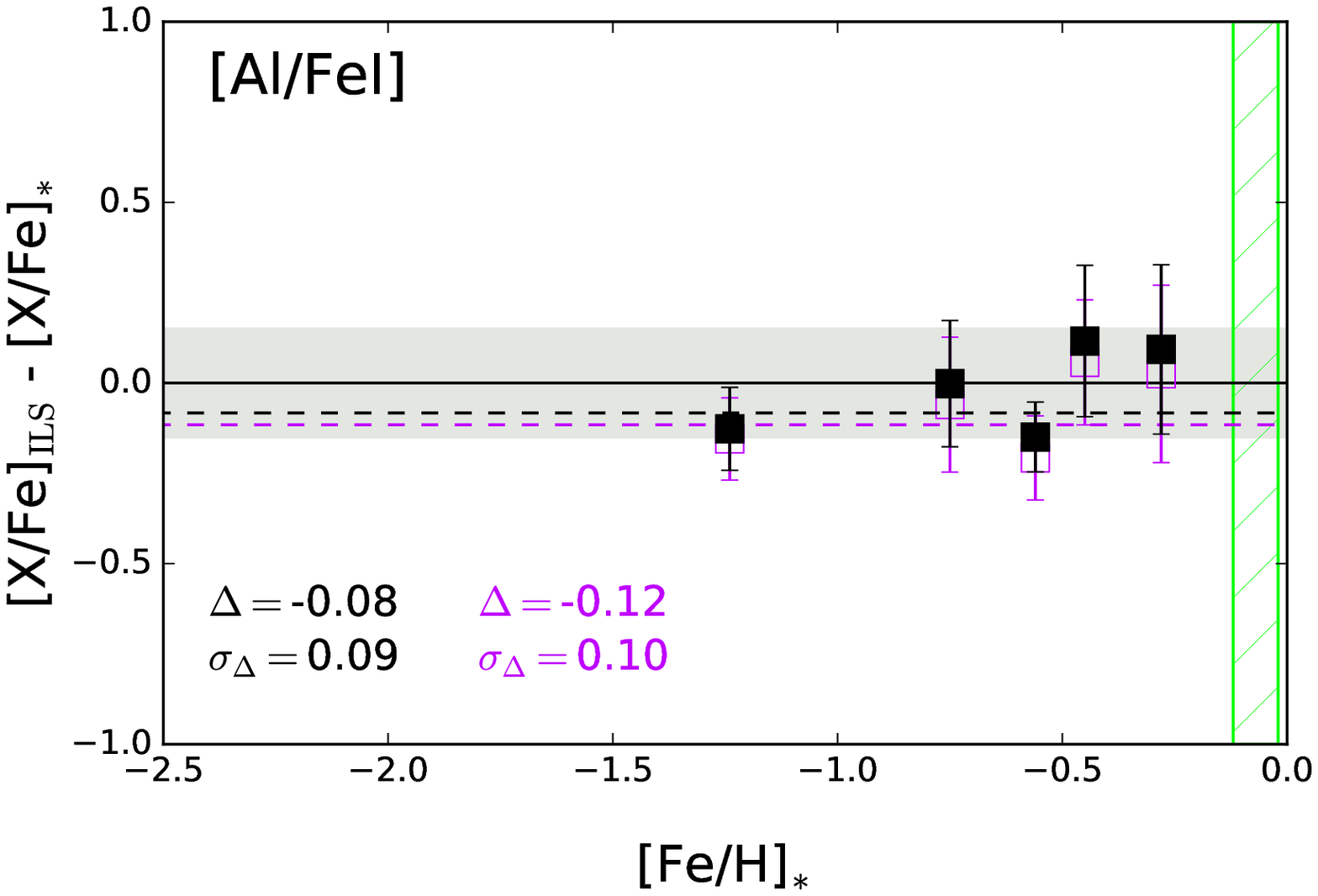}
\includegraphics[scale=0.4]{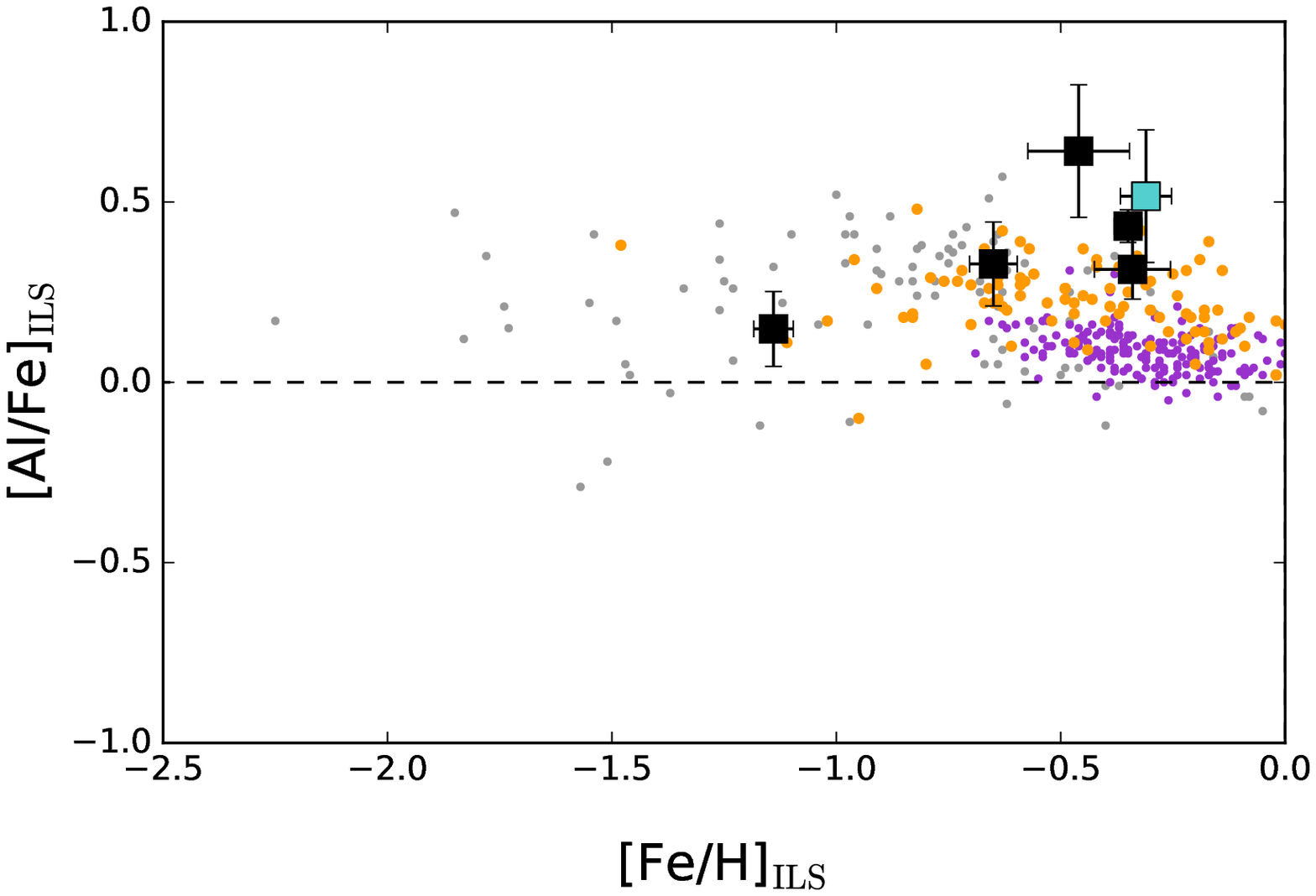}

\includegraphics[scale=0.4]{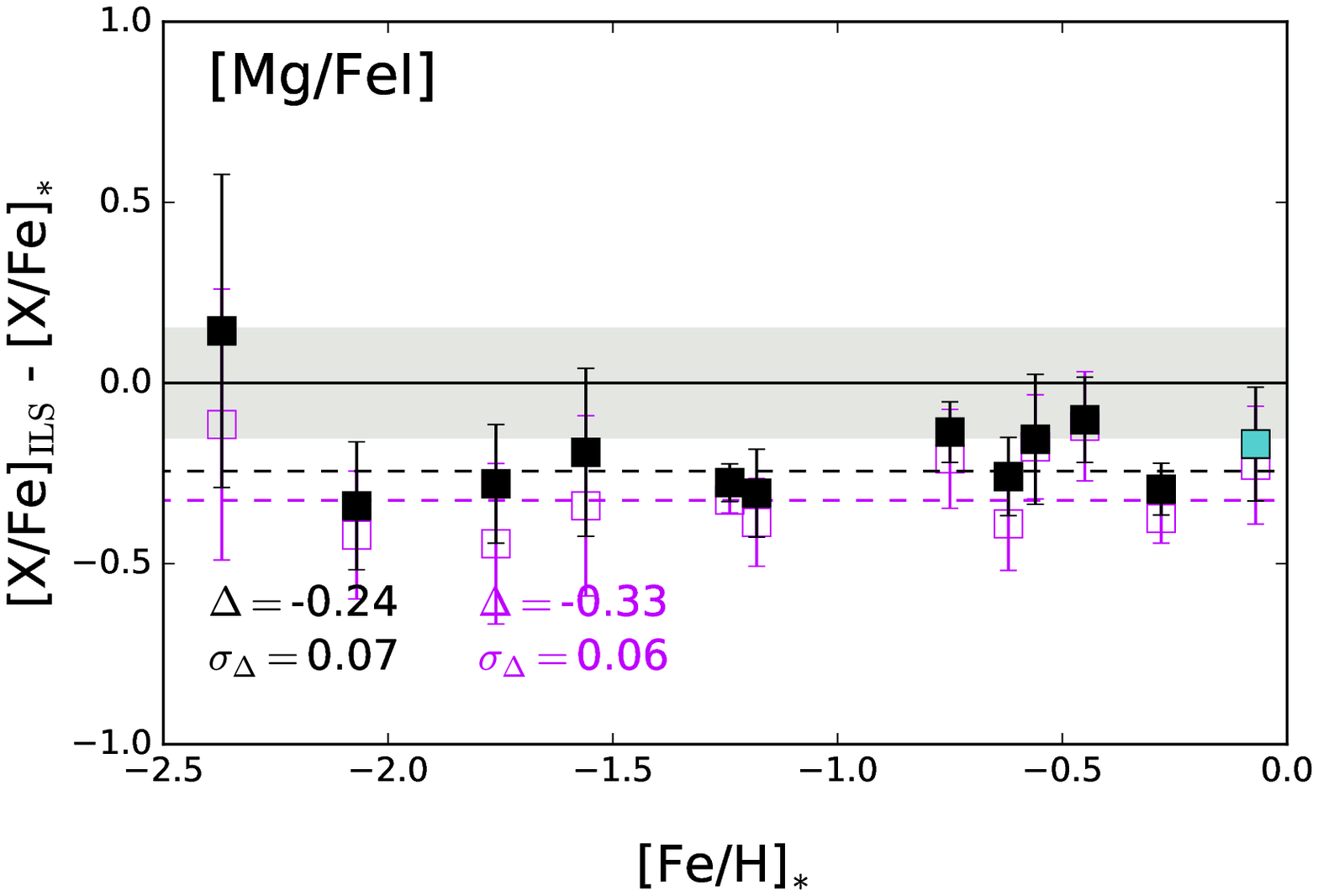}
\includegraphics[scale=0.4]{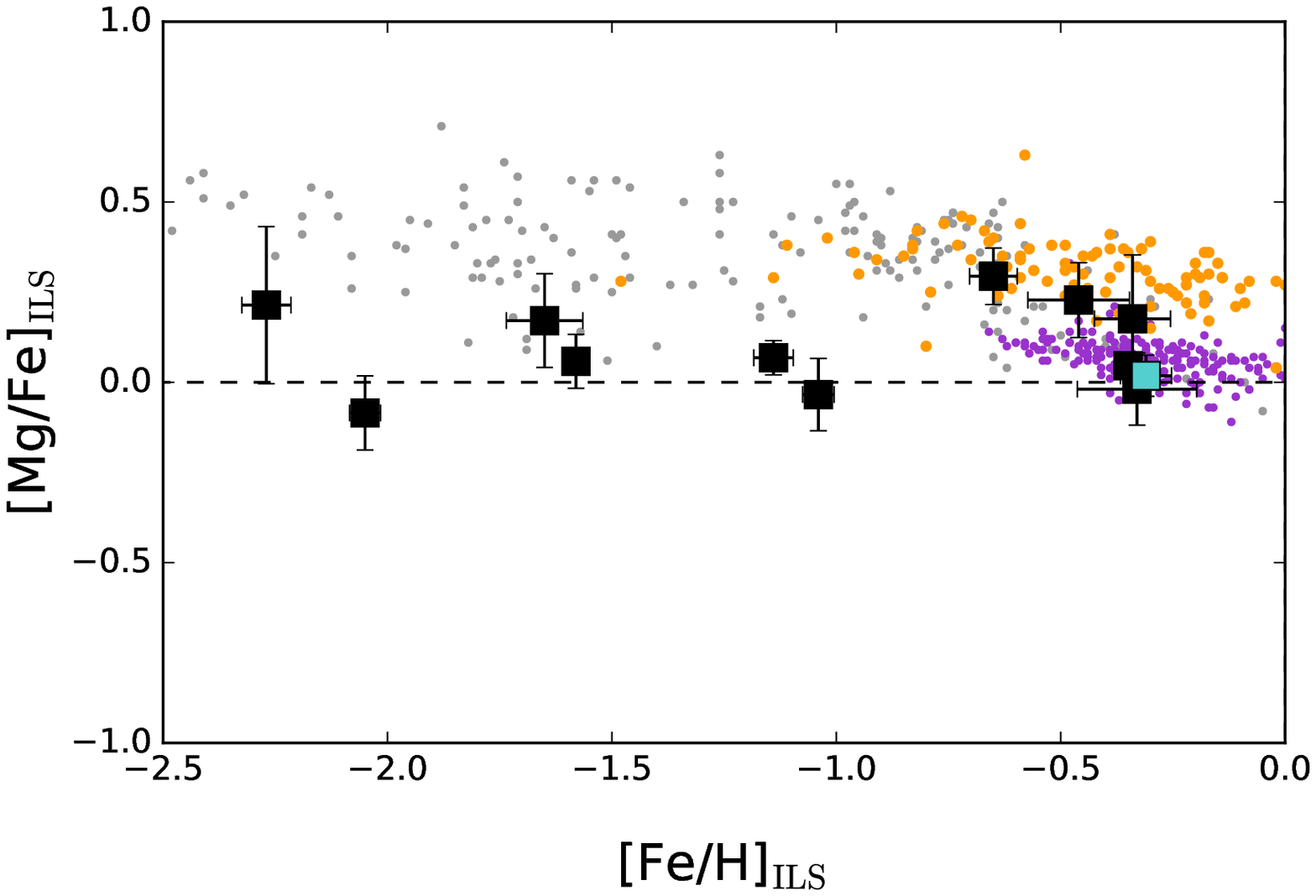}

\caption{ The same as Figure \ref{fig:alpha} for Na I, Al I and Mg I.}
\label{fig:light} 
\end{figure*}

For the comparison to abundances from individual stars,    for each abundance ratio we compute an unweighted mean  ratio from the values given in the references, as well as an uncertainty equal to the standard deviation of the reference measurements to provide a sense of the spread in abundances measured from different authors. For most elements the spread in reference abundances is between 0.05 to 0.15 dex. Most of the element ratios are only measured in a subset of the references used for the [Fe/H] comparison for each cluster.   In many cases we find only one reference value for a given element and in these cases we assign a default  uncertainty of 0.1 dex, which is in the middle of the range  we find for other elements.    The reference abundance ratios we use for comparisons in this section are given in Table \ref{tab:biglit}, and we explicitly list the number of references available for each abundance ratio.  We note that we have used the abundances as given in each work and have not attempted to account for differences in analyses such as the solar abundance distribution used, different log {\it gf} values, model atmospheres, etc.  These analysis differences can potentially lead to systematic offsets, particularly for elements where the reference value is from a single work.  

 For each element we calculate the difference of the IL abundance ratio and the reference ratio and display the residuals for all GCs as a function of the  reference [Fe/H]    
 in the corresponding figures.  The residuals are shown with error bars that are equal to the IL measurement uncertainty  and the reference uncertainty added in quadrature. We show the absolute abundances in purple and the differential abundances in black. For each element we next calculate a systematic offset, which we label $\Delta$,   for the entire sample of 12 GCs, which corresponds to the weighted mean of the residuals.  This value is shown for both absolute and differential abundances in each panel and also corresponds to the  purple and black dashed  lines in each figure, respectively.    The weighted standard deviation of the residuals, which we label $\sigma_{\Delta}$, are also shown in the panel.

 We  have chosen to calculate the systematic offsets without including measurements for NGC 6528, which is often an outlier in the comparisons, as it was for [Fe/H].   Therefore, the systematic offsets are valid for where the method is best applied according to this work, which is generally -2.4$<$[Fe/H]$<$-0.3, or for the metallicity range noted for the particular element.  The measurements for NGC 6528 are still shown for comparison in the figures, where they are highlighted as cyan squares.  
 
A third  goal in this section is to determine if the best agreement is found for the ionized species when the abundance is calculated  with respect to ionized Fe, as is standard practice in stellar abundance studies. Taking the ratios with respect to Fe II may not be the ideal method for the IL analysis because Fe II lines are more difficult and sometimes not possible to measure in the IL spectra.  To investigate this we compare each of the ionized species (Ti II, Sc II, Y II, Ba II, La II, Nd II, Eu II) in two panels, one where the IL ratio is taken with respect to Fe I and one where the IL ratio is taken with respect to Fe II.

{  There are several caveats to keep in mind in this analysis of the performance of the GC IL technique.    First,  many elements are only measured in a subset of the 12 GCs, so the cluster by cluster comparison may only be valid for a part of the metallicity range that the entire sample covers. Second, for many of the elements we do not have reference abundances for every GC for which we measure an abundance, particularly for NGC 6440 where we only have references for Mg, Al, Si, Ca, and Ti from \cite{origlia08}.  This means that this analysis could potentially be expanded if additional stellar reference abundances are obtained, in many cases increasing the range in metallicity for which the comparisons could be made (Co, for example). Therefore, in the figures we have highlighted areas in [Fe/H] in green where the GC IL measurement exists but our analysis is ultimately  limited by a lack of reference abundances, and for which further progress could be made in the future. In addition, we find it helpful to compare the IL abundances to the abundance patterns observed in MW field stars, to get a sense if the measurements for these additional GCs are reasonable.   A third caveat is that there are instances where we have only a single reference abundance and have reasons from the reference work to believe that there could be issues with the reference measurement, which could impact the comparisons made for the IL abundances (V, for example).   To highlight these areas, in the Figures we show the abundance for the GC affected with a cross instead of a square, and comment on how the calculated offsets would change if the GC is not included in the comparison.  We also use a cross  where there may be an issue with the IL abundance measurement (Eu, for example).  A final caveat is that there are elements for which the lowest metallicity measurements are particularly difficult or not possible, and even though our results are applicable to a more limited metallicity range, this is not necessarily a failure because we have not or don't expect to make measurements in extragalactic GCs in this regime.

\begin{figure*}
\centering
\includegraphics[scale=0.4]{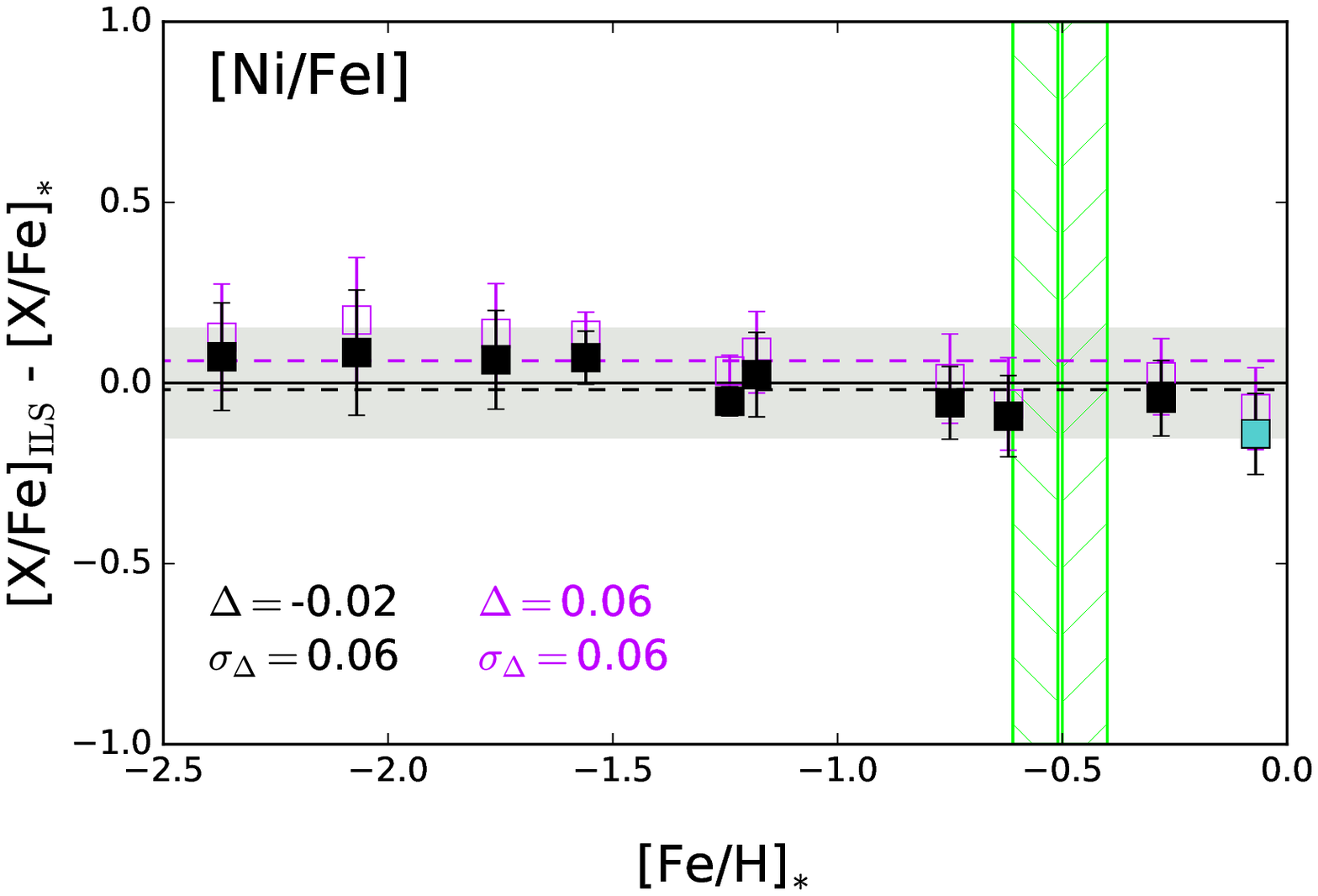}
\includegraphics[scale=0.4]{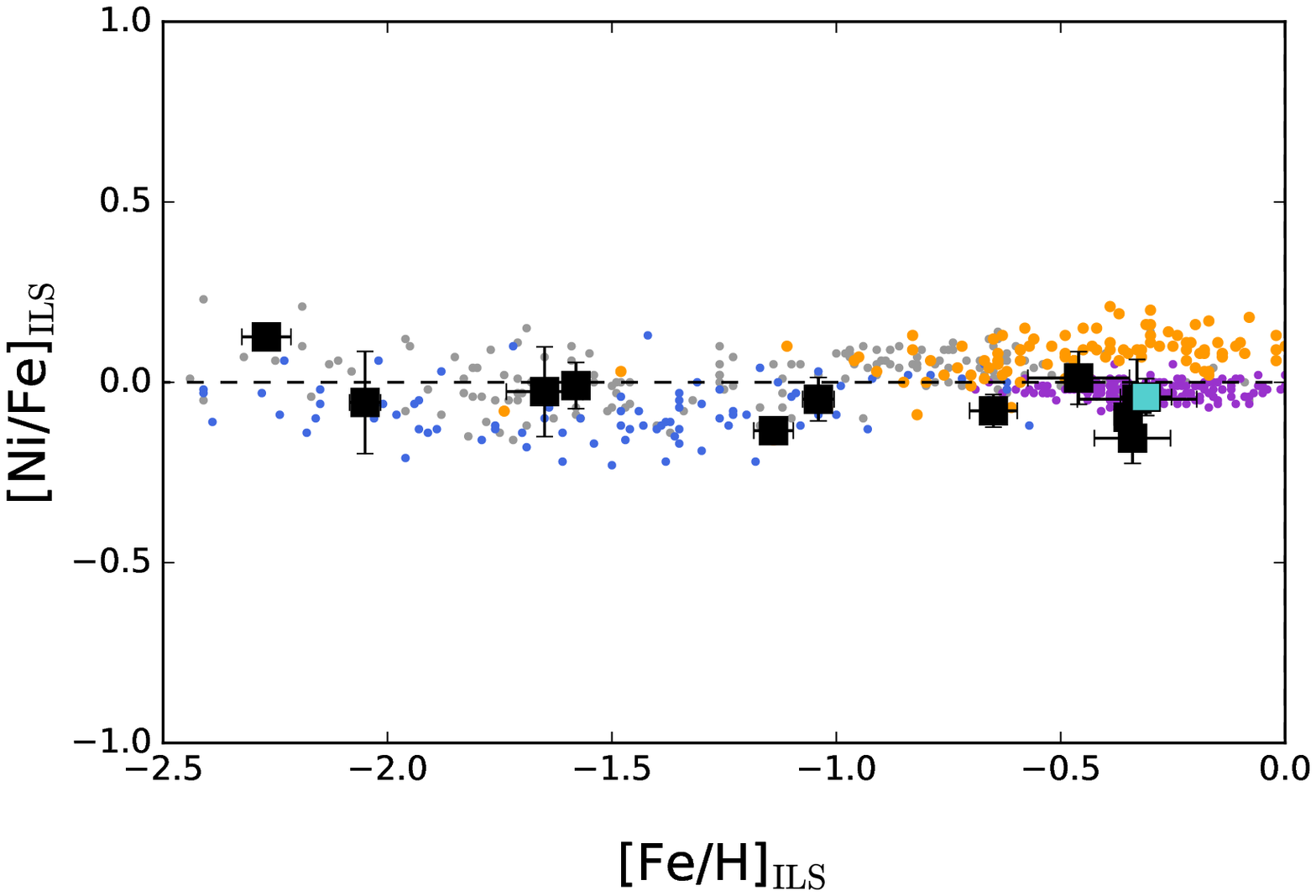}

\includegraphics[scale=0.4]{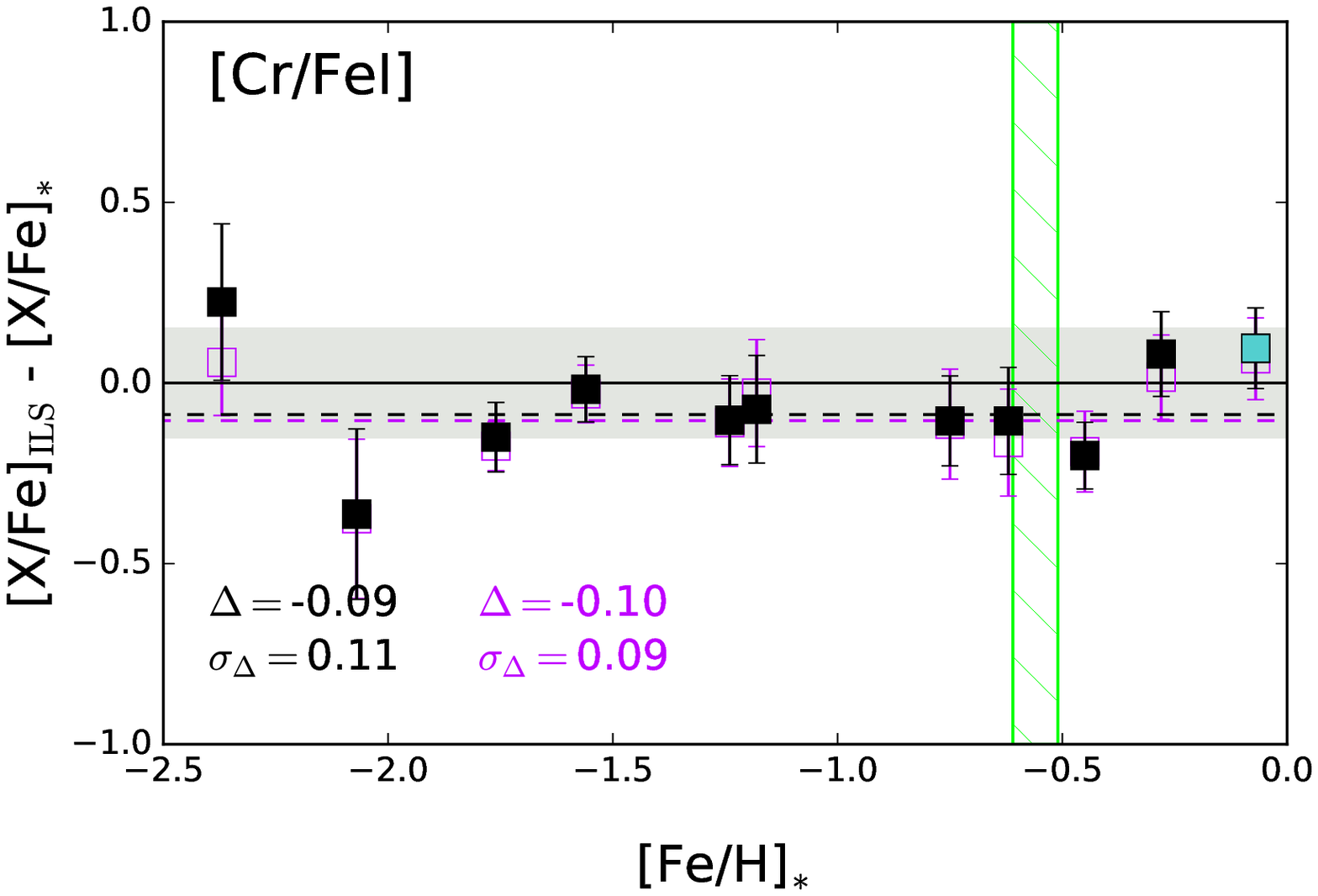}
\includegraphics[scale=0.4]{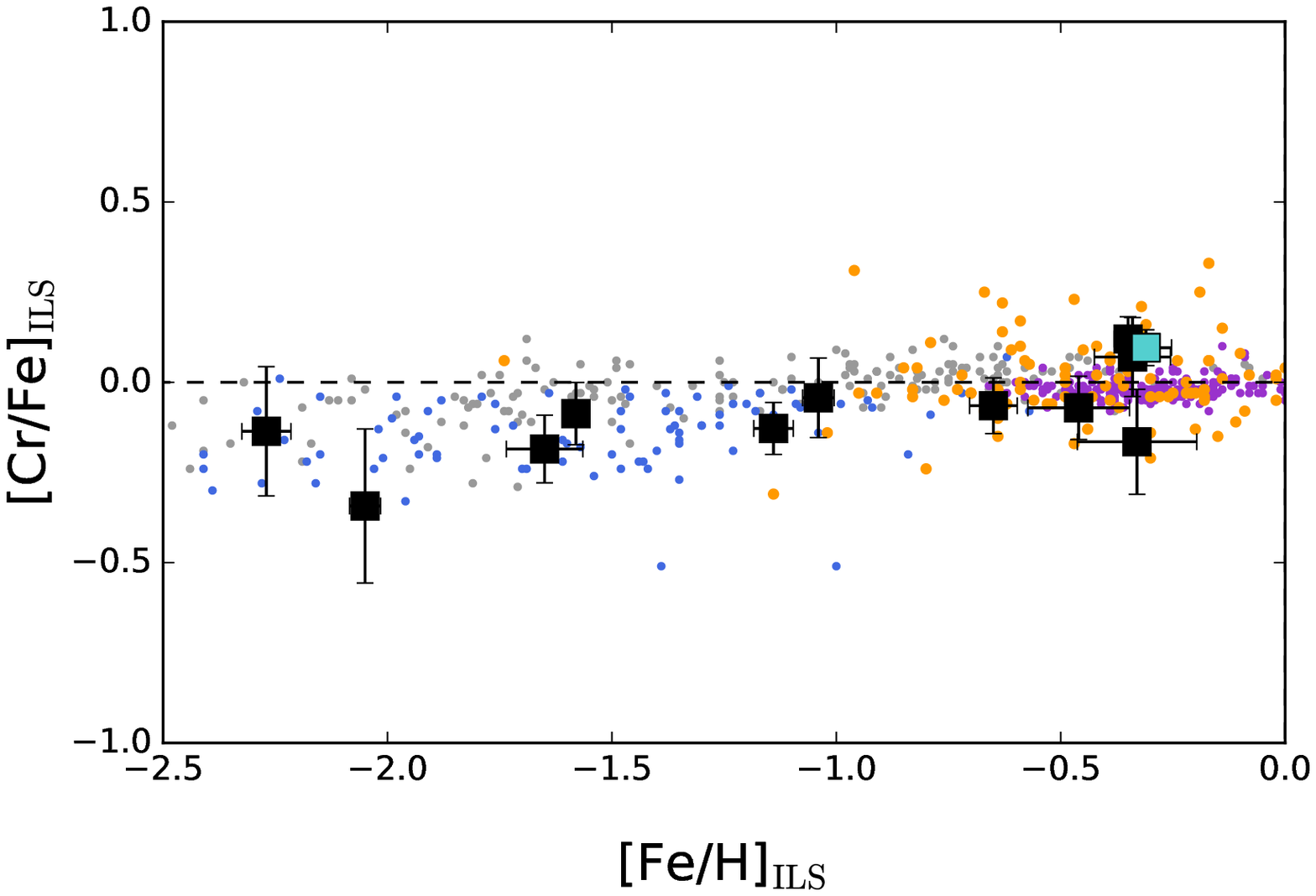}

\includegraphics[scale=0.4]{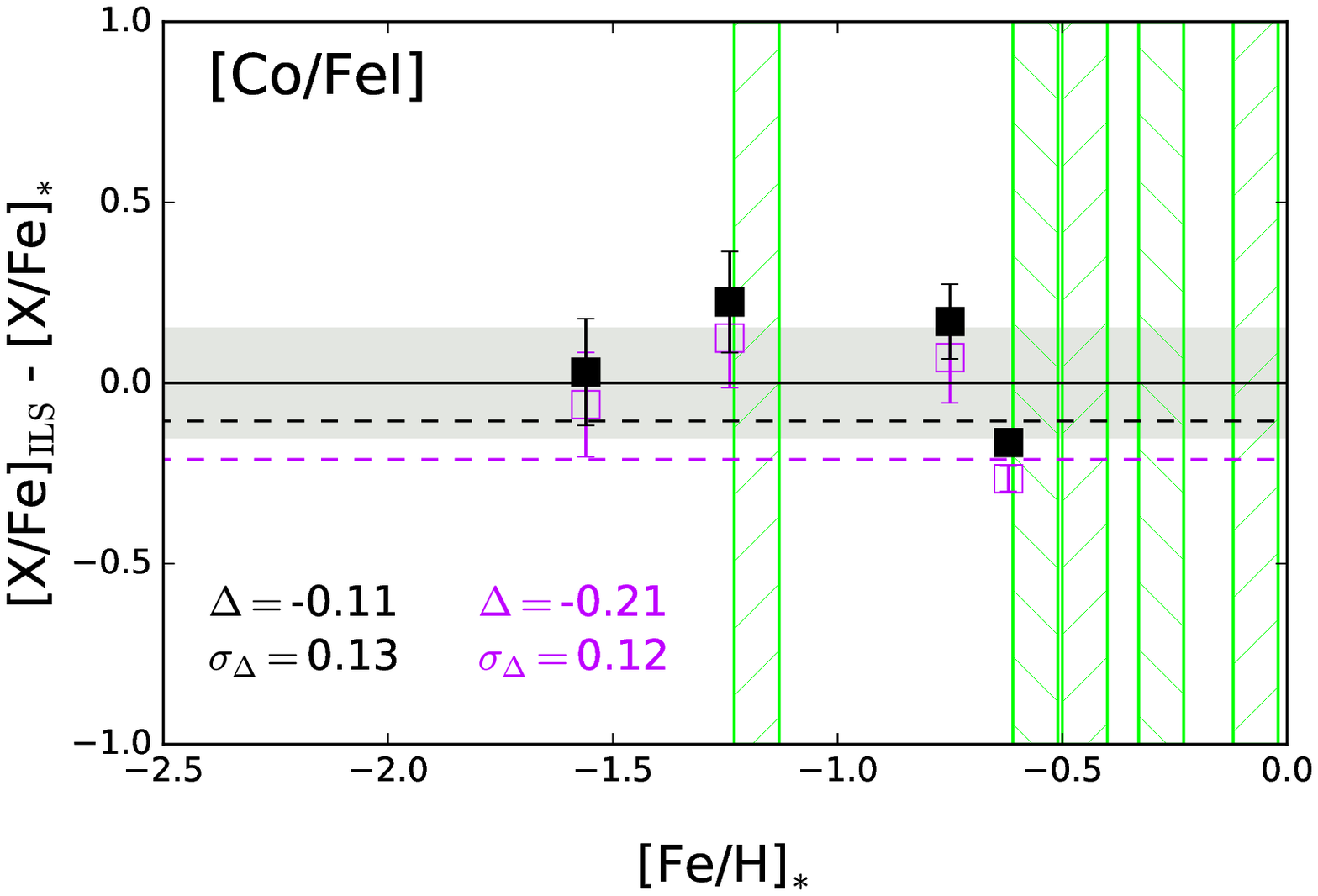}
\includegraphics[scale=0.4]{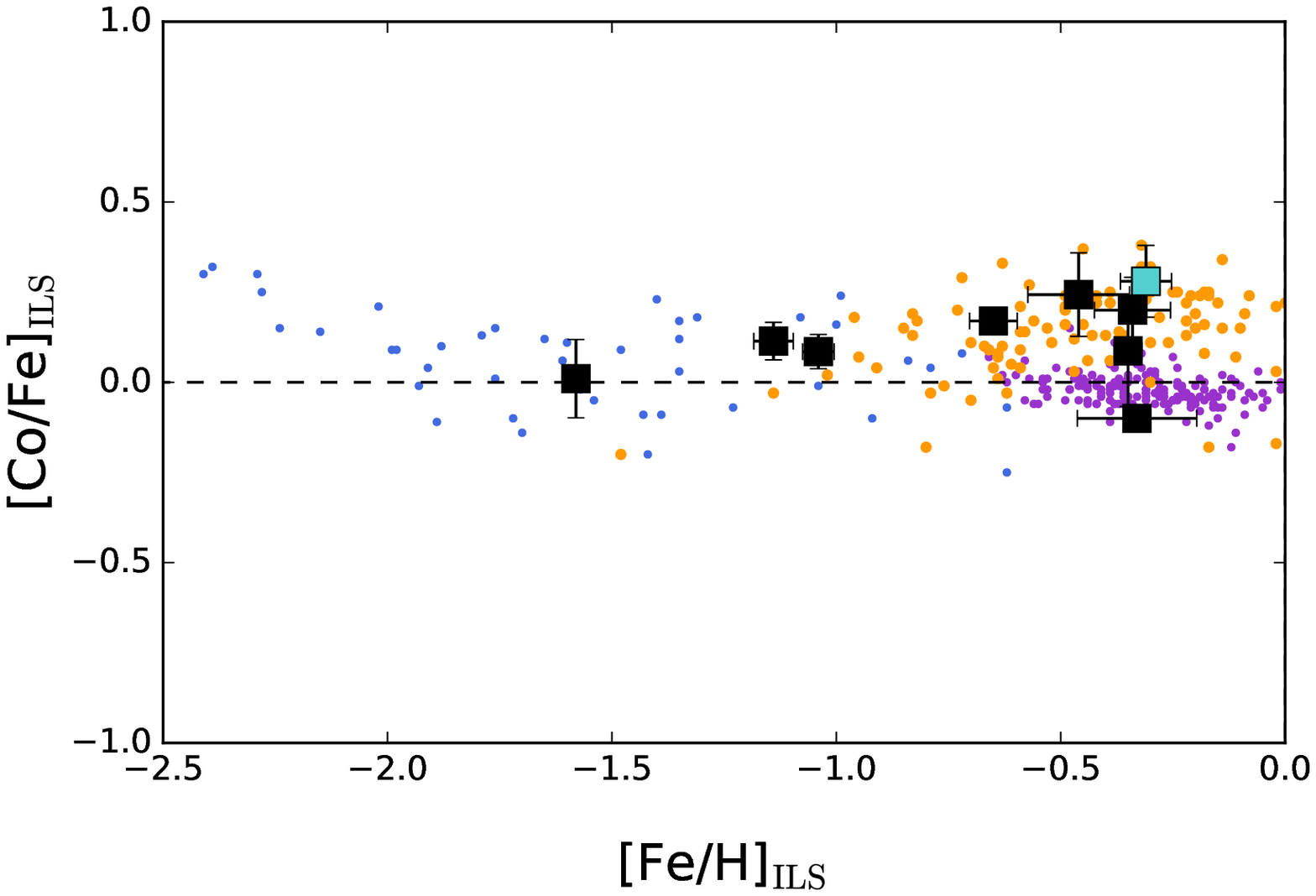}

\caption{ The same as Figure \ref{fig:alpha} for Ni I, Cr I, and Co I. }
\label{fig:fepeak} 
\end{figure*}

\begin{figure*}
\centering
\includegraphics[scale=0.4]{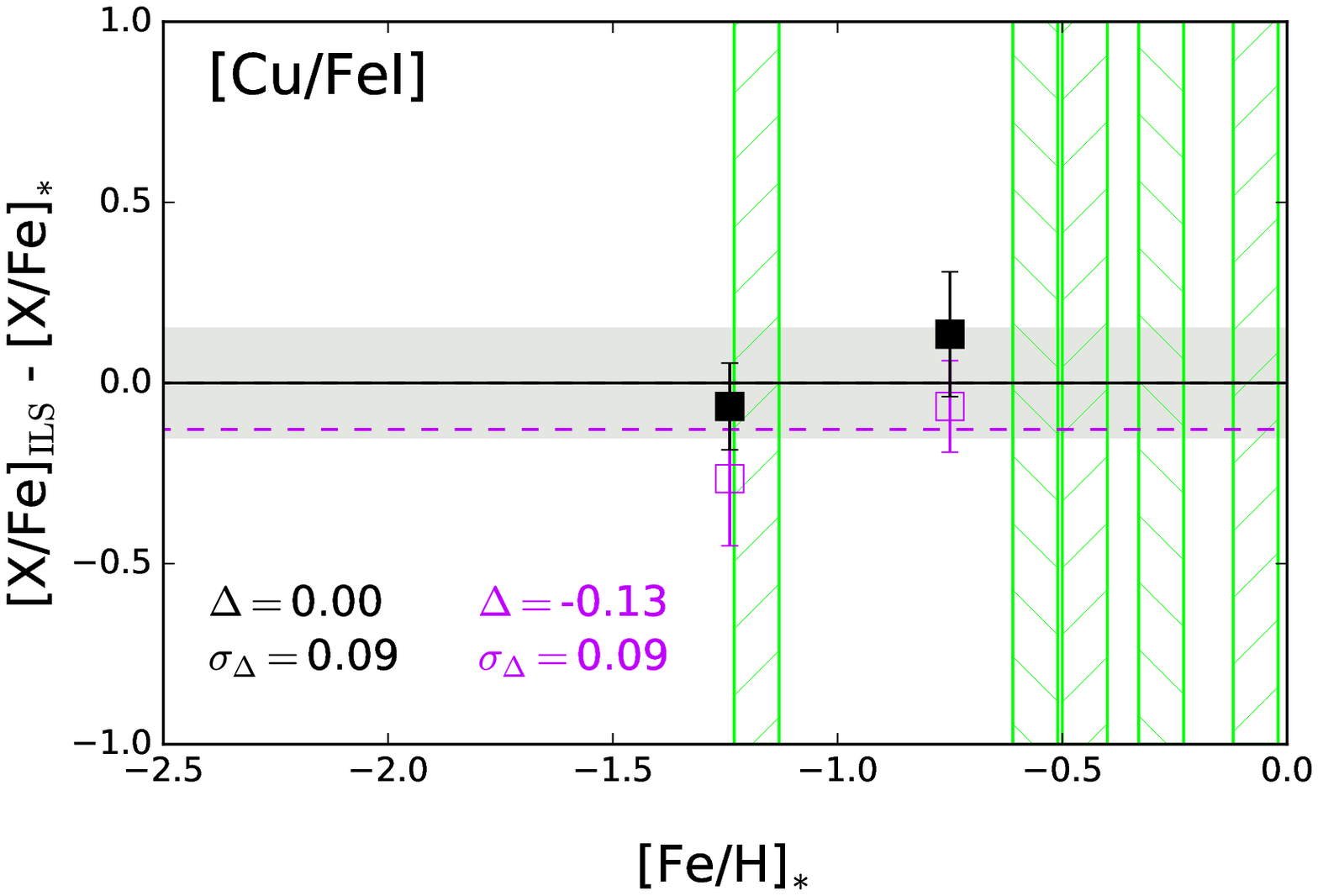}
\includegraphics[scale=0.4]{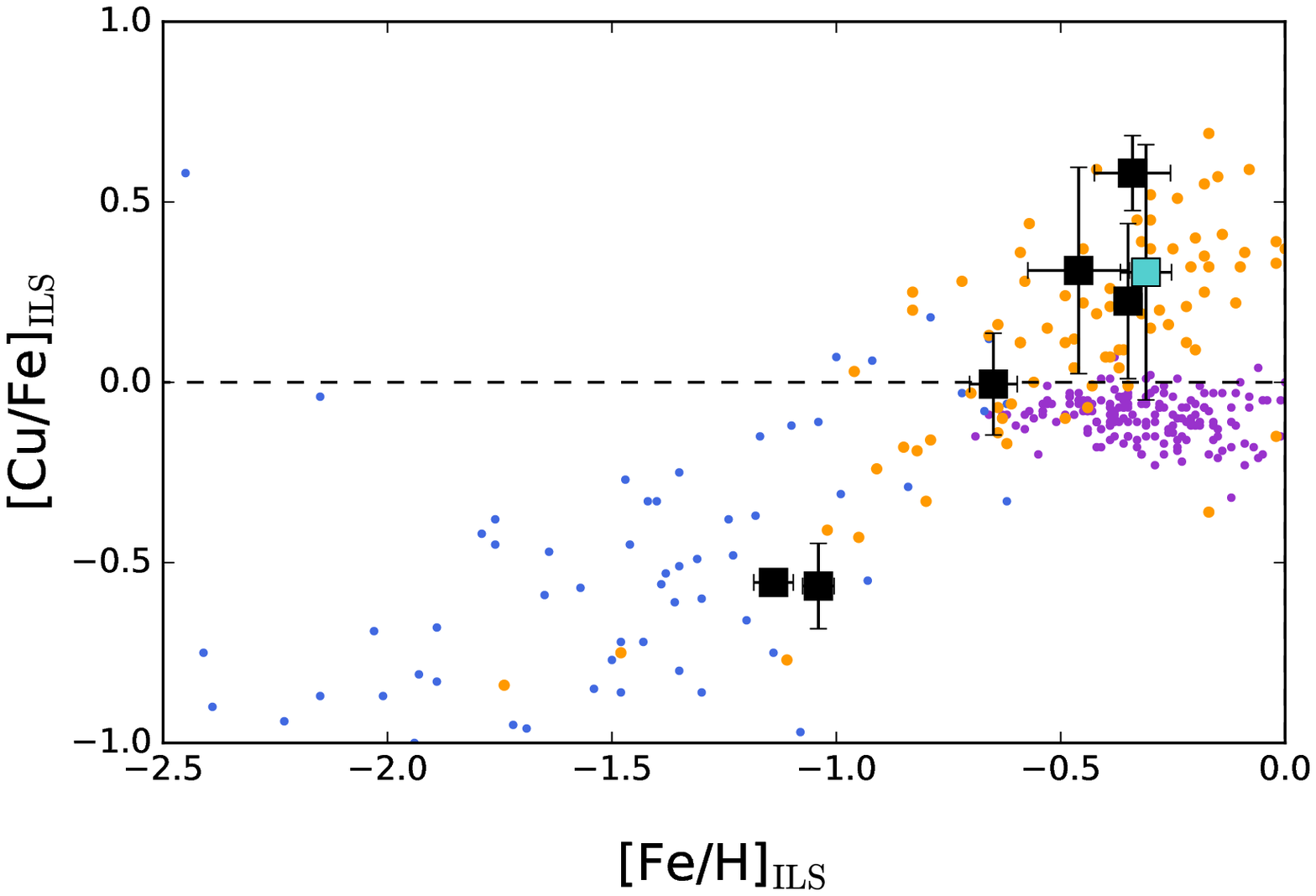}

\includegraphics[scale=0.4]{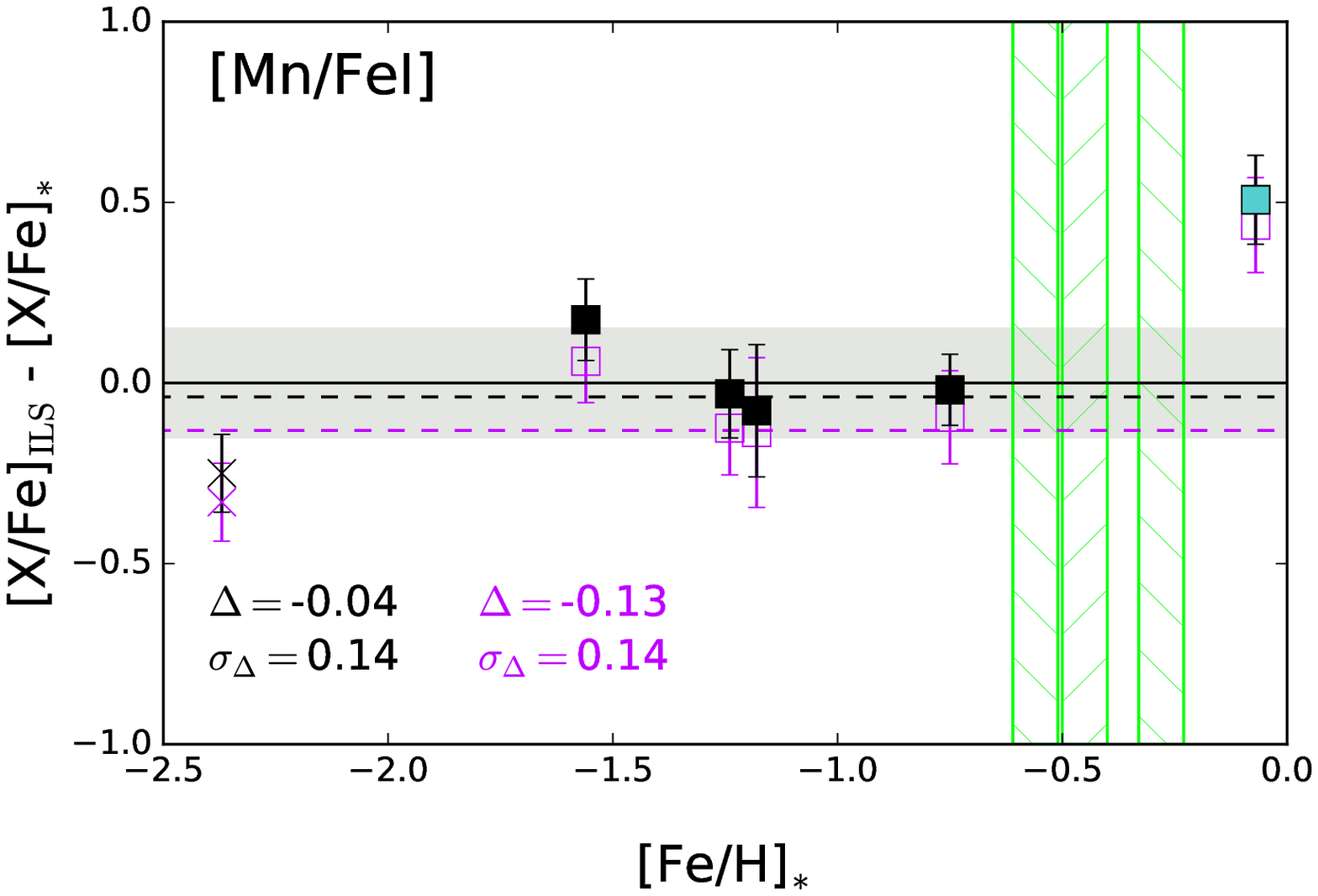}
\includegraphics[scale=0.4]{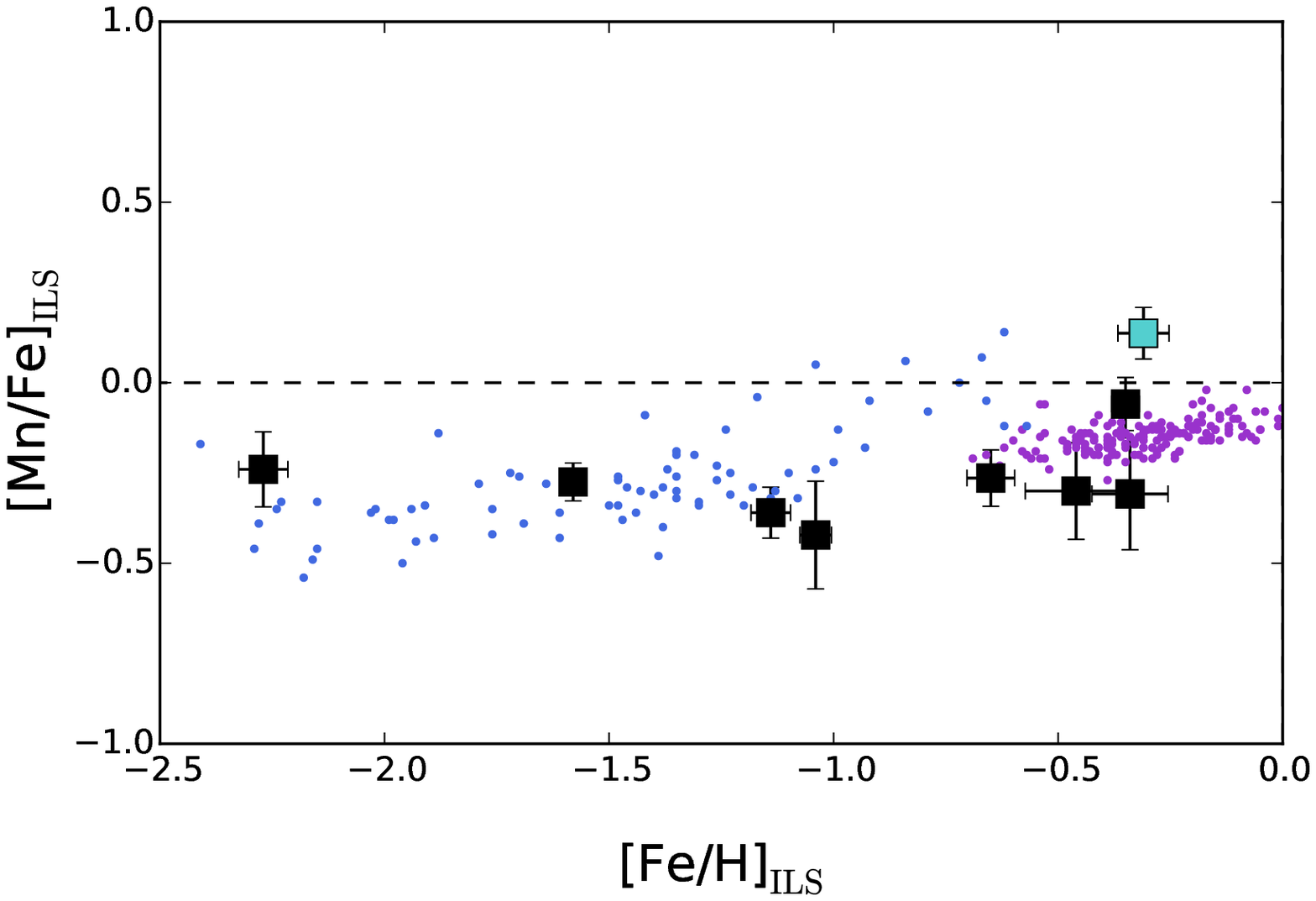}

\includegraphics[scale=0.4]{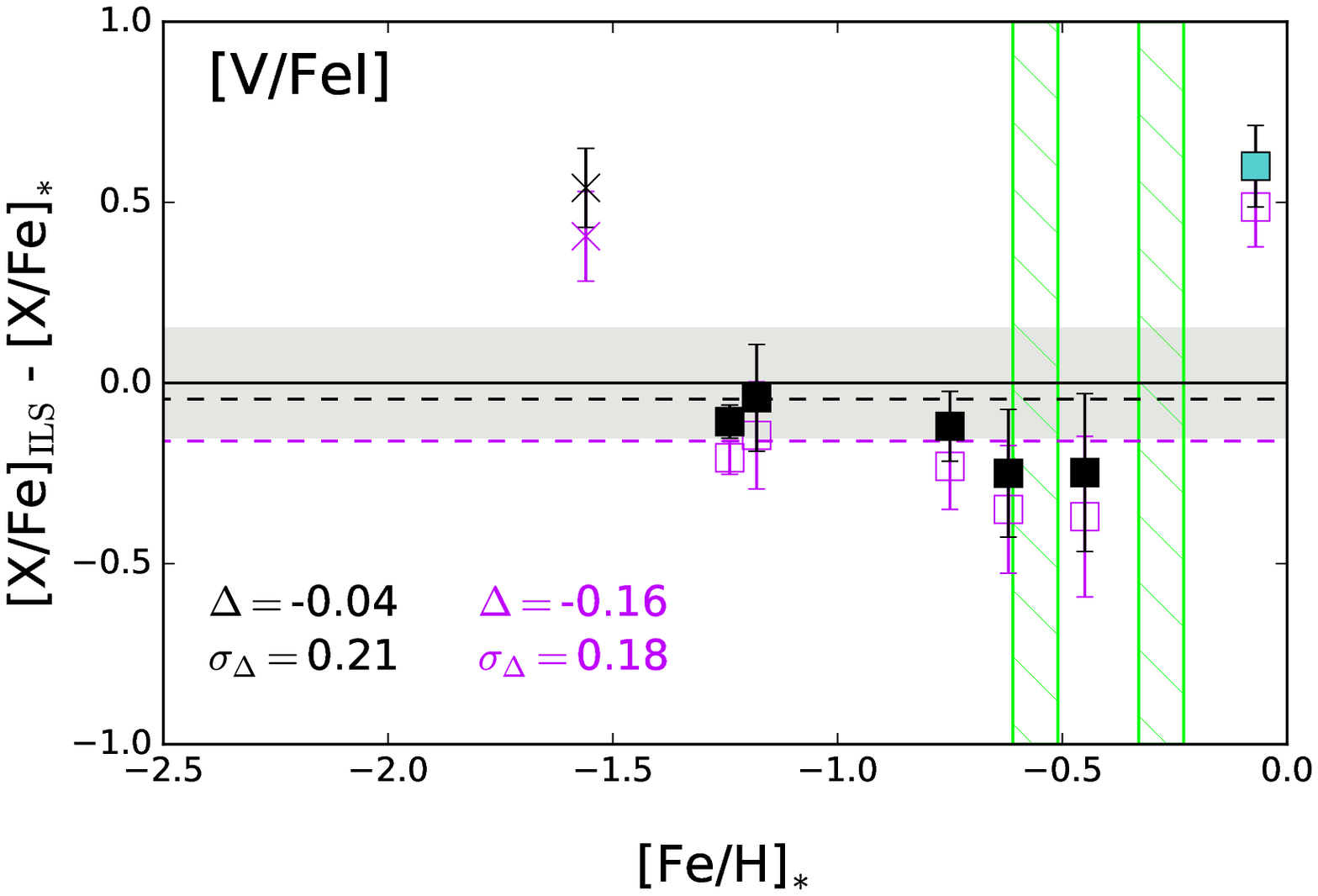}
\includegraphics[scale=0.4]{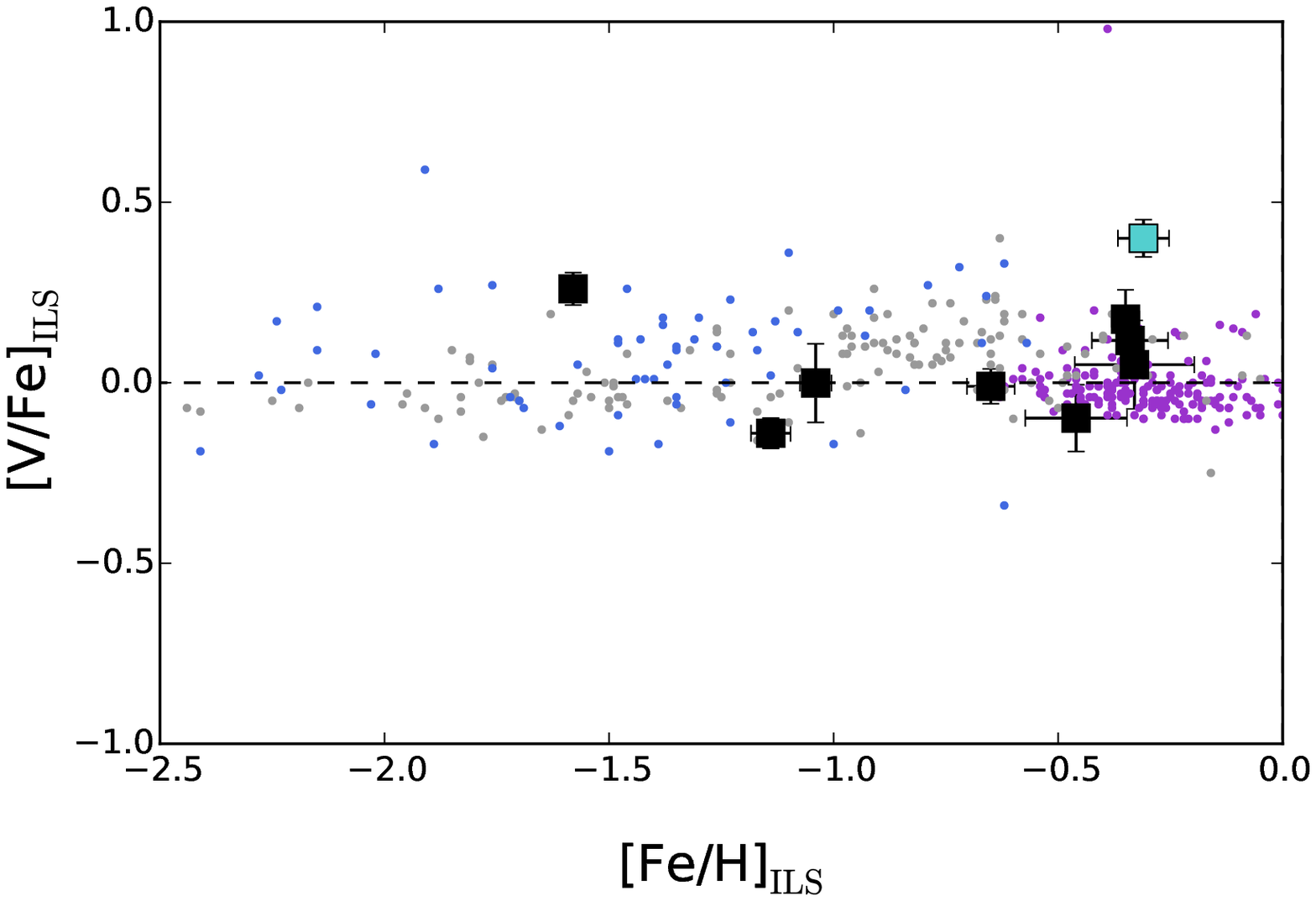}

\caption{ The same as Figure \ref{fig:alpha} for Cu I, Mn I and V I.  For Mn we show the measurement for Fornax 3 with a cross to keep in mind that there are possible issues with the reference abundance. The same is done for the NGC 6752 measurement of V.  Note that these measurements are still included in the calculated offsets shown on the panels.}
\label{fig:fepeak2} 
\end{figure*}

\subsection{Alpha Elements: Ca, Si, Ti}

 Abundances of alpha elements are important for chemical evolution studies to investigate the star formation rate using the relative  timescales of  Type Ia supernovae  (SNIa) and Type II supernovae (SNII), as well as the stellar initial mass function  \citep{1979ApJ...229.1046T,1992AJ....103.1621S}. 
The results for Ca I, Si I, and Ti I are shown in Figure \ref{fig:alpha} and the results for Ti II with respect to neutral and ionized Fe are shown in Figure \ref{fig:ti2}. [Ca/Fe] and [Si/Fe] show the best agreement; the mean residual in [Ca/Fe] is $\Delta=-0.00$ with a spread of $\sigma_{\Delta}=0.12$ for differential abundances, and for absolute abundances   $\Delta=-0.05$ and $\sigma_{\Delta}=0.10$. For  [Si/Fe] differential abundances result in  $\Delta=0.00$ with $\sigma_{\Delta}=0.07$, and absolute abundances result in $\Delta=+0.09$ with $\sigma_{\Delta}=0.07$.   Measurable Si I features in the IL are generally fewer in number and weaker than the Ca I features,  and so are more difficult to measure.  Because of this we were unable to measure Si I  in the two GCs with [Fe/H]$<-2$, and so we can only evaluate the agreement in [Si/Fe] for GCs with [Fe/H]$>-2,$ whereas [Ca/Fe] can be more easily measured over the entire range of $-2.4< $[Fe/H] $<-0.1$.   We note that in our studies of extragalactic GCs, we usually do not measure Si I lines in GCs with [Fe/H]$<-2$ \citepalias{paper4,m31p2}, so the range tested here will be the most applicable for IL work.   We find that for both Ca I and Si I the differential abundances remove the small systematics seen with absolute abundances.   In the right panels of Figure \ref{fig:alpha} we compare the IL measurements with MW field star abundances from the literature, so that the behavior of the abundances can be put into context.  Both [Ca/Fe] and [Si/Fe] agree well with the halo-like plateau for [Fe/H]$<-0.7$.   At the highest metallicities we find that the [Ca/Fe] and [Si/Fe] of the bulge GCs  are more consistent with the alpha element ratios observed in the MW bulge stars of \cite{johnson14} than with the disk stars of \cite{reddy03}.

The results for Ti I are shown in the bottom panel of Figure \ref{fig:alpha}, where we find  $\Delta=+0.04$ with $\sigma_{\Delta}=0.23$ for differential abundances and $\Delta=+0.16$ with $\sigma_{\Delta}=0.20$ for absolute abundances.   The differential abundances improve the overall comparison, although the scatter is still fairly large. 
 We note that if we had included NGC 6528, both the offsets and scatter would  be larger.  Both the IL measurements and the reference measurements appear robust for the more metal poor GCs;  the IL measurements are from more than 4 Ti I features per GC and the reference measurements are all averages from more than one work.

We show the abundance ratio results for Ti II where the ratio is shown with respect to both Fe I and Fe II in Figure \ref{fig:ti2}. The IL Ti II measurements themselves tend to have larger uncertainties than Ca I or Si I, which may be because most of the Ti II features are in the bluer, lower S/N region of the data ($\lambda <$ 5000 \AA).  
For differential abundances we find $\Delta=-0.07$ with $\sigma_{\Delta}=0.12$, and for absolute abundances we find a significantly higher offset of $\Delta=+0.14$.  The offsets are both a bit smaller when the ratios are taken with respect to Fe II, but the  $\sigma_{\Delta}$ is a little larger.   The Ti II measurement for NGC 6752 is the outlier in the comparison, however we note that there may be an issue with the single reference abundance from \cite{cavallo2004}.  These authors note that their Ti II abundance is 0.24 dex higher than their Ti I abundance, and caution that the Ti II abundances may be suffering from an unknown systematic error.  Our Ti II measurement is lower than the value in \cite{cavallo2004}, so better agreement would indeed be found if the \cite{cavallo2004} value was systematically high.  If we calculate the offset without the NGC 6752 measurement the corresponding  improvement for Fe I ratios would be $\Delta=-0.05$ with $\sigma_{\Delta}=0.10$.    At high metallicity both the Ti I and Ti II ratios tend to be higher than what is observed in the MW disk population, and show more similarity to bulge stars \citep[e.g.][]{gonzalez11,fulbright07,alves10}.

In conclusion, we find that Ca I and Si I  produce the most accurate IL alpha element abundances with respect to previously measured abundances in GC stars, with no overall systematic offset when differential abundances are used.   Individual measurements of Ti II often have larger line-to-line scatter than Ca I or Si I, but overall match the reference values well, with the exception of NGC 6752.
 Ti I   abundance ratios have an overall offset that is small, however the scatter is  larger than for Ca I, Si I, and Ti II, so we conclude that alpha element ratios are preferably done with the Ca I, Si I and Ti II.   
\subsection{Light Elements: Na, Al, Mg}

In Figure \ref{fig:light} we present the results for Na I, Al I, and Mg I.  Note that  these elements have been shown to {\it not} be mono-metallic in GCs \citep[e.g.][and references therein]{grattonrev}, and so there is sometimes a large spread in the reference abundance values, particularly for Na.   Despite this the IL abundances show reasonable  agreement with the mean reference values for Na and Al, with  offsets of $\Delta=-0.05$  for Na I,  and $-0.08$ and $-0.12$  for Al I when differential and absolute abundances are used, respectively.   We find that the high metallicity GCs have [Al/Fe]  significantly higher than solar ratios, which is  consistent with the MW bulge stars shown in Figure \ref{fig:light}.

We are not able to make comparisons for the lowest metallicity GCs since the few features of these elements are weak in IL spectra.   However, this is not necessarily problematic, because  the metallicity ranges evaluated for the  Na and Al lines in this work correspond to the same metallicity ranges for which we were able to measure these elements in M31 GCs in \citetalias{m31p2}.    We note that the weighted mean and scatter are dominated by the GCs with smaller dispersions in the mean reference abundance, which can be misleading in the case of GCs with significant star-to-star abundance variations. For example, the weighted offset in [Na/Fe] is very small, but both the residuals and the dispersion in stellar reference abundances for NGC 2808 and NGC 6388 are large.

The agreement for [Mg/Fe] is the worst  of all of the elements considered in this work.  We have shown in our previous work that IL [Mg/Fe] in GCs do not track the IL abundances of  alpha elements \citepalias{m31p1,paper4,m31p2}.  Similar results for IL spectra analyses were  found by \cite{sakari} and \cite{larsen12}, although the systematic offsets in \cite{sakari} were not as severe as we find.   In this work we find that the offset in [Mg/Fe] when compared to stellar abundances is $\Delta=-0.24$, with a small scatter of $\sigma_{\Delta}=0.07$ when differential abundances are used and a larger offset of  $\sigma_{\Delta}$=-0.33 for absolute abundances. Like \cite{sakari} we find that the abundances are closer to literature values when differential abundances are used, but a significant discrepancy still remains, particularly for the metal poor clusters.  As an experiment to see if strong line wings were affecting the abundance we re-synthesized the spectra using Barklem damping constants, however this change results in even lower [Mg/Fe] values.  It is possible that the IL Mg I abundances could be particularly affected by a mismatch in the dwarf to giant ratio in the CMDs, but a simple test  of removing the CMD boxes of dwarfs with T$_{\rm{eff}}<$5500 raises the derived abundance by $\sim0.05-0.10$ dex, which is not enough to overcome the discrepancy.  
   
While we can expect that the Mg  ratios might not match the stellar abundances due  to  star-to-star  abundance differences, the Mg star-to-star differences are usually smaller than observed for Na or Al, so it is puzzling that the agreement is consistently much worse for Mg than for Na or Al.   It is possible that large abundance differences exist in a small fraction of luminous cool giants, and that these are dominating the IL measurement but are not represented in the samples of individual stars that have been observed in each GC.   There could also be significant NLTE effects in some fraction of the stars \citep[e.g][]{mgnlte}, so it would be interesting to see if an in depth analysis of  NLTE effects on the IL measurements solves the discrepancy for Mg.

\begin{figure}
\centering
\includegraphics[scale=0.4]{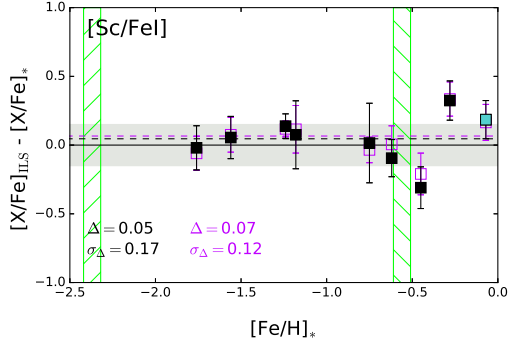}
\includegraphics[scale=0.4]{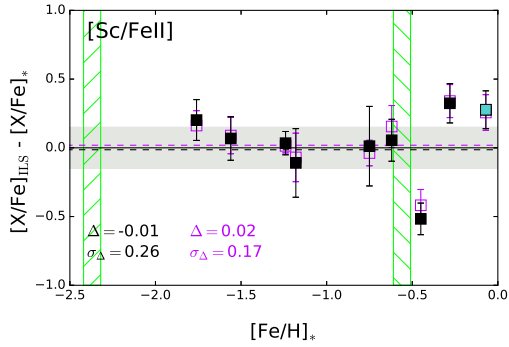}
\includegraphics[scale=0.4]{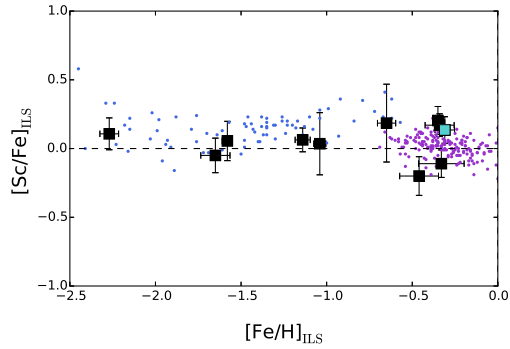}

\caption{ The same as Figure \ref{fig:ti2} for Sc II.}
\label{fig:sc} 
\end{figure}

\begin{figure}
\centering

\includegraphics[angle=90,scale=0.35]{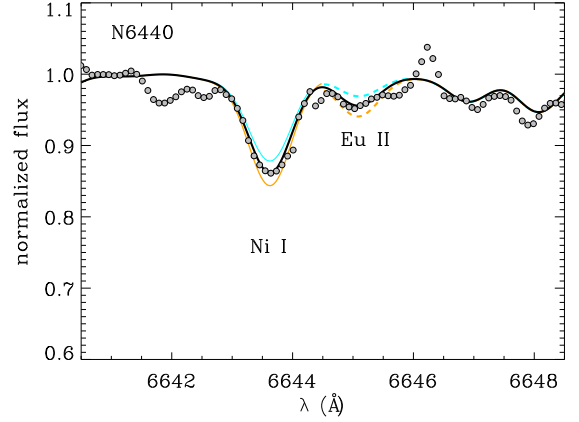}
\includegraphics[angle=90,scale=0.35]{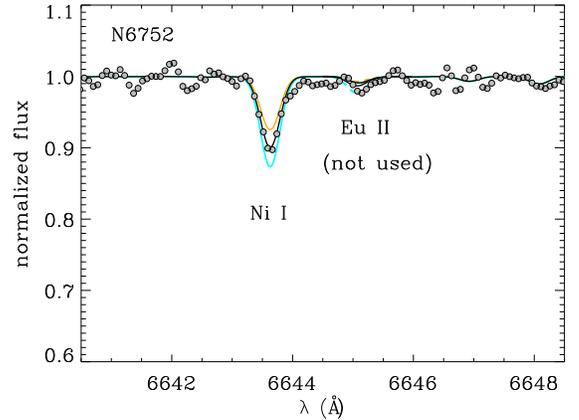}

\caption{Examples of data and synthesis for Ni (6643 \AA) and Eu (6645 \AA). Data is shown as gray points and is smoothed by 3 pixels.  The top panel shows NGC 6440, where a measurement was obtained for Eu II, and the bottom panel shows NGC 6752, for which the Eu II line was too weak to be measured accurately.  In each panel the solid black line corresponds to synthesized spectra with the mean Ni and Eu abundance (assumed to be [Eu/Fe]=$+0.3$ for NGC 6752). The cyan  and orange solid lines show $-0.3$ dex and $+0.3$ dex in [Ni/Fe] from the mean value, respectively.  The cyan and orange dotted lines show $-0.3$ dex and $+0.3$ dex in [Eu/Fe] from the mean value, respectively.     }
\label{fig:eu_synth1} 
\end{figure}

We conclude that on average the IL measurements of Na and Al are  comparable to the mean and/or range of abundances from  the reference studies, even though these elements can vary significantly within the stars in a given GC.   While the IL abundances of these light elements are not straightforward to interpret, they can perhaps be used in the future to constrain the presence of star-to-star abundance variations in unresolved extragalactic globular clusters.
Mg abundances show poor agreement, which reinforces our previous conclusions that IL Mg is a poor indicator of alpha element abundance, which may be caused by true flux weighted variation in the cluster stars that is not reflected in the stellar comparison samples or may be caused by a failure in the IL technique as applied in this work.   Although [Mg/Fe] ratios are often used for [$\alpha$/Fe] constraints at low to moderate spectral resolution for extragalactic systems, we conclude that Mg measurements of individual lines from GC IL spectra using our current technique do not reproduce results from standard stellar analyses.

\subsection{Fe Peak Elements: Ni, Cr, Co, Cu, Mn, V, Sc}

We measure abundances for 7 Fe peak elements; the results for Ni I, Cr I, and Co I are shown in Figure \ref{fig:fepeak}, Cu I, Mn I and V I are shown in Figure \ref{fig:fepeak2}, and Sc II is shown in Figure \ref{fig:sc}.   

 We are able to measure Ni I and Cr I over the entire metallicity range.  Ni I measurements agree well with reference abundances over the entire range, and Cr I agrees best for GCs with [Fe/H]$>$-2.  For both elements  the smallest offsets are for differential abundances, which result in $\Delta=-0.02$ with $\sigma_{\Delta}=0.06$ and $\Delta=-0.09$ with $\sigma_{\Delta}=0.11$ for Ni and Cr, respectively.   The [Ni/Fe] and [Cr/Fe]  measurements also agree well with the abundance patterns of MW field stars, as can be seen in the right panels of Figure \ref{fig:fepeak}.   Examples of high S/N Ni measurements are shown in Figure \ref{fig:eu_synth1} for NGC 6440 and NGC 6752.  As seen in Figure \ref{fig:fepeak}, we  find that [Cr/Fe] ratios for the more metal-poor GCs ([Fe/H]$<-1.5$) tend to be lower than those of higher metallicity GCs.  While the difference is small and not statistically significant due to the measurement uncertainties and the  sample size,  this is consistent with other LTE analyses of MW RGB stars at these metallicities \citep[e.g.][]{ishigaki13,2004ApJ...612.1107C,1994AA...285..440N}, and  also seen in IL analyses of M31 GCs \citep{m31p1,m31inprep}.

We measure Co I in 9 GCs, but can only compare to stellar abundances for  4 GCs  with $-1.6<$[Fe/H]$<-0.7$.  The agreement is not as good as that of Ni or Cr; we find a mean offset for differential abundances of  $\sigma_{\Delta}$-0.11, and  $\sigma_{\Delta}=0.13$.  The differential abundances lower the systematic offset by 0.1 dex over the absolute abundances.   We note that Co I could be evaluated more thoroughly if we had stellar reference values available for the GCs with [Fe/H]$>-0.7$;  the results for the remaining 5 GCs with IL measurements are shown in the bottom right panel of Figure \ref{fig:fepeak}.  

 [Cu/Fe] is useful for study of SNII nucleosynthesis and the time delay of SNIa \citep{2016arXiv160705299M}.    We show Cu I  in Figure \ref{fig:fepeak2}. Cu I is difficult to measure as we use only 2 features (5105\AA, 5782\AA), and these are both weak transitions,  somewhat blended and affected by hyperfine splitting.   We are very limited in our evaluation of Cu because we have only two GCs for which we can make comparisons to stellar values.   We find that the differential abundances result in an offset of $\Delta=0.00$, while the absolute abundances result in  a larger offset of $\sigma_{\Delta}$=-0.13. In total we measure Cu I for 7 of the GCs, most of which are in the higher metallicity range of our sample, as can be seen in the upper right panel of Figure \ref{fig:fepeak2}. The GCs with [Fe/H]$<-0.5$ lie in the lower range of the [Cu/Fe] observed in MW field stars.  The  GCs with [Fe/H]$>-0.5$ have  higher [Cu/Fe] that are $\gtrsim0.2$ dex  above solar ratios. Although the measurement uncertainty is large in these cases, enhanced [Cu/Fe] has also been seen in MW bulge stars \citep{johnson14}.  With similar findings for the alpha elements Ca, Ti, Si, and Al, the Cu results at high metallicity provide more evidence that the bulge GCs have bulge-like stellar abundances.       We conclude that Cu has potential for extragalactic studies if the behavior at high metallicity could be more thoroughly evaluated  with additional stellar measurements.

 Mn is an interesting element for chemical evolution studies because it may provide constraints on the ratio of SNIa to SN II, as well as the mechanism and progenitors of SNIa \citep[see review by ][]{2016arXiv160705299M}, although NLTE effects may change this interpretation \citep[e.g.][]{2008A&A...492..823B}. 
Here, IL Mn I is measured in 9 GCs, but can only be compared to stellar results for 6 GCs.  The GCs missing in the comparison are those with [Fe/H]$>-0.6$, so Mn is an element for which the IL comparison could be revisited with additional stellar abundance data to evaluate this element in a larger range in [Fe/H] than we are able to do in this work.   The offset calculated for differential abundances is $\Delta=-0.04$  with $\sigma_{\Delta}$=0.14 and for absolute abundances is $\Delta=-0.13$  with $\sigma_{\Delta}$=0.14. Mn I is one of the elements for which the agreement for NGC 6528 is particularly poor.  However, since we do not include NGC 6528 in the comparison,  the scatter is  driven by  an  IL abundance for Fornax 3 that is  $\sim$0.3 dex lower than the stellar result. \cite{letarte} measure [Mn/Fe]=0.01 for Fornax 3 from two stars,  however their Mn abundances may be more uncertain than other elements, as they could only measure Mn in 3 of the 9 stars in their sample in Fornax GCs. A  solar [Mn/Fe] is somewhat surprising because it is higher than what is seen in the MW and dwarf galaxies at [Fe/H]$\sim -2$ \citep{2012A&A...541A..45N}, but the Mn measurement or its implication are not discussed further by \cite{letarte}.  The IL measurement of [Mn/Fe]$=-0.3$ is  more consistent with the other lower metallicity GCs in our sample, as well as the MW field stars shown in the middle right panel of Figure \ref{fig:fepeak2}, which could mean that IL [Mn/Fe]  behave as expected despite the disagreement for Fornax 3.  If we exclude the comparison for Fornax 3 we  would derive $\Delta=+0.03$  with $\sigma_{\Delta}$=0.09,  which would improve the agreement for Mn.  In light of this, we give an optimal range for [Mn/Fe] of $-1.7<$[Fe/H]$<-0.8$ until more comparisons can be made at high and low [Fe/H].

We measure V I in 9 GCs, 7 of which can be compared to stellar references.  The offset for the  GCs is $\Delta=-0.04$ when differential abundances are used, with a large spread of $\sigma_{\Delta}=0.21$.  The offset is larger when absolute abundances are used; we derive $\Delta=-0.16$ with  $\sigma_{\Delta}$=0.18.
NGC 6528 again has the most discrepant ratio.  The weighted scatter is driven by the [V/Fe] of NGC 6752, which is  significantly higher than the stellar comparison.  In the 5 references in Table \ref{tab:source} for NGC 6752 there is only one [V/Fe] measurement, which is  found in \cite{yong2005}. These authors discuss that the [V/Fe]$=-0.28$ that they find is low in comparison to a previous measurement by \cite{1995ApJ...441L..81N} (which is not included in our reference list for NGC 6752), and suggest that their measurement error of V I may be underestimated because of the large line to line scatter in their V abundances.   If we compare the [V/Fe] results for all of the GCs with the MW field stars in the bottom right panel of Figure \ref{fig:fepeak2}, we find that the IL measurements overlap with the MW field star abundances, although the  majority of the GCs fall in the lower range. This is consistent with the offset we would obtain if we eliminated NGC 6752 from the comparison,  suggesting that the  [V/Fe] measured in IL is systematically $\sim0.1$ dex low.

\begin{figure}
\centering
\includegraphics[scale=0.4]{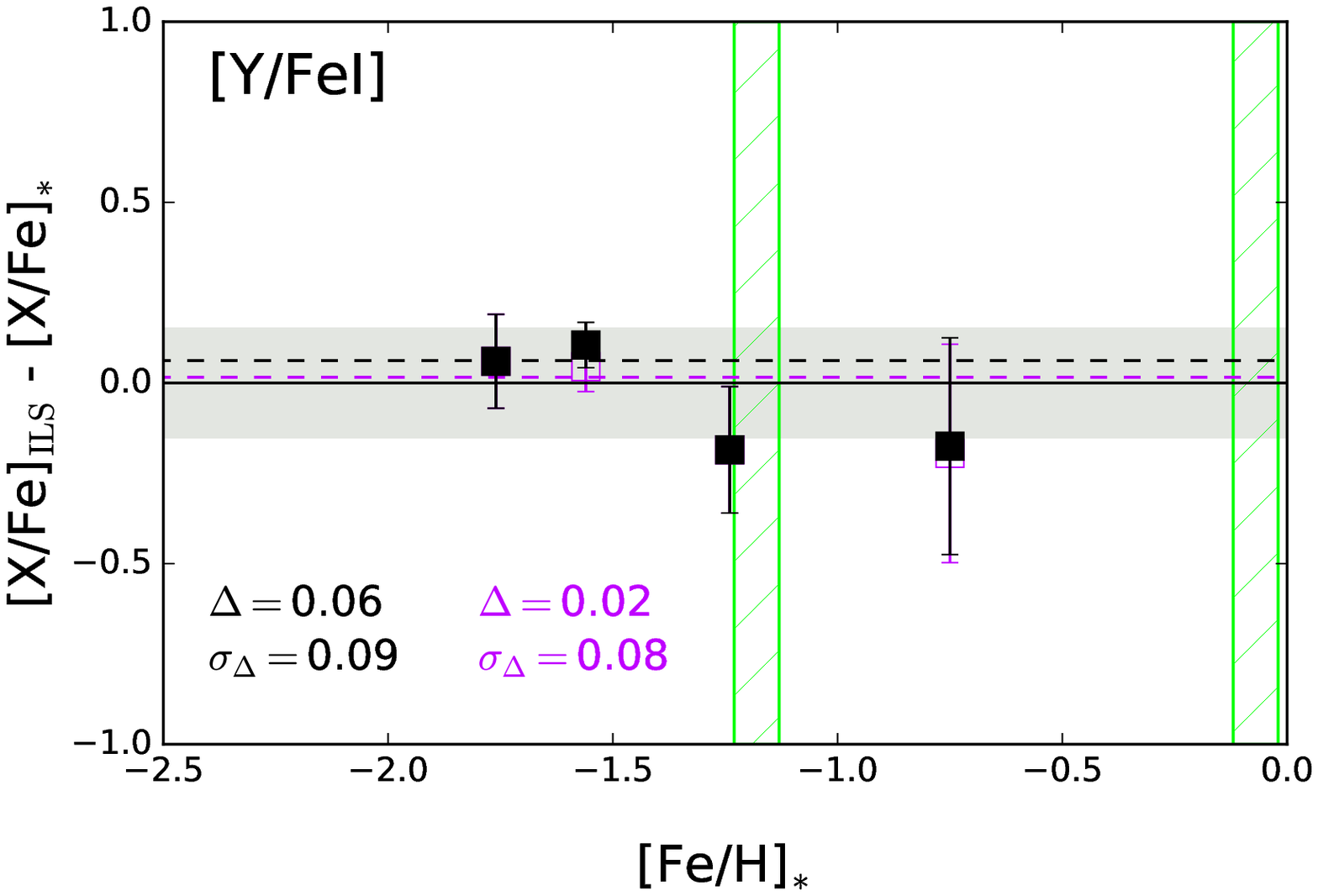}
\includegraphics[scale=0.4]{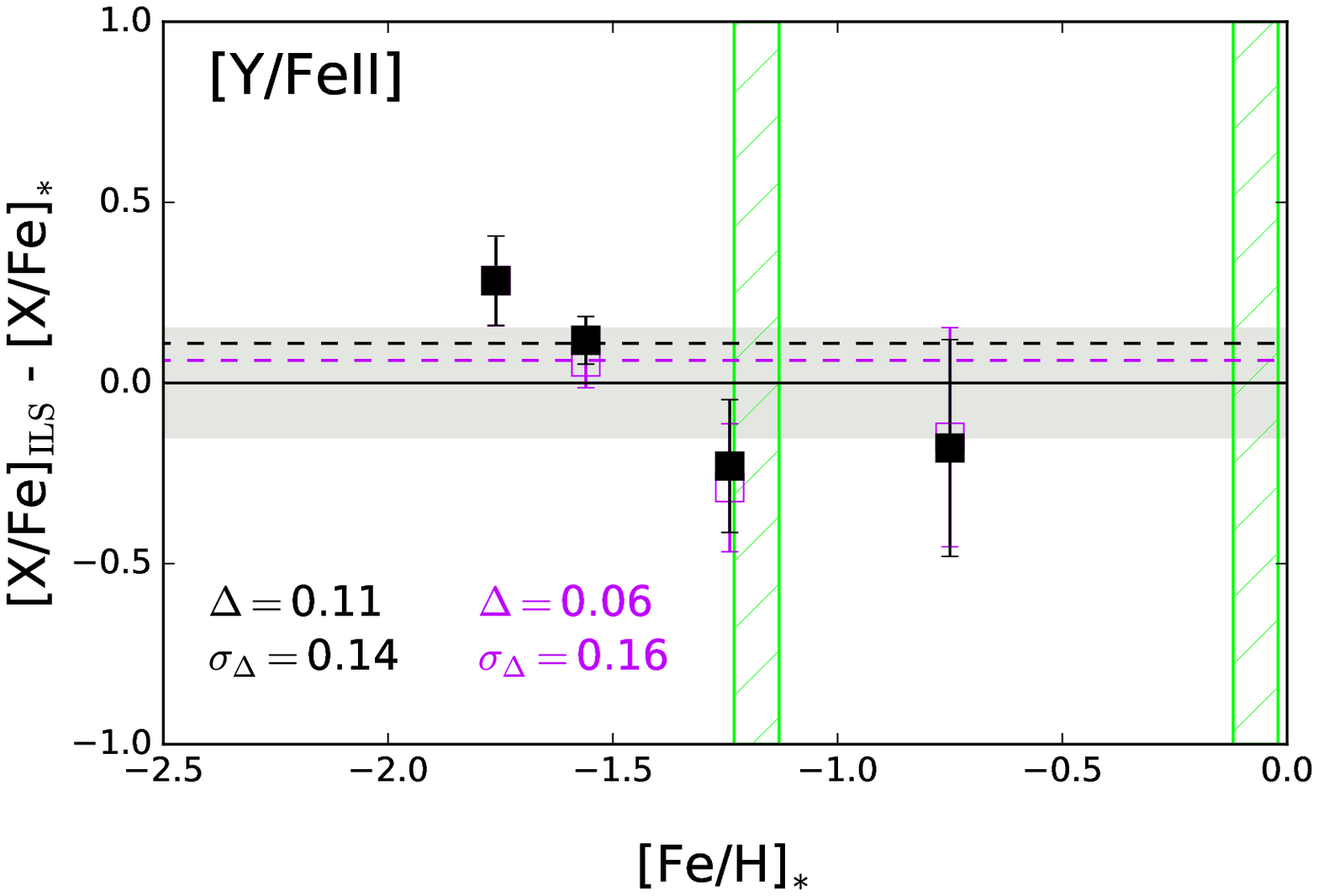}
\includegraphics[scale=0.4]{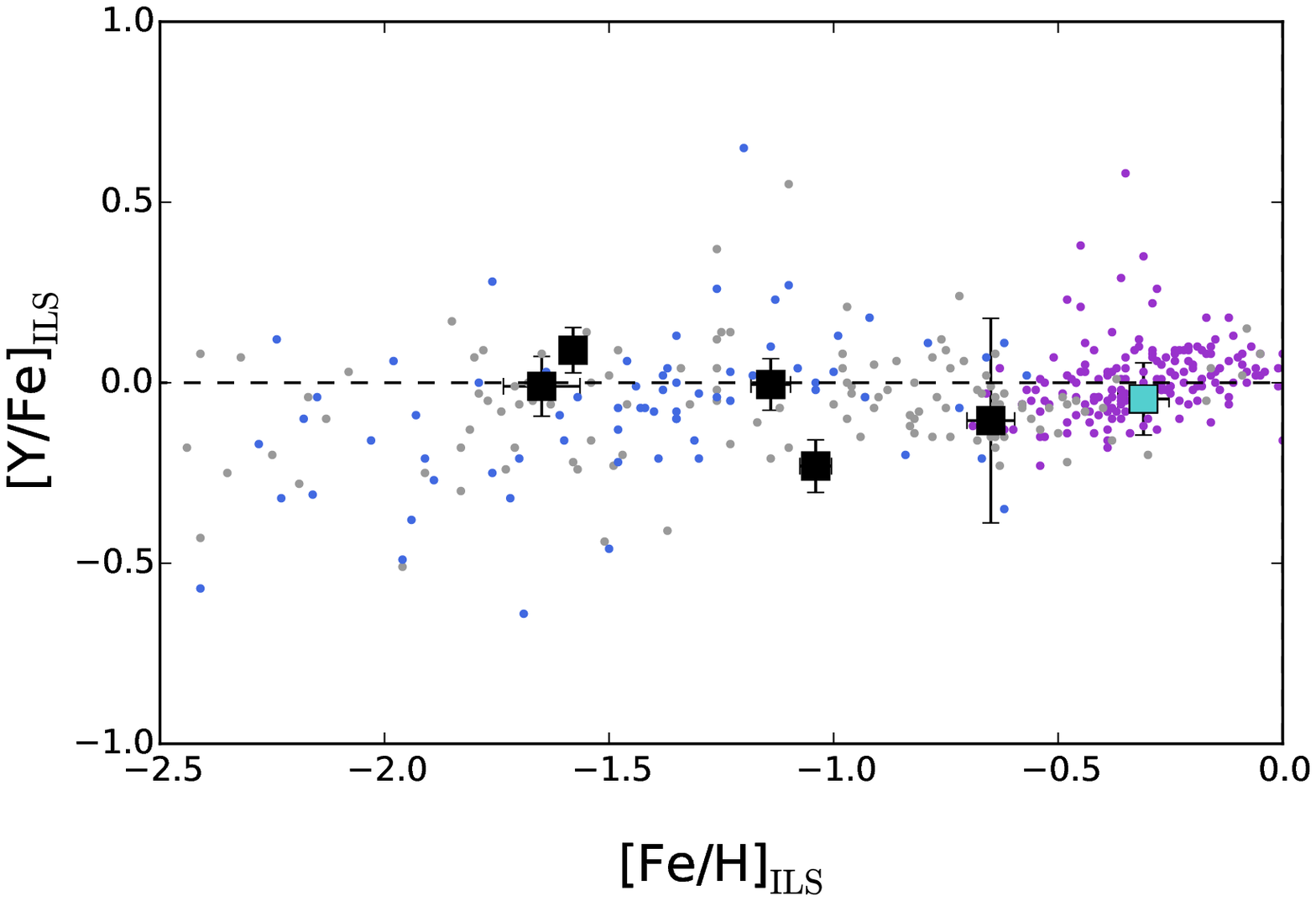}

\caption{ The same as Figure \ref{fig:ti2} for Y II.}
\label{fig:y} 
\end{figure}

\begin{figure}
\centering
\includegraphics[scale=0.4]{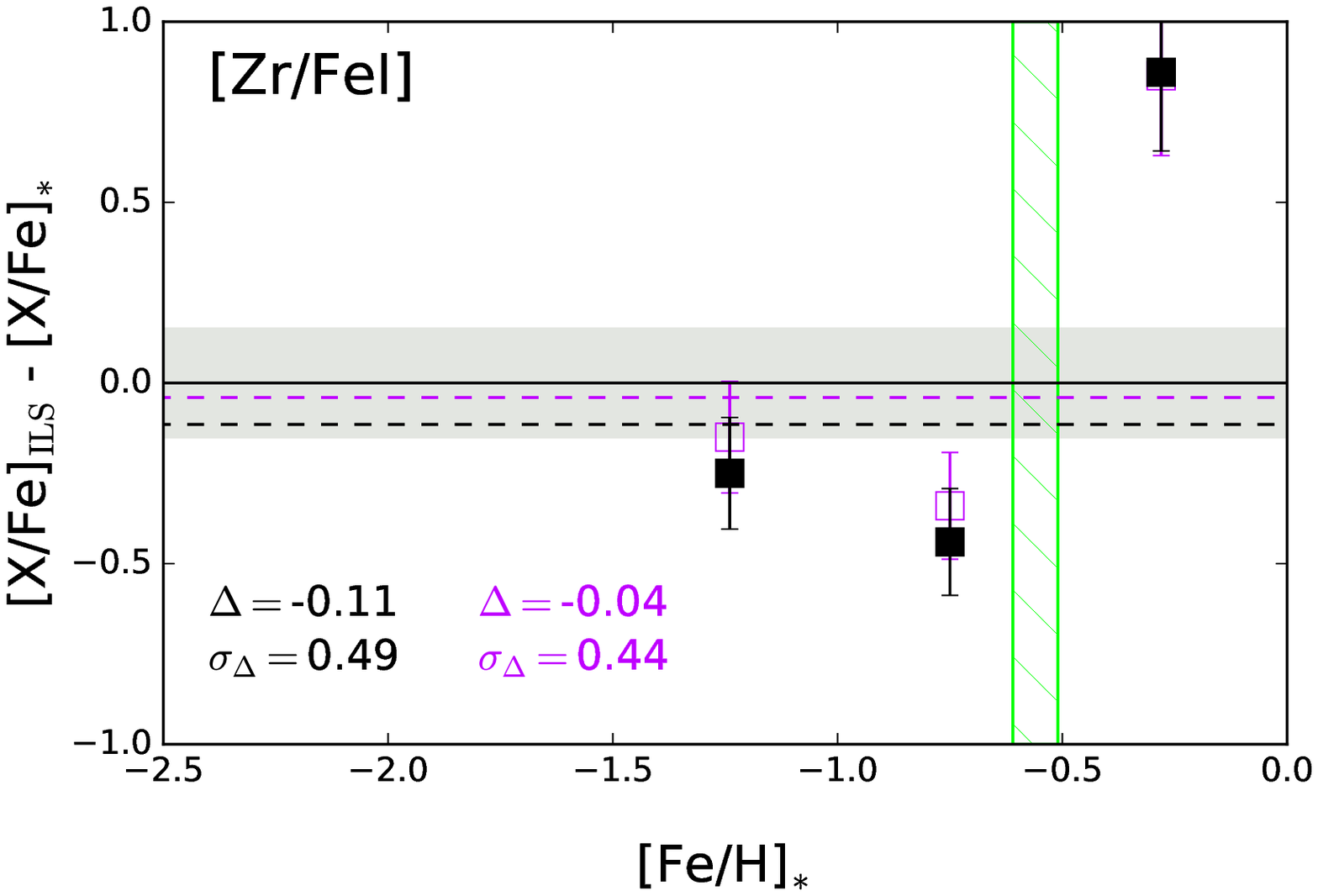}
\includegraphics[scale=0.4]{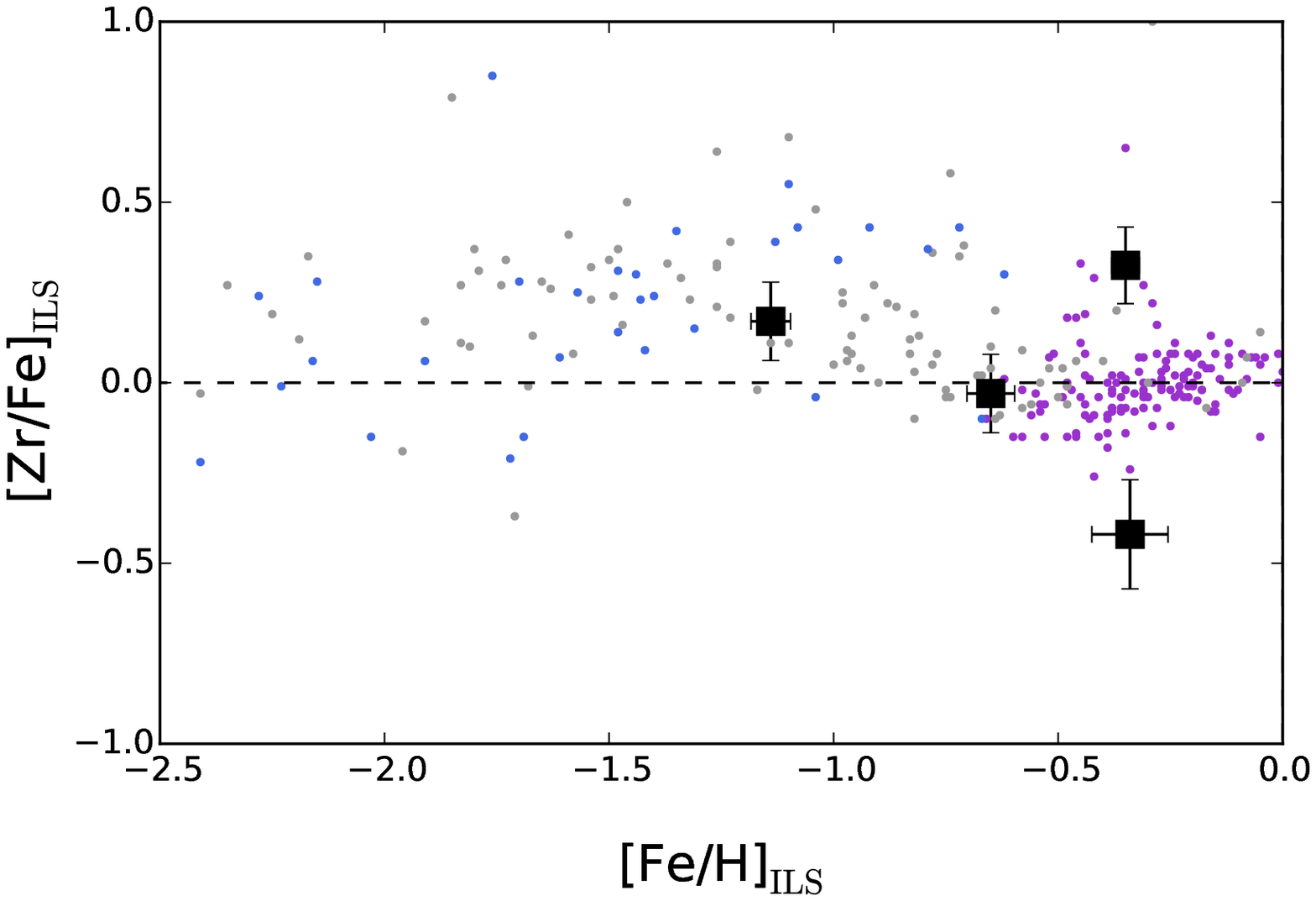}

\caption{The same as Figure \ref{fig:alpha} for Zr I. }
\label{fig:zr} 
\end{figure}

\begin{figure}
\centering
\includegraphics[scale=0.4]{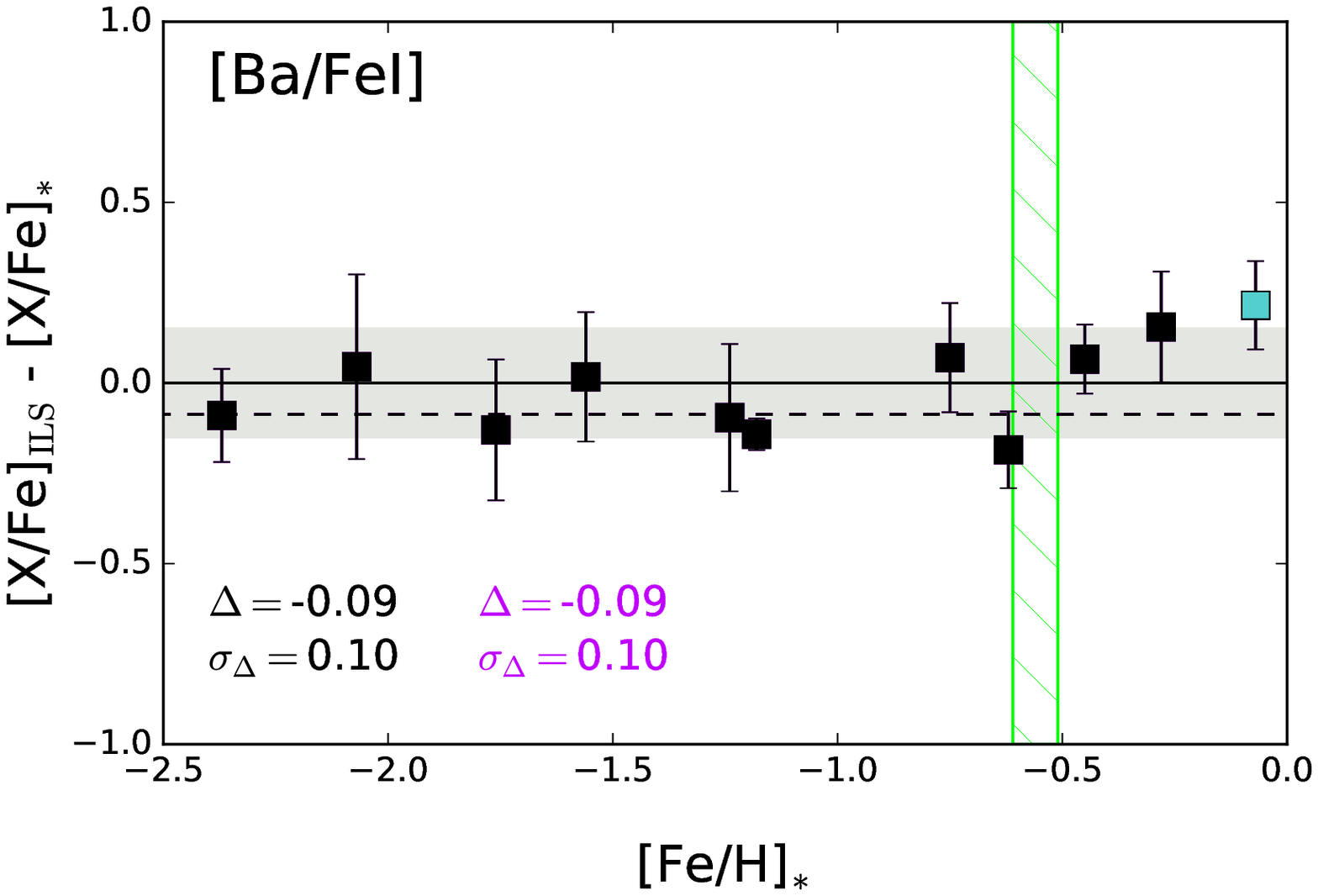}
\includegraphics[scale=0.4]{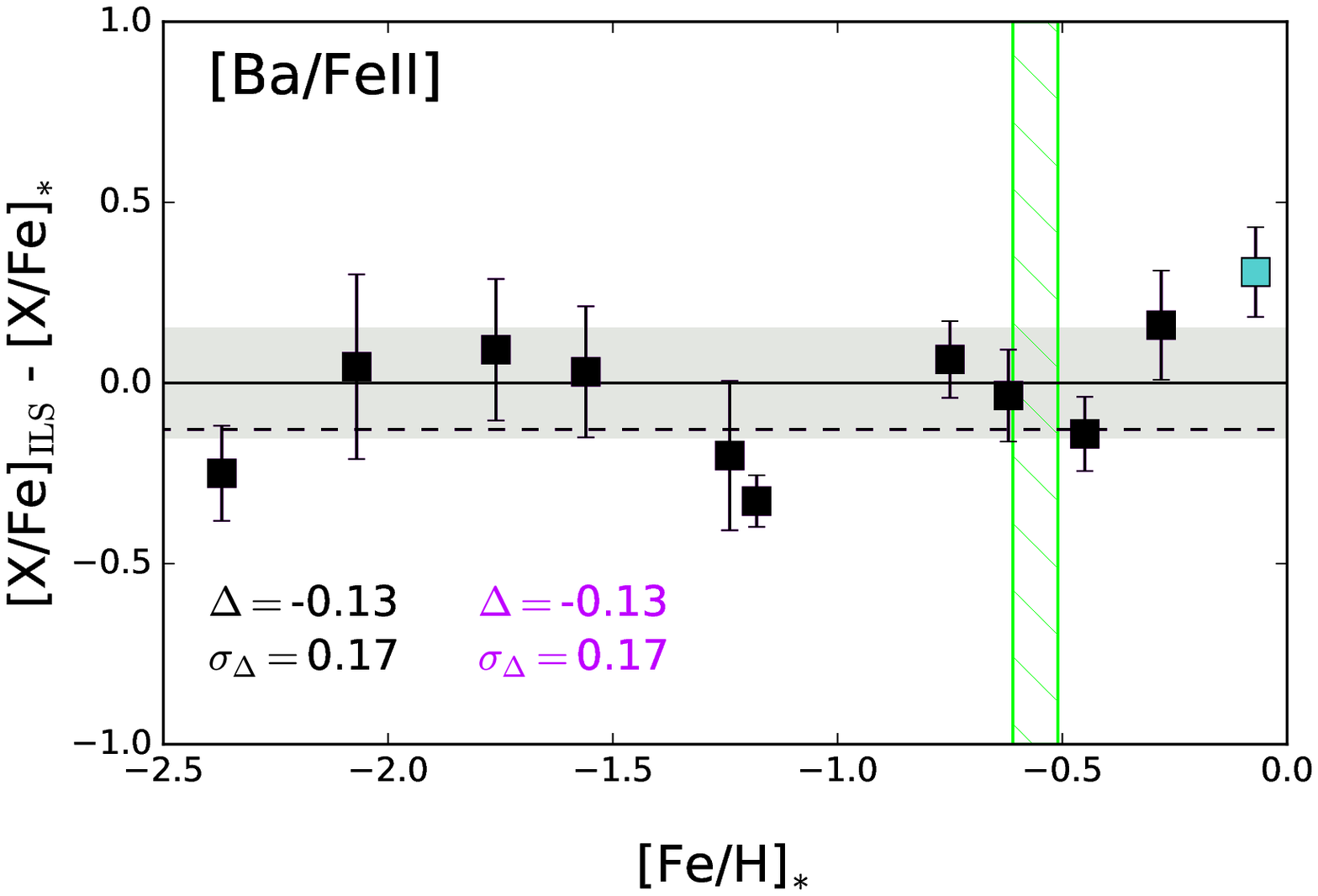}
\includegraphics[scale=0.4]{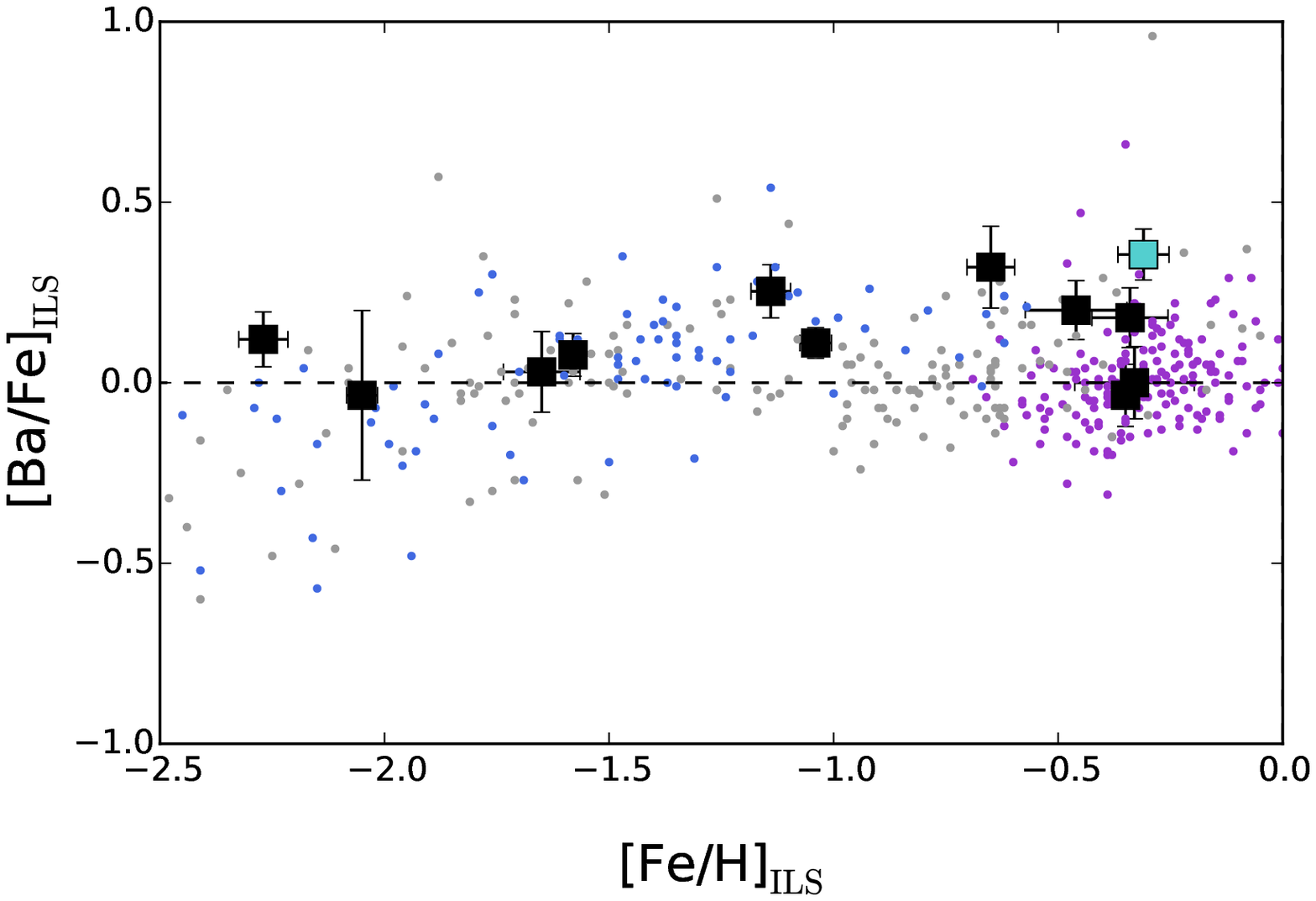}

\caption{ The same as Figure \ref{fig:ti2} for Ba II.}
\label{fig:ba} 
\end{figure}

In Figure \ref{fig:sc} we show the results for Sc II for ratios taken with respect to both Fe I and Fe II.  
The offsets are a bit smaller when the ratios are taken with respect to Fe II; in the differential case $\sigma_{\Delta}=0.05$ for Fe I and   $\sigma_{\Delta}=-0.01$ for Fe II, however the scatter is much larger when the ratio is taken with respect to Fe II.  
The differential and absolute abundances give similar results to each other in each case.   The outliers in the comparison are the higher metallicity GCs NGC 6441 and NGC 6553. Both reference values are from one study each, which means the comparison between analyses is more susceptible to systematics from differences in model atmospheres, stellar parameters, and line lists. For NGC 6441 we only measure 1 line, and the GC has one of the larger velocity dispersions of 18 \kms, which can make abundance measurements more uncertain.   For NGC 6553, \cite{cohen99} do not use HFS   in  analysis of Sc II, which could cause a systematic offset, and we note that their  abundance was  determined from a single line, although a different line than the ones we measure for this GC.   With these potential issues in mind,  we suggest an optimal range for Sc II to be $-1.7 < $[Fe/H] $<-0.6$, although we note this range could be extended to [Fe/H]=$-2.4$ with additional stellar measurements for Fornax 3.

 In summary, of the Fe-peak elements, Ni I and Cr I are the most easily measurable across the whole metallicity range of the sample and show the best agreement with stellar reference abundances. Sc II compares well for a smaller range of [Fe/H] of $-1.7 < $[Fe/H] $<-0.6$.   Cu I shows potential if the comparison to reference abundances could be expanded at high metallicity. 
The comparisons for Co I and Mn I are also limited by a lack of reference abundances at high metallicity; Mn I shows good agreement for intermediate metallicities.   V I can  be measured consistently for the more metal rich GCs, but tends to give abundance ratios that are systematically low.

\subsection{Neutron Capture Elements: Y, Zr, Ba, La, Nd, Eu}

We present IL measurements for Y II, Zr I, Ba II, La II, Nd II and Eu II in Figures \ref{fig:y}-\ref{fig:eu}.    The neutron capture  elements are generally more difficult to measure in IL spectra because there are fewer transitions, they are often weak features in the IL, and are often blended with other transitions.    With the exception of Ba II, we are only able to measure these heavy elements in a few clusters each, and there are often no comparison values in the references in Table \ref{tab:source}.    Although we do not do so in this work, we can in principle  put upper limits on some of these ratios using the IL spectral synthesis, which could prove interesting in extragalactic systems where neutron capture element information is scarce.

In Figure \ref{fig:y} we show the results for Y II, which we measure in 6 GCs and compare to stellar values for 4 GCs.  Y is useful in chemical evolution studies for study of the s-process, and with Ba can provide ratios useful for evaluating the relative contributions to two s-process peaks \citep{1999ARA&A..37..239B}.  For Y II we find that the absolute abundances give smaller offsets, although in both cases the offsets are less than 0.1 dex. When the ratio is taken with respect to Fe I the absolute ratios give $\Delta$=0.02 and  $\sigma_{\Delta}$=0.08, and the differential abundances result in $\Delta$=0.06 and  $\sigma_{\Delta}$=0.09.  When the ratios are taken with respect to Fe II abundances the systematic offsets and scatter increase for both absolute and differential abundances.   The Y II comparison is for a smaller range in [Fe/H] than Ba II, although we note that we have generally  been unable to measure Y II in GCs more metal poor than those tested here \citepalias{m31p1,paper4,m31inprep}. 
  As seen in Figure \ref{fig:y}, the [Y/Fe] of the  6 GCs  compares well with MW field star ratios.

Zr I is one of the most difficult elements to measure in the IL spectra, usually blended, and we typically must measure the abundance from a single line.  We measure [Zr/Fe] in 4 of the more metal rich GCs but only have comparison values for 3 GCs: NGC 104, NGC 362 and NGC 6553.  The agreement with stellar abundances is  poor, and both the absolute and differential abundances have the largest scatter for any element tested in this work, with $\sigma_{\Delta}>$0.4 in both cases.  
 The [Zr/Fe] of all 4 GCs are shown in the bottom panel of Figure \ref{fig:zr}, where they appear to overlap with MW field stars, but there is large scatter in both the IL GC measurements and the stellar measurements.  Given the poor comparison we conclude that Zr will not be useful for GC IL abundance work. 

By contrast, Ba II is generally the easiest heavy element to measure in the IL spectra because of several strong transitions.   We show an example of the data and synthesis for a Ba II line in NGC 6752 in Figure \ref{fig:ba_synth1}.  Ba II is extensively used in chemical evolution studies as a probe of the s-process \citep[e.g][]{1999ARA&A..37..239B,2008ARA&A..46..241S, 2016arXiv160705299M}.   We are able to measure Ba II in all 12 GCs in the sample, 11 of which have reference values from individual stars.   The [Ba/Fe] results are shown in Figure \ref{fig:ba}, and we find that both the offset and scatter are best when [Ba/Fe] is calculated with respect to Fe I abundance rather than Fe II abundance, Both the absolute and differential abundances give the same result since we do not find variations between lines in our solar abundance analysis.   For ratios taken with respect to Fe I we find $\Delta=-0.09$ and $\sigma_{\Delta}=0.10$.

\begin{figure}
\centering

\includegraphics[scale=0.4]{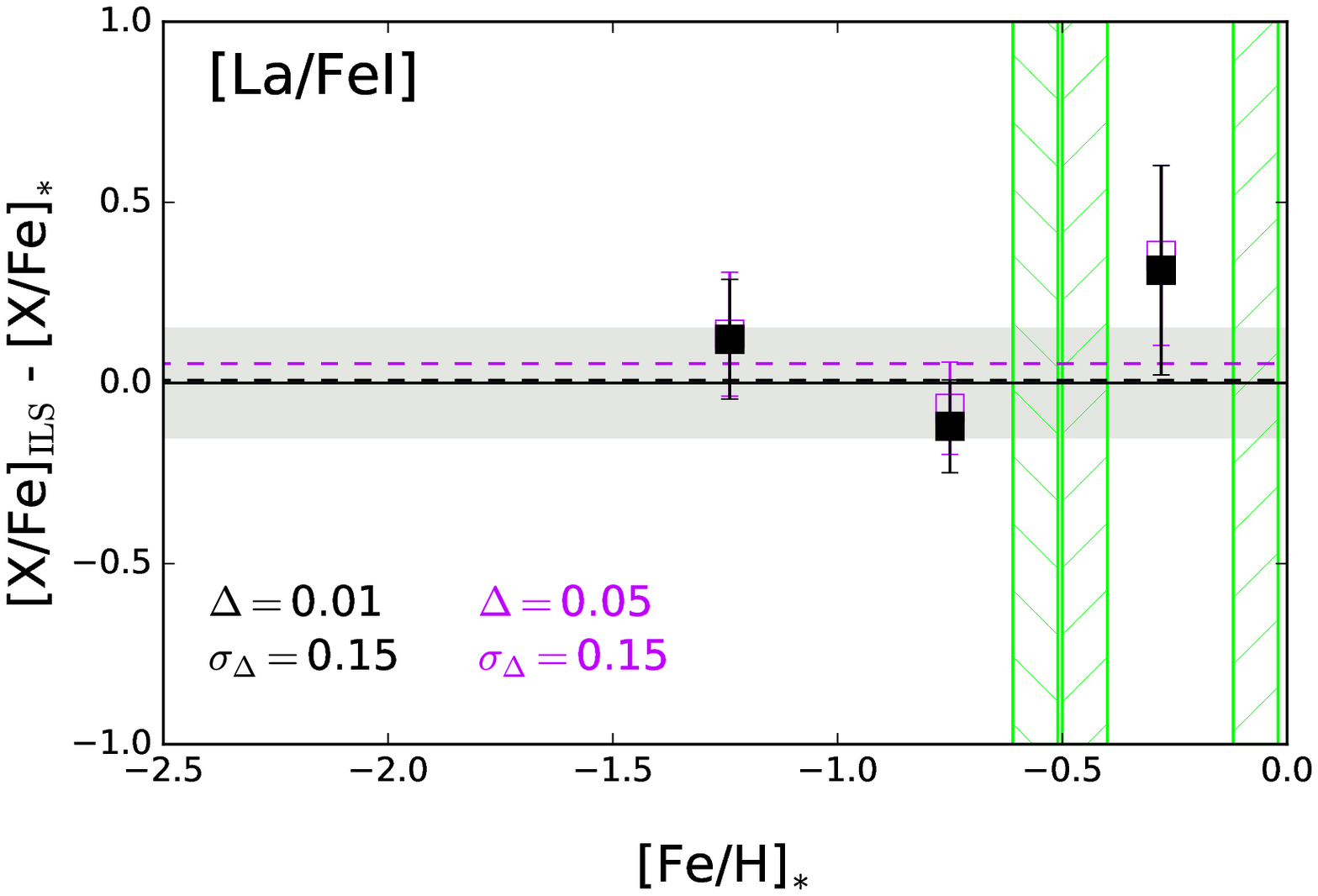}
\includegraphics[scale=0.4]{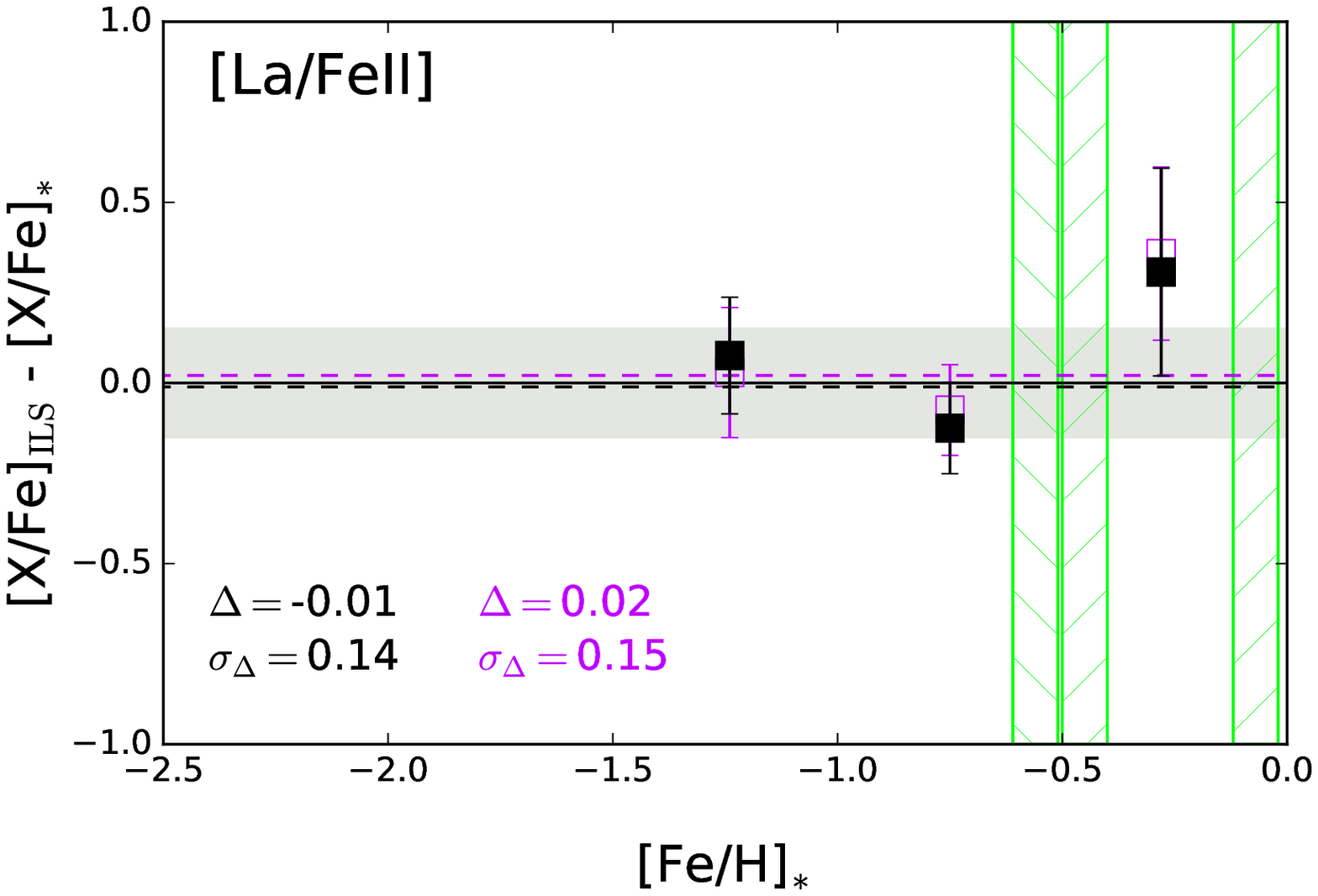}
\includegraphics[scale=0.4]{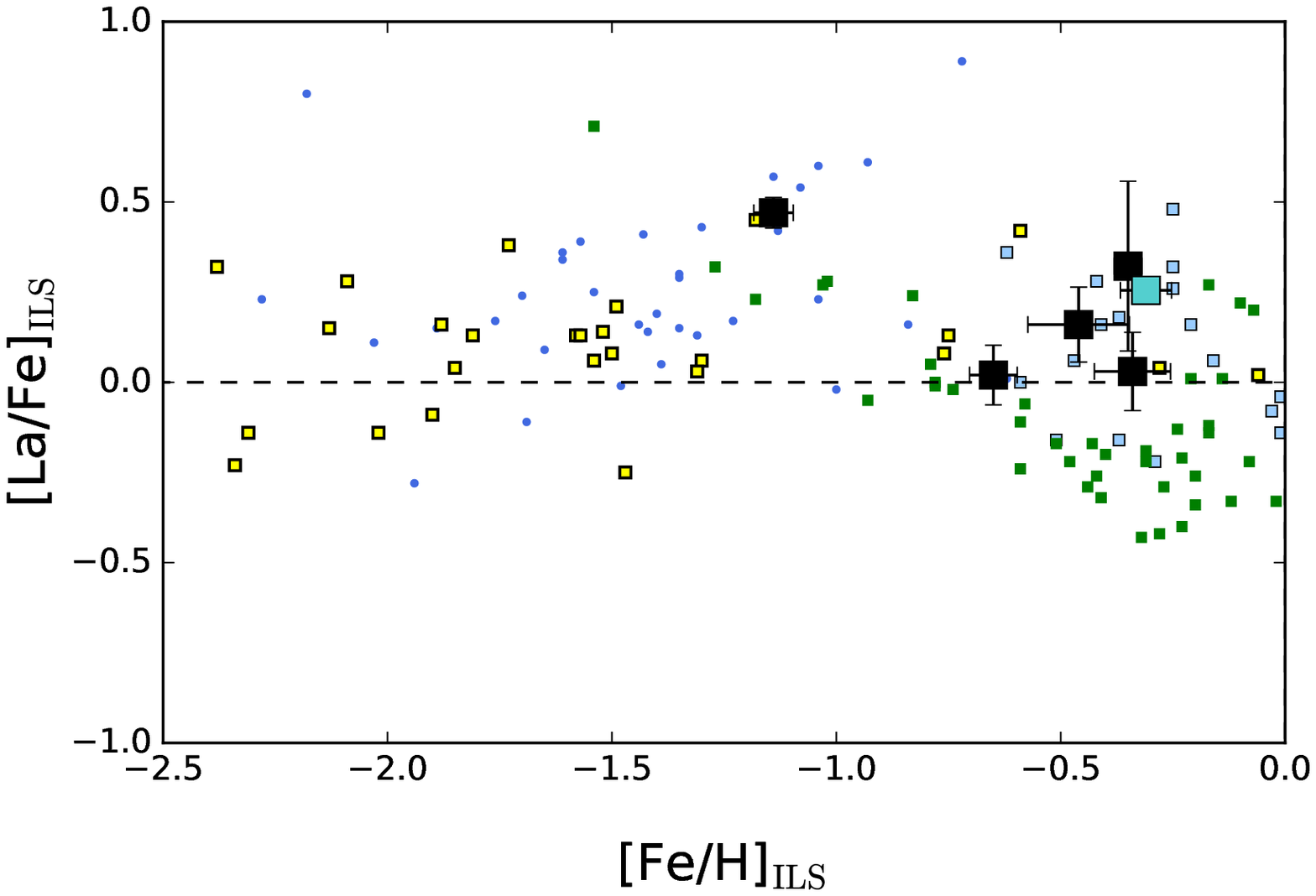}

\caption{ The same as Figure \ref{fig:ti2} for La II.   Additional stellar reference abundances correspond to bulge stars of  \cite{2016A&A...586A...1V} (light blue squares) and \cite{2012ApJ...749..175J} (green squares), as well as GC stellar abundances from \cite{pritzl05}.  }
\label{fig:la} 
\end{figure}

\begin{figure}
\centering
\includegraphics[scale=0.4]{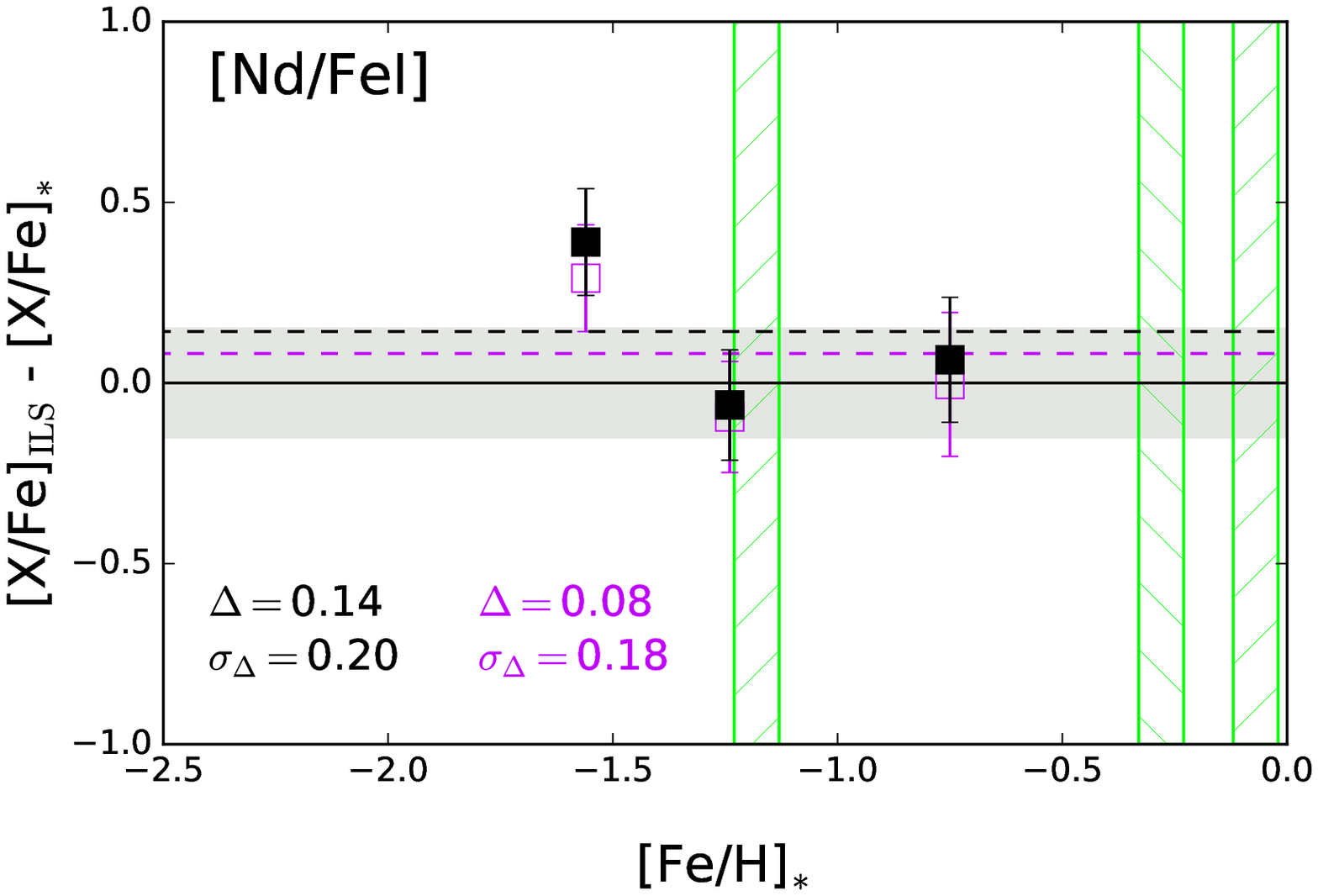}
\includegraphics[scale=0.4]{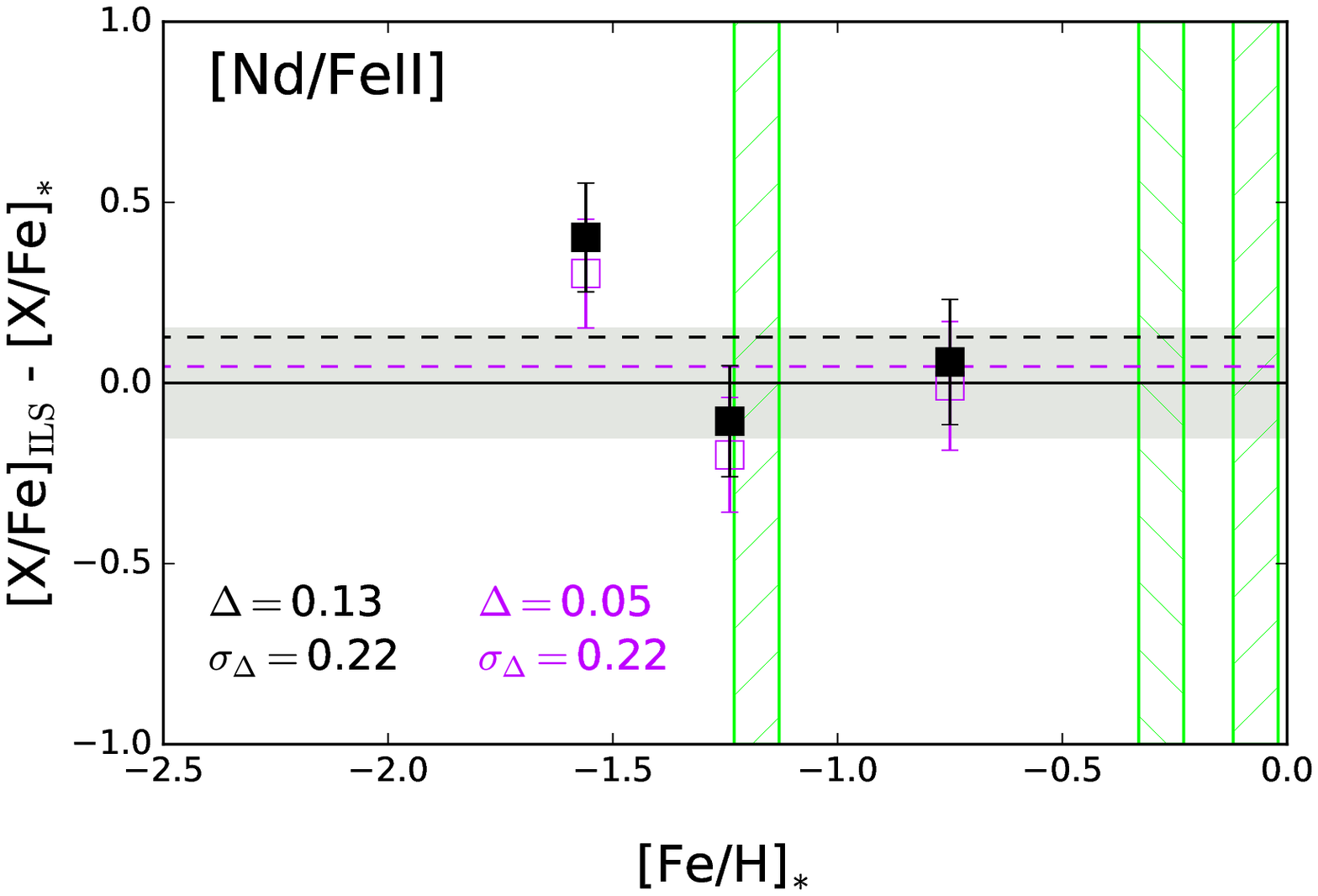}
\includegraphics[scale=0.4]{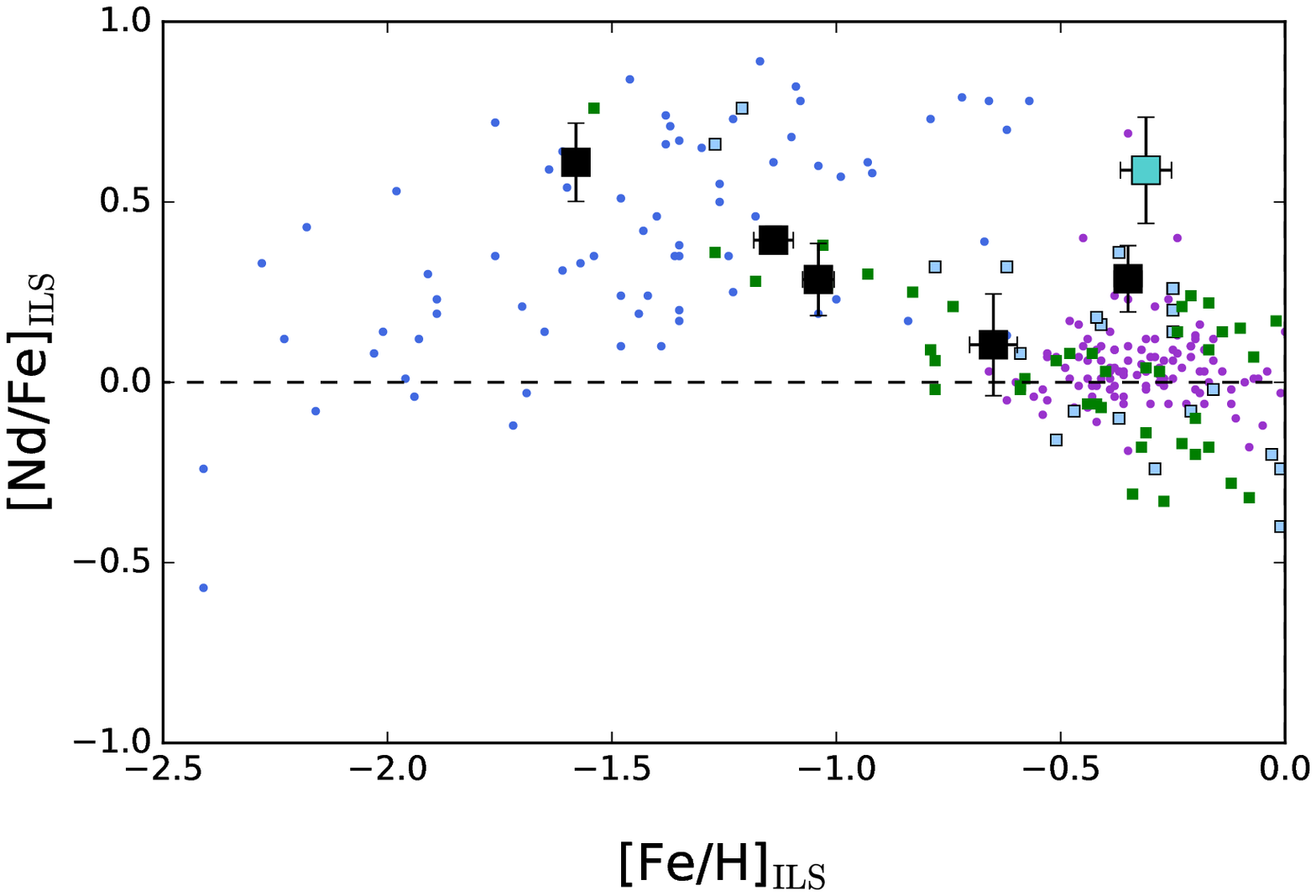}

\caption{The same as Figure \ref{fig:ti2} for Nd II. Additional stellar reference abundances are listed in Figure \ref{fig:la}.}
\label{fig:nd} 
\end{figure}

\begin{figure}
\centering
\includegraphics[scale=0.4]{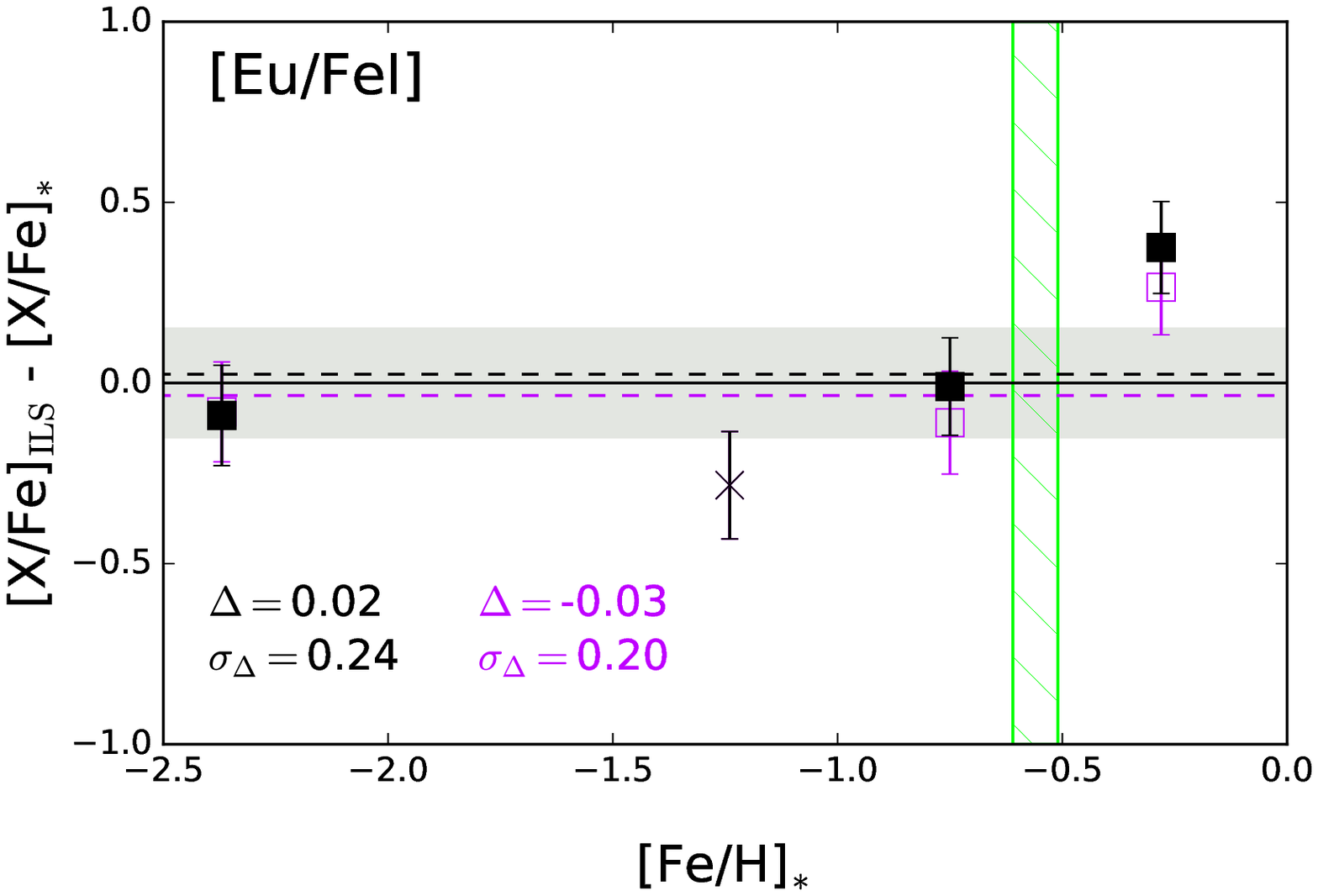}
\includegraphics[scale=0.4]{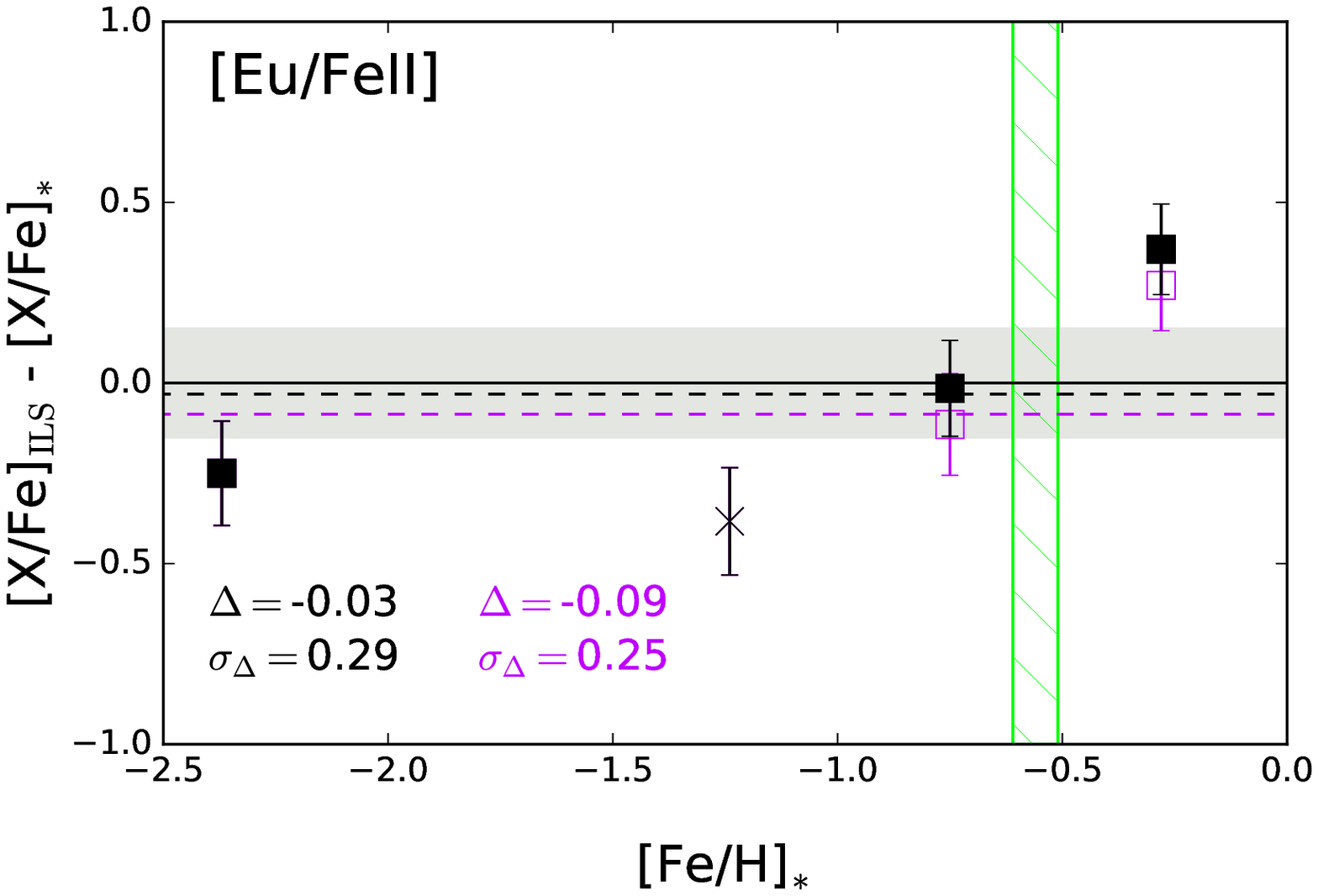}
\includegraphics[scale=0.4]{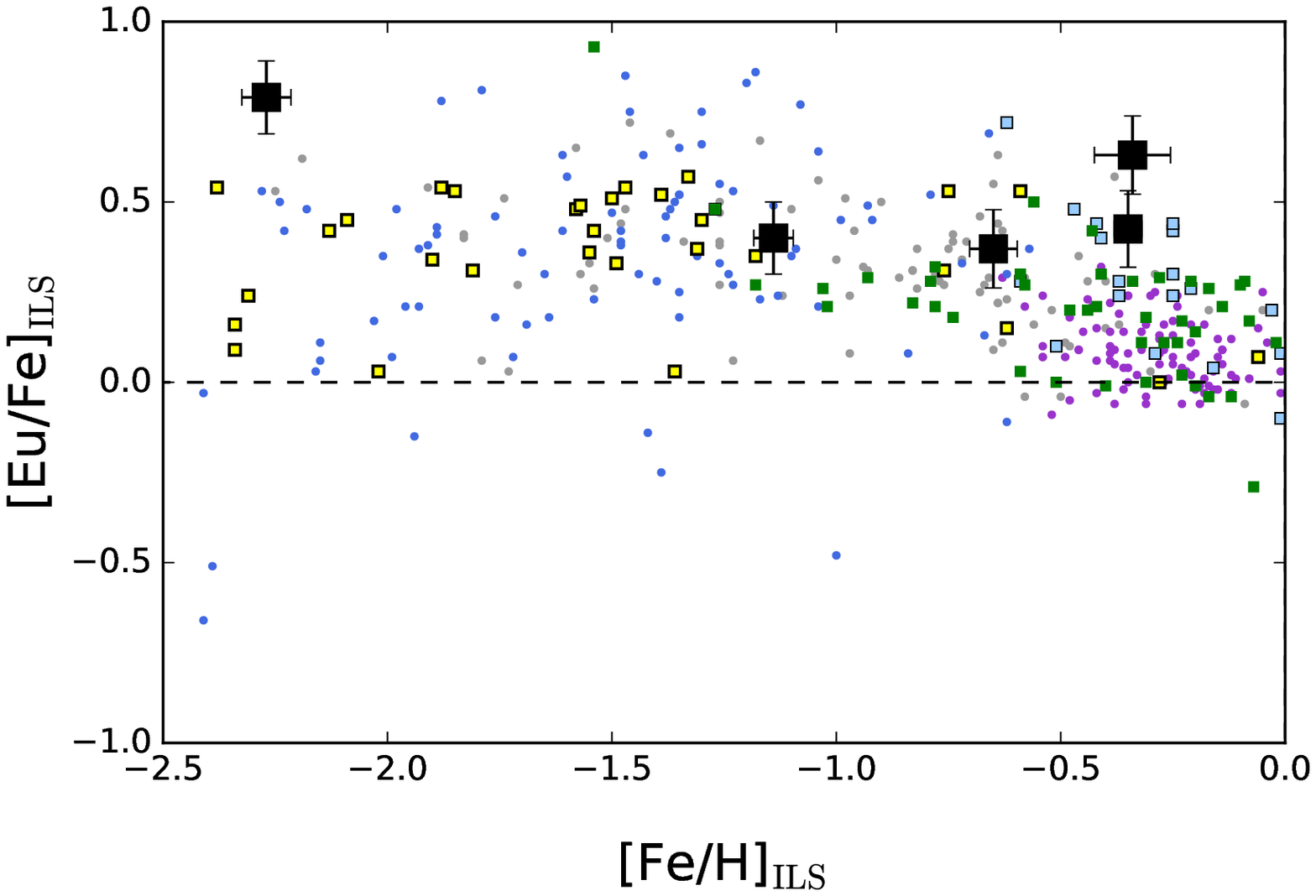}

\caption{The same as Figure \ref{fig:ti2} for Eu II. Additional stellar reference abundances are listed in Figure \ref{fig:la}. }
\label{fig:eu} 
\end{figure}

\begin{deluxetable*}{lrrrr}
\tablecolumns{5}
\tablewidth{0pc}
\tablecaption{Weighted Offsets\label{tab:offset_final}}
\tablehead{ \colhead{Ratio}  &\colhead{$\Delta$} &\colhead{$\sigma_{\Delta}$}  &\colhead{N$_{GC}$} &\colhead{[Fe/H]}\\ & & & & \colhead{Range Analyzed} \\ & (dex) & (dex) }

\startdata
\cutinhead{Offsets Including All GCs}

$[$Fe I/H$]_{Harris}$ & $-$0.02  & 0.11 & 11 & $-2.1<$[Fe/H]$<-0.1$ \\
$[$Fe I/H$]_{*}$ &$+$0.07  & 0.08 & 12 & $-2.4<$[Fe/H]$<-0.1$  \\

 &&&&Optimal $-2.4<$[Fe/H]$<-0.3$\\
\cutinhead{Offsets without NGC 6528}

$[$Fe I/H$]_{Harris}$ & $-$0.01  & 0.10 & 10 & $-2.1<$[Fe/H]$<-0.3$\\
$[$Fe I/H$]_{*}$ &$+$0.08  & 0.07 & 11 & $-2.4<$[Fe/H]$<-0.3$\\

\hline
\\

$[$Na I/Fe I$]$ & $-$0.05  & 0.13& 7 & $-1.7<$[Fe/H]$<-0.3$  \\
\hline
\\
$[$Mg I/Fe I$]$ & $-$0.24  & 0.07& 11 & $-2.4<$[Fe/H]$<-0.3$ \\
\hline
\\
$[$Al I/Fe I$]$ & $-$0.08 &  0.09 & 5 & $-1.2<$[Fe/H]$<-0.3$  \\
\hline
\\
$[$Si I/Fe I$]$ &$+$0.00  & 0.07 & 9  &$ -1.8<$[Fe/H]$<-0.3$ \\
\hline
\\

$[$Ca I/Fe I$]$ &  $+$0.00  & 0.12 &11 &$ -2.4<$[Fe/H]$<-0.3$  \\
\hline
\\
$[$Ti I/Fe I$]$ & $+$0.04  & 0.23 & 10  &$ -2.4<$[Fe/H]$<-0.3$  \\
\hline
\\
$[$Ti II/Fe I$]$ &$-$0.07  & 0.12 & 9 &$ -2.4<$[Fe/H]$<-0.3$  \\
$[$Ti II/Fe II$]$ & $-$0.06  & 0.13 & 9 &$-2.4<$[Fe/H]$<-0.3$ \\
\hline
\\
$[$Sc II/Fe I$]$ & $+$0.05  & 0.17 & 8 &$ -1.7<$[Fe/H]$<-0.3$ \\
$[$Sc II/Fe II$]$ &$-$0.01  & 0.26& 8 &$ -1.7<$[Fe/H]$<-0.3$   \\
&&&&Optimal $-1.7<$[Fe/H]$<-0.6$ \\

\hline
\\
$[$V I/Fe I$]$ &$-$0.04  & 0.21 & 6 &$ -1.6<$[Fe/H]$<-0.4$  \\

\hline
\\
$[$Cr I/Fe I$]$ & $-$0.09  & 0.11 & 10 &$ -2.4<$[Fe/H]$<-0.3$ \\
&&&&Optimal $-1.8<$[Fe/H]$<-0.3$ \\
\hline
\\
$[$Mn I/Fe I$]$ &$-$0.04 &  0.14 & 5 &$ -2.4<$[Fe/H]$<-0.8$\\
&&&&Optimal  $-1.7<$[Fe/H]$<-0.8$\\
\hline
\\
$[$Co I/Fe I$]$ & $-$0.11 &  0.13 & 4 &$ -1.6<$[Fe/H]$<-0.4$ \\
\hline
\\
$[$Ni I/Fe I$]$ & $-$0.02 &  0.06 & 9 &$ -2.4<$[Fe/H]$<-0.3$ \\
\hline
\\
$[$Cu I/Fe I$]$ & $+$0.00 &  0.09 & 2 &$ -1.2<$[Fe/H]$<-0.4$ \\
\hline
\\
$[$Y II/Fe I$]$ & $+$0.06 &  0.09 & 4 &$ -1.8<$[Fe/H]$<-0.8$ \\
$[$Y II/Fe II$]$ &$+$0.11  & 0.14 & 4 &$ -1.8<$[Fe/H]$<-0.8$ \\
\hline
\\
$[$Zr I/Fe I$]$ & $-$0.11  & 0.49& 3 &$ -1.2<$[Fe/H]$<-0.3$ \\
\hline
\\

$[$Ba II/Fe I$]$ & $-$0.09  & 0.10 & 10 &$ -2.4<$[Fe/H]$<-0.3$ \\
$[$Ba II/Fe II$]$ & $-$0.13 &  0.17 & 10 &$ -2.4<$[Fe/H]$<-0.3$ \\
\hline
\\
$[$La II/Fe I$]$ & $+$0.01  & 0.15 & 3 &$ -1.2<$[Fe/H]$<-0.3$  \\
$[$La II/Fe II$]$ &$-$0.01  & 0.14 & 3 &$ -1.2<$[Fe/H]$<-0.3$ \\
\hline
\\

$[$Nd II/Fe I$]$ & $+$0.14 &  0.20 & 3 &$ -1.6<$[Fe/H]$<-0.8$ \\
$[$Nd II/Fe II$]$ & $+$0.13 &  0.22 & 3 &$ -1.6<$[Fe/H]$<-0.8$ \\
\hline
\\
$[$Eu II/Fe I$]$ &$+$0.02  & 0.24 & 4 &$ -2.4<$[Fe/H]$<-0.3$ \\
$[$Eu II/Fe II$]$ & $-$0.03  & 0.29 & 4 &$ -2.4<$[Fe/H]$<-0.3$ \\

\enddata
\tablerefs{ Final weighted offsets ($\Delta$) and weighted scatter ($\sigma_{\Delta}$) for each element ratio, corresponding to the abundance comparisons where line-by-line differential abundances were calculated.   N$_{GC}$ shows the number of GCs for which a comparison to stellar reference abundances was available. With the exception of the first two rows, all offsets are calculated after excluding the comparison for NGC 6528. For ionized species we show the offsets both when ratios are taken with respect to Fe I and Fe II abundances. In column 5 we explicitly list the range in [Fe/H] over which the offset and scatter were calculated. When the accuracy of the IL abundances for an element is found to be best over a limited range in [Fe/H] we give an optimal metallicity range below the tested range in [Fe/H].  See text for details.  }
\end{deluxetable*}

\begin{figure}
\centering

\includegraphics[angle=90,scale=0.35]{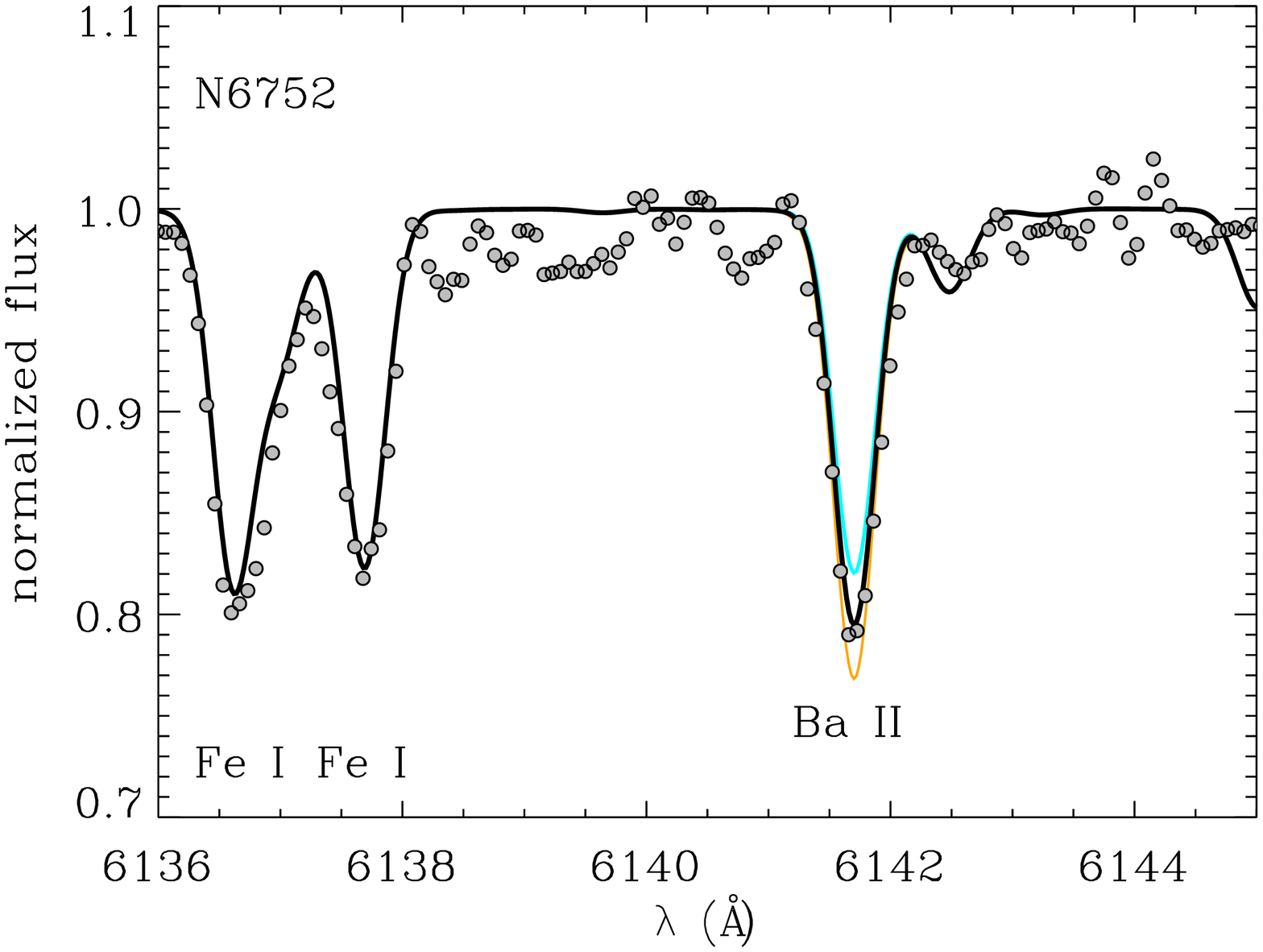}

\caption{Examples of data and synthesis for Ba (6141 \AA) with nearby Fe I lines (6136, 6137 \AA).  Ba II lines are strong and can be measured throughout the entire metallicity range of the sample. Data for NGC 6752 is shown as gray points and is smoothed by 3 pixels.   The solid black line corresponds to synthesized spectra with the mean Ba abundance. The cyan  and orange solid lines show $-0.3$ dex and $+0.3$ dex in [Ba/Fe] from the mean value, respectively.     }
\label{fig:ba_synth1} 
\end{figure}

\begin{figure}
\centering

\includegraphics[angle=90,scale=0.35]{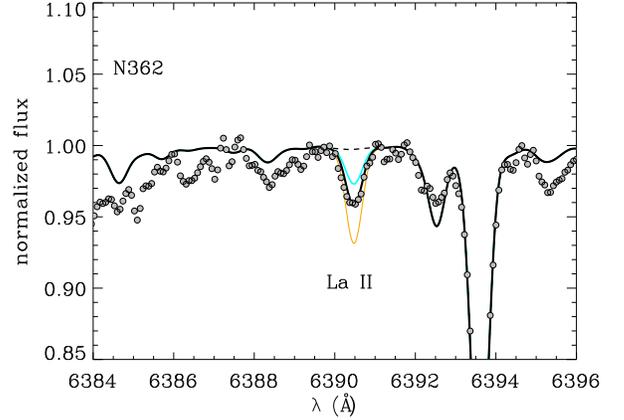}
\includegraphics[angle=90,scale=0.35]{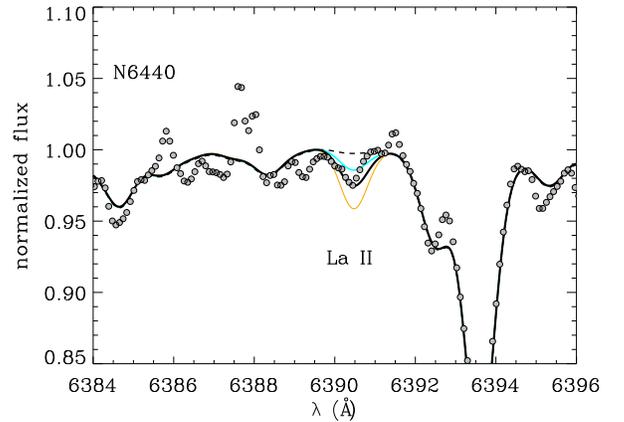}

\caption{Examples of data and synthesis for La (6390 \AA).   La II lines are weak and can only be measured where S/N of the data is high. The top panel shows NGC 362, where the La II measurement is reasonably clean and the bottom panel shows NGC 6440, for which the data is lower S/N but we still make a measurement.  Data is shown as gray points and is smoothed by 3 pixels.   In each panel the solid black line corresponds to synthesized spectra with the mean La abundance. The cyan  and orange solid lines show $-0.4$ dex and $+0.4$ dex in [La/Fe] from the mean value, respectively.   As a consistency check, with the dashed black line we show the region synthesized without the La II line  to check for the influence of underlying blends.     }
\label{fig:la_synth1} 
\end{figure}

\begin{figure}
\centering

\includegraphics[angle=90,scale=0.35]{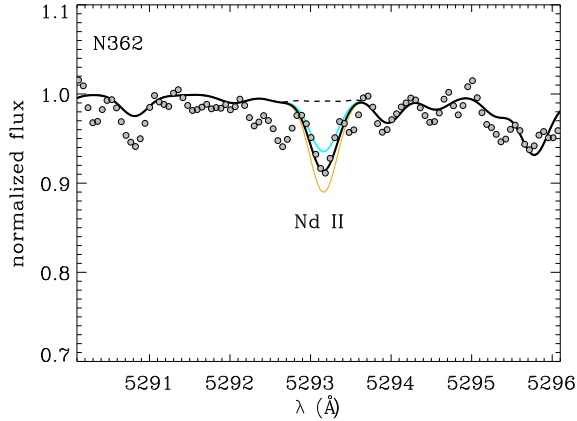}

\caption{Examples of data and synthesis for Nd (5293 \AA).  Nd II lines are weak and can only be measured where S/N of the data is high. Data for NGC 362 is shown as gray points and is smoothed by 3 pixels.   The solid black line corresponds to synthesized spectra with the mean Nd abundance. The cyan  and orange solid lines show $-0.3$ dex and $+0.3$ dex in [Nd/Fe] from the mean value, respectively. As a consistency check, with the dashed black line we show the region synthesized without the Nd II line  to check for the influence of underlying blends.        }
\label{fig:nd_synth1} 
\end{figure}

\begin{figure}
\centering
\includegraphics[angle=90,scale=0.35]{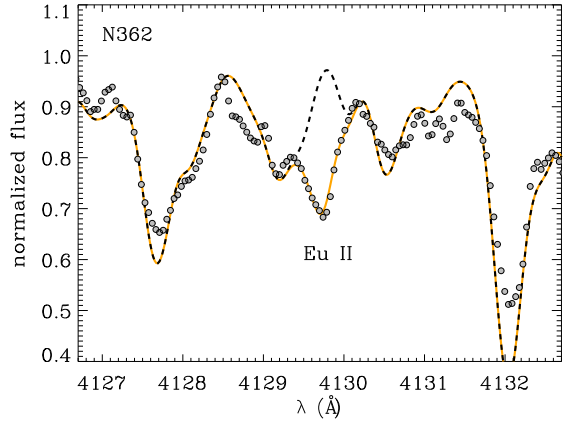}
\includegraphics[angle=90,scale=0.35]{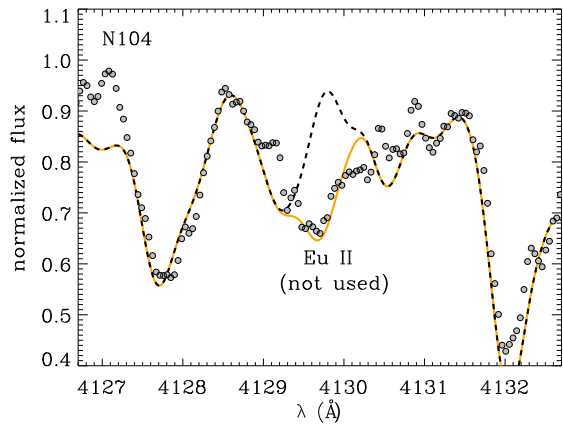}
\includegraphics[angle=90,scale=0.35]{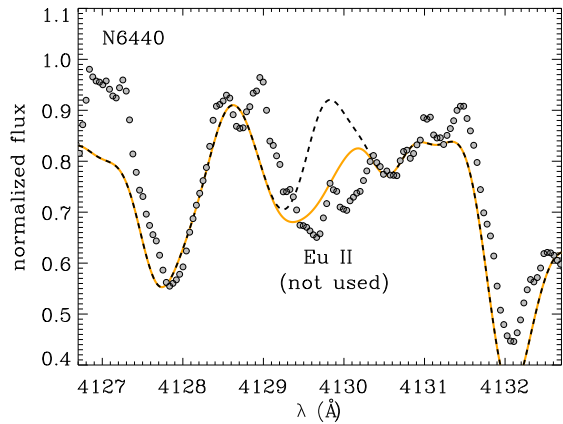}

\caption{Examples of the Eu II feature at 4129 \AA.  From top to bottom the GCs have increasing metallicity (NGC 362=-1.1, NGC 104=-0.7, and NGC 6440=-0.4.) Data is shown by gray points and is smoothed by 3 pixels. The solid orange line shows spectra synthesized with the mean [Eu/Fe] for each GC, and the dashed black line shows spectra synthesized without the Eu II feature included.  The region has strongly blended features, but a Eu II measurement can be made for more metal poor GCs with high enough S/N, as in the case of NGC 362.  With the increasing metallicity of NGC 104 and NGC 6440 the spectra become more blended, the synthesis is a poorer match to the data in the entire region, and the pseudo-continuum placement is more uncertain.    }
\label{fig:eu_synth2} 
\end{figure}

\begin{figure}
\centering
\includegraphics[scale=0.55]{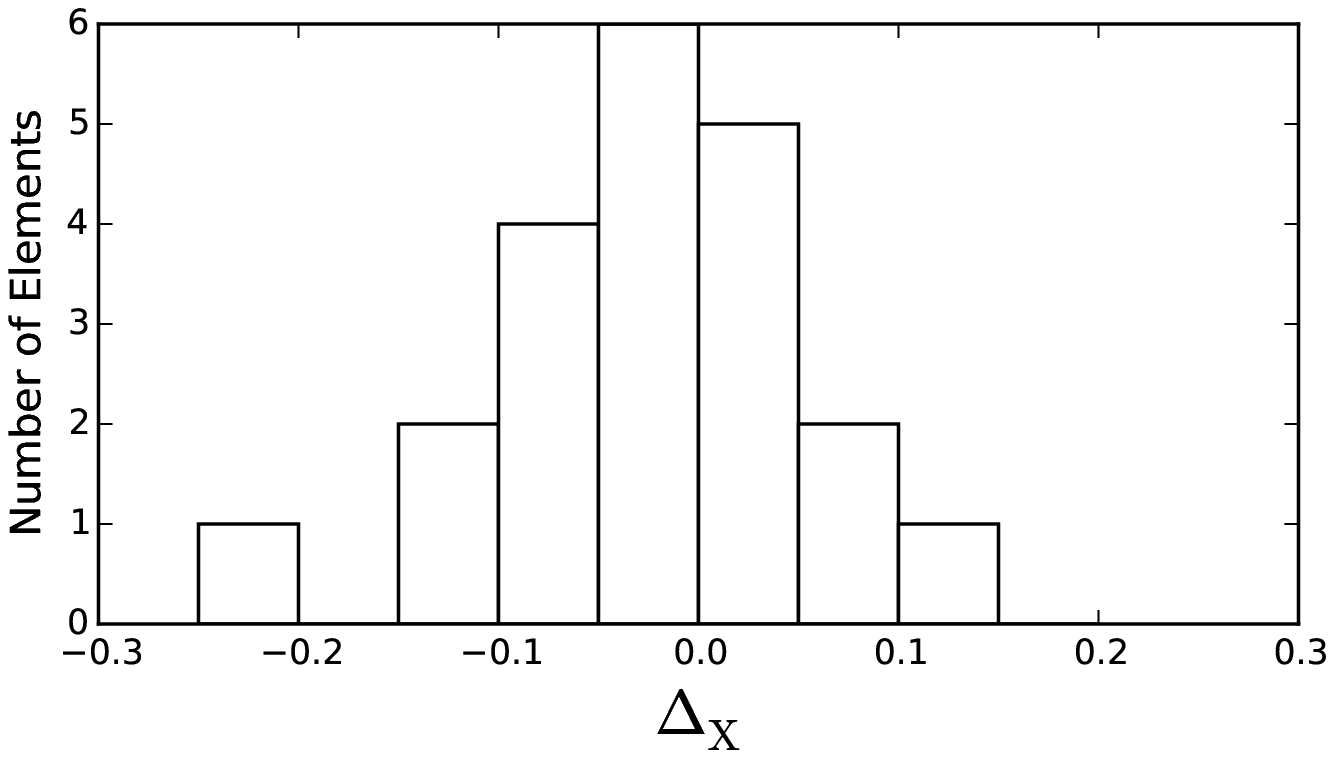}
\includegraphics[scale=0.55]{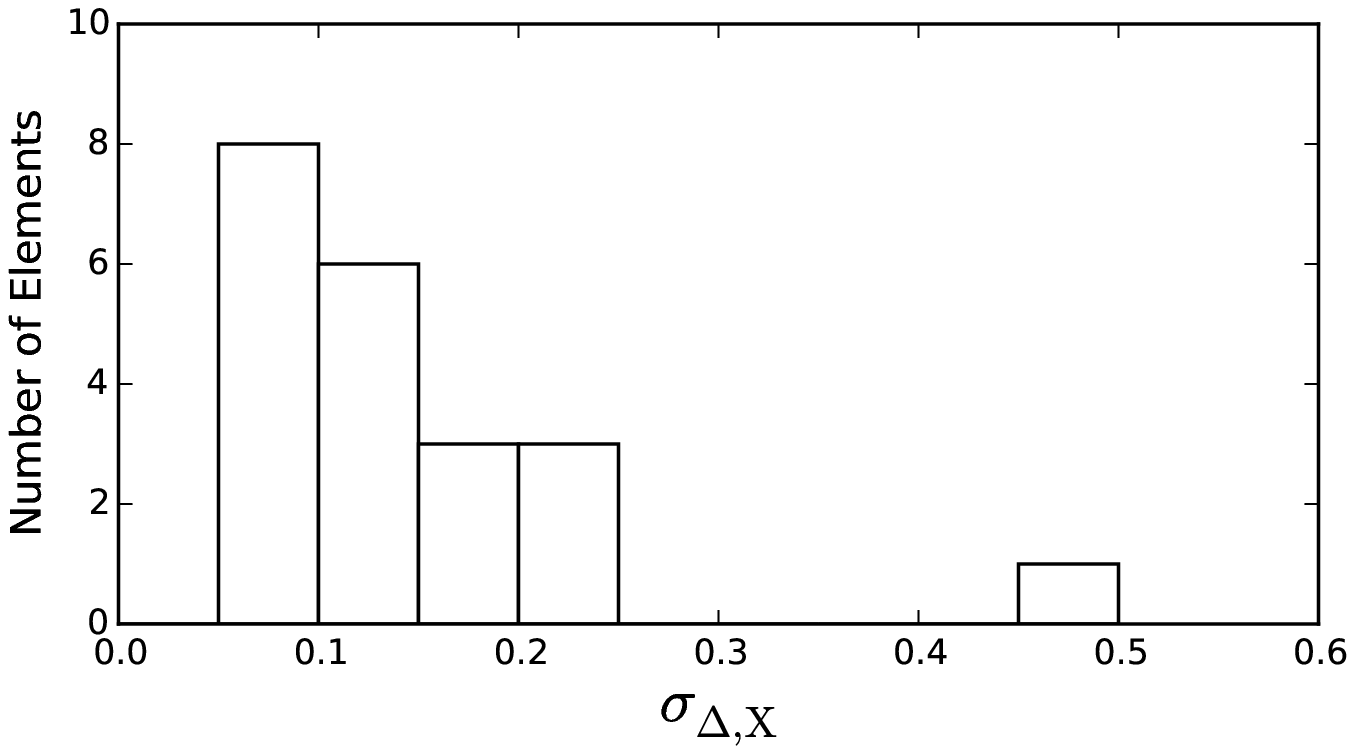}

\caption{ Distribution of values of the  systematic offset $\Delta$  (top) and the  scatter around the offset $\sigma_{\Delta}$ (bottom) for  elements summarized in Table \ref{tab:offset_final},  for which IL abundances   have been compared to stellar abundances. For Fe I  the values used are calculated for the stellar comparison, [Fe/H]$_{*}$, without NGC 6528.  For ionized species the values taken with respect to Fe I are used. }
\label{fig:hist_results} 
\end{figure}

Figure \ref{fig:la} shows the comparisons for La II.   In the IL spectra we find 3 measurable La II features, but they are difficult to measure because they are weak lines and sensitive to nearby  blends.  Like Ba, La  is useful for study of the s-process \citep[e.g][]{1999ARA&A..37..239B,2008ARA&A..46..241S}.   Examples of two La measurements, one with higher S/N and one with lower S/N, are shown in Figure \ref{fig:la_synth1}.   While  we measure La II in 6 GCs we can only compare to stellar references abundances in 3 GCs.  The agreement is similar both when [La/Fe] ratios are taken with respect to Fe I and Fe II and for differential and absolute abundances. The differential abundances offsets are $\Delta=+0.01$ with $\sigma_{\Delta}=0.15$ for Fe I ratios and $\Delta=-0.01$ with $\sigma_{\Delta}=0.14$ with Fe II ratios.    We find that two out of the three GCs compare well to references, while the measurement for the most metal-rich GC out of the three, NGC 6553, is offset to higher abundances, but with larger uncertainty.  The large uncertainty is due both to a large line-to-line scatter for the 2 lines in IL and a large difference in the 2 reference abundances. It would be interesting to determine if the larger uncertainties in both the stellar and IL measurements  are particular to this GC or a result of the difficulty in measurements at high metallicity.  While we do have 3 additional measurements of La II at [Fe/H]$>-0.6$, we are limited by a lack of stellar reference abundances, so this is an issue that could be resolved in the future with additional stellar measurements.     We conclude that La II is promising for heavy element constraints in extragalactic GCs, although more stellar abundances are needed to better evaluate La at higher metallicities.

 At solar composition, Nd is produced roughly half in the s-process and half in the r-process \citep{2008ARA&A..46..241S}. Like La II, Nd II is difficult to measure because the features are generally weak and sensitive to blends.   Spectra and synthesis of the 5293 \rAA line in NGC 362 are presented in Figure \ref{fig:nd_synth1}. 
 Nd II is measured in 6 GCs, but can only be compared to stellar results in 3 GCs. When comparing Fe I and Fe II ratios we find that the differential abundances give similar results to each other and the absolute abundances have a smaller offset but higher scatter for Fe II ratios. For the differential abundances we find $\Delta=+0.14$ and $\sigma_{\Delta}=0.20$ for Fe I ratios and $\Delta=+0.13$ and $\sigma_{\Delta}=0.22$ for Fe II ratios.   Nd II is particularly difficult to measure in IL spectra of  metal poor GCs, and the largest disagreement is for NGC 6752 ([Fe/H]$\sim-1.6$), for which we measure a high abundance of [Nd/Fe]$\sim=+0.6$  from a single weak line, compared to the reference value of [Nd/Fe]$=+0.22$ measured by \cite{yong2005}.   Given this discrepancy, Nd measurements from IL spectra for GCs with [Fe/H]$<-1.5$ should be interpreted cautiously.   We are limited  in evaluating Nd II for [Fe/H]$>-0.5$ by a lack of stellar results, so it will be interesting to revisit this element if more stellar results become available. 
 In summary, Nd II is promising for extragalactic studies for moderate to high metallicity GCs, but more stellar comparisons are needed at high metallicity.  Measurements in low metallicity GCs will be very difficult in IL and may result in significant systematic offsets.

Eu II measurements can be especially useful for chemical tagging studies because Eu is primarily produced in the {\it r}-process \citep[e.g.][ and references therein]{2008ARA&A..46..241S}, however Eu II is particularly difficult to measure in IL spectra because of the line broadening caused by the intrinsic velocity dispersions of the GCs and line blending for the bluest features.  The Eu II transition at 6645 \rAA  is the most easily accessible and least blended feature for metal-rich GCs, and is the feature we are able to measure in 3 of the more metal-rich GCs in our sample.  For metal-poor GCs the 4129 \rAA transition is the cleanest strong line we can measure, and which we have used for abundance measurements in 2  metal-poor GCs in our sample.  Unfortunately we cannot make  simultaneous clean measurements of Eu II at 6645 \rAA and 4129 \rAA in any of these GCs.  Examples of this difficulty are shown in Figure \ref{fig:eu_synth1} and Figure \ref{fig:eu_synth2}.  In the more metal poor GCs the 6645 \rAA line is too weak to make a clean measurement, and in the more metal rich GCs the 4129 \rAA line becomes too blended and the pseudo-continuum normalization too uncertain.   

We compare our Eu II results for 4 of the GCs in Figure \ref{fig:eu}; we find that Fe I abundances give similar or better results than Fe II abundances.  The differential abundances give a smaller offset than the absolute abundances.  We find $\Delta=+0.02$ and $\sigma_{\Delta}=0.24$ for differential abundances and  $\Delta=-0.03$ and $\sigma_{\Delta}=0.29$ for absolute abundances. From the bottom panel in Figure \ref{fig:eu} we see that the [Eu/Fe] we measure for the two GCs with [Fe/H]$>-0.5$ tends to be higher than what it is seen in MW stars, but the values for GCs with [Fe/H]$<-0.5$ agree better with MW stars.   Two of the GCs show good agreement with references, but the overall scatter is large due to the the IL abundance for NGC 362 being underestimated and the NGC 6553 abundance being overestimated. Both the comparison values for these GCs come from more than one reference. Out of the IL abundances  the measurement for NGC 362 is likely the most uncertain because it is made from the 4129 \rAA line at the metallicity  limit of where this line may be starting to be significantly affected by blending, as seen in Figure \ref{fig:eu_synth2}.  Consequently, we recommend that the 4129 \rAA line only be used for the most metal-poor GCs with smaller velocity dispersions.     We  conclude that Eu measurements for individual GCs will be difficult to use for precise chemical tagging because of systematic uncertainties at the 0.25 dex level, however, in the IL very enhanced or strict upper limits on  [Eu/Fe] may still be meaningful, as well as the average [Eu/Fe] observed for a large number of GCs at similar metallicity.

 In summary, we find that the easiest and most accurate neutron capture element to measure across the entire metallicity range is Ba II, and Y II compares reasonably well within the uncertainties for the smaller range in [Fe/H] that can be tested. La II and Nd II can be measured but the metallicity range we can evaluate is limited without additional stellar reference abundances.   Eu II can also be measured, although usually with larger uncertainties, so it will be difficult to use for precise chemical tagging, although perhaps a useful aggregate value of [Eu/Fe] can be obtained from group of IL measurements. Zr compares badly to references and will not be useful for IL studies.

\section{Summary and Conclusions}
\label{sec:summary}

Detailed abundance analysis based on high resolution integrated light spectra  was performed for 11 MW GCs and 1 Fornax GC, with the primary goal of a comprehensive comparison of IL abundances to abundances from spectra of individual stars  in those same GCs.  The effects of stochastic population sampling and horizontal branch morphology were accounted for in the final solutions.  
We present comparisons of the [Fe/H] from IL spectra to catalogued [Fe/H] from \citep[][2010 edition]{harris}, as well as to the mean [Fe/H] from recent abundance studies of individual stars by different authors for each GC.     In addition to Fe we compare abundances for 19 elements measured in the IL spectra to the mean abundances from studies of individual stars, when available. The elements tested  include light, alpha, Fe-peak and neutron capture elements.  We calculate a systematic offset and standard deviation of the abundance comparison for each element. We have performed both an absolute abundance analysis where solar ratios are taken relative to the solar abundance distribution of \cite{asplundreview} and a line-by-line differential abundance analysis where we have calculated abundances with solar abundances we have measured for each line, when possible. For ionized species we perform two comparisons, one in which the abundance ratio is taken with respect to the abundance from Fe I lines, and one in which the abundance ratio is taken with respect to the abundance of  Fe II lines. 

A summary of the final systematic offsets is given in Table \ref{tab:offset_final}, where we have given the values calculated with line-by-line differential abundances. We note that the most metal-rich GC, NGC 6528,  is often an outlier in the comparisons, as it is in the comparison for [Fe/H], and so we don't include it in the calculated offsets.    With the exception of [Mg/Fe], the calculated offsets for each species are all $<$0.15 dex, with most of the offsets $<$0.1 dex, as can be seen in the histogram of our results in Figure \ref{fig:hist_results}.  The  scatter around the systematic offset can vary considerably depending on the element and the number of GCs that can be compared; the distribution of the scatter is shown in the bottom panel of Figure \ref{fig:hist_results}.   Significant outliers that dominate the scatter for particular elements are discussed further in the text.

A summary of our results:

\begin{itemize}
  \item   We find that the abundance of Fe is best constrained using Fe I lines, but that including Fe II lines can marginally improve the solutions when enough Fe II lines ($>$10) can be measured.

  \item For the [Fe/H] comparison of 11 GCs to  the \citetalias{harris}, we find a total  offset of $+0.07$ with a standard deviation of 0.08. For the comparison of 12 GCs to averaged stellar abundances we  find an offset of $-0.02$ dex with a standard deviation of 0.11, which demonstrates that the IL abundance method can reliably determine [Fe/H] to the level of $\sim$0.1 dex.
   
  \item  In both [Fe/H] comparisons the worst agreement is found for clusters with [Fe/H]$>-0.3$.   The most significant outlier in both comparisons is the most metal rich GC in the sample, NGC 6528.  We conclude that the IL method is most accurately applied for [Fe/H] in the metallicity range of $-2.4<$[Fe/H]$<-0.3$.  A larger sample of training set GCs with solar metallicity and with less background contamination is necessary to more fully understand the [Fe/H] underestimation at high metallicity.

  \item For nearly all elements we find that the differential abundances result in smaller systematic offsets, although in most cases the difference in the systematic offset between the two methods  is less than 0.1 dex.  Because the systematic offsets are smaller we choose to use the differential abundance ratios as our final results and in our future work.

\item   We do not  find a significant difference in the  calculated offsets between Fe I and Fe II ratios, as they  all agree to within 0.06 dex, however the scatter in the offset is somewhat larger  for Fe II ratios for the elements with the greatest number of comparisons (Ti II, Sc II, Y II, Ba II).  Since the abundance of Fe II can  be difficult to measure in the IL spectra, we recommend that  for consistency ionized abundance ratios are taken with respect to Fe I abundances in the future.  

\item The best results are typically found for elements that can be easily measured across all or most of the metallicity range in both the IL and stellar analysis:  Fe I, Ca I, Si I, Ni I, and Ba II.

\item Cr I provides  good results for GCs with metallicities of   [Fe/H]$>-2$.

\item Ti II provides good results for 9 GCs, with one outlier, although the individual IL measurements for GCs can have higher line-to-line scatter than measurements for Ca I or Si I. 

\item While Ti I can be measured across the entire range in metallicity, the results show a larger scatter compared to reference abundances, therefore we find  Ca I, Si I, and Ti II to be preferable for studies of [$\alpha$/Fe] ratios.

\item Sc II provides good results for GCs with [Fe/H]$< -0.5$, and it's performance at [Fe/H]$<-2$ could be evaluated if stellar comparison abundances were available in that regime.

\item Elements that show promising agreement but the metallicity range of our analysis is limited  due to  a lack of stellar results for comparison include: Co I, Cu I, Mn I, Nd II, and La II.

\item Al I provides good results where it can be measured in IL, which is for [Fe/H]$>-1.3$, and is unlikely to be measured in more metal poor GCs from the 6696 \rAA doublet, which is weak in IL spectra.  

\item Na I provides mixed results where it can be measured in IL, which is for [Fe/H]$>-2$, but is usually consistent with the range in Na I that is observed in GCs with star-to-star variations in Na.

\item Eu II is difficult to measure and we recommend using the 4129 \rAA line for only the most metal poor GCs.  The 6645 \rAA can only be measured in more metal rich GCs with high S/N. The 0.25 dex scatter in the comparison suggests that individual measurements should not be used for precise chemical tagging, however we are optimistic that an aggregate measure of [Eu/Fe] would be possible in a large sample of extragalactic GCs.   

\item Mg I measurements here are consistently lower than from individual stars, with the largest systematic offset we observe.  It is unclear from this sample if there is a consistent offset.   A better understanding of the issues with the IL measurement must be gained before Mg can be used in study of extragalactic GCs. We suggest a more thorough analysis of NLTE effects on the IL abundances as a starting point in the future.

\item  Zr I shows poor results and could only be measured in a subset of GCs, therefore we recommend against using the IL abundances in the future.

\item  The bulge GCs analyzed here show abundance patterns similar to MW bulge stars for Ca, Si, Ti, Al, and Cu.

\end{itemize}

 In conclusion, IL abundances of certain key elements measured in GCs can be used to accurately study chemical evolution of distant galaxies for most of the metallicity range spanned by GCs.  More progress could be made on other elements with additional stellar data. We find poor agreement for GCs with the highest metallicities ([Fe/H]$>-0.3$). Unfortunately, for these metallicities there are very few resolved GCs with reliable abundances in the literature from single RGB stars that also have high enough central surface brightnesses to obtain high quality IL spectra, so it will be difficult to further test the IL analysis at these metallicities in the future.  IL  abundances from very high metallicity GCs ([Fe/H]$>-0.3$) should be interpreted cautiously,  particularly if the derived  abundance ratios are outliers with respect to lower metallicity GCs in the system.

\acknowledgements
The authors thank the anonymous referee, whose helpful comments improved the paper. J.E.C. is supported by an NSF Astronomy and Astrophysics Postdoctoral Fellowship under award AST-1302710.  The authors thank D. Zaritsky and P. Pessev for the observations of metal rich bulge clusters made with the Magellan Clay Telescope.

\bibliographystyle{apj}
\bibliography{master_ref}

\end{document}